%% file: main.tex
\documentclass[journal,comsoc, 12pt,onecolumn,draftclsnofoot]{IEEEtran}

\usepackage{textcomp}
\usepackage{verbatim}
\usepackage{dsfont}

\usepackage{tikz}
\usetikzlibrary{patterns, arrows,backgrounds,fit,tikzmark,positioning,calc,decorations,snakes,shapes,matrix}
\usetikzlibrary{decorations.pathreplacing}
\usetikzlibrary{calc,arrows.meta,positioning}

\usepackage[T1]{fontenc}
\usepackage{lmodern}

\usepackage{cite}

\usepackage{amsfonts}
\usepackage{amsmath,bm, amssymb}
\usepackage{mathtools,hyperref}
\usepackage{amsthm}
\usepackage{mathrsfs}

\usepackage{xcolor}

\usepackage{subcaption} 
\usepackage{nicefrac} 
\usepackage{array}

\usepackage[subnum]{cases} 
\usepackage[inline]{enumitem}
\usepackage{lscape}
\usepackage{cleveref}
\usepackage{multicol}
\usepackage{multirow}	

\usepackage{pgfplots}
\let\OLD=\pgfnodeparttextbox
\usepackage{signalflowdiagram}
\let\pgfnodeparttextbox=\OLD

\usepackage{colortbl}
\usepackage{adjustbox}

\theoremstyle{remark}

\newtheorem{proposition}{{Proposition}}

\definecolor{RedBlue}{rgb}{0.8,0,0.5}
\definecolor{RedBlueGreen}{rgb}{0.8,0.6,0.5}
\definecolor{YellowOrange}{rgb}{0.4,0.4,0}
\definecolor{OliveGreen}{rgb}{0,0.6,0}

\DeclareMathOperator*{\argmin}{argmin}
\DeclareMathOperator*{\argmax}{argmax}
\DeclareMathOperator{\sign}{sign}

\ifCLASSINFOpdf
\else
\fi
\usepackage{amsmath}
\usepackage[cmintegrals]{newtxmath}
\begin{document}
	
\title{Robust Interference Management for SISO Systems with Multiple Over-the-Air Computations}

\author{\IEEEauthorblockN{\small{Jaber~Kakar$^{*}$,~\IEEEmembership{\small Student~Member,~IEEE} and~Aydin~Sezgin$^{*}$,~\IEEEmembership{\small Senior~Member,~IEEE}\\}
	\IEEEauthorblockA{$^{*}$Institute of Digital Communication Systems,
		Ruhr-Universit{\"a}t Bochum, Germany \\
		Email: \{jaber.kakar, aydin.sezgin\}@rub.de}
	}        
}  

\markboth{Draft}%
{Shell \MakeLowercase{\textit{et al.}}: Bare Demo of IEEEtran.cls for IEEE Communications Society Journals}

\makeatletter
\newcommand*{\rom}[1]{\expandafter\@slowromancap\romannumeral #1@}
\makeatother

\maketitle

\begin{abstract} 
	Over-the-air computation (AirComp) represents a promising concept that leverages on the superposition property of wireless multiple access channels (MAC). This property facilitates the computation of sums $s=\sum_{k=1}^{K}x_k$ of real-valued, distributed sensor (Tx) data $x_k$ for a fusion center (Rx). In today's context, where spectrum is scarce, we may wish to not only compute one, but rather (for any mutually exclusive collection of sensor index sensor sets $\mathcal{D}_m$) $M$, $M\geq 2$, sums $s_m=\sum_{k\in\mathcal{D}m}x_k$ over a shared complex-valued MAC at once with minimal mean-squared error ($\mathsf{MSE}$). Finding appropriate Tx-Rx scaling factors balance between a low error in the computation of $s_n$ and the interference induced by it in the computation of other sums $s_m$, $m\neq n$. In this paper, we are interested in designing an optimal Tx-Rx scaling policy that minimizes the mean-squared error $\max_{m\in[1:M]}\mathsf{MSE}_m$ subject to a Tx power constraint with maximum power $P$. We show that an optimal design of the Tx-Rx scaling policy $\left(\bar{\bm a},\bar{\bm b}\right)$ involves optimizing (a) their phases and (b) their absolute values in order to (i) decompose the computation of $M$ sums into, respectively, $M_R$ and $M_I$ ($M=M_R+M_I$) calculations over real and imaginary part of the Rx signal and (ii) to minimize the computation over each part -- real and imaginary -- individually. The primary focus of this paper is on (b). We derive conditions (i) on the feasibility of the optimization problem and (ii) on the Tx-Rx scaling policy of a local minimum for $M_w=2$ computations over the real ($w=R$) or the imaginary ($w=I$) part. 
	Extensive simulations over one receiving chain for $M_w=2$ show that the level of interference in terms of $\Delta D=|\mathcal{D}_2|-|\mathcal{D}_1|$ plays an important role on the ergodic worst-case $\mathsf{MSE}$. At very high $\mathsf{SNR}$, typically only the sensor with the weakest channel transmits with full power while all remaining sensors transmit with less to limit the interference. Interestingly, we observe that due to residual interference, the ergodic worst-case $\mathsf{MSE}$ is not vanishing; rather, it converges to $\frac{|\mathcal{D}_1||\mathcal{D}_2|}{K}$ as $\mathsf{SNR}\rightarrow\infty$.  
\end{abstract}


\IEEEpeerreviewmaketitle

\section{Introduction}
\label{sec:intro}

In the era of Big Data and Internet-of-Things (IoT), enormous quantities of data are exchanged among a staggering number of mobile devices (e.g., sensors). According to DOMO, the global internet population grew by $500$ million from 2017 to 2018 and reached now $4.3$ billions \cite{DOMO} creating around $1.7$ MB of data per second. Simultaneously, the number of IoT devices is exponentially growing and forecasts predict 125 billion IoT devices by 2030 \cite{vxchnge}. These devices are a key contributing factor in the massive growth of data.

In IoT applications involving massive amount of data, wireless data aggregation (WDA) represents a promising solution for data collection from sensors with limited spectrum bandwidth \cite{Abari_2016}. WDA is of particular relevance when there are latency restrictions on the processing of sensor data. In AirComp -- a novel WDA technique that leverages on the superposition property of the wireles multiple access channel (MAC) -- data signals can be combined in both a linear and non-linear manner. Specifically, a fusion center (FC) receives a linear combination of sensor signals weighted by the channels' coefficients. This allows realizing the summation, or averaging, of sensor signals through over-the-air transmissions. Through appropriate pre-processing functions $\psi_k\left(\cdot\right)$ at sensor $k=1,\ldots,K$, and post-scaling function $\varphi\left(\cdot\right)$ at the FC not only averaging but more complex target functions $\phi\left(\cdot\right)$ on the sensor data $\left(x_1,\ldots,x_K\right)$ from the class of so-called \emph{nomographic functions} (e.g., geometric mean) which omit the representation 
\begin{align*}
\phi\left(x_1,\ldots,x_K\right)=\varphi\left(\sum_{k=1}^{K}\psi_k\left(x_k\right)\right)
\end{align*} can be attained \cite{Buck_1979}. For AirComp systems, the idea in \cite{Nazer_2007, Goldenbaum_SP_2013, Goldenbaum_TCom_2013} is to let each sensor process its own data $x_k$ according to $\psi_k\left(x_k\right)$, such that the MAC generates the intermediate result $\sum_{k=1}^{K}\psi_k\left(x_k\right)$ as an input of $\varphi\left(\cdot\right)$ which in return gives the desired function $\phi$ evaluated on the sensor data $\left(x_1,\ldots,x_K\right)$.\footnote{We emphasize that the idea of over-the-air computation has also been applied very early by the information theory community in the construction of the so-called compute-and-forward relaying strategy that harnesses from the interference caused by the simultaneous transmission in a MAC \cite{Nazer_2011}.} The advantage of the decomposition principle in AirComp is two-fold. On the one hand, the computation task is decomposed into $K+1$ subtasks which are, respectively, $\psi_k\left(\cdot\right)$, $k=1,\ldots,K$, assigned to the $k$-th sensor and $\varphi\left(\cdot\right)$ allotted to the FC. On the other hand, completing the computation is limited to a single time slot rather than a $K$ slot TDMA scheme. Specific use cases of AirComp are, amongst others, distributed machine learning \cite{Tandon_2017, Amiri_2019} and over-the air consensus \cite{Molinari_2018}. 

\subsection{Related Work}
\label{subsec:rel_work}

An important aspect of AirComp research is about its system design. More detailed, the design of pre- and postprocessing functions for different functions $\phi(\cdot)$ (e.g., geometric mean, maximum) have been analyzed in references \cite{Goldenbaum_SP_2013, Goldenbaum_TCom_2013, Abari_2016}. Robust designs that account for synchronization offsets between sensors \cite{Goldenbaum_TCom_2013, Abari_2015} and imperfect or lack of channel state information \cite{Goldenbaum_2014_CSI, Dong_2020} are also studied. More recently, two research groups \cite{Cao_2019, Liu_2019} have independently developed for an averaging target function, the jointly global-optimal pre- and postprocessing scalars that minimize the (non-convex) mean-squared error ($\mathsf{MSE}$) subject to a per-sensor, peak transmit power constraint. The authors make the observation that the optimal pre-processing is a mixture of the channel-inversion and energy-greedy policy. In \cite{Zhu_2019}, \emph{Zhu et al.} consider the optimization problem that minimizes the computation distortion of a MIMO AirComp system with multi-modal sensors by zero-forcing precoding and aggregation beamforming design. This setup facilitates a multiplexing gain in the sense that at most $\min\left(N_T,N_R\right)$\footnote{$N_T$ and $N_R$ are, respectively, the number of transmit (sensor) and receive (FC) antennas.} functions $\phi_m$ can be computed in a single slot. The system model of \cite{Zhu_2019} is extended in \cite{Li_2019} to a wirelessly-powered, MIMO AirComp system.     

To the best of the authors' knowledge, the aspect of interference management in AirComp systems -- particularly for SISO systems with no multiplexing gain -- is largely unstudied. As part of this study, the goal is (i) to better understand how multiple computations over a shared MAC influence the computation distortion and (ii) deduce an interference management policy for low, medium and high $\mathsf{SNR}$ which minimizes the worst $\mathsf{MSE}$.

\subsection{Contributions}
\label{subsec:contributions}

In this paper, we consider a single-antenna AirComp system with which we seek to compute, for any mutually exclusive collection of sensor index sensor sets $\mathcal{D}_m$ of arbitrary cardinality, $M$ sums $s_m=\sum_{k\in\mathcal{D}m}x_k$, $m=1,\ldots,M$, with real-valued sensor inputs $x_k$ over a shared MAC. As opposed to a MIMO system considered in \cite{Zhu_2019}, where there is a multiplexing gain, for a SISO system there is none. For this system under study, the main contributions of this paper are the following.        
\begin{itemize}
	\item We cast the AirComp problem as an optimization problem that minimizes the worst-case $\mathsf{MSE}$ -- $\max_{m\in[1:M]}\mathsf{MSE}_m$ -- over all possible transmit-receive scaling (Tx-Rx) policies. In this optimization problem, the per-sensor prescaling policy is subjected to a power constraint. Due to the coupling of Tx and Rx-scaling, this problem is non-convex.   
	\item An orthogonalization principle that decomposes the computation of $M$ sums into $M_R$ computations over the real and $M_I$ computations over the imaginary part of the receiving chain is suggested ($M=M_R+M_I$). Not only for the special case, where $M=2$, but also larger $M>2$, we show the optimality of this decomposition rule. 
	When this orthgonalization principle is applied, the optimization of the worst-case $\mathsf{MSE}$ is separated for real and imaginary part.  
	\item The worst-case $\mathsf{MSE}$ optimization along one part -- say without loss of generality the real part -- for $M_{R}=2$ computations is considered. For this case, the $\mathsf{MSE}$ minimization problem is reformulated to a fractional program. We study its feasibility and derive conditions on the maximum tolerable noise variance. Through means of the Karush-Kuhn-Tucker (KKT) condition \cite{Boyd_Book}, we determine a close-form expression which satisfies the second-order sufficient condition of relative matrix inertias \cite{Han_1985} of local minima. The solution resembles global optimal solutions of interference-free scenarios \cite{Cao_2019, Liu_2019} in the sense that sensors $k\in\mathcal{P}_m\subseteq\mathcal{D}_m$, $m\in[1:M_R]$, with stronger channels transmit with less-than full power while all remaining sensors operate at peak power. The cardinality of these sets are of utmost importance for optimal interference management.  
	\item We consider in our simulations a single reeiving chain -- say without loss of generality the real part -- for $M_R=2$ computations.\footnote{In the simulations, $K=|\mathcal{D}_1|+|\mathcal{D}_2|$ holds.} They show that the level of interference in terms of $\Delta D=|\mathcal{D}_2|-|\mathcal{D}_1|$ plays an important role on the ergodic worst-case $\mathsf{MSE}$. At very high $\mathsf{SNR}$, typically only the sensor with the weakest channel transmits with full power while all remaining sensors transmit with less to limit the interference. Interestingly, we observe that due to residual interference, the ergodic worst-case $\mathsf{MSE}$ is not vanishing; rather, it converges to $\frac{|\mathcal{D}_1||\mathcal{D}_2|}{K}$ as $\mathsf{SNR}\rightarrow\infty$. This result gives us an approximate design guideline on deciding which pair of computation indices $\left(m_{1,w},m_{2,w}\right)$ shall be computed along the real ($w=R$) and imaginary ($w=I$) processing chain.  
\end{itemize}

\subsection{Paper Organization}
\label{subsec:paper_org}

The remainder of this paper is organized as follows. Section \ref{sec:sym_mod} deals with the AirComp system model and its respective $\mathsf{MSE}$ optimization problem. In section \ref{sec:orthogonal_computation}, the orthogonalization principle for muliplexing the real-valued computation along real and imaginary part of the receiving chain is described. The optimization problem after orthogonalization is formulated and solved in the section \ref{sec:sol_opt_orthgonal_problem}. The main subject of section \ref{sec:num_res} is the discussion of the simulation results. Finally, section \ref{sec:conclusion} concludes this work.         

\textbf{Notations:} For a complex number $z$, $\Re\{z\}$ and $\Im\{z\}$ denote, respectively, the real and imaginary part. $z^{*}$ is the complex conjugate of $z$. Throughout this paper, we denote sets by calligraphic letters (e.g., $\mathcal{S}$), vectors by bold, lower-case letters (e.g., $\bm b$) and matrices by capitalized, bold-face letters (e.g., $\bm B$). $\bm x_{[i]}$ represents the $i$-th largest component in $\bm x$ and $\bm x_{\mathcal{S}}$ is the collection of elements of $\bm x=(x_1,x_2,\ldots, x_N)^{T}$ indexed by $\mathcal{S}\subseteq[1:N]$. $\sign\left\{x\right\}$ is the signum function that extracts the sign of a real number $x$. Finally, we use $[x]^{+}$ as a shorthand notation for $\max\left(0,x\right)$.

\section{System Model and Problem Formulation}
\label{sec:sym_mod}

\subsection{System Model}

\begin{figure}[h]
	\begin{center}
		\begin{tikzpicture}[node distance=0.5cm]
			\begin{scope}
				\node[input]      (in11)                 {\footnotesize $x_1$};
				\node[modulator] (mul11)  [node distance=1cm, right from=in11] {};
				\node[input] (in12)  [above from=mul11]  {\nodepart{above}{\footnotesize $\bar{b}_1$}};
				\node[modulator] (mul12)  [node distance=1cm, right from=mul11] {};
				\node[input] (in13)  [above from=mul12]  {\nodepart{above}{\footnotesize $\bar{h}_1$}};
				\node[] (bend1)  [node distance=0.85cm, right from=mul12] {};
				
				\path[r>] (in11) -- (mul11);
				\path[r>] (in12) -- (mul11);
				\path[r>] (mul11) -- (mul12);
				\path[r>] (in13) -- (mul12);
			\end{scope}
			\begin{scope}[yshift=-1.5cm]
				\node[input]      (in21)                 {\footnotesize $x_2$};
				\node[modulator] (mul21)  [node distance=1cm, right from=in21] {};
				\node[input] (in22)  [above from=mul21]  {\nodepart{above}{\footnotesize $\bar{b}_2$}};
				\node[modulator] (mul22)  [node distance=1cm, right from=mul21] {};
				\node[input] (in23)  [above from=mul22]  {\nodepart{above}{\footnotesize $\bar{h}_2$}};
				 \node[] (bend2)  [node distance=0.85cm, right from=mul22] {};
				
				\path[r>] (in21) -- (mul21);
				\path[r>] (in22) -- (mul21);
				\path[r>] (mul21) -- (mul22);
				\path[r>] (in23) -- (mul22);
			\end{scope}
			\begin{scope}[yshift=-2.625cm]
				\node[]      (ink1)                 {$\vdots$};
				\node[] (mulk1)  [node distance=0.85cm, right from=ink1] {$\vdots$};
				\node[] (mulk2)  [node distance=0.85cm, right from=mulk1] {$\vdots$};
			\end{scope}		
			\begin{scope}[yshift=-4.25cm]
				\node[input]      (inK1)                 {\footnotesize $x_K$};
				\node[modulator] (mulK1)  [node distance=1cm, right from=inK1] {};
				\node[input] (inK2)  [above from=mulK1]  {\nodepart{above}{\footnotesize $\bar{b}_K$}};
				\node[modulator] (mulK2)  [node distance=1cm, right from=mulK1] {};
				\node[input] (inK3)  [above from=mulK2]  {\nodepart{above}{\footnotesize $\bar{h}_K$}};
				\node[] (bendK)  [node distance=4.5cm, right from=inK1] {};
				
				\path[r>] (inK1) -- (mulK1);
				\path[r>] (inK2) -- (mulK1);
				\path[r>] (mulK1) -- (mulK2);
				\path[r>] (inK3) -- (mulK2);
			\end{scope}		
				\node[adder]  (add1)  [below right=1.85cm and 0.5cm of mul12] {};
				\node[adder]  (add2)  [node distance=0.5cm, right from=add1] {};
				\node[output] (out) [node distance=1cm, right from=add2] {\footnotesize $y=\sum_{k=1}^{K}\bar{h}_k \bar{b}_k x_k+n$};
				\node[input] (inadd2)  [above from=add2]  {\nodepart{above}{\footnotesize $n$}};
				\path[r>] (mul12) -| (add1);
				\path[r>] (mul22) |- (add1);
				\path[r>] (mulK2) -| (add1);
				\path[r>] (add1) -- (add2);
				\path[r>] (inadd2) -- (add2);
				\path[r] (add2) -- (out);
				
				 Annotate the MAC channel
				\node[] (left_top)  [above right=1.5 cm and 0.5cm of mul11] {};
				\node[] (right_top)  [above right=3.55 cm and 0.35cm of add2] {};
				\node[] (left_bot)  [node distance=7cm, below from=left_top] {};
				\draw[dashed] (left_top) -- (left_bot);
				\node[] (right_bot)  [node distance=7cm, below from=right_top] {};			
				\draw[dashed] (right_top) -- (right_bot);
				
				\node[align=center] (channel)  [above left=0 cm and 0cm of right_bot] {\footnotesize Multiple-access\\[-2.25ex] \footnotesize channel (MAC)};				
				\node[align=center] (Tx)  [above left=0 cm and 0cm of left_bot] {\footnotesize Sensors\\[-2.25ex] \footnotesize (Tx)};	
				\node[align=center] (Tx)  [above right=0 cm and 0cm of right_bot] {\footnotesize Receiver\\[-2.25ex] \footnotesize (Rx)};	
				
				\node[filter] (fil_Re) [above right=1.25cm and 0.75cm of out] {\footnotesize $\Re\left\{\cdot\right\}$};	
				\node[filter] (fil_Im) [below right=1.25cm and 0.75cm of out] {\footnotesize $\Im\left\{\cdot\right\}$};	
				
				\path[r>] (out) |- (fil_Re);
				\path[r>] (out) |- (fil_Im);	
				
				\node[modulator] (mulRx_Re)  [node distance=1cm, right from=fil_Re] {};
				\node[modulator] (mulRx_Im)  [node distance=1cm, right from=fil_Im] {};
				\node[input] (inRx_Re)  [above from=mulRx_Re]  {\nodepart{above}{\footnotesize $\bar{a}_{m,R}$}};
				\node[input] (inRx_Im)  [above from=mulRx_Im]  {\nodepart{above}{\footnotesize $\bar{a}_{m,I}$}};	
				
				\path[r>] (fil_Re) |- (mulRx_Re);
				\path[r>] (fil_Im) |- (mulRx_Im);	
				\path[r>] (inRx_Re) -- (mulRx_Re);	
				\path[r>] (inRx_Im) -- (mulRx_Im);
				
				\node[adder]  (add_est)  [below right=1.5cm and 0.75cm of mulRx_Re] {};	
				
				\path[r>] (mulRx_Re) -| (add_est);
				\path[r>] (mulRx_Im) -| (add_est);
				
				\node[output]  (out_est)  [node distance=0.5cm, right from=add_est] {\footnotesize $\hat{s}_m$};
				
				\path[r>] (add_est) -- (out_est);
															
		\end{tikzpicture}
	\end{center}
	\caption{\small Illustration of the AirComp system. Note that $x_k$, $\bar{a}_{m,R}$, $\bar{a}_{m,I}$ and $\hat{s}_m$ are real, while the remaining variables are typically complex.}
	\label{fig:sym_mod_aircomp}
\end{figure}
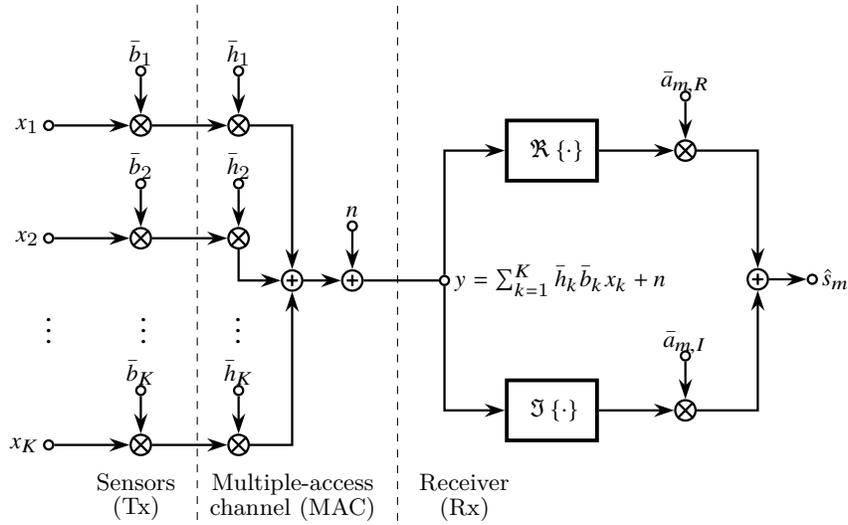

We consider a $K$-sensor, single-antenna AirComp multiple access channel (MAC) system as shown in Fig. \ref{fig:sym_mod_aircomp}. In this system, each sensor's pre-processed signal $x_k\in\mathbb{R}$, $\forall k\in[1:K]$, is scaled by its scaling factor $\bar{b}_k\in\mathbb{C}$ and conveyed to the receiver (FC) through the MAC. The collection of all $K$ Tx-scaling factors are denoted by $\bar{\bm{b}}\triangleq\left(\bar{b}_k\right)_{k\in[1:K]}$. Thus, the received signal $y$ becomes
\begin{align}
y=\sum_{k=1}^{K}\bar{h}_k \bar{b}_k x_k+n,\label{eq:rx_sig}
\end{align} where $h_k\in\mathbb{C}$ is the channel coefficient of sensor $k$ and $n\sim\mathcal{CN}\left(0,2\sigma^{2}\right)$ is additive white Gaussian noise. We assume that the channel coefficients $\bar{\bm h}\triangleq\left(\bar{h}_k\right)_{k\in[1:K]}$ are both known by the sensors and the receiver. Additionally, the sensors transmissions are assumed to be perfectly synchronized. The pre-processed sensor signals $x_k$, $\forall k\in[1:K]$, are independent of each other with each of them being zero mean and unit variance. 

Under these assumptions, the goal of the AirComp problem with \emph{multiple simultaneous computations} is to compute $M$, $M\geq 2$, desired sums $\left(s_m\right)_{m\in[1:M]}$ given by
\begin{align}
s_m=\sum_{k\in\mathcal{D}_m}x_k \label{eq:desired_sums_to_be_computed}
\end{align} over the MAC at once with the lowest possible computation distortion. The indexing sets $\mathcal{D}_m$\footnote{We call $\mathcal{D}_m$ the $m$-th \emph{computation sensor} index set.} denote which sensors collaborate in the computation of the $m$-th sum $s_m$. We do not make any assumptions on the realizations of $\mathcal{D}_m$ other than $\mathcal{D}_m\neq\emptyset$, and $\mathcal{D}_m\cap\mathcal{D}_n=\emptyset$, $m\neq n$, $\forall m,n\in[1:M]$. 

The computation distortion is measured by the mean-squared error ($\mathsf{MSE}$)
\begin{align}
\mathsf{MSE}_m=\mathbb{E}\left[|\hat{s}_m-s_m|^{2}\right],\label{eq:MSE_general}
\end{align} where $\hat{s}_m$ is a linear estimate of $s_m$ given by
\begin{align}
\hat{s}_m=\bar{a}_{m,R}\:\Re\left\{y\right\}+\bar{a}_{m,I}\:\Im\left\{y\right\}=\Re\left\{\bar{a}_m^{*}y\right\}
\end{align} with $\bar{a}_m^{*}$ being the complex conjugate of the $m$-th Rx-scaling factor $\bar{a}_m=\bar{a}_{m,R}+j\bar{a}_{m,I}\in\mathbb{C}$ of the vector $\bar{\bm a}\triangleq\left(\bar{a}_m\right)_{m\in[1:M]}$. By the assumption that $x_k$ is of zero mean and unit variance, the power consumption of sensor $k$ is $\mathbb{E}\left[|\bar{b}_k x_k|^{2}\right]=|\bar{b}_k|^{2}$. In the remainder of this paper, we denote the absolute values of $\bar{b}_k$ and $\bar{h}_k$ by $b_k$ and $h_k$, respectively.  

\subsection{Problem Formulation}

A more explicit representation of $\mathsf{MSE}_m$ as a function of $\bar{\bm a}$ and $\bar{\bm b}$ using $\tilde{b}_k=\bar{h}_k\bar{b}_k$ is
\begin{align}
\mathsf{MSE}_m&=\sum_{k\in\mathcal{D}_m}\big\lvert \Re\left\{\bar{a}_m^{*}\tilde{b}_k\right\}-1\big\rvert^{2}+\sum_{\ell\in\mathcal{D}_m^{C}}\big\lvert\Re\left\{\bar{a}_m^{*}\tilde{b}_\ell\right\}\big\rvert^{2}+\sigma^{2}|\bar{a}_m|^{2},\label{eq:MSE_detailed}
\end{align} 
where $\mathcal{D}_m^{C}\triangleq [1:K]\setminus\mathcal{D}_m$. Alternatively, we may rewrite \eqref{eq:MSE_detailed} in its polar form using $\phi_k=\arg\left\{\tilde{b}_k\right\}$ and $\alpha_m=\arg\left\{\bar{a}_m\right\}$. 
\begin{align}
\mathsf{MSE}_m&=\sum_{k\in\mathcal{D}_m}\Big\lvert \big\lvert\bar{a}_m\tilde{b}_k\big\rvert\cos\left(\phi_k-\alpha_m\right)-1\Big\rvert^{2}\nonumber\\
&\qquad+\sum_{\ell\in\mathcal{D}_m^{C}}\Big\lvert \big\lvert\bar{a}_m\tilde{b}_\ell\big\rvert\cos\left(\phi_\ell-\alpha_m\right)\Big\rvert^{2}+\sigma^{2}|\bar{a}_m|^{2}
\label{eq:MSE_detailed_trig}
\end{align} Now, for given channel realizations $\bar{\bm h}$, a \emph{robust} MSE-minimization problem in terms of a combined Tx-Rx policy, i.e., designing $\left(\bar{\bm a}, \bar{\bm b}\right)$ \emph{jointly}, can be formulated according to
\begin{subequations}
	\begin{alignat}{2}
	&\!\min_{\bar{\bm a},\bar{\bm b}}\max_{m\in[1:M]}        &\:\:& \mathsf{MSE}_m\label{eq:optProb_Original}\\
	&\text{subject to} &      &b_k^{2}\leq P,\forall k\in[1:K].\label{eq:pwr_constraint}
	\end{alignat}
\end{subequations}

\section{Orthogonalization over Real and Imaginary Parts}
\label{sec:orthogonal_computation}

From Eq. \eqref{eq:MSE_detailed} one can infer that when optimizing an individual $\mathsf{MSE}$, say $\mathsf{MSE}_m$, it is preferrable to choose for $\big\lvert\bar{a}_{m}\tilde{b}_k\big\rvert\neq 0$, $\forall k\in[1:K]$ 
\begin{align*}
\Im\left\{\bar{a}_m^{*}\tilde{b}_k\right\}&=0,\quad\forall k\in\mathcal{D}_m,\\
\Re\left\{\bar{a}_m^{*}\tilde{b}_\ell\right\}&=0,\quad\forall \ell\in\mathcal{D}_m^{C}. 
\end{align*} In \eqref{eq:MSE_detailed_trig} this translates to setting the phase differences to $\phi_k-\alpha_m=0$, $\phi_\ell-\alpha_m=\pm\nicefrac{\pi}{2}$, $\forall k\in\mathcal{D}_m$, $\forall \ell\in\mathcal{D}_m^{C}$. As a result, $\mathsf{MSE}_m$ is interference-free and corresponds to the point-to-point $\mathsf{MSE}$, denoted by $\mathsf{MSE}^{(\mathsf{P2P})}_m$\footnote{The optimization of $\mathsf{MSE}^{(\mathsf{P2P})}_m$ with respect to parameters $a_m$ and $b_k$, $k\in\mathcal{D}_m$ is discussed in detail in \cite{Cao_2019, Liu_2019}.}
\begin{align}
\mathsf{MSE}^{(\mathsf{P2P})}_m=\sum_{k\in\mathcal{D}_m}\big\lvert |\bar{a}_m\tilde{b}_k|-1\big\rvert^{2}+\sigma^{2}|\bar{a}_m|^{2},\label{eq:P2P_MSE}
\end{align} while the remaining $\mathsf{MSE}$s -- $\mathsf{MSE}_n$, $\forall n\in[1:M],n\neq m$ -- equal
\begin{align}
\mathsf{MSE}_n&=\sum_{k\in\mathcal{D}_n}\Big\lvert |\bar{a}_n\tilde{b}_k|\cos\left(\alpha_m-\alpha_n\pm\frac{\pi}{2}\right)-1\Big\rvert^{2}+\sum_{\ell\in\mathcal{D}_m}\Big\lvert |\bar{a}_n\tilde{b}_\ell|\cos\left(\alpha_m-\alpha_n\right)\Big\rvert^{2}\nonumber\\
&\quad+\sum_{\ell\in\mathcal{D}_n^{C}\setminus\mathcal{D}_m}\Big\lvert |\bar{a}_n\tilde{b}_\ell|\cos\left(\alpha_m-\alpha_n\pm\frac{\pi}{2}\right)\Big\rvert^{2}+\sigma^{2}|\bar{a}_n|^{2}.\label{eq:remaining_MSEs}
\end{align} 
For the special case $M=2$, where $\mathcal{D}_n^{C}\setminus\mathcal{D}_m=\emptyset$, it is \emph{optimal} to choose $\alpha_m-\alpha_n=\mp\nicefrac{\pi}{2}$ in \eqref{eq:remaining_MSEs} such that $\mathsf{MSE}_n=\mathsf{MSE}_n^{(\mathsf{P2P})}$. For this special case, a simple choice that satisfies all phase difference conditions is $\left(\alpha_m,\alpha_n\right)=\left(\phi_k,\phi_\ell\right)=(0,\nicefrac{\pi}{2})$ $\forall k\in\mathcal{D}_m$, $\forall \ell\in\mathcal{D}_n$, $m\neq n$. Simply said, this strategy \emph{orthogonalizes} the computation of $s_1$ and $s_2$, i.e., $s_m$ is either solely computed along the real $(\bar{a}_{m,I}=0)$ or imaginary $(\bar{a}_{m,R}=0)$ Rx processing chain of Fig. \ref{fig:sym_mod_aircomp}.

We can extend this orthgonalization strategy to the case where $M>2$. To this end, we define real and imaginary \emph{computation index} sets $\mathcal{C}_R\subseteq[1:M]$ and $\mathcal{C}_I\subseteq[1:M]$, 
$\mathcal{C}_R\cup\mathcal{C}_I=[1:M]$, that assign which computation is delegated to real and imaginary processing chains. The union $\mathcal{C}_R\cap\mathcal{C}_I$\footnote{We denote the cardinalites by $M_w=|\mathcal{C}_w|$, $w\in\{R,I\}$. Note that these cardinalities in its most general form satisfy $M\leq M_R+M_I$.} specifies the computations that are computed along both chains. In this paper, we assume that $\mathcal{C}_R\cap\mathcal{C}_I=\emptyset$ such that $M=M_R+M_I$. 
Then, for a given computation index set pair $\left(\mathcal{C}_R,\mathcal{C}_I\right)$ 
\begin{align}
\mathsf{MSE}_{m,w}^{\perp}\left(\mathcal{C}_w\right)&=\sum_{k\in\mathcal{D}_m}\Big\lvert |\bar{a}_{m,w}|h_k b_k-1\Big\rvert^{2}\nonumber\\
&\qquad+\sum_{\ell\in \mathcal{D}_{\mathcal{C}_{w}}\setminus\mathcal{D}_m}\Big\lvert |\bar{a}_{m,w}|h_\ell b_\ell\Big\rvert^{2}+\sigma^{2}|\bar{a}_{m,w}|^{2},\label{eq:MSE_Rx_processing_chain}\\
\mathsf{MSE}_{w}^{\perp}\left(\mathcal{C}_w\right)&=\min_{\substack{\left(|\bar{a}_{m,w}|\right)_{m\in\mathcal{C}_{w}},\\\left(b_k\right)_{k\in\mathcal{D}_{\mathbb{C}_w}}\leq\sqrt{P}\bm 1}}\max_{m\in\mathcal{C}_{w}}\left(\mathsf{MSE}_{m,\mathcal{C}_{w}}^{\perp}\right),\label{eq:opt_worst_case_MSE_Rx_processing_chain}
\end{align} denote, respectively, the associated $\mathsf{MSE}$ of $s_m$ and the optimized, worst-case $\mathsf{MSE}$ of all computations along Rx processing chain $w\in\{R,I\}$. Overall, the worst-case $\mathsf{MSE}$ for this pair then becomes 
\begin{align}
\mathsf{MSE}^{\perp}\left(\mathcal{C}_R,\mathcal{C}_I\right)=\max\left(\mathsf{MSE}_{R}^{\perp}\left(\mathcal{C}_R\right),\mathsf{MSE}_{I}^{\perp}\left(\mathcal{C}_I\right)\right).\label{eq:opt_worst_case_MSE_given_comp_idx_pair} 
\end{align} 
Note that \emph{symmetry} applies, i.e., processing computations with indices in $\mathcal{C}_R$ ($\mathcal{C}_I$) can be processed either along the real and imaginary Rx processing chain; thus, $\mathsf{MSE}^{\perp}_{w}\left(\mathcal{C}_w\right)=\mathsf{MSE}^{\perp}_{\bar{w}}\left(\mathcal{C}_w\right)$ for $w\neq\bar{w}$. Solving \eqref{eq:opt_worst_case_MSE_given_comp_idx_pair} over all possible pairs $\left(\mathcal{C}_R,\mathcal{C}_I\right)$ gives the optimal $\mathsf{MSE}$ of the orthogonalization scheme. 
\begin{align}
\mathsf{MSE}^{\perp,\star}\triangleq\mathsf{MSE}^{\perp}\left(\mathcal{C}_R^{\star},\mathcal{C}_I^{\star}\right)=\min_{\substack{\left(\mathcal{C}_R^{\star},\mathcal{C}_I^{\star}\right)\in\mathbb{N}^{2}_{++}:\\\mathcal{C}_R\cup\mathcal{C}_I=[1:M]}}\mathsf{MSE}^{\perp}\left(\mathcal{C}_R,\mathcal{C}_I\right)\label{eq:optimal_MSE_orthogonal_scheme}
\end{align} In general, $\mathsf{MSE}^{\perp,\star}\geq\mathsf{MSE}^{\star}$, with $\mathsf{MSE}^{\star}$ being the optimum of problem $(7)$. In fact, in Appendix \ref{apdx:comp_MSEs}, we show that equality holds. The basic operation needed to perform the optimization in \eqref{eq:opt_worst_case_MSE_given_comp_idx_pair} and \eqref{eq:optimal_MSE_orthogonal_scheme} is solving \eqref{eq:opt_worst_case_MSE_Rx_processing_chain}. To this end, the next section addresses the solution of \eqref{eq:opt_worst_case_MSE_Rx_processing_chain} for the special case of $M_w=2$.     

\section{Solution to Optimization Problem \eqref{eq:opt_worst_case_MSE_Rx_processing_chain}}
\label{sec:sol_opt_orthgonal_problem}

In this section, we outline the main ideas and concepts of our locally optimal solution to problem \eqref{eq:opt_worst_case_MSE_Rx_processing_chain}. Rigorous proofs are appended to the appendix and referred to wherever necessary.    

\subsection{Preliminaries}
\label{subsec:prelim}

Due to the symmetry property, we consider without loss of generality the optimization problem \eqref{eq:opt_worst_case_MSE_Rx_processing_chain} for the real Rx processing chain when $M_R=2$. In the sequel of this paper, for ease of presentation, we simplify some notation\footnote{We denote $|\bar{a}_{m,R}|=c_{m}$ and pretend that $\mathcal{C}_I=\emptyset$ such that $\cup_{m=1}^{2}\mathcal{D}_m=[1:K]$ for $\mathcal{C}_R=[1:M]=\{1,2\}$. We implicitly assume in this section that $b_k$, $c_m$ and $h_k$ are non-negative. Throughout the remaining part of this paper, we omit using the subscript '$R$' and the superscript '$\perp$'.}. With this simplification in notation, $\mathsf{MSE}_{m,\mathcal{C}_R}^{\perp}$ becomes
\begin{align}
\mathsf{MSE}_m\left(c_m,\bm b\right)&=c_m^{2}\underbrace{\left(\sigma^{2}+\sum_{k=1}^{K}h_k^{2}b_k^{2}\right)}_{\triangleq A}-2c_m\underbrace{\left(\sum_{k\in\mathcal{D}_m}h_k b_k\right)}_{\triangleq B_{\mathcal{D}_m}}+|\mathcal{D}_m|.\label{eq:MSE_m_orthogonal_Rx_real_special_case}		
\end{align} Note that $A=\sigma^{2}+\sum_{k=1}^{K}h_k^{2}b_k^{2}=\sigma^{2}+\sum_{m=1}^{2}C_{\mathcal{D}_m}$, where $C_{\mathcal{D}_m}=\sum_{k\in\mathcal{D}_m}h_k^{2}b_k^{2}$. From the Cauchy-Schwarz inequality and $A\geq C_{\mathcal{D}_m}$, we know that $\nicefrac{B_{\mathcal{D}_m}^{2}}{A}\leq\nicefrac{B_{\mathcal{D}_m}^{2}}{C_{\mathcal{D}_m}}\leq |\mathcal{D}_m|$ and thus $A|\mathcal{D}_m|-B_{\mathcal{D}_m}^{2}\geq 0$. The comparison of the different $\mathsf{MSE}$s for non-negative Rx-scaling factors $c_m$ in Appendix \ref{apdx:lb_bound} allows us to reformulate the optimization problem \eqref{eq:opt_worst_case_MSE_Rx_processing_chain} to the following fractional program for $\Delta D\triangleq |\mathcal{D}_2|-|\mathcal{D}_1|$. 
\begin{subequations}
	\begin{alignat}{2}
	&\!\min_{\bm b\geq\bm 0}        &\:\:& -\frac{B_{\mathcal{D}_1}^2}{A}\label{eq:revised_optProb}\\
	&\text{subject to} &      &b_k-\sqrt{P}\leq 0,\forall k\in[1:K],\label{eq:revised_pwr_constraint}\\
	&&& \frac{B_{\mathcal{D}_1}^{2}-B_{\mathcal{D}_2}^{2}}{A}+\Delta D=0.\label{eq:equal_MSE}
	\end{alignat}
\end{subequations} Throughout this paper, we denote, respectively, the $k$-th \emph{inequality} constraint function by $g_k\left(\bm b\right)\triangleq b_k-\sqrt{P}$ and the \emph{equality} constraint function by $h\left(\bm b\right)=\nicefrac{B_{\mathcal{D}_1}^2}{A}-\nicefrac{B_{\mathcal{D}_2}^2}{A}+\Delta D$.  

\subsection{Feasability of Problem (15)}
\label{subsec:feas}

In this subsection, we would like to know when problem (15) is infeasible. Clearly, for $b_k\in[0,\sqrt{P}]$, $\forall k\in[1:K]$, there exists no solution if the equality constraint \eqref{eq:equal_MSE} is not satisfied. Interestingly, in the case that $\Delta D=0$, the optimization problem is always feasible. This is since \eqref{eq:equal_MSE} reduces to the linear condition $B_{\mathcal{D}_2}-B_{\mathcal{D}_1}=0$ for which there incurs no requirement on the noise variance $\sigma^2$; thus, one can always find a feasible vector $\bm b$ satisfying that particular equality constraint. Henceforth, we assume that $\Delta D\neq 0$. Then, \eqref{eq:equal_MSE} is not satisfied for $\bm 0\leq \bm b\leq \sqrt{P}\bm 1$ and $\sigma^2>0$ if  
\begin{align}
\sigma^{2}>\tilde{\sigma}^{2}\triangleq\left[\max_{\bm 0\leq \bm b\leq \sqrt{P}\bm 1}\left(\frac{1}{\Delta D}\left(B_{\mathcal{D}_2}^{2}-B_{\mathcal{D}_1}^{2}\right)-C_{\mathcal{D}_1}-C_{\mathcal{D}_2}\right)\right]^{+}.\label{eq:feas_cond_1} 
\end{align} In other words, the problem has no solution if the noise variance $\sigma^{2}$ 
exceeds the threshold noise variance $\tilde{\sigma}^{2}$. This threshold can be further simplified by  
\begin{align}
\tilde{\sigma}^{2}=
\begin{cases}
\left[\max_{\substack{0\leq b_k\leq \sqrt{P}:\\k\in\mathcal{D}_2}}\left(\frac{1}{\Delta D}B_{\mathcal{D}_2}^{2}-C_{\mathcal{D}_2}\right)\right]^{+}&\text{ if }\Delta D>0\\
\left[\max_{\substack{0\leq b_k\leq \sqrt{P}:\\k\in\mathcal{D}_1}}\left(-\frac{1}{\Delta D}B_{\mathcal{D}_1}^{2}-C_{\mathcal{D}_1}\right)\right]^{+}&\text{ if }\Delta D<0			
\end{cases}.\label{eq:feas_cond_2} 
\end{align} 

\subsubsection{Lower Bound $\underline{\tilde{\sigma}}^{2}$}

In the following, we establish a lower bound $\underline{\tilde{\sigma}}^{2}$ on $\tilde{\sigma}^{2}$. A lower bound on $\tilde{\sigma}^{2}$ is to choose a mixture of full power transmission, i.e., $b_k=\sqrt{P}$ for $k\in\mathcal{G}_i^{C}\subseteq\mathcal{D}_i$ and less-than full power transmission, i.e., $b_j=\frac{H_{\mathcal{G}_i}}{h_j}<\sqrt{P}$ for $j\in\mathcal{G}_i\subseteq\mathcal{D}_i$, where $\mathcal{G}_i\cup\mathcal{G}_i^{C}=\mathcal{D}_i$. Specifically, we choose
\begin{align*}
H_{\mathcal{G}_i}=\frac{\sqrt{P}\left(\sum_{k\in\mathcal{G}_i^{C}}h_k\right)}{(-1)^{i}\Delta D-|\mathcal{G}_i|}
\end{align*} for $(-1)^{i}\Delta D-|\mathcal{G}_i|>0$ such that
\begin{align*}
h_k\sqrt{P}\leq H_{\mathcal{G}_i}\leq h_j\sqrt{P}
\end{align*} for $j\in\mathcal{G}_i$ and $k\in\mathcal{G}_i^{C}$. Above inequality suggests that the indices of the smallest $|\mathcal{G}_i^{C}|$ channel coefficients are attributed to the set $\mathcal{G}_i^{C}$. The remaining $|\mathcal{G}_i|$ indices construct $\mathcal{G}_i$. For this choice of $b_k$, $k\in\mathcal{D}_i$, we get
\begin{align}
&\underline{\tilde{\sigma}}^{2}=\nonumber\\
&\begin{cases}
\left[\frac{1}{\Delta D}\left(|\mathcal{G}_2|H_{\mathcal{G}_2}+\sqrt{P}\left(\sum_{k\in\mathcal{G}_2^{C}}h_k\right)\right)^{2}-\left(|\mathcal{G}_2|H_{\mathcal{G}_2}^{2}+P\left(\sum_{k\in\mathcal{G}_2^{C}}h_k^{2}\right)\right)\right]^{+}&\text{ if }\Delta D>0\\
\left[-\frac{1}{\Delta D}\left(|\mathcal{G}_1|H_{\mathcal{G}_1}+\sqrt{P}\left(\sum_{k\in\mathcal{G}_1^{C}}h_k\right)\right)^{2}-\left(|\mathcal{G}_1|H_{\mathcal{G}_1}^{2}+P\left(\sum_{k\in\mathcal{G}_1^{C}}h_k^{2}\right)\right)\right]^{+}&\text{ if }\Delta D<0\\
\end{cases}.\label{eq:lb_bound_noie_th}
\end{align}  

However, it may often be cumbersome to determine $H_{\mathcal{G}_i}$. To this end, we seek to find an upper bound $\overline{\tilde{\sigma}}^{2}$ which ultimately allows us to approximate $\tilde{\sigma}^2$. 

\subsubsection{Upper Bound $\overline{\tilde{\sigma}}^{2}$}

We can verify that $B_{\mathcal{D}_m}^{2}=C_{\mathcal{D}_m}+F_{\mathcal{D}_m}$, where
\begin{align*}
F_{\mathcal{D}_m}=\sum_{k\in \mathcal{D}_m}\sum_{\substack{j\in \mathcal{D}_m:\\ j\neq k}}h_k h_j b_k b_j.
\end{align*} Since $\tilde{\sigma}^2$ in \eqref{eq:feas_cond_2} depends on both $C_{\mathcal{D}_i}$ and $F_{\mathcal{D}_i}$, we find an upper bound on $F_{\mathcal{D}_i}$ as a function of  $C_{\mathcal{D}_i}$.
\begin{align*}
F_{\mathcal{D}_i}&=\sum_{k\in \mathcal{D}_i}\sum_{\substack{j\in \mathcal{D}_i:\\ j\neq k}}h_k h_j b_k b_j\stackrel{(a)}{=}\sum_{k\in \mathcal{D}_i}\left(h_k b_k \bm 1^{T}_{|\mathcal{D}_i|-1}\bm q_{\mathcal{D}_i\setminus\{k\}}\right)\\
&\stackrel{(b)}{\leq}\sum_{k\in\mathcal{D}_i}\left(\sqrt{|\mathcal{D}_i|-1}h_k b_k\sqrt{\sum_{j\in\mathcal{D}_i\setminus\{k\}}h_j^{2}b_j^{2}}\right)\\
&\leq\sqrt{|\mathcal{D}_i|-1}\sqrt{C_{\mathcal{D}_i}}\underbrace{\sum_{k\in\mathcal{D}_i} h_k b_k}_{=B_{\mathcal{D}_i}=\sqrt{C_{\mathcal{D}_i}+F_{\mathcal{D}_i}}}, 
\end{align*} where step $(a)$ uses $\bm q_{\mathcal{D}_i\setminus\{k\}}=\left(h_j b_j\right)_{j\in\mathcal{D}_i\setminus\{k\}}$ such that $\bm 1^{T}_{|\mathcal{D}_i|-1}\bm q_{\mathcal{D}_i\setminus\{k\}}=\sum_{j\in\mathcal{D}_i\setminus\{k\}}h_j b_j$. Step $(b)$ follows from the Cauchy-Schwarz inequality $|\bm u^{T}\bm v|\leq ||\bm u||_2||\bm v||_2$ with $\bm u=h_k b_k\bm 1^{T}_{|\mathcal{D}_i|-1}$, $\bm v=\bm q_{\mathcal{D}_i\setminus\{k\}}$, $||\bm u||_2=\sqrt{|\mathcal{D}_i|-1}h_k b_k$ and $||\bm v||_2=\sqrt{\sum_{j\in\mathcal{D}_i\setminus\{k\}}h_j^{2}b_j^{2}}=\sqrt{C_{\mathcal{D}_i}-h_k^{2} b_k^{2}}$\footnote{In the next step, we use the upper bound $\sqrt{C_{\mathcal{D}_i}-h_k^{2} b_k^{2}}\leq\sqrt{C_{\mathcal{D}_i}}$ instead of the tighter bound $\sqrt{C_{\mathcal{D}_i}-h_k^{2} b_k^{2}}\leq\sqrt{C_{\mathcal{D}_i}}\left(1-\frac{h_k^{2} b_k^{2}}{2C_{\mathcal{D}_i}}\right)$ for more compact bounding expressions.}. From $F_{\mathcal{D}_i}\geq 0$ and above inequality, one can derive that
\begin{align}
0\leq F_{\mathcal{D}_i}\leq \frac{C_{\mathcal{D}_i}}{2}\left(|\mathcal{D}_i|-1+\underbrace{\sqrt{\left(|\mathcal{D}_i|-1\right)\left(|\mathcal{D}_i|+3\right)}}_{\triangleq\zeta_{\mathcal{D}_i}}\right).\label{eq:up_bound_F_D_i}
\end{align} Using \eqref{eq:up_bound_F_D_i} along with $C_{\mathcal{D}_i}\leq P\left(\sum_{k\in\mathcal{D}_i}h_k^{2}\right)$ in Eq. \eqref{eq:feas_cond_2}, we get 
\begin{align}
\overline{\tilde{\sigma}}^{2}\triangleq
\begin{cases}
\frac{P\left(\sum_{k\in\mathcal{D}_2}h_k^{2}\right)}{2}\left(\frac{1-2\Delta D+|\mathcal{D}_2|+\zeta_{\mathcal{D}_2}}{\Delta D}\right)\quad&\text{ if }\Delta D>0\\
-\frac{P\left(\sum_{k\in\mathcal{D}_1}h_k^{2}\right)}{2}\left(\frac{1+2\Delta D+|\mathcal{D}_1|+\zeta_{\mathcal{D}_1}}{\Delta D}\right)\quad&\text{ if }\Delta D<0
\end{cases}.\label{eq:ub_bound_noie_th} 
\end{align} 
\subsubsection{Approximation on $\tilde{\sigma}^{2}$}
Since $\zeta_{\mathcal{D}_i}\approx|\mathcal{D}_i|-1$, we approximate $\tilde{\sigma}^{2}$ by
\begin{align*}
\tilde{\sigma}^{2}\approx
\begin{cases}
\frac{P|\mathcal{D}_1|\left(\sum_{k\in\mathcal{D}_2}h_k^{2}\right)}{\Delta D}\quad&\text{ if }\Delta D>0\\
\frac{P|\mathcal{D}_2|\left(\sum_{k\in\mathcal{D}_1}h_k^{2}\right)}{-\Delta D}\quad&\text{ if }\Delta D<0
\end{cases}.
\end{align*} As far as the existence of a feasible solution to the optimization problem is concerned, the approximation on $\tilde{\sigma}^2$ suggests the following main influencing factors on the non-emptiness of the feasible set. These are (i) the $\mathsf{SNR}=\nicefrac{P}{\sigma^{2}}$, (ii) the ratio $\nicefrac{|\mathcal{D}_1|}{\Delta D}$ ($\nicefrac{|\mathcal{D}_2|}{-\Delta D}$) for $\Delta D>0$ ($\Delta D<0$) and  (iii)  the channel statistics, i.e., mean and variance of $h_k$. 
As any one of these three factors increases, it is less likely that the feasible set is empty.  

\subsection{Solution through KKT-Conditions}
\label{subsec:sol_KKT}

We determine a solution to the optimization problem by considering the KKT-conditions given in Appendix \ref{apdx:KKT_cond}. The complementary slackness condition \eqref{eq:comp_slack} suggests that there are two sets of sensors. On the one hand, there are sensors $k\in\mathcal{P}_m$, $\mathcal{P}_m\subseteq\mathcal{D}_m$, $\forall m\in[1:2]$, that do not transmit with full power, i.e., $b_k<\sqrt{P}$, and thus the Lagrange multiplier being $\lambda_k=0$. The remaining sensors $\mathcal{P}_m^{C}=\mathcal{D}_m\setminus\mathcal{P}_m$, $\forall m\in[1:2]$, on the other hand, transmit with full power, i.e., $b_k=\sqrt{P}$ for $k\in\mathcal{P}_m^{C}$. In Appendix \ref{apdx:KKT_cond_sol}, we use these two sets to determine the KKT-point $\bm b'=\left(b_1',\ldots, b_K'\right)^T$ with its $k$-th element being either $b_k'=\sqrt{P}$ if $k\in\bigcup_{m=1}^{2}\mathcal{P}_m^C$ and $b_k'=\frac{E_{\mathcal{P}_m}}{h_k}$ if $k\in\bigcup_{m=1}^{2}\mathcal{P}_m$. We show that the extreme cases where $a)$ all sensors transmit with less-than full power, i.e., $\left(|\mathcal{P}_1|,|\mathcal{P}_2|\right)=\left(|\mathcal{D}_1|,|\mathcal{D}_2|\right)$, and $b)$ all sensors transmit with full power, i.e., $\left(|\mathcal{P}_1|,|\mathcal{P}_2|\right)=\left(0,0\right)$, do not give us a feasible KKT-point. Instead, the only viable KKT-solutions occur at intermediate cardinality cases of $a)$ and $b)$ where $0<|\mathcal{P}_1|+|\mathcal{P}_2|<K$ or $0<|\mathcal{P}_1^C|+|\mathcal{P}_2^C|<K$. These give us the cases $c)$ $\left(|\mathcal{P}_1|,|\mathcal{P}_2|\right)\geq\left(1,1\right)$ (excluding case $a)$) or $d)$ $|\mathcal{P}_m|\geq 1,|\mathcal{P}_n|=0$, $m\neq n$. For cases $c)$ and $d)$, we derive conditions on $E_{\mathcal{P}_m}$ specified in Eqs. \eqref{eq:cond_satisfy_dual_prim_feas_case_c}, \eqref{eq:cond_satisfy_dual_prim_feas_case_d}, such that $\bm b'$ and its respective (Lagrange) dual vector $\bm \lambda'=\left(\lambda_1',\ldots,\lambda_K'\right)^{T}$ produce a KKT-solution that is both primal feasible (cf. Eqs. \eqref{eq:prim_feas_ineq_const}, \eqref{eq:prim_feas_eq_const}) and dual feasible (cf. Eq. \eqref{eq:dual_feas}). Further, from the conditions \eqref{eq:cond_satisfy_dual_prim_feas_case_c}, \eqref{eq:cond_satisfy_dual_prim_feas_case_d}, we can also retrieve the design principle of the sets $\mathcal{P}_m$ and $\mathcal{P}_m^C$, $\forall m\in[1:2]$. That is, for $m\in[1:2]$, the sensors of the strongest $|\mathcal{P}_m|$ channels $h_k$ of the vector $\bm h_{\mathcal{D}_m}\triangleq\left(h_k\right)_{k\in\mathcal{D}_m}$ are attributed to $\mathcal{P}_m$, while the remaining $|\mathcal{P}_m^{C}|=|\mathcal{D}_m|-|\mathcal{P}_m|$ sensors with weaker channels in $\bm h_{\mathcal{D}_m}$ are accumulated in the set $\mathcal{P}_m^C$. Interestingly, this design choice is in agreement with intuition. As one would assume, it is important to exploit every sensors computation to keep the $\mathsf{MSE}$ as low as possible. To this end, one seeks to balance out the effective power surplus $|h_k b_k-h_\ell b_\ell|$ of sensors $k\in\mathcal{P}_m$ with stronger channels $h_k$ against sensors $\ell\in\mathcal{P}_m^C$ with weaker channels $h_\ell$. Naturally, the cardinality of $\mathcal{P}_m$ categorizes the relative level of weak to strong channels and thus plays a crucial role in the achievable worst-case $\mathsf{MSE}$.

\subsection{Linear Independence Constraint Qualification (LICQ)}
\label{subsec:LICQ}

We recall that in order for a minimum point $\bm b'$ to satisfy the KKT-conditions of Appendix \ref{apdx:KKT_cond}, the problem should satisfy some regularity conditions. One common condition, which we use here is the LICQ. The LICQ is satisfied if the gradients of the active inequality constraints, i.e., $\nabla g_k\left(\bm b'\right)$, $\forall k\in\bigcup_{m=1}^{2}\mathcal{P}_m^{C}$, and the gradient of the equality constraint $\nabla h\left(\bm b'\right)$ are linearly independent at $\bm b'$; or, in other words, the $K\times \left(|\mathcal{P}_1^C|+|\mathcal{P}_2^C|+1\right)$ matrix
\begin{align}
\bm J=\begin{bmatrix}
\bm G\left(\bm b'\right) & \nabla h\left(\bm b'\right)
\end{bmatrix},\label{eq:jacobian} 
\end{align} where $\bm G\left(\bm b'\right)=\left(\nabla g_k\left(\bm b'\right)\right)_{k\in\bigcup_{m=1}^{2}\mathcal{P}_m^{C}}$ has to be of full rank. Thus $\mathsf{rank}\left(\bm J\right)=|\mathcal{P}_1^C|+|\mathcal{P}_2^C|+1$. In Appendix \ref{apdx:LICQ}, we show that the LICQ is satisfied for the KKT-point $\bm b'$ of Appendix \ref{apdx:KKT_cond_sol}.    

\subsection{Second-Order Sufficient Condition}

Consider the Lagrangian function 
\begin{align*}
\mathcal{L}(\bm b,\bm \lambda,\mu)=-\frac{B_{\mathcal{D}_1}^{2}}{A}+\sum_{k=1}^{K}\lambda_k\left(b_k-\sqrt{P}\right)+\mu\left(\frac{B_{\mathcal{D}_1}^{2}}{A}-\frac{B_{\mathcal{D}_2}^{2}}{A}+\Delta D\right)
\end{align*} of the optimization problem $(15)$. From optimization theory, the second-order sufficient condition of optimality is known to be following.  

\begin{proposition}[Second-order sufficient condition]
	If $\nabla_{\bm b}\:\mathcal{L}(\bm b',\bm \lambda',\mu')=\bm 0$, if $\bm b'$ feasible, if strict complementarity holds, i.e., $\lambda_k'>0$, $\forall k\in \bigcup_{m=1}^{2}\mathcal{P}_m^{C}$ and if
	\begin{align}
	&\bm s^{T}\bm H\bm s>0,\forall \bm s\in\mathcal{S}\triangleq\left\{\bm s\in\mathbb{R}^{K}:\bm s\neq\bm 0,\:\bm s^{T}\nabla h\left(\bm b'\right)=0,\:\bm s^{T}\nabla g_k\left(\bm b'\right)=0,\forall k\in\cup_{m=1}^{2}\mathcal{P}_m^{C}\right\},\label{eq:cond_SOC}
	\end{align} where $\bm H=\nabla_{\bm b}^{2}\:\mathcal{L}(\bm b',\bm \lambda',\mu')$, then it follows that $\bm b'$ is a local minimizer.
\end{proposition} Comparing $\mathcal{S}$ in \eqref{eq:cond_SOC} with $\bm J$ in \eqref{eq:jacobian}, we see that $\mathcal{S}$ is the kernel of $\bm J^{T}$, i.e., $\mathcal{S}=\mathsf{ker}\left(\bm J^{T}\right)$. Note that $\mathsf{dim}\left(\mathcal{S}\right)=K-\mathsf{rank}\left(\bm J\right)$ which simplifies to $\mathsf{dim}\left(\mathcal{S}\right)=K-|\mathcal{P}_1^C|-|\mathcal{P}_2^C|-1$,
if the LICQ is satisfied. Recall from linear algebra, that the \emph{matrix inertia} $\pi\left(\bm H\right)$ of a symmetric $K\times K$ real matrix is defined to be the triple $\left(\rho,\eta,\theta\right)$, where $\rho$, $\eta$ and $\theta$ are, respectivly, the numbers of positive, negative and zero eingevalues of the matrix $\bm H$ with multiplicities counted \cite{Horn_2012}. \emph{Han and Fujiwara} introduce the notion of \emph{relative} inertia $\pi\left(\bm H/\mathcal{S}\right)$ for a symmetric $K\times K$ real matrix \cite{Han_1985}. They follow from Sylvester's law of inertia that
\begin{align}
\pi\left(\bm H/\mathcal{S}\right)=\pi\left(\bm S^{T}\bm H\bm S\right),\label{eq:relative_matrix_inertia}
\end{align} where $\bm S$ is a matrix whose columns form a basis of $\mathcal{S}$. One can infer that the second-order sufficient condition \eqref{eq:cond_SOC} is equivalent to the relative matrix inertia being $\pi\left(\bm H/\mathcal{S}\right)=\left(\mathsf{dim}\left(\mathcal{S}\right),0,0\right)$. However, examining the relative inertia may be tedious. A more practical approach is to check directly for the matrix inertia of the KKT-matrix
\begin{align}
\bm K\triangleq\begin{bmatrix}
\bm H & \bm J \\
\bm J^{T} & \bm 0
\end{bmatrix}.\label{eq:KKT_mat}
\end{align} To this end, \emph{Han and Fujiwara} establishen in \cite[Theorem 3.4]{Han_1985} for $\theta\left(\bm H/\mathcal{S}\right)=0$\footnote{In Theorem 3.1, they show that $\theta\left(\bm K\right)=\theta\left(\bm H/\mathcal{S}\right)+\mathsf{dim}\left(\mathsf{ker}\left(\bm J\right)\right)$. In other words, if $\theta\left(\bm K\right)=0$ $\implies$ $\theta\left(\bm H/\mathcal{S}\right)=0$.}, the direct relationship 
\begin{align*}
\pi\left(\bm K\right)=\pi\left(\bm H/\mathcal{S}\right)+\left(\mathsf{rank}\left(\bm J\right),\mathsf{rank}\left(\bm J\right),|\mathcal{P}_1^C|+|\mathcal{P}_2^C|+1-\mathsf{rank}\left(\bm J\right)\right).
\end{align*} For the KKT-point $\bm b'$ which satisfies the LICQ such that $\mathsf{rank}\left(\bm J\right)=|\mathcal{P}_1^C|+|\mathcal{P}_2^C|+1$ gives 
\begin{align}
\pi\left(\bm K\right)=\left(K,|\mathcal{P}_1^C|+|\mathcal{P}_2^C|+1,0\right)\label{eq:KKT_mat_inertia_LICQ_satisfied}
\end{align} if $\pi\left(\bm H/\mathcal{S}\right)=\left(K-|\mathcal{P}_1^C|-|\mathcal{P}_2^C|-1,0,0\right)$. Thus, in our simulation, to check for the second-order sufficient condition, we verify if \eqref{eq:KKT_mat_inertia_LICQ_satisfied} is satisfied.    

\section{Simulation Results}
\label{sec:num_res}

In this section, we provide simulation results for validation of our proposed solution of optimization problem \eqref{eq:opt_worst_case_MSE_Rx_processing_chain}. Thus, we implicitly assume that the orthogonalization principle in computation described in \ref{sec:orthogonal_computation} is deployed. Then, without loss of generality, we can focus on the real processing chain. We stick to the notation used in section \ref{sec:sol_opt_orthgonal_problem}. In the simulation, we model the absolute values of the channel ceofficients $h_k$ (which is actually $|h_k|$) by i.i.d. Rayleigh fading, i.e., $h_k\sim\mathsf{Rayl}\left(\sigma_R\right)$, where $\sigma_R$ is the scale parameter of the Rayleigh distribution. Then the mean and variance of $h_k$, $\forall k\in[1:K]$, are, respectively, $\mu_{h}=\sigma_R\sqrt{\nicefrac{\pi}{2}}$ and $\sigma^{2}_h=\frac{(4-\pi)}{2}\sigma_R^2$. If not otherwise specified, we choose $\sigma^{2}_h=1$, while we set the noise variance to $\sigma^{2}=1$. For a \emph{fixed} realization of $\left(\mathcal{D}_1,\mathcal{D}_2\right)$, where $\mathcal{D}_m\subseteq[1:K]$, $\forall m\in[1:2]$, with $K=30$, we compute the normalized average $\mathsf{MSE}$ -- $\nicefrac{\mathbb{E}_{\bm h}\left[\mathsf{MSE}\right]}{\max\left(|\mathcal{D}_1|,|\mathcal{D}_2|\right)}$ -- for different $\mathsf{SNR}=\nicefrac{P}{\sigma^{2}}$ in the range of $-5$ dB up to $50$ dB in $5$ dB increments. Note that for a fixed realization of $\left(\mathcal{D}_1,\mathcal{D}_2\right)$, we approximate $\mathbb{E}_{\bm h}\left[\mathsf{MSE}\right]$ as an average $\mathsf{MSE}$ over the number of \emph{feasible} realizations. Clearly, this number is always less or equal to the number of channel realizations, which we fix to $7500$. 

\subsection{Feasibility}
\label{subsec:feas_sim_res}

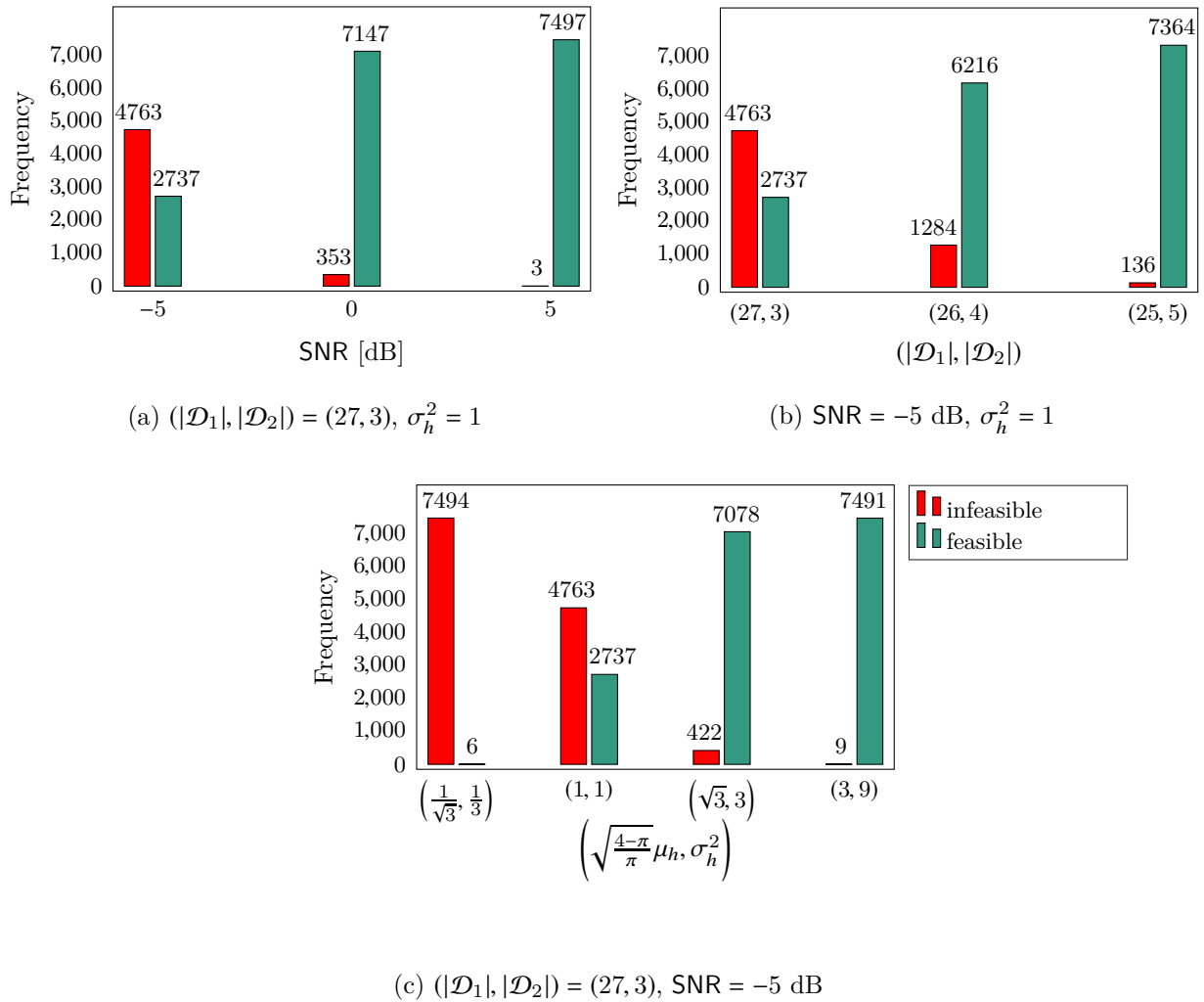
\begin{figure}
	\centering
	\begin{subfigure}{.5\textwidth}
		\input{plots/Feas_Anal_SNR_Impact_Plot.tex}
		\caption{$\left(|\mathcal{D}_1|,|\mathcal{D}_2|\right)=\left(27,3\right)$, $\sigma^{2}_{h}=1$}
		\label{fig:feas_1}
	\end{subfigure}%
	\begin{subfigure}{.5\textwidth}
		\input{plots/Feas_Anal_Set_Card_Impact_Plot.tex}
		\caption{$\mathsf{SNR}=-5$ dB, $\sigma^{2}_{h}=1$}
		\label{fig:feas_2}
	\end{subfigure}\vskip\floatsep%
	\begin{subfigure}{0.5\textwidth}
		\centering
		\input{plots/Feas_Anal_Channel_Stat_Impact_Plot.tex}
		\caption{$\left(|\mathcal{D}_1|,|\mathcal{D}_2|\right)=\left(27,3\right)$, $\mathsf{SNR}=-5$ dB}
		\label{fig:feas_3}
	\end{subfigure}
	\caption{\small Histogram of the number of infeasible (red) and feasible (green) realizations for varying (a) $\mathsf{SNR}$, (b) cardinality vector $\left(|\mathcal{D}_1|,|\mathcal{D}_2|\right)$ and (c) channel statistics $\left(\mu_h,\sigma^2_h\right)$. Recall that the noise variance is $\sigma^2=1$ and that there are in total $7500$ realizations.}
	\label{fig:sim_feas}
\end{figure}

In this subsection, we discuss how (a) the $\mathsf{SNR}$, (b) the cardinality vector $\left(|\mathcal{D}_1|,|\mathcal{D}_2|\right)$ and (c) the channel statistics of $h_k$ affect the feasibility of the optimization problem $(15)$. To this end, we count the number of all feasible and infeasible realizations as we either increase (a) the $\mathsf{SNR}$ from $-5$ dB to $5$ dB in $5$ dB (additive) increments, (b) the cardinality ratio $\nicefrac{|\mathcal{D}_2|}{|\Delta D|}$ or (c) the channel statistics $\left(\mu_h,\sigma^2_h\right)$ of $h_k$, $\forall k$. We can see in Fig. \ref{fig:sim_feas} that an increase of any of those parameters has a positive impact on the feasibility of the optimization problem. This is in accordance with the discussion of subsection \ref{subsec:feas}. Optimization problems that are parametrized by either a low $\mathsf{SNR}$, large cardinality imbalances $|\Delta D|$ or Rayleigh distributions with a low mean and a low standard deviation are prone to suffer from infeasibility. However, the plots in \ref{fig:feas_1}-\ref{fig:feas_3} show that sufficiently large/small values of these parameters, e.g., $\mathsf{SNR}\geq 5$ dB or $|\Delta D|\leq 20$, make the optimization problem almost always feasible.  

\begin{figure}
	\centering
	\begin{subfigure}{.5\textwidth}
		\input{plots/Revised_Best_Card_Set_delta_d_4_SNR_-5.tex}
		\caption{$\mathsf{SNR}=-5$ dB}
		\label{fig:best_set_delta_d_4_SNR_-5}
	\end{subfigure}%
	\begin{subfigure}{.5\textwidth}
		\input{plots/Revised_Best_Card_Set_delta_d_4_SNR_10.tex}
		\caption{$\mathsf{SNR}=10$ dB}
		\label{fig:best_set_delta_d_4_SNR_10}
	\end{subfigure}\vskip\floatsep%
	\begin{subfigure}{0.5\textwidth}
		\input{plots/Revised_Best_Card_Set_delta_d_4_SNR_30.tex}
		\caption{$\mathsf{SNR}=30$ dB}
		\label{fig:best_set_delta_d_4_SNR_30}
	\end{subfigure}%
	\begin{subfigure}{0.5\textwidth}
		\input{plots/Revised_Best_Card_Set_delta_d_4_SNR_50.tex}
		\caption{$\mathsf{SNR}=50$ dB}
		\label{fig:best_set_delta_d_4_SNR_50}
	\end{subfigure}
	\caption{\small Histogram of the optimal cardinality set $\left(|\mathcal{P}_1^{\star}|,|\mathcal{P}_2^{\star}|\right)$ for $\left(|\mathcal{D}_1|,|\mathcal{D}_2|\right)=\left(13,17\right)$ and varying $\mathsf{SNR}\in\left\{-5,10,30,50\right\}$ dB.}
	\label{fig:sim_best_set_delta_d_4}
\end{figure}
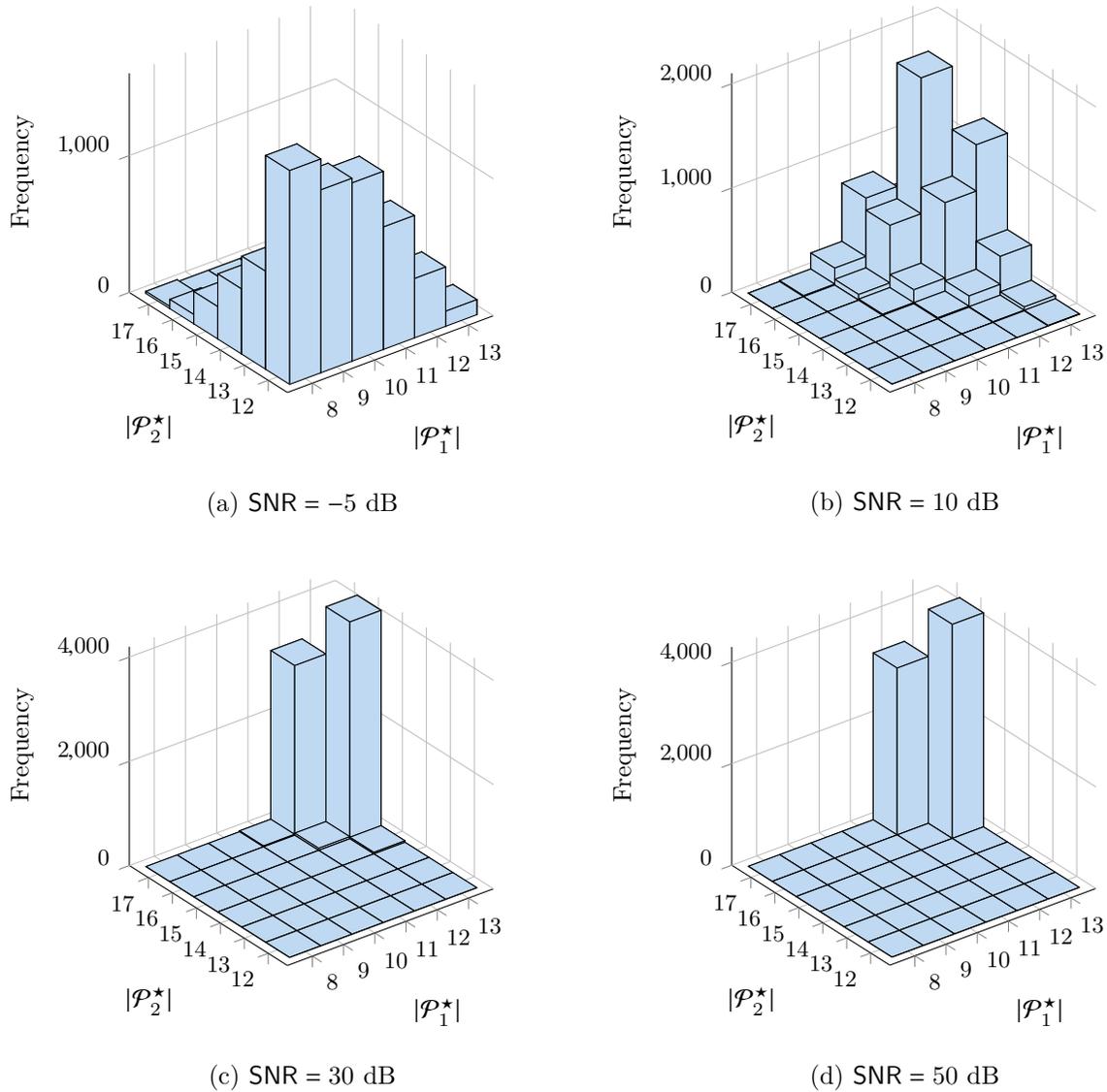

\begin{figure}
	\centering
	\begin{subfigure}{.5\textwidth}
		\input{plots/Revised_Best_Card_Set_delta_d_-14_SNR_-5.tex}
		\caption{$\mathsf{SNR}=-5$ dB}
		\label{fig:best_set_delta_d_-14_SNR_-5}
	\end{subfigure}%
	\begin{subfigure}{.5\textwidth}
		\input{plots/Revised_Best_Card_Set_delta_d_-14_SNR_10.tex}
		\caption{$\mathsf{SNR}=10$ dB}
		\label{fig:best_set_delta_d_-14_SNR_10}
	\end{subfigure}\vskip\floatsep%
	\begin{subfigure}{0.5\textwidth}
		\input{plots/Revised_Best_Card_Set_delta_d_-14_SNR_30.tex}
		\caption{$\mathsf{SNR}=30$ dB}
		\label{fig:best_set_delta_d_-14_SNR_30}
	\end{subfigure}%
	\begin{subfigure}{0.5\textwidth}
		\input{plots/Revised_Best_Card_Set_delta_d_-14_SNR_50.tex}
		\caption{$\mathsf{SNR}=50$ dB}
		\label{fig:best_set_delta_d_-14_SNR_50}
	\end{subfigure}
	\caption{\small Histogram of the optimal cardinality set $\left(|\mathcal{P}_1^{\star}|,|\mathcal{P}_2^{\star}|\right)$ for $\left(|\mathcal{D}_1|,|\mathcal{D}_2|\right)=\left(22,8\right)$ and varying $\mathsf{SNR}\in\left\{-5,10,30,50\right\}$ dB.}
	\label{fig:sim_best_set_delta_d_-14}
\end{figure}
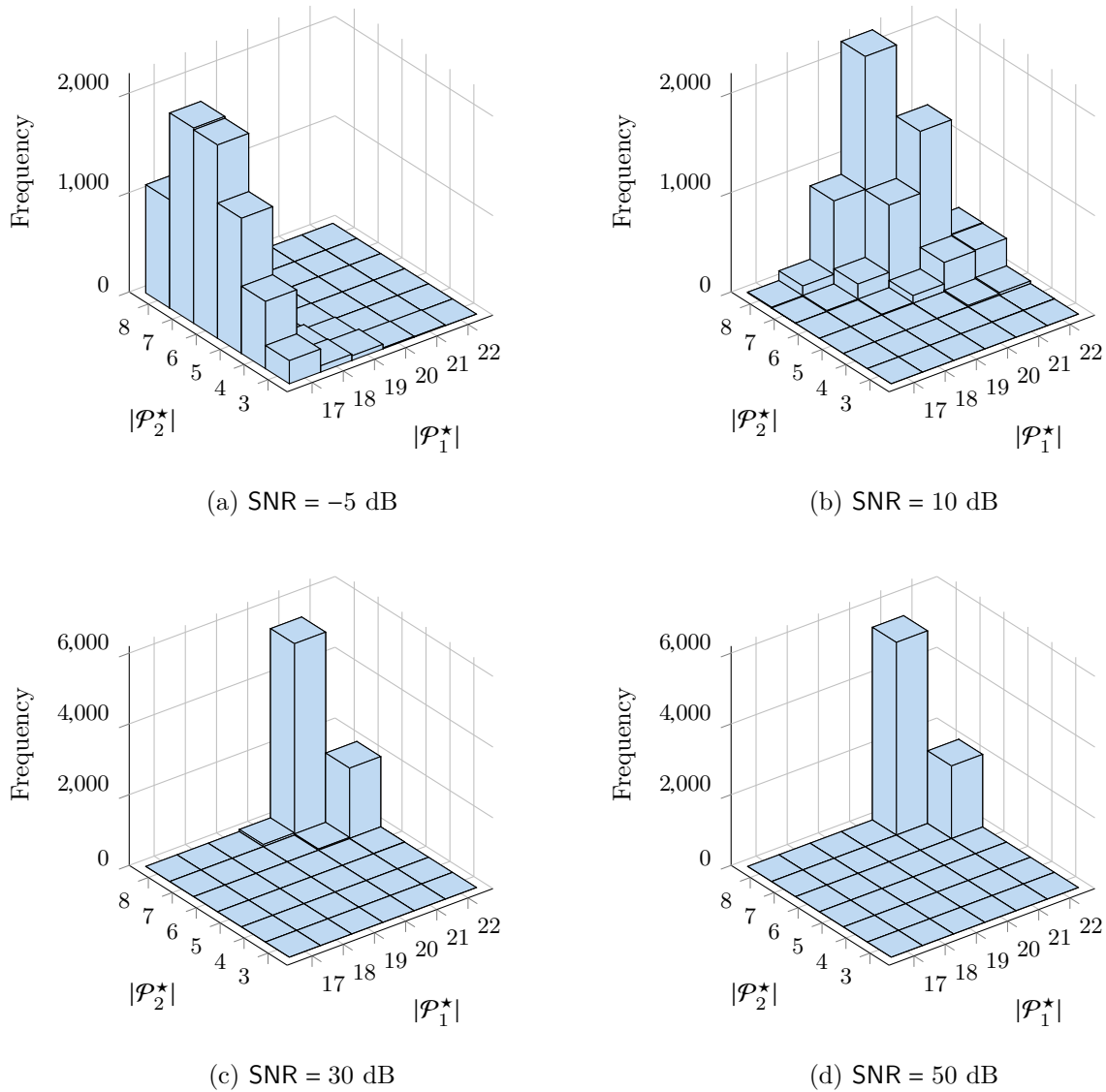

\subsection{Optimal Cardinality Set $\left(|\mathcal{P}_1^{\star}|,|\mathcal{P}_2^{\star}|\right)$}
\label{subsec:best_card_set}

In this section, we discuss the influence of the $\mathsf{SNR}$ and $\left(|\mathcal{D}_1|,|\mathcal{D}_2|\right)$ on the optimal cardinality vector $\left(|\mathcal{P}_1^{\star}|,|\mathcal{P}_2^{\star}|\right)$. To this end, we plot histograms of $\left(|\mathcal{P}_1^{\star}|,|\mathcal{P}_2^{\star}|\right)$ for $\left(|\mathcal{D}_1|,|\mathcal{D}_2|\right)=\left(13,17\right)$ (Fig. \ref{fig:sim_best_set_delta_d_4}) and $\left(|\mathcal{D}_1|,|\mathcal{D}_2|\right)=\left(22,8\right)$ (Fig. \ref{fig:sim_best_set_delta_d_-14}) for $\mathsf{SNR}\in\{-5,10,30,50\}$ dB. Qualitatively, at very high $\mathsf{SNR}$, the distortion attributed to the interfering computation is dominant over the noise. For this case, letting all sensors transmit with full power is often detrimental for the accuracy in computation as it imposes significant interference. Rather, to limit the interference, we let only one single sensor -- namely sensor $\tilde{k}$ with its channel $h_{\tilde{k}}$ matching the overall weakest channel $h_{\min}\triangleq\min_{k\in[1:K]} h_k$-- transmit with full power; in other words, either $\left(|\mathcal{P}_1^{\star}|,|\mathcal{P}_2^{\star}|\right)=\left(|\mathcal{D}_1|,|\mathcal{D}_2|-1\right)$ or $\left(|\mathcal{P}_1^{\star}|,|\mathcal{P}_2^{\star}|\right)=\left(|\mathcal{D}_1|-1,|\mathcal{D}_2|\right)$. The probability that $\tilde{k}\in\mathcal{D}_m$ is $\mathbb{P}\left(\tilde{k}\in\mathcal{D}_m\right)=\nicefrac{|\mathcal{D}_m|}{K}$. At $\mathsf{SNR}=50$ dB, we see in Fig. \ref{fig:best_set_delta_d_4_SNR_50} (similarly for Fig. \ref{fig:best_set_delta_d_-14_SNR_50})) that $\left(|\mathcal{P}_1^{\star}|,|\mathcal{P}_2^{\star}|\right)=\left(13,16\right)$ or $\left(|\mathcal{P}_1^{\star}|,|\mathcal{P}_2^{\star}|\right)=\left(12,17\right)$ are the only optimal cardinality vectors which occur, respectively, with relative frequencies $\nicefrac{4215}{7500}$ ($\approx \mathbb{P}\left(\tilde{k}\in\mathcal{D}_2\right)=\nicefrac{17}{30}$) and $\nicefrac{3285}{7500}$ ($\approx \mathbb{P}\left(\tilde{k}\in\mathcal{D}_1\right)=\nicefrac{13}{30}$). 
As we decrease the $\mathsf{SNR}$, irrespective of $\left(|\mathcal{D}_1|,|\mathcal{D}_2|\right)$, we observe that $\left(|\mathcal{P}_1^{\star}|,|\mathcal{P}_2^{\star}|\right)$ becomes more dispersive. This observation implies that it is often better to let more sensors transmit with full power. Particularly, the lower the $\mathsf{SNR}$, the more dispersion we observe in the histogram. For instance, while for $\mathsf{SNR}=30$ dB, the dispersion is almost non-existent, this effect is more prevalent for the histograms at $\mathsf{SNR}\in\{-5,10\}$ dB. The histogram of $\mathsf{SNR}=10$ dB remains of similar shape as the ones for $\mathsf{SNR}\in\{30,50\}$ dB. This is not the case for $\mathsf{SNR}=-5$ dB where the effect of the noise is more dominant over the interference which allows more sensors to transmit with full power than at $\mathsf{SNR}\in\left\{10,30,50\right\}$ dB. This reflects in a drop of the cardinalities $|\mathcal{P}_1^{\star}|$ and $|\mathcal{P}_2^{\star}|$ (cf. Figs. \ref{fig:best_set_delta_d_4_SNR_-5} and \ref{fig:best_set_delta_d_-14_SNR_-5}).  

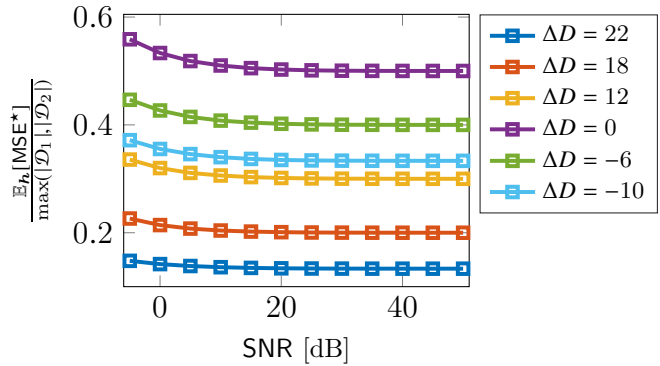
\begin{figure}
	\centering
	\input{plots/MSE_over_SNR_Plot.tex}
	\caption{\small{The normalized, average $\mathsf{MSE}$ -- $\nicefrac{\mathbb{E}_{\bm h}\left[\mathsf{MSE}\right]}{\max\left(|\mathcal{D}_1|,|\mathcal{D}_2|\right)}$ -- versus $\mathsf{SNR}$ for different $\Delta D$ realizations with $K=30$ sensors.}}
	\label{fig:avg_MSE_over_SNR}
\end{figure}

\begin{figure}
	\centering
	\input{plots/MSE_over_SNR_Plot_with_Benchmark_Comp.tex}
	\caption{\small{Performance comparison between the robust $\mathsf{MSE}$ scheme of this paper with the full-power scheme ($b_k=\sqrt{P}$, $\forall k\in[1:K]$).}}
	\label{fig:avg_MSE_over_SNR_benchmark}
\end{figure}
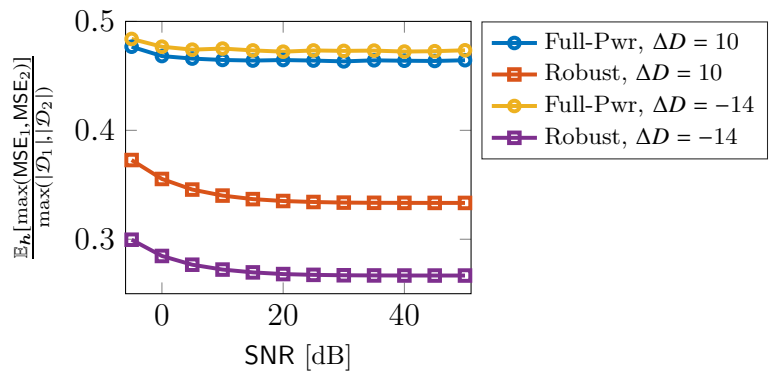

\subsection{Achievable Average MSE}
\label{subsec:ach_avg_MSE}

Now, we elaborate on the behavior of the achievable, normalized $\mathsf{MSE}$ given by  $\nicefrac{\mathbb{E}_{\bm h}\left[\mathsf{MSE}\right]}{\max\left(|\mathcal{D}_1|,|\mathcal{D}_2|\right)}$. 

Fig. \ref{fig:avg_MSE_over_SNR} shows the average, normalized $\mathsf{MSE}$ over $\mathsf{SNR}$. We see that irrespective of $\Delta D$, the average $\mathsf{MSE}$ is monotonously decreasing in $\mathsf{SNR}$. However, for almost all channel realizations, the smaller $|\Delta D|$, the more interference is imposed on the calculation of $s_m$ through the simultaneous computation of $s_n$, $m\neq n$. In Fig. \ref{fig:avg_MSE_over_SNR}, this reflects on a decreasing behavior of the average $\mathsf{MSE}$ (independent of $\mathsf{SNR}$) as we increase $|\Delta D|$ from $0$ to $22$. In our simulations, we observe a symmetric behavior, i.e., for $\Delta D$ and $-\Delta D$, the average $\mathsf{MSE}$s are almost identical. As the $\mathsf{SNR}$ rises, we see in Fig. \ref{fig:avg_MSE_over_SNR} that for all values of $\Delta D$, the normalized, average $\mathsf{MSE}$ converges. Interestingly, for $\mathsf{SNR}\rightarrow\infty$, we infer from our simulation that the convergence limit becomes $\mathbb{E}_{\bm h}\left[\mathsf{MSE}^{\star}\right]\rightarrow\frac{|\mathcal{D}_1||\mathcal{D}_2|}{K}$. The normalized, average $\mathsf{MSE}$ for medium $\mathsf{SNR}$ ($\mathsf{SNR}\approx 10$ dB) is already close to this limit. 

In Fig. \ref{fig:avg_MSE_over_SNR_benchmark}, we compare the robust $\mathsf{MSE}$ scheme with a benchmark scheme, namely the full-power scheme, where $\forall k\in[1:K]$, $b_k=\sqrt{P}$, or in cardinality-sense $\left(|\mathcal{P}_1|,|\mathcal{P}_2|\right)=\left(0,0\right)$. As already discussed, our scheme outperforms the full-power scheme. More detailed, the relative gain at $\mathsf{SNR}=-5$ dB ($\mathsf{SNR}=50$ dB) for $\Delta D=10$ and $\Delta D=-14$ are, respectively, $28$\% and $61$\% ($39$\% and $77$\%). This increase in the relative gain from $\mathsf{SNR}=-5$ dB to $\mathsf{SNR}=50$ dB is since the robust scheme much more resembles the full-power scheme at low $\mathsf{SNR}$ than at high $\mathsf{SNR}$. This resemblance can be quantified by comparing the optimal cardinalities $\left(|\mathcal{P}_1^{\star}|,|\mathcal{P}_2^{\star}|\right)$ at low and high $\mathsf{SNR}$ with $\left(|\mathcal{P}_1|,|\mathcal{P}_2|\right)=\left(0,0\right)$ of the benchmark scheme (cf. subsection \ref{subsec:best_card_set}).     

\section{Concluding Remarks}
\label{sec:conclusion}  

In this work, we consider a multiple access channel (MAC) with $K$ sensors as transmitters and a single receiver. For $M$ mutually exclusive sensor index sets $\mathcal{D}_m\subseteq[1:K]$, $\forall m\in[1:M]$ of arbitrary cardinality, the MAC is used as a medium to compute the sums $s_m=\sum_{k\in\mathcal{D}_m}x_k$, $\forall m\in[1:M]$, of real-valued sensor observations $x_k$ simultaneously. The goal is to minimize the worst-case, mean-sqaured error, i.e., $\max_{m\in[1:M]}\mathsf{MSE}_m$, over all feasible Tx-Rx scaling policies $(\bm a,\bm b)$ subject to a Tx-power constraint. We show that an optimal design of the Tx-Rx scaling policy involves optimizing (a) their phases and (b) their absolute values to orthgonalize and minimize the computation over both real and imaginary part. The primary focus of this paper is on (b). We derive conditions (i) on the feasibility of the optimization problem and (ii) on the Tx-Rx scaling policy of a local minimum for $M_R=2$ computations over the real or the imaginary part. Extensive simulations show that the level of interference in terms of $\Delta D=|\mathcal{D}_2|-|\mathcal{D}_1|$ plays an important role on the ergodic worst-case $\mathsf{MSE}$. Interestingly, we observe that the ergodic worst-case $\mathsf{MSE}$ is not vanishing; rather, it converges to $\frac{|\mathcal{D}_1||\mathcal{D}_2|}{K}$ as $\mathsf{SNR}\rightarrow\infty$.

\appendices

\section{Comparison of $\mathsf{MSE}_m$ and $\mathsf{MSE}_{m,w}^{\perp}\left(\mathcal{C}_w\right)$}
\label{apdx:comp_MSEs}

Recall from section \ref{sec:orthogonal_computation} that
\begin{align*}
\mathsf{MSE}_m&=\sum_{k\in\mathcal{D}_m}\Big\lvert \big\lvert\bar{a}_m\tilde{b}_k\big\rvert\cos\left(\phi_k-\alpha_m\right)-1\Big\rvert^{2}+\sum_{\ell\in\mathcal{D}_m^{C}}\Big\lvert \big\lvert\bar{a}_m\tilde{b}_\ell\big\rvert\cos\left(\phi_\ell-\alpha_m\right)\Big\rvert^{2}+\sigma^{2}|\bar{a}_m|^{2}.
\end{align*} 
Naturally, the phase difference $\phi_k-\alpha_m\in\left(-\nicefrac{\pi}{2},\nicefrac{\pi}{2}\right)$ so that $0<\cos\left(\phi_k-\alpha_m\right)<1$ and $\mathsf{MSE}_m<|\mathcal{D}_m|$. Next, we define $v_{km}\triangleq\Big\lvert \big\lvert\bar{a}_m\tilde{b}_k\big\rvert\cos\left(\phi_k-\alpha_m\right)-1\Big\rvert$ with its range being $0\leq v_{km}<1$. The \emph{smallest} $|\tilde{b}_k|$\footnote{Another -- but larger in magnitude -- solution is $|\tilde{b}_k|=\frac{1+v_{km}}{|\bar{a}_m|\cos\left(\phi_k-\alpha_m\right)}$.}, $k\in\mathcal{D}_m$, that attains $v_{km}$ is
\begin{align*}
|\tilde{b}_k|=\frac{1-v_{km}}{|\bar{a}_m|\cos\left(\phi_k-\alpha_m\right)}.
\end{align*} Due to the power constraint $\big\lvert\tilde{b}_k\big\rvert\leq h_k\sqrt{P}$, or equivalently
\begin{align*}
v_{km}\geq 1-|\bar{a}_m|h_k\cos\left(\phi_k-\alpha_m\right)\sqrt{P},
\end{align*} we can refine the range of $v_{km}$ to be
\begin{align}
\left[1-|\bar{a}_m|h_k\cos\left(\phi_k-\alpha_m\right)\sqrt{P}\right]^{+}\leq v_{km}\leq 1.\label{eq:range_Xkm}
\end{align} Now, we may rewrite $\mathsf{MSE}_m$ in terms of $v_{km}$ as follows.
\begin{align}
\widetilde{\mathsf{MSE}}_m&=\sum_{k\in\mathcal{D}_m}v_{km}^{2}+\sum_{\substack{n=1\\n\neq m}}^{M}\left(\sum_{\ell\in\mathcal{D}_n}\frac{\big\lvert\bar{a}_{m}\big\rvert^{2}\left(1-v_{\ell n}\right)^{2}}{\big\lvert\bar{a}_{n}\big\rvert^{2}}\frac{\cos^{2}\left(\phi_\ell-\alpha_m\right)}{\cos^{2}\left(\phi_\ell-\alpha_n\right)}\right)+\sigma^{2}|\bar{a}_m|^{2}.\label{eq:MSE_m_comparison_orth_one}
\end{align} We infer that the initial optimization problem $(7)$ is equivalent to
\begin{subequations}
	\begin{alignat}{2}
	&\!\min_{\substack{\bm \alpha, \bm \phi,\\\bm a',\bm v}}\:\max_{m\in[1:M]}        &\:\:& \widetilde{\mathsf{MSE}}_m\label{eq:optProb_Original_modifie}\\
	&\text{subject to} &      &\eqref{eq:range_Xkm},\qquad\forall k\in\mathcal{D}_m,\forall m\in[1:M].\label{eq:X_kn}
	\end{alignat}
\end{subequations} for $\bm\alpha=[\alpha_1,\ldots,\alpha_M]^{T}$, $\bm\phi=[\phi_1,\ldots,\phi_K]^{T}$, $\bm a'=[|\bar{a}_1|,\ldots,|\bar{a}_M|]^{T}$ and $\bm v=\left(v_{km}\right)_{k\in\mathcal{D}_{m},m\in[1:M]}$. Choosing $\phi_{\ell}-\alpha_n=u\pi$, $u\in\mathbb{Z}$, $\forall \ell\in\mathcal{D}_n$\footnote{This suggests that for $u=0$ and distinct $\ell_1$, $\ell_2\in\mathcal{D}_n$, $\alpha_n=\phi_{\mathcal{D}_n}=\phi_{\ell_1}=\phi_{\ell_2}$.}  has the following two positive effects. Namely, (i) we find a tight lower bound $\widetilde{\mathsf{MSE}}'_m$, $\forall m\in[1:M]$ given by
\begin{align}
\widetilde{\mathsf{MSE}}'_m&=\sum_{k\in\mathcal{D}_m}v_{km}^{2}+\sum_{\substack{n=1\\n\neq m}}^{M}\left(\frac{\big\lvert\bar{a}_{m}\big\rvert^{2}\cos^{2}\left(\alpha_n-\alpha_m\right)}{\big\lvert\bar{a}_{n}\big\rvert^{2}}\sum_{\ell\in\mathcal{D}_n}\left(1-v_{\ell n}\right)^{2}\right)+\sigma^{2}|\bar{a}_m|^{2},\label{eq:MSE_m_comparison_orth_one_lb}
\end{align} since $	\frac{\cos^{2}\left(\phi_\ell-\alpha_m\right)}{\cos^{2}\left(\phi_\ell-\alpha_n\right)}\geq \cos^{2}\left(\phi_\ell-\alpha_m\right)$
and (ii) the range of $v_{\ell n}$ (cf. \eqref{eq:range_Xkm}) is maximized because $1-|\bar{a}_n|h_\ell\cos\left(\phi_\ell-\alpha_n\right)\sqrt{P}\geq 1-|\bar{a}_n|h_\ell\sqrt{P}$
so that
\begin{align}
\left[1-|\bar{a}_n|h_\ell\sqrt{P}\right]^{+}\leq v_{\ell n}\leq 1.\label{eq:range_Xkm2}
\end{align} Note that this choice can, if anything, improve upon the optimal $\mathsf{MSE}$. We observe that \eqref{eq:MSE_m_comparison_orth_one_lb} depends on $\cos^{2}\left(\Delta \alpha_{nm}\right)$, where $\Delta\alpha_{nm}\triangleq\alpha_n-\alpha_m$ is a phase difference. Due to the symmetric behavior of $\cos^{2}\left(\Delta \alpha_{nm}\right)$, we can assume without loss of generality $\forall m,n\in[1:M]$ that $\Delta\alpha_{nm}\in[-\nicefrac{\pi}{2},\nicefrac{\pi}{2}]$. 
This allows us represent each phase difference by a convex combination
\begin{align}
\Delta\alpha_{nm}=
\begin{cases}
\lambda_{nm}\cdot 0+\left(1-\lambda_{nm}\right)\cdot\frac{\pi}{2}=\frac{\pi\left(1-\lambda_{nm}\right)}{2}\quad&\text{ if }\Delta a_{nm}\geq 0\\
\lambda_{nm}\cdot 0-\left(1-\lambda_{nm}\right)\cdot\frac{\pi}{2}=-\frac{\pi\left(1-\lambda_{nm}\right)}{2}\quad&\text{ if }\Delta a_{nm}\leq 0\\
\end{cases}
\label{eq:convex_combinnation}
\end{align} for $\lambda_{nm}\in[0,1]$. Exploiting \eqref{eq:convex_combinnation} and the \emph{concavity} of $\cos^{2}\left(x\right)$ in $x\in[-\nicefrac{\pi}{2},\nicefrac{\pi}{2}]$, we conclude that
\begin{align}
\cos^{2}\left(\Delta\alpha_{nm}\right)\geq \lambda_{nm}\cos^{2}\left(0\right)+\left(1-\lambda_{nm}\right)\cos^{2}\left(\frac{\pi}{2}\right)=\lambda_{nm}.\label{eq:concavity}
\end{align} To exploit \eqref{eq:concavity}, we need an understanding of the mapping from $\Delta\alpha_{nm}$ to $\lambda_{nm}$, i.e., $\Delta\alpha_{nm}\rightarrow\lambda_{nm}$ for $\Delta\alpha_{nm}\in[-\nicefrac{\pi}{2},\nicefrac{\pi}{2}]$ and $\lambda_{nm}\in[0,1]$. To this end, due to \eqref{eq:convex_combinnation} we observe the following mapping on $\lambda_{nm}$ for $m,n,o\in[1:M]$
\begin{align}
&\Delta\alpha_{mn}+\Delta\alpha_{nm}=0\implies\lambda_{mn}-\lambda_{nm}=0,\label{eq:equivalence1}\\
&\Delta\alpha_{mm}=0\implies\lambda_{mm}=1,\label{eq:equivalence2}\\
&\Delta\alpha_{nm}=\Delta\alpha_{no}+\Delta\alpha_{om}\implies \lambda_{nm}=\chi_{nm}\left(\lambda_{no},\lambda_{om},\sign{\left\{\Delta\alpha_{no}\cdot\Delta\alpha_{om}\right\}}\right)\label{eq:equivalence4},
\end{align} where for $\lambda_{no}$, $\lambda_{om}\in[0,1]$, $z\in\{-1,0,1\}$
\begin{align}
\chi_{nm}\left(\lambda_{no},\lambda_{om},z\right)\triangleq
\begin{cases}
\chi_{nm}^{(z\geq 0)}\left(\lambda_{no},\lambda_{om}\right)\quad&\text{ for }z\in\{0,1\}\\
\chi_{nm}^{(z\leq 0)}\left(\lambda_{no},\lambda_{om}\right)\quad&\text{ for }z\in\{-1,0\}
\end{cases}\label{eq:chi_nm}
\end{align} with
\begin{align*}
\chi_{nm}^{(z\geq 0)}\left(\lambda_{no},\lambda_{om}\right)&\triangleq
\begin{cases}
1-\lambda_{no}-\lambda_{om}\quad&\text{ if }\:0\leq\lambda_{no}+\lambda_{om}\leq 1\\
-1+\lambda_{no}+\lambda_{om}\quad&\text{ if }\:1\leq\lambda_{no}+\lambda_{om}\leq 2
\end{cases},
\\
\chi_{nm}^{(z\leq 0)}\left(\lambda_{no},\lambda_{om}\right)&\triangleq
\begin{cases}
1+\lambda_{no}-\lambda_{om}\quad&\text{ if }\:-1\leq\lambda_{no}-\lambda_{om}\leq 0\\
1-\lambda_{no}+\lambda_{om}\quad&\text{ if }\:0\leq\lambda_{no}-\lambda_{om}\leq 1\\		
\end{cases}.
\end{align*} Wherever unnecessary, we omit the variable $z$ of the function $\chi_{nm}$. Note that this function has the following properties: 
\begin{itemize}
	\item[(a)] range: $0\leq\chi_{nm}\leq 1$,
	\item[(b)] symmetry: $\chi_{nm}\left(\lambda_{no},\lambda_{om},z\right)=\chi_{nm}\left(\lambda_{om},\lambda_{no},z\right)$,
	\item[(c)] optimum:
	\begin{itemize}
		\item $\argmin\chi_{nm}^{(z\geq 0)}\left(\lambda_{no},\lambda_{om}\right)=\left\{\left(\lambda_{no},\lambda_{om}\right)\in[0,1]^{2}:\lambda_{no}+\lambda_{om}=1\right\}$, 
		\item $\argmin\chi_{nm}^{(z\leq 0)}\left(\lambda_{no},\lambda_{om}\right)=\left\{\left(0,1\right),\left(1,0\right)\right\}$, 		
		\item $\argmax\chi_{nm}^{(z\geq 0)}\left(\lambda_{no},\lambda_{om}\right)=\left\{\left(0,0\right),\left(1,1\right)\right\}$, 		
		\item $\argmax\chi_{nm}^{(z\leq 0)}\left(\lambda_{no},\lambda_{om}\right)=\left\{\left(\lambda_{no},\lambda_{om}\right)\in[0,1]^{2}:\lambda_{no}-\lambda_{om}=0\right\}$,
	\end{itemize}   
	\item[(d)] extreme points:
	\begin{itemize}
		\item $\chi_{nm}\left(\lambda_{no},0\right)=1-\lambda_{no}$,
		\item $\chi_{nm}\left(\lambda_{no},1\right)=\lambda_{no}$, 
		\item $\chi_{nm}\left(0,0\right)=\chi_{nm}\left(1,1\right)=1$,
		\item $\chi_{nm}\left(0,1\right)=\chi_{nm}\left(1,0\right)=0$.
	\end{itemize}
\end{itemize} Ultimately, using Eqs. \eqref{eq:concavity}, \eqref{eq:equivalence1},\eqref{eq:equivalence4} and \eqref{eq:chi_nm}, we can lower bound $\widetilde{\mathsf{MSE}}_m'$, $\forall m\in[1:M]$ and some $o\in[1:M]\setminus\{m\}$ by 
\begin{align}
\widetilde{\mathsf{MSE}}''_m&=\sum_{k\in\mathcal{D}_m}v_{km}^{2}+\sum_{\substack{n=1\\n\neq m}}^{M}\left(\frac{\big\lvert\bar{a}_{m}\big\rvert^{2}}{\big\lvert\bar{a}_{n}\big\rvert^{2}}\chi_{nm}\left(\lambda_{on},\lambda_{om}\right)\sum_{\ell\in\mathcal{D}_n}\left(1-v_{\ell n}\right)^{2}\right)+\sigma^{2}|\bar{a}_m|^{2}.
\end{align} Due to the optimum property of $\chi_{nm}$, it is always best to choose $\left(\lambda_{on},\lambda_{om}\right)\in\left\{\left(0,1\right),\left(1,0\right)\right\}$. Recall that $\lambda_{on}=0$ ($\lambda_{on}=1$) represents a phase difference of $\Delta\alpha_{on}=\nicefrac{\pi}{2}$ ($\Delta\alpha_{on}=0$). Ultimately, this leads to the orthgonalization strategy and the exhaustive search in \eqref{eq:optimal_MSE_orthogonal_scheme} of the optimal computation index sets. Thus, in the optimum $\mathsf{MSE}_m$ and $\mathsf{MSE}_{m,w}^{\perp}\left(\mathcal{C}_w\right)$ match. 

\section{Lower Bound on $\mathsf{MSE}_{m,\mathcal{C}_R}^{\perp}$}
\label{apdx:lb_bound}

Note that $\mathsf{MSE}_j\geq|\mathcal{D}_j|-\nicefrac{B_{\mathcal{D}_j}^{2}}{A}$. Thus for $c_m\in\mathbb{R}_{+}$, $|\mathcal{D}_1|-\nicefrac{B_{\mathcal{D}_1}^{2}}{A}$ and $|\mathcal{D}_2|-\nicefrac{B_{\mathcal{D}_2}^{2}}{A}$ function as lower bounds on the worst-case $\mathsf{MSE}$. Above observation implies that ideally, we would like to ensure that the lower bounds on the MSEs are tight, i.e., 
\begin{align}
\mathsf{MSE}_j=\mathsf{MSE}_k=|\mathcal{D}_k|-\nicefrac{B_{\mathcal{D}_k}^{2}}{A}= |\mathcal{D}_j|-\nicefrac{B_{\mathcal{D}_j}^{2}}{A}.\label{eq:optimal}
\end{align}

\section{KKT-Conditions}
\label{apdx:KKT_cond}

The Lagrangian for the optimization problem of subsection \ref{subsec:prelim} with Lagrange multipliers $\bm \lambda=\left(\lambda_1,\ldots,\lambda_K\right)^{T}$ and $\mu$ is
\begin{align*}
\mathcal{L}(\bm b,\bm \lambda,\mu)=\underbrace{-\frac{B_{\mathcal{D}_1}^{2}}{A}}_{\triangleq f_0\left(\bm b\right)}+\sum_{k=1}^{K}\lambda_k\underbrace{\left(b_k-\sqrt{P}\right)}_{\triangleq g_k\left(\bm b\right)}+\mu\underbrace{\left(\frac{B_{\mathcal{D}_1}^{2}}{A}-\frac{B_{\mathcal{D}_2}^{2}}{A}+\Delta D\right)}_{\triangleq h\left(\bm b\right)}.
\end{align*} To derive the KKT-conditions, we first determine the partial derivative
\begin{align}
\frac{\partial \mathcal{L}}{\partial b_k}&=
\begin{cases}
-\frac{2B_{\mathcal{D}_1}h_k}{A^{2}}\left(A-h_k b_k B_{\mathcal{D}_1} \right)(1-\mu)+\frac{2B_{\mathcal{D}_2}^{2}}{A^{2}}h_k^{2}b_k\mu+\lambda_k\quad\text{ if }k\in\mathcal{D}_1\\
\frac{2B_{\mathcal{D}_1}^{2}}{A^{2}}h_k^{2}b_k(1-\mu)-\frac{2B_{\mathcal{D}_2}h_k\mu}{A^{2}}\left(A-h_k b_k B_{\mathcal{D}_2} \right)+\lambda_k\quad\text{ if }k\in\mathcal{D}_2
\end{cases}.
\label{eq:partial_derivate_lagrange_function} 	
\end{align} Then, the KKT-conditions are as follows.
\begin{align}
\nabla_{\bm b}\:\mathcal{L}&=\bm 0\qquad&\text{(stationarity)}\label{eq:stationarity}\\
g_k\left(\bm b\right)&\leq 0,\forall k\in[1:K]\qquad&\text{(inequality feasibility constraint)}\label{eq:prim_feas_ineq_const}\\
h\left(\bm b\right)&=0\qquad&\text{(equality feasibility constraint)}\label{eq:prim_feas_eq_const}\\
\bm \lambda&\geq \bm 0\qquad&\text{(dual feasibility)}\label{eq:dual_feas}\\
\lambda_k g_k\left(\bm b\right)&=0,\forall k\in[1:K]&\text{(complementary slackness)}\label{eq:comp_slack}
\end{align}	

\section{Solution to KKT-Conditions \eqref{eq:stationarity}--\eqref{eq:comp_slack}}
\label{apdx:KKT_cond_sol}

Note that the complementary slackness condition \eqref{eq:comp_slack} suggests that a KKT-point $\bm b'$ has entries $b_k<\sqrt{P}$ and $b_k=\sqrt{P}$. The subsets $\mathcal{P}_m$ and $\mathcal{P}_m^{C}$ of $\mathcal{D}_m$, $\forall m\in[1:2]$, indicate which of the sensors in $\mathcal{D}_m$ transmit with full power and less-than full power. Further, for the latter type of sensors, their respective Lagrange multipliers are $\lambda_k=0$, $\forall k\in\mathcal{P}_m$. If $\mathcal{P}_m\neq\emptyset$, then we define for $j\in\mathcal{P}_m$
\begin{align}
E_{\mathcal{P}_m}=h_j b_j,\label{eq:E_Pm}
\end{align} such that
\begin{align}
B_{\mathcal{D}_m}&=|\mathcal{P}_m|E_{\mathcal{P}_m}+\sqrt{P}h_m',\label{eq:B_Dm_E_Pi}\\
A&=\sigma^{2}+\sum_{m=1}^{2}\left(|\mathcal{P}_m|E_{\mathcal{P}_m}^{2}+Ph_{sm}\right),\label{eq:A_E_Pi}
\end{align} where $h_m'=\sum_{k\in\mathcal{P}_m^{C}}h_k$ and $h_{sm}=\sum_{k\in\mathcal{P}_m^{C}}h_k^{2}$. We emphasize that in the case of $\mathcal{P}_m=\emptyset$ $\left(\mathcal{P}_m^C=\emptyset\right)$, $E_{\mathcal{P}_m}=0$ $\left(h_{m}'=h_{sm}=0,\forall m\in[1:2]\right)$. The equality feasibility constraint \eqref{eq:prim_feas_eq_const} in terms of $\left(E_{\mathcal{P}_1},E_{\mathcal{P}_2}\right)$ becomes the following.
\begin{align}
\frac{B_{\mathcal{D}_2}^{2}}{A}&-\frac{B_{\mathcal{D}_1}^{2}}{A}=\Delta D\stackrel{\eqref{eq:E_Pm}, \eqref{eq:B_Dm_E_Pi}, \eqref{eq:A_E_Pi}}{\iff}\nonumber\\
&\begin{cases}
|\mathcal{P}_2|E_{\mathcal{P}_2}-|\mathcal{P}_1|E_{\mathcal{P}_1}+\sqrt{P}\left(h_2'-h_1'\right)=0&\text{ if }\Delta D=0\\
\left(|\mathcal{P}_2|E_{\mathcal{P}_2}+\sqrt{P}h_2'\right)^{2}-\left(|\mathcal{P}_1|E_{\mathcal{P}_1}+\sqrt{P}h_1'\right)^{2}&\text{ }\\
\qquad-\Delta D\left(\sigma^{2}+\sum_{m=1}^{2}\left(|\mathcal{P}_m|E_{\mathcal{P}_m}^{2}+Ph_{sm}\right)\right)=0&\text{ if }\Delta D\neq 0
\end{cases}.\label{eq:primal_feas_eq_E_Pi}
\end{align} The condition \eqref{eq:stationarity} gives us the following expressions on $\mu$. 
\begin{align}
\mu=
\begin{cases}
\frac{B_{\mathcal{D}_1}\left(A-h_k b_k B_{\mathcal{D}_1}\right)}{A\left(h_k b_k \Delta D+B_{\mathcal{D}_1}\right)}-\frac{\lambda_k A}{2 h_k\left(h_k b_k\Delta D+B_{\mathcal{D}_1}\right)}&\text{ if }k\in\mathcal{D}_1\\
\frac{h_k b_k\left(B_{\mathcal{D}_2}^{2}-A\Delta D\right)}{A\left(B_{\mathcal{D}_2}-h_k b_k \Delta D\right)}+\frac{\lambda_k A}{2 h_k\left(B_{\mathcal{D}_2}-h_k b_k \Delta D\right)}&\text{ if }k\in\mathcal{D}_2			
\end{cases}
\label{eq:mu_expression}
\end{align} We need to ensure that the two expressions on $\mu$ for $k\in\mathcal{D}_1$ and $k\in\mathcal{D}_2$ in \eqref{eq:mu_expression} are of the same value. With Eqs. \eqref{eq:E_Pm}, \eqref{eq:B_Dm_E_Pi}, \eqref{eq:A_E_Pi}, the definitions on $\mathcal{P}_m$ and $\mathcal{P}_m^{C}$, this is the case for $\left(E_{\mathcal{P}_1},E_{\mathcal{P}_2}\right)$ and $\lambda_k$, $\forall k\in\mathcal{P}_m^{C}$, $m\in[1:2]$, if the following set of equations are satisfied. 
\begin{align}
&\frac{2B_{\mathcal{D}_m}B_{\mathcal{D}_n}^{2}}{A^{2}\left(B_{\mathcal{D}_m}+(-1)^{m+1}E_{\mathcal{P}_m}\Delta D\right)}\left(h_k \sqrt{P}-E_{\mathcal{P}_m}\right)=-\frac{\lambda_k}{h_k}\text{ for }j\in\mathcal{P}_m,k\in\mathcal{P}_m^{C},m\neq n,\label{eq:mu_identical_1}\\
&\frac{2B_{\mathcal{D}_m}B_{\mathcal{D}_n}^{2}}{A^{2}}\sqrt{P}\left(h_k-h_j\right)=B_{\mathcal{D}_m}\left(\frac{\lambda_j}{h_j}-\frac{\lambda_k}{h_k}\right)\nonumber\\
&\qquad+(-1)^{m+1}\Delta D\sqrt{P}\left(\frac{\lambda_j}{h_j}h_k -\frac{\lambda_k}{h_k}h_j\right)\text{ for }j,k\in\mathcal{P}_m^{C},m\neq n\label{eq:mu_identical_2}\\
&E_{\mathcal{P}_1}B_{\mathcal{D}_1}+E_{\mathcal{P}_2}B_{\mathcal{D}_2}=A\text{ for }j\in\mathcal{P}_1,k\in\mathcal{P}_2,\label{eq:mu_identical_3}\\
	&\frac{2B_{\mathcal{D}_1}B_{\mathcal{D}_2}}{A^{2}}\left(\frac{A-h_j\sqrt{P}B_{\mathcal{D}_m}-E_{\mathcal{P}_n}B_{\mathcal{D}_n}}{B_{\mathcal{D}_n}+(-1)^{i}E_{\mathcal{P}_n}\Delta D}\right)=\frac{\lambda_j}{h_j}\text{ for }j\in\mathcal{P}_m^{C},k\in\mathcal{P}_{n},m\neq n,\label{eq:mu_identical_4}\\
&\frac{2B_{\mathcal{D}_1}B_{\mathcal{D}_2}}{A^{2}}\left(A-\sqrt{P}\left(h_j  B_{\mathcal{D}_1}+h_k B_{\mathcal{D}_2}\right)\right)=\frac{\lambda_j}{h_j}B_{\mathcal{D}_2}+\frac{\lambda_k}{h_k}B_{\mathcal{D}_1}\nonumber\\
&\qquad-\Delta D\sqrt{P}\left(\frac{\lambda_j}{h_j}h_k-\frac{\lambda_k}{h_k}h_j\right)\text{ for }j\in\mathcal{P}_1^{C},k\in\mathcal{P}_2^{C}.\label{eq:mu_identical_5}
\end{align} Now, we iterate through all possible cardinality vectors $\left(|\mathcal{P}_1|,|\mathcal{P}_2|\right)$, where every cardinality $|\mathcal{P}_m|,|\mathcal{P}_m^C|\geq 0$ satisfies
\begin{align*}
|\mathcal{P}_m|+|\mathcal{P}_m^C|&=|\mathcal{D}_m|.
\end{align*} For each of those choices, we design $\mathcal{P}_m$ and determine $\left(E_{\mathcal{P}_1},E_{\mathcal{P}_2}\right)$ as well as the Lagrange multipliers $\lambda_k$, $k\in\mathcal{P}_m^C$, $m\in[1:2]$, from Eqs. \eqref{eq:primal_feas_eq_E_Pi}, \eqref{eq:mu_identical_1}, \eqref{eq:mu_identical_2}, \eqref{eq:mu_identical_3}, \eqref{eq:mu_identical_4} and \eqref{eq:mu_identical_5}. We start with the two extreme cases, where $a)$ no sensor transmits with full power, i.e.,  $\left(|\mathcal{P}_1|,|\mathcal{P}_2|\right)=\left(|\mathcal{D}_1|,|\mathcal{D}_2|\right)$ and $b)$ all sensors transmit with full power, i.e.,  $\left(|\mathcal{P}_1|,|\mathcal{P}_2|\right)=\left(0,0\right)$. Then, we study intermediate cases of $a)$ and $b)$. Namely, we consider $c)$ cardinalities $\left(|\mathcal{P}_1|,|\mathcal{P}_2|\right)\geq\left(1,1\right)$ (excluding the extreme case $a)$) and finally $d)$ $|\mathcal{P}_m|\geq 1$ and $|\mathcal{P}_n|=0$ for $m\neq n$. 

\paragraph{a) $\left(|\mathcal{P}_1|,|\mathcal{P}_2|\right)=\left(|\mathcal{D}_1|,|\mathcal{D}_2|\right)$:} Since $\sigma^{2}>0$, \eqref{eq:mu_identical_3} is never satisfied for this case. Hence, irrespective of $\Delta D$, this case does not produce a KKT-point.       

\paragraph{b) $\left(|\mathcal{P}_1|,|\mathcal{P}_2|\right)=\left(0,0\right)$:} We can check for $\left(E_{\mathcal{P}_1},E_{\mathcal{P}_2}\right)=\left(0,0\right)$ in \eqref{eq:primal_feas_eq_E_Pi} that only degenerate channel conditions satisfy the equality constraint. For channel coefficients drawn from a continuous distributions, these degenerate conditions occur with zero probability. 

\paragraph{c) $\left(|\mathcal{P}_1|,|\mathcal{P}_2|\right)\geq\left(1,1\right)$:} For this case, we determine $\left(E_{\mathcal{P}_1},E_{\mathcal{P}_2}\right)$ by solving the system of equations \eqref{eq:primal_feas_eq_E_Pi} and \eqref{eq:mu_identical_3} which is either linear if $\Delta D=0$ or quadratic otherwise ($\Delta D\neq 0$). For the special case $\Delta D=0$, we find a rather short, real-valued expression on $E_{\mathcal{P}_m}$ for $m\neq n$ and $\Delta h=h_2'-h_1'$.  
\begin{align}
E_{\mathcal{P}_m,c)}^{(\Delta D=0)}&=\frac{|\mathcal{P}_n|\left(\sigma^{2}+P\left(h_{s1}+h_{s2}\right)\right)+(-1)^{n}P\left(\Delta h\right) h_n'}{\sqrt{P}\left(|\mathcal{P}_1|h_2'+|\mathcal{P}_2|h_1'\right)}\label{eq:E_Pi_sol_DeltaD_0_case_c}
\end{align} However, for $\Delta D\neq 0$, the solution is lengthy and omitted here for ease of presentation. We remind the reader that only real and non-negative solutions are allowed. Assuming the existence of a real solution vector $\left(E_{\mathcal{P}_1},E_{\mathcal{P}_2}\right)$, which only depends on $\bm h$, $\sigma^{2}$, $P$, $\Delta D$ and $|\mathcal{P}_m|$, we can infer from \eqref{eq:E_Pm}, \eqref{eq:mu_identical_1}, \eqref{eq:mu_identical_2}, \eqref{eq:mu_identical_4}, \eqref{eq:mu_identical_5} that   
\begin{align}
b_j&=\frac{E_{\mathcal{P}_m}}{h_j},\label{eq:bj_sol_case_c}\\
\lambda_k&=\frac{2B_{\mathcal{D}_m}B_{\mathcal{D}_n}^{2}}{A^{2}\left(B_{\mathcal{D}_m}+(-1)^{m+1}E_{\mathcal{P}_m}\Delta D\right)}h_k\left(E_{\mathcal{P}_m}-h_k \sqrt{P}\right),\label{eq:lambdak_sol_case_c}
\end{align} for $j\in\mathcal{P}_m$ and $k\in\mathcal{P}_m^{C}$, $\forall m\in[1:2]$.

\paragraph{d) $|\mathcal{P}_m|\geq 1,|\mathcal{P}_n|=0$:} For this case, we solve for $E_{\mathcal{P}_m}$ in \eqref{eq:primal_feas_eq_E_Pi} when $E_{\mathcal{P}_n}=0$. Again, in the interest of simplicity, we only state the explicit expression for $\Delta D=0$ (and omit the one for $\Delta D\neq 0$).
\begin{align}
E_{\mathcal{P}_m,d)}^{(\Delta D=0)}&=\frac{(-1)^{n}\sqrt{P}\left(\Delta h\right)}{|\mathcal{P}_m|}\label{eq:E_Pi_sol_DeltaD_0_case_d}
\end{align} A given $E_{\mathcal{P}_m}$ allows us to compute the primal and dual variables for $j\in\mathcal{P}_m$, $k\in\mathcal{P}_m^{C}$ and $\ell\in\mathcal{P}_n^C$ according to
\begin{align}
b_j&=\frac{E_{\mathcal{P}_m}}{h_j},\nonumber\\
\lambda_k&=\frac{2B_{\mathcal{D}_m}B_{\mathcal{D}_n}^{2}}{A^{2}\left(B_{\mathcal{D}_m}+(-1)^{m+1}E_{\mathcal{P}_m}\Delta D\right)}h_k\left(E_{\mathcal{P}_m}-h_k \sqrt{P}\right),\nonumber\\
\lambda_\ell&=\frac{2B_{\mathcal{D}_1}B_{\mathcal{D}_2}}{A^{2}}\left(\frac{A-E_{\mathcal{P}_m}B_{\mathcal{D}_m}-h_{\ell}\sqrt{P}B_{\mathcal{D}_n}}{B_{\mathcal{D}_m}+(-1)^{m+1}E_{\mathcal{P}_m}\Delta D}\right).\label{eq:lambdal_sol_case_d}
\end{align} 

So far, we have computed $E_{\mathcal{P}_m}$ and $\lambda_k$ for cases $c)$ and $d)$. Wherever necessary, to distinguish their values, we use the subscripts $c)$ and $d)$ for $E_{\mathcal{P}_m}$. For $j\in\mathcal{P}_m$, we know from \eqref{eq:prim_feas_ineq_const} that $b_j\in[0,\sqrt{P})$. This suggests that for $j\in\mathcal{P}_m$, and cases $c)$ and $d)$ that
\begin{align}
0\leq E_{\mathcal{P}_m}<h_j\sqrt{P}.\label{eq:primal_feas_case_c_and_d}
\end{align} Simultaneously, the dual feasibility constraint $\bm\lambda\geq\bm 0$ confines the range of $E_{\mathcal{P}_m}$ further. The Lagrange multipliers of cases $c)$ and $d)$ (cf. \eqref{eq:lambdak_sol_case_c} and \eqref{eq:lambdal_sol_case_d}) are fractions $\nicefrac{\mathsf{num}}{\mathsf{den}}$ with the same common denominator $\mathsf{den}$ but potentially a different numerator $\mathsf{num}$. Clearly, $\lambda_k>0$ is equivalent to either $\left\{\mathsf{num}>0,\mathsf{den}>0\right\}$ or $\left\{\mathsf{num}< 0,\mathsf{den}<0\right\}$. Without going into details, this gives us ultimately the following conditions on $E_{\mathcal{P}_m}$ for $k\in\mathcal{P}_m^{C}$ and
\begin{itemize}
	\item case $c)$, $\forall m,n\in[1:2]$, $m\neq n$:
	\begin{align}
	\begin{cases}
	h_k\sqrt{P}< E_{\mathcal{P}_{m,c)}}&\text{ if }|\mathcal{D}_n|\geq|\mathcal{P}_m^C|\\
	h_k\sqrt{P}< E_{\mathcal{P}_{m,c)}}<\frac{\sqrt{P}h_m'}{|\mathcal{P}_m^C|-|\mathcal{D}_n|}&\text{ if }|\mathcal{D}_n|<|\mathcal{P}_m^C|			
	\end{cases},\label{eq:cond_satisfy_dual_feas_case_c}
	\end{align}
	\item case $d)$, $\ell\in\mathcal{P}_n^C$, $m\neq n$:
	\begin{align}
	\begin{cases}
	h_k\sqrt{P}< E_{\mathcal{P}_{m,d)}}<\frac{\sigma^{2}+P\left(h_{s1}+h_{s2}-h_{\ell}h_n'\right)}{\sqrt{P}h_m'}&\text{ if }|\mathcal{D}_n|\geq|\mathcal{P}_m^C|\\
	h_k\sqrt{P}< E_{\mathcal{P}_{m,d)}}<\min\left(\frac{\sigma^{2}+P\left(h_{s1}+h_{s2}-h_{\ell}h_n'\right)}{\sqrt{P}h_m'},\frac{\sqrt{P}h_m'}{|\mathcal{P}_m^C|-|\mathcal{D}_n|}\right)&\text{ if }|\mathcal{D}_n|<|\mathcal{P}_m^C|			
	\end{cases}.\label{eq:cond_satisfy_dual_feas_case_d}
	\end{align} 
\end{itemize} We would like to highlight to the reader that typically \emph{strict complementarity}, i.e., $\lambda_k>0$, $\forall k\in\mathcal{P}_m^C$, $m\in[1:2]$, holds, since the sets in \eqref{eq:cond_satisfy_dual_feas_case_c} and \eqref{eq:cond_satisfy_dual_feas_case_d} constraining $E_{\mathcal{P}_m}$ are usually open. The case where $E_{\mathcal{P}_m}$ matches with the right-hand or left-hand side of the inequalities (which causes $\lambda_k=0$) happes only for degenerate channel conditions which have zero probability. This observation is needed when we consider the second-order sufficient condition. 

Combining \eqref{eq:primal_feas_case_c_and_d}, \eqref{eq:cond_satisfy_dual_feas_case_c} and \eqref{eq:cond_satisfy_dual_feas_case_d} gives us the final conditions on $E_{\mathcal{P}_m}$ such that the vector $\bm b'=\left(b_1',\ldots, b_K'\right)^T$ with its $k$-th element corresponding to
\begin{align}
b_k'=
\begin{cases}
\sqrt{P}\qquad&\text{ if }k\in\bigcup_{m=1}^{2}\mathcal{P}_m^C\\
\frac{E_{\mathcal{P}_m}}{h_k}\qquad&\text{ if }k\in\bigcup_{m=1}^{2}\mathcal{P}_m\\
\end{cases},
\end{align} becomes a feasible KKT-point. Further, the combination of these conditions suggests that the largest $|\mathcal{P}_m|\leq |\mathcal{D}_m|$ channel coefficients of $\bm h_{\mathcal{D}_m}\triangleq\left(h_k\right)_{k\in\mathcal{D}_m}$ generate the vector $\bm h_{\mathcal{P}_m}\triangleq\left(h_k\right)_{k\in\mathcal{P}_m}$, i.e., 
\begin{align*}
\bm h_{\mathcal{P}_m}=\left(\bm h_{\mathcal{D}_{m_{[k]}}}\right)_{k=1}^{|\mathcal{P}_m|},
\end{align*} where $\bm h_{\mathcal{D}_{m_{[k]}}}$ is the $k$-th largest component of $\bm h_{\mathcal{D}_m}$ and $\mathcal{P}_m$ the respective index set of $\bm h_{\mathcal{P}_m}$ from $[1:K]$. The remaining $|\mathcal{P}_m^C|$ channel coefficients of $\bm h_{\mathcal{D}_m}$ form $\bm h_{\mathcal{P}_m^C}$. Defining $\overline{v}_m=\bm h_{\mathcal{D}_{m_{[|\mathcal{P}_m|]}}}$, $\underline{v}_m=\bm h_{\mathcal{D}_{m_{[|\mathcal{P}_m|+1]}}}$ and $\overline{w}_n=\bm h_{\mathcal{D}_{n_{[1]}}}$ allows us to compactly write the condition on $E_{\mathcal{P}_m}$ such that primal and dual feasibility hold. This gives us for
\begin{itemize}
	\item case $c)$, $\forall m,n\in[1:2]$, $m\neq n$:
	\begin{align}
	\begin{cases}
	\underline{v}_m\sqrt{P}< E_{\mathcal{P}_{m,c)}}<\overline{v}_m\sqrt{P}&\text{ if }|\mathcal{D}_n|\geq|\mathcal{P}_m^C|\\
	\underline{v}_m\sqrt{P}< E_{\mathcal{P}_{m,c)}}<\min\left(\overline{v}_m\sqrt{P},\frac{\sqrt{P}h_m'}{|\mathcal{P}_m^C|-|\mathcal{D}_n|}\right)&\text{ if }|\mathcal{D}_n|<|\mathcal{P}_m^C|			
	\end{cases},\label{eq:cond_satisfy_dual_prim_feas_case_c}
	\end{align}
	\item case $d)$, $m\neq n$:
	\begin{align}
	\begin{cases}
		\underline{v}_m\sqrt{P}< E_{\mathcal{P}_{m,d)}}<\min\left(\overline{v}_m\sqrt{P},\frac{\sigma^{2}+P\left(h_{s1}+h_{s2}-\overline{w}_n h_n'\right)}{\sqrt{P}h_m'}\right)&\text{ if }|\mathcal{D}_n|\geq|\mathcal{P}_m^C|\\
	\underline{v}_m\sqrt{P}< E_{\mathcal{P}_{m,d)}}<\min\left(\overline{v}_m\sqrt{P},\frac{\sigma^{2}+P\left(h_{s1}+h_{s2}-\overline{w}_n h_n'\right)}{\sqrt{P}h_m'},\frac{\sqrt{P}h_m'}{|\mathcal{P}_m^C|-|\mathcal{D}_n|}\right)&\text{ if }|\mathcal{D}_n|<|\mathcal{P}_m^C|			
	\end{cases}.\label{eq:cond_satisfy_dual_prim_feas_case_d}
	\end{align}  
\end{itemize}

\section{Verification of the Linear Independence Constraint Qualification (LICQ)}
\label{apdx:LICQ}

Recall that the LICQ holds at $\bm b'$, iff $\nabla g_k\left(\bm b'\right)$, $\forall k\in\bigcup_{m=1}^{2}\mathcal{P}_{m}^{C}$ and $\nabla h\left(\bm b'\right)$ are linearly independent. The gradient of the $k$-th inequality constraint is simply the standard unit vector along the $k$-th coordinate axis, i.e., $\nabla g_k\left(\bm b'\right)=\bm e_j$. Thus, naturally, the vectors of all active inequality constraints are all linearly independent. The partial derivative of a feasible $\bm b'$ satisfying \eqref{eq:prim_feas_eq_const} is 
\begin{align*}
\frac{\partial h}{\partial b_k}&=
\begin{cases}
\frac{2h_k}{A}\left(B_{\mathcal{D}_1}+h_k b_k \Delta D\right)&\quad\text{ if }k\in\mathcal{D}_1\\
-\frac{2h_k}{A}\left(B_{\mathcal{D}_2}-h_k b_k \Delta D\right)&\quad\text{ if }k\in\mathcal{D}_2	
\end{cases}		
\end{align*} such that
\begin{align*}
\nabla h\left(\bm b'\right)&=\frac{2}{A\left(\bm b'\right)}\left(\sum_{k\in\mathcal{D}_1}h_k\left(B_{\mathcal{D}_1}\left(\bm b'\right)+h_k b_k'\Delta D\right)\bm e_k\right)-\frac{2}{A\left(\bm b'\right)}\left(\sum_{k\in\mathcal{D}_2}h_k\left(B_{\mathcal{D}_2}\left(\bm b'\right)-h_k b_k'\Delta D\right)\bm e_k\right).
\end{align*} Note that in general $\frac{2h_k}{A\left(\bm b'\right)}\neq 0$, $\forall k\in[1:K]$, and
\begin{align*}
&B_{\mathcal{D}_m}+(-1)^{m+1}h_k b_k'\Delta D\stackrel{\eqref{eq:B_Dm_E_Pi}}{=}|\mathcal{P}_m|E_{\mathcal{P}_m}+\sqrt{P}h_m'+(-1)^{m+1}h_k b_k'\Delta D\\
&\quad=
\begin{cases}
E_{\mathcal{P}_m}\left(|\mathcal{D}_{n}|-|\mathcal{P}_m^{C}|\right)+\sqrt{P}h_m'\quad&\text{ if }k\in\mathcal{P}_m\\
|\mathcal{P}_m|E_{\mathcal{P}_m}+\sqrt{P}\left(h_m'+(-1)^{m+1}h_k\Delta D\right)\quad&\text{ if }k\in\mathcal{P}_m^{C}
\end{cases}\quad\neq 0
\end{align*} for $m\neq n$ and $m\in[1:2]$ in case of non-degenerate channel realizations. However, since for 
\begin{itemize}
	\item case $c)$: $|\bigcup_{m=1}^{2}\mathcal{P}_m^{C}|<K$,
	\item case $d)$: $|\mathcal{P}_{m}^{C}|<|\mathcal{D}_m|<K$, 
\end{itemize} it follows that the span of gradient vectors of active constraints is independent of $\nabla h\left(\bm b'\right)$. This establishes the LICQ for KKT-points of cases $c)$ and $d)$.\footnote{Naturally, the LICQ does not hold for Case $b)$ (all sensors transmit with full power), since the collection of gradient vectors of active constraints spans $\mathbb{R}^{K}$.}

\ifCLASSOPTIONcaptionsoff
  \newpage
\fi

\bibliographystyle{IEEEtran}
\bibliography{Citations}

\end{document}

%% file: plots/Feas_Anal_SNR_Impact_Plot.tex
\definecolor{mycolor1}{rgb}{0.20000,0.60000,0.50000}%
	\begin{tikzpicture}
	\begin{axis}[scale=0.875,
	ybar,
	bar shift auto,
	height=6cm,
	width=9cm,
	tick align=inside,
	tick style={draw=none},
	ymajorgrids,
	major grid style={draw=white},
	xtick={1,2,3},
	xticklabels={$-5$, $0$,$5$},
	xticklabel style = {font=\footnotesize},
	xlabel style = {font=\small},
	yticklabel style = {font=\footnotesize},
	ylabel style = {font=\small},	
	ymin=-150,
	ymax=8500,
	ytick={0,1000,2000,3000,4000,5000,6000,7000},
	ylabel={Frequency},
	xlabel={$\mathsf{SNR}$ [dB]},
	legend style={
		at={(.09,-0.15)},
		anchor=north west,
		legend columns=-1,
		/tikz/every even column/.append style={column sep=1.0cm}
	},
	] 
	                      
	\addplot[fill=red,draw=black] coordinates{
		(1,4763)
		(2,353)
		(3, 3)}; 
	\addplot[fill=mycolor1, draw=black,] coordinates {
		(1, 2737)
		(2, 7147)
		(3, 7497)}; 
	
	\node[above, align=center]
	at (axis cs:0.93,4763) {\footnotesize$4763$};
	\node[above, align=center]
	at (axis cs:1.91,353) {\footnotesize$353$};
	\node[above, align=center]
	at (axis cs:2.93,3) {\footnotesize$3$};
	\node[above, align=center]
	at (axis cs:1.12,2737) {\footnotesize$2737$};
	\node[above, align=center]
	at (axis cs:2.07,7147) {\footnotesize$7147$};
	\node[above, align=center]
	at (axis cs:3.06,7497) {\footnotesize$7497$};
	\end{axis} 
	\end{tikzpicture}

%% file: plots/Feas_Anal_Set_Card_Impact_Plot.tex
\definecolor{mycolor1}{rgb}{0.20000,0.60000,0.50000}%
	\begin{tikzpicture}
	\begin{axis}[scale=0.875,
	ybar,
	bar shift auto,
	height=6cm,
	width=9.0cm,
	tick align=inside,
	tick style={draw=none},
	ymajorgrids,
	major grid style={draw=white},
	xtick={1,2,3},
	xticklabels={{$(27,3)$},{$(26,4)$},{$(25,5)$}},
	xticklabel style = {font=\footnotesize},
	xlabel style = {font=\small},
	yticklabel style = {font=\footnotesize},
	ylabel style = {font=\small},	
	ymin=-150,
	ymax=8500,
	ytick={0,1000,2000,3000,4000,5000,6000,7000},
	ylabel={Frequency},
	xlabel={$\left(|\mathcal{D}_1|,|\mathcal{D}_2|\right)$},
	legend style={
		at={(.09,-0.15)},
		anchor=north west,
		legend columns=-1,
		/tikz/every even column/.append style={column sep=1.0cm}
	},
	]                         
	\addplot[fill=red,draw=black] coordinates{
		(1,4763)
		(2,1284)
		(3, 136)}; 
	\addplot[fill=mycolor1, draw=black,] coordinates {
		(1, 2737)
		(2, 6216)
		(3, 7364)}; 
	
	\node[above, align=center]
	at (axis cs:0.94,4763) {\footnotesize$4763$};
	\node[above, align=center]
	at (axis cs:1.87,1284) {\footnotesize$1284$};
	\node[above, align=center]
	at (axis cs:2.90,136) {\footnotesize$136$};
	\node[above, align=center]
	at (axis cs:1.12,2737) {\footnotesize$2737$};
	\node[above, align=center]
	at (axis cs:2.08,6216) {\footnotesize$6216$};
	\node[above, align=center]
	at (axis cs:3.07,7364) {\footnotesize$7364$};
	\end{axis} 
	\end{tikzpicture}

%% file: plots/Feas_Anal_Channel_Stat_Impact_Plot.tex
\definecolor{mycolor1}{rgb}{0.20000,0.60000,0.50000}%
	\begin{tikzpicture}
	\begin{axis}[scale=0.875,
	ybar,
	bar shift auto,
	height=6cm,
	width=9cm,
	tick align=inside,
	tick style={draw=none},
	ymajorgrids,
	major grid style={draw=white},
	xtick={1,2,3,4},
	xticklabels={{$\left(\frac{1}{\sqrt{3}},\frac{1}{3}\right)$}, {$\left(1,1\right)$},{$\left(\sqrt{3},3\right)$},{$\left(3,9\right)$}},
	xlabel={$\left(\sqrt{\frac{4-\pi}{\pi}}\mu_h,\sigma^{2}_{h}\right)$},
	xticklabel style = {font=\footnotesize},
	xlabel style = {font=\small},
	yticklabel style = {font=\footnotesize},
	ylabel style = {font=\small},	
	ymin=-150,
	ymax=8500,
	ytick={0,1000,2000,3000,4000,5000,6000,7000},
	ylabel={Frequency},
	legend style={legend cell align=left, align=left, draw=white!15!black}, legend pos=outer north east,
	]                     
	\addplot[fill=red,draw=black] coordinates{
		(1, 7494)
		(2,4763)
		(3,422)
		(4, 9)}; \addlegendentry[text width=45pt, text depth=]{\footnotesize{infeasible}}
	\addplot[fill=mycolor1, draw=black,] coordinates {
		(1, 6)
		(2, 2737)
		(3, 7078)
		(4, 7491)}; \addlegendentry[text width=65pt, text depth=]{\footnotesize{feasible}}
	\node[above, align=center]
	at (axis cs:0.92,7494) {\footnotesize$7494$};		
	\node[above, align=center]
	at (axis cs:1.87,4763) {\footnotesize$4763$};
	\node[above, align=center]
	at (axis cs:2.87,422) {\footnotesize$422$};
	\node[above, align=center]
	at (axis cs:3.90,9) {\footnotesize$9$};
	\node[above, align=center]
	at (axis cs:1.12,6) {\footnotesize$6$};	
	\node[above, align=center]
	at (axis cs:2.18,2737) {\footnotesize$2737$};
	\node[above, align=center]
	at (axis cs:3.11,7078) {\footnotesize$7078$};
	\node[above, align=center]
	at (axis cs:4.07,7491) {\footnotesize$7491$};
	\end{axis} 
	\end{tikzpicture}

%% file: plots/Revised_Best_Card_Set_delta_d_4_SNR_-5.tex
%
%
\begin{tikzpicture}

\begin{axis}[%
scale=0.6, width=3.284in,
height=3.566in,
at={(0.758in,0.481in)},
scale only axis,
unbounded coords=jump,
colormap={patchmap}{[1pt] rgb(0pt)=(0.75,0.85,0.95); rgb(899pt)=(0.75,0.85,0.95)},
xmin=7.2,
xmax=13.8,
xtick={ 8,  9, 10, 11, 12, 13},
xticklabel style = {font=\footnotesize},
tick align=outside,
xlabel style={font=\color{white!15!black}},
xlabel style = {font=\small},
xlabel={$|\mathcal{P}_{1}^{\star}|$},
ymin=11.2,
ymax=17.8,
yticklabel style = {font=\footnotesize},
ytick={12, 13, 14, 15, 16, 17},
ylabel style={font=\color{white!15!black}},
ylabel style = {font=\small},
ylabel={$|\mathcal{P}_{2}^{\star}|$},
zmin=0,
zmax=1600,
zlabel style={font=\color{white!15!black}},
zlabel style = {font=\small},
zticklabel style = {font=\footnotesize},
zlabel={Frequency},
view={-37.5}{30},
axis background/.style={fill=white},
title style={font=\bfseries},
axis x line*=bottom,
axis y line*=left,
axis z line*=left,
xmajorgrids,
ymajorgrids,
zmajorgrids,
legend style={at={(1.03,1)}, anchor=north west, legend cell align=left, align=left, draw=white!15!black}
]

\addplot3[%
surf,
shader=flat corner, draw=black, mesh/rows=30]
table[row sep=crcr, colormap name=surfmap, point meta=\thisrow{c}] {%
	x	y	z	c\\
	7.501	11.501	0	0\\
	7.501	11.501	0	1\\
	8.499	11.501	0	2\\
	8.499	11.501	0	3\\
	nan	nan	0	4\\
	8.501	11.501	0	5\\
	8.501	11.501	0	6\\
	9.499	11.501	0	7\\
	9.499	11.501	0	8\\
	nan	nan	0	9\\
	9.501	11.501	0	10\\
	9.501	11.501	0	11\\
	10.499	11.501	0	12\\
	10.499	11.501	0	13\\
	nan	nan	0	14\\
	10.501	11.501	0	15\\
	10.501	11.501	0	16\\
	11.499	11.501	0	17\\
	11.499	11.501	0	18\\
	nan	nan	0	19\\
	11.501	11.501	0	20\\
	11.501	11.501	0	21\\
	12.499	11.501	0	22\\
	12.499	11.501	0	23\\
	nan	nan	0	24\\
	12.501	11.501	0	25\\
	12.501	11.501	0	26\\
	13.499	11.501	0	27\\
	13.499	11.501	0	28\\
	nan	nan	0	29\\
	7.501	11.501	0	30\\
	7.501	11.501	1561	31\\
	8.499	11.501	1561	32\\
	8.499	11.501	0	33\\
	nan	nan	0	34\\
	8.501	11.501	0	35\\
	8.501	11.501	1335	36\\
	9.499	11.501	1335	37\\
	9.499	11.501	0	38\\
	nan	nan	0	39\\
	9.501	11.501	0	40\\
	9.501	11.501	1301	41\\
	10.499	11.501	1301	42\\
	10.499	11.501	0	43\\
	nan	nan	0	44\\
	10.501	11.501	0	45\\
	10.501	11.501	894	46\\
	11.499	11.501	894	47\\
	11.499	11.501	0	48\\
	nan	nan	0	49\\
	11.501	11.501	0	50\\
	11.501	11.501	433	51\\
	12.499	11.501	433	52\\
	12.499	11.501	0	53\\
	nan	nan	0	54\\
	12.501	11.501	0	55\\
	12.501	11.501	110	56\\
	13.499	11.501	110	57\\
	13.499	11.501	0	58\\
	nan	nan	0	59\\
	7.501	12.499	0	60\\
	7.501	12.499	1561	61\\
	8.499	12.499	1561	62\\
	8.499	12.499	0	63\\
	nan	nan	0	64\\
	8.501	12.499	0	65\\
	8.501	12.499	1335	66\\
	9.499	12.499	1335	67\\
	9.499	12.499	0	68\\
	nan	nan	0	69\\
	9.501	12.499	0	70\\
	9.501	12.499	1301	71\\
	10.499	12.499	1301	72\\
	10.499	12.499	0	73\\
	nan	nan	0	74\\
	10.501	12.499	0	75\\
	10.501	12.499	894	76\\
	11.499	12.499	894	77\\
	11.499	12.499	0	78\\
	nan	nan	0	79\\
	11.501	12.499	0	80\\
	11.501	12.499	433	81\\
	12.499	12.499	433	82\\
	12.499	12.499	0	83\\
	nan	nan	0	84\\
	12.501	12.499	0	85\\
	12.501	12.499	110	86\\
	13.499	12.499	110	87\\
	13.499	12.499	0	88\\
	nan	nan	0	89\\
	7.501	12.499	0	90\\
	7.501	12.499	0	91\\
	8.499	12.499	0	92\\
	8.499	12.499	0	93\\
	nan	nan	0	94\\
	8.501	12.499	0	95\\
	8.501	12.499	0	96\\
	9.499	12.499	0	97\\
	9.499	12.499	0	98\\
	nan	nan	0	99\\
	9.501	12.499	0	100\\
	9.501	12.499	0	101\\
	10.499	12.499	0	102\\
	10.499	12.499	0	103\\
	nan	nan	0	104\\
	10.501	12.499	0	105\\
	10.501	12.499	0	106\\
	11.499	12.499	0	107\\
	11.499	12.499	0	108\\
	nan	nan	0	109\\
	11.501	12.499	0	110\\
	11.501	12.499	0	111\\
	12.499	12.499	0	112\\
	12.499	12.499	0	113\\
	nan	nan	0	114\\
	12.501	12.499	0	115\\
	12.501	12.499	0	116\\
	13.499	12.499	0	117\\
	13.499	12.499	0	118\\
	nan	nan	0	119\\
	nan	nan	0	120\\
	nan	nan	0	121\\
	nan	nan	0	122\\
	nan	nan	0	123\\
	nan	nan	0	124\\
	nan	nan	0	125\\
	nan	nan	0	126\\
	nan	nan	0	127\\
	nan	nan	0	128\\
	nan	nan	0	129\\
	nan	nan	0	130\\
	nan	nan	0	131\\
	nan	nan	0	132\\
	nan	nan	0	133\\
	nan	nan	0	134\\
	nan	nan	0	135\\
	nan	nan	0	136\\
	nan	nan	0	137\\
	nan	nan	0	138\\
	nan	nan	0	139\\
	nan	nan	0	140\\
	nan	nan	0	141\\
	nan	nan	0	142\\
	nan	nan	0	143\\
	nan	nan	0	144\\
	nan	nan	0	145\\
	nan	nan	0	146\\
	nan	nan	0	147\\
	nan	nan	0	148\\
	nan	nan	0	149\\
	7.501	12.501	0	150\\
	7.501	12.501	0	151\\
	8.499	12.501	0	152\\
	8.499	12.501	0	153\\
	nan	nan	0	154\\
	8.501	12.501	0	155\\
	8.501	12.501	0	156\\
	9.499	12.501	0	157\\
	9.499	12.501	0	158\\
	nan	nan	0	159\\
	9.501	12.501	0	160\\
	9.501	12.501	0	161\\
	10.499	12.501	0	162\\
	10.499	12.501	0	163\\
	nan	nan	0	164\\
	10.501	12.501	0	165\\
	10.501	12.501	0	166\\
	11.499	12.501	0	167\\
	11.499	12.501	0	168\\
	nan	nan	0	169\\
	11.501	12.501	0	170\\
	11.501	12.501	0	171\\
	12.499	12.501	0	172\\
	12.499	12.501	0	173\\
	nan	nan	0	174\\
	12.501	12.501	0	175\\
	12.501	12.501	0	176\\
	13.499	12.501	0	177\\
	13.499	12.501	0	178\\
	nan	nan	0	179\\
	7.501	12.501	0	180\\
	7.501	12.501	715	181\\
	8.499	12.501	715	182\\
	8.499	12.501	0	183\\
	nan	nan	0	184\\
	8.501	12.501	0	185\\
	8.501	12.501	193	186\\
	9.499	12.501	193	187\\
	9.499	12.501	0	188\\
	nan	nan	0	189\\
	9.501	12.501	0	190\\
	9.501	12.501	72	191\\
	10.499	12.501	72	192\\
	10.499	12.501	0	193\\
	nan	nan	0	194\\
	10.501	12.501	0	195\\
	10.501	12.501	12	196\\
	11.499	12.501	12	197\\
	11.499	12.501	0	198\\
	nan	nan	0	199\\
	11.501	12.501	0	200\\
	11.501	12.501	1	201\\
	12.499	12.501	1	202\\
	12.499	12.501	0	203\\
	nan	nan	0	204\\
	12.501	12.501	0	205\\
	12.501	12.501	0	206\\
	13.499	12.501	0	207\\
	13.499	12.501	0	208\\
	nan	nan	0	209\\
	7.501	13.499	0	210\\
	7.501	13.499	715	211\\
	8.499	13.499	715	212\\
	8.499	13.499	0	213\\
	nan	nan	0	214\\
	8.501	13.499	0	215\\
	8.501	13.499	193	216\\
	9.499	13.499	193	217\\
	9.499	13.499	0	218\\
	nan	nan	0	219\\
	9.501	13.499	0	220\\
	9.501	13.499	72	221\\
	10.499	13.499	72	222\\
	10.499	13.499	0	223\\
	nan	nan	0	224\\
	10.501	13.499	0	225\\
	10.501	13.499	12	226\\
	11.499	13.499	12	227\\
	11.499	13.499	0	228\\
	nan	nan	0	229\\
	11.501	13.499	0	230\\
	11.501	13.499	1	231\\
	12.499	13.499	1	232\\
	12.499	13.499	0	233\\
	nan	nan	0	234\\
	12.501	13.499	0	235\\
	12.501	13.499	0	236\\
	13.499	13.499	0	237\\
	13.499	13.499	0	238\\
	nan	nan	0	239\\
	7.501	13.499	0	240\\
	7.501	13.499	0	241\\
	8.499	13.499	0	242\\
	8.499	13.499	0	243\\
	nan	nan	0	244\\
	8.501	13.499	0	245\\
	8.501	13.499	0	246\\
	9.499	13.499	0	247\\
	9.499	13.499	0	248\\
	nan	nan	0	249\\
	9.501	13.499	0	250\\
	9.501	13.499	0	251\\
	10.499	13.499	0	252\\
	10.499	13.499	0	253\\
	nan	nan	0	254\\
	10.501	13.499	0	255\\
	10.501	13.499	0	256\\
	11.499	13.499	0	257\\
	11.499	13.499	0	258\\
	nan	nan	0	259\\
	11.501	13.499	0	260\\
	11.501	13.499	0	261\\
	12.499	13.499	0	262\\
	12.499	13.499	0	263\\
	nan	nan	0	264\\
	12.501	13.499	0	265\\
	12.501	13.499	0	266\\
	13.499	13.499	0	267\\
	13.499	13.499	0	268\\
	nan	nan	0	269\\
	nan	nan	0	270\\
	nan	nan	0	271\\
	nan	nan	0	272\\
	nan	nan	0	273\\
	nan	nan	0	274\\
	nan	nan	0	275\\
	nan	nan	0	276\\
	nan	nan	0	277\\
	nan	nan	0	278\\
	nan	nan	0	279\\
	nan	nan	0	280\\
	nan	nan	0	281\\
	nan	nan	0	282\\
	nan	nan	0	283\\
	nan	nan	0	284\\
	nan	nan	0	285\\
	nan	nan	0	286\\
	nan	nan	0	287\\
	nan	nan	0	288\\
	nan	nan	0	289\\
	nan	nan	0	290\\
	nan	nan	0	291\\
	nan	nan	0	292\\
	nan	nan	0	293\\
	nan	nan	0	294\\
	nan	nan	0	295\\
	nan	nan	0	296\\
	nan	nan	0	297\\
	nan	nan	0	298\\
	nan	nan	0	299\\
	7.501	13.501	0	300\\
	7.501	13.501	0	301\\
	8.499	13.501	0	302\\
	8.499	13.501	0	303\\
	nan	nan	0	304\\
	8.501	13.501	0	305\\
	8.501	13.501	0	306\\
	9.499	13.501	0	307\\
	9.499	13.501	0	308\\
	nan	nan	0	309\\
	9.501	13.501	0	310\\
	9.501	13.501	0	311\\
	10.499	13.501	0	312\\
	10.499	13.501	0	313\\
	nan	nan	0	314\\
	10.501	13.501	0	315\\
	10.501	13.501	0	316\\
	11.499	13.501	0	317\\
	11.499	13.501	0	318\\
	nan	nan	0	319\\
	11.501	13.501	0	320\\
	11.501	13.501	0	321\\
	12.499	13.501	0	322\\
	12.499	13.501	0	323\\
	nan	nan	0	324\\
	12.501	13.501	0	325\\
	12.501	13.501	0	326\\
	13.499	13.501	0	327\\
	13.499	13.501	0	328\\
	nan	nan	0	329\\
	7.501	13.501	0	330\\
	7.501	13.501	475	331\\
	8.499	13.501	475	332\\
	8.499	13.501	0	333\\
	nan	nan	0	334\\
	8.501	13.501	0	335\\
	8.501	13.501	59	336\\
	9.499	13.501	59	337\\
	9.499	13.501	0	338\\
	nan	nan	0	339\\
	9.501	13.501	0	340\\
	9.501	13.501	9	341\\
	10.499	13.501	9	342\\
	10.499	13.501	0	343\\
	nan	nan	0	344\\
	10.501	13.501	0	345\\
	10.501	13.501	2	346\\
	11.499	13.501	2	347\\
	11.499	13.501	0	348\\
	nan	nan	0	349\\
	11.501	13.501	0	350\\
	11.501	13.501	0	351\\
	12.499	13.501	0	352\\
	12.499	13.501	0	353\\
	nan	nan	0	354\\
	12.501	13.501	0	355\\
	12.501	13.501	0	356\\
	13.499	13.501	0	357\\
	13.499	13.501	0	358\\
	nan	nan	0	359\\
	7.501	14.499	0	360\\
	7.501	14.499	475	361\\
	8.499	14.499	475	362\\
	8.499	14.499	0	363\\
	nan	nan	0	364\\
	8.501	14.499	0	365\\
	8.501	14.499	59	366\\
	9.499	14.499	59	367\\
	9.499	14.499	0	368\\
	nan	nan	0	369\\
	9.501	14.499	0	370\\
	9.501	14.499	9	371\\
	10.499	14.499	9	372\\
	10.499	14.499	0	373\\
	nan	nan	0	374\\
	10.501	14.499	0	375\\
	10.501	14.499	2	376\\
	11.499	14.499	2	377\\
	11.499	14.499	0	378\\
	nan	nan	0	379\\
	11.501	14.499	0	380\\
	11.501	14.499	0	381\\
	12.499	14.499	0	382\\
	12.499	14.499	0	383\\
	nan	nan	0	384\\
	12.501	14.499	0	385\\
	12.501	14.499	0	386\\
	13.499	14.499	0	387\\
	13.499	14.499	0	388\\
	nan	nan	0	389\\
	7.501	14.499	0	390\\
	7.501	14.499	0	391\\
	8.499	14.499	0	392\\
	8.499	14.499	0	393\\
	nan	nan	0	394\\
	8.501	14.499	0	395\\
	8.501	14.499	0	396\\
	9.499	14.499	0	397\\
	9.499	14.499	0	398\\
	nan	nan	0	399\\
	9.501	14.499	0	400\\
	9.501	14.499	0	401\\
	10.499	14.499	0	402\\
	10.499	14.499	0	403\\
	nan	nan	0	404\\
	10.501	14.499	0	405\\
	10.501	14.499	0	406\\
	11.499	14.499	0	407\\
	11.499	14.499	0	408\\
	nan	nan	0	409\\
	11.501	14.499	0	410\\
	11.501	14.499	0	411\\
	12.499	14.499	0	412\\
	12.499	14.499	0	413\\
	nan	nan	0	414\\
	12.501	14.499	0	415\\
	12.501	14.499	0	416\\
	13.499	14.499	0	417\\
	13.499	14.499	0	418\\
	nan	nan	0	419\\
	nan	nan	0	420\\
	nan	nan	0	421\\
	nan	nan	0	422\\
	nan	nan	0	423\\
	nan	nan	0	424\\
	nan	nan	0	425\\
	nan	nan	0	426\\
	nan	nan	0	427\\
	nan	nan	0	428\\
	nan	nan	0	429\\
	nan	nan	0	430\\
	nan	nan	0	431\\
	nan	nan	0	432\\
	nan	nan	0	433\\
	nan	nan	0	434\\
	nan	nan	0	435\\
	nan	nan	0	436\\
	nan	nan	0	437\\
	nan	nan	0	438\\
	nan	nan	0	439\\
	nan	nan	0	440\\
	nan	nan	0	441\\
	nan	nan	0	442\\
	nan	nan	0	443\\
	nan	nan	0	444\\
	nan	nan	0	445\\
	nan	nan	0	446\\
	nan	nan	0	447\\
	nan	nan	0	448\\
	nan	nan	0	449\\
	7.501	14.501	0	450\\
	7.501	14.501	0	451\\
	8.499	14.501	0	452\\
	8.499	14.501	0	453\\
	nan	nan	0	454\\
	8.501	14.501	0	455\\
	8.501	14.501	0	456\\
	9.499	14.501	0	457\\
	9.499	14.501	0	458\\
	nan	nan	0	459\\
	9.501	14.501	0	460\\
	9.501	14.501	0	461\\
	10.499	14.501	0	462\\
	10.499	14.501	0	463\\
	nan	nan	0	464\\
	10.501	14.501	0	465\\
	10.501	14.501	0	466\\
	11.499	14.501	0	467\\
	11.499	14.501	0	468\\
	nan	nan	0	469\\
	11.501	14.501	0	470\\
	11.501	14.501	0	471\\
	12.499	14.501	0	472\\
	12.499	14.501	0	473\\
	nan	nan	0	474\\
	12.501	14.501	0	475\\
	12.501	14.501	0	476\\
	13.499	14.501	0	477\\
	13.499	14.501	0	478\\
	nan	nan	0	479\\
	7.501	14.501	0	480\\
	7.501	14.501	236	481\\
	8.499	14.501	236	482\\
	8.499	14.501	0	483\\
	nan	nan	0	484\\
	8.501	14.501	0	485\\
	8.501	14.501	8	486\\
	9.499	14.501	8	487\\
	9.499	14.501	0	488\\
	nan	nan	0	489\\
	9.501	14.501	0	490\\
	9.501	14.501	0	491\\
	10.499	14.501	0	492\\
	10.499	14.501	0	493\\
	nan	nan	0	494\\
	10.501	14.501	0	495\\
	10.501	14.501	0	496\\
	11.499	14.501	0	497\\
	11.499	14.501	0	498\\
	nan	nan	0	499\\
	11.501	14.501	0	500\\
	11.501	14.501	0	501\\
	12.499	14.501	0	502\\
	12.499	14.501	0	503\\
	nan	nan	0	504\\
	12.501	14.501	0	505\\
	12.501	14.501	0	506\\
	13.499	14.501	0	507\\
	13.499	14.501	0	508\\
	nan	nan	0	509\\
	7.501	15.499	0	510\\
	7.501	15.499	236	511\\
	8.499	15.499	236	512\\
	8.499	15.499	0	513\\
	nan	nan	0	514\\
	8.501	15.499	0	515\\
	8.501	15.499	8	516\\
	9.499	15.499	8	517\\
	9.499	15.499	0	518\\
	nan	nan	0	519\\
	9.501	15.499	0	520\\
	9.501	15.499	0	521\\
	10.499	15.499	0	522\\
	10.499	15.499	0	523\\
	nan	nan	0	524\\
	10.501	15.499	0	525\\
	10.501	15.499	0	526\\
	11.499	15.499	0	527\\
	11.499	15.499	0	528\\
	nan	nan	0	529\\
	11.501	15.499	0	530\\
	11.501	15.499	0	531\\
	12.499	15.499	0	532\\
	12.499	15.499	0	533\\
	nan	nan	0	534\\
	12.501	15.499	0	535\\
	12.501	15.499	0	536\\
	13.499	15.499	0	537\\
	13.499	15.499	0	538\\
	nan	nan	0	539\\
	7.501	15.499	0	540\\
	7.501	15.499	0	541\\
	8.499	15.499	0	542\\
	8.499	15.499	0	543\\
	nan	nan	0	544\\
	8.501	15.499	0	545\\
	8.501	15.499	0	546\\
	9.499	15.499	0	547\\
	9.499	15.499	0	548\\
	nan	nan	0	549\\
	9.501	15.499	0	550\\
	9.501	15.499	0	551\\
	10.499	15.499	0	552\\
	10.499	15.499	0	553\\
	nan	nan	0	554\\
	10.501	15.499	0	555\\
	10.501	15.499	0	556\\
	11.499	15.499	0	557\\
	11.499	15.499	0	558\\
	nan	nan	0	559\\
	11.501	15.499	0	560\\
	11.501	15.499	0	561\\
	12.499	15.499	0	562\\
	12.499	15.499	0	563\\
	nan	nan	0	564\\
	12.501	15.499	0	565\\
	12.501	15.499	0	566\\
	13.499	15.499	0	567\\
	13.499	15.499	0	568\\
	nan	nan	0	569\\
	nan	nan	0	570\\
	nan	nan	0	571\\
	nan	nan	0	572\\
	nan	nan	0	573\\
	nan	nan	0	574\\
	nan	nan	0	575\\
	nan	nan	0	576\\
	nan	nan	0	577\\
	nan	nan	0	578\\
	nan	nan	0	579\\
	nan	nan	0	580\\
	nan	nan	0	581\\
	nan	nan	0	582\\
	nan	nan	0	583\\
	nan	nan	0	584\\
	nan	nan	0	585\\
	nan	nan	0	586\\
	nan	nan	0	587\\
	nan	nan	0	588\\
	nan	nan	0	589\\
	nan	nan	0	590\\
	nan	nan	0	591\\
	nan	nan	0	592\\
	nan	nan	0	593\\
	nan	nan	0	594\\
	nan	nan	0	595\\
	nan	nan	0	596\\
	nan	nan	0	597\\
	nan	nan	0	598\\
	nan	nan	0	599\\
	7.501	15.501	0	600\\
	7.501	15.501	0	601\\
	8.499	15.501	0	602\\
	8.499	15.501	0	603\\
	nan	nan	0	604\\
	8.501	15.501	0	605\\
	8.501	15.501	0	606\\
	9.499	15.501	0	607\\
	9.499	15.501	0	608\\
	nan	nan	0	609\\
	9.501	15.501	0	610\\
	9.501	15.501	0	611\\
	10.499	15.501	0	612\\
	10.499	15.501	0	613\\
	nan	nan	0	614\\
	10.501	15.501	0	615\\
	10.501	15.501	0	616\\
	11.499	15.501	0	617\\
	11.499	15.501	0	618\\
	nan	nan	0	619\\
	11.501	15.501	0	620\\
	11.501	15.501	0	621\\
	12.499	15.501	0	622\\
	12.499	15.501	0	623\\
	nan	nan	0	624\\
	12.501	15.501	0	625\\
	12.501	15.501	0	626\\
	13.499	15.501	0	627\\
	13.499	15.501	0	628\\
	nan	nan	0	629\\
	7.501	15.501	0	630\\
	7.501	15.501	68	631\\
	8.499	15.501	68	632\\
	8.499	15.501	0	633\\
	nan	nan	0	634\\
	8.501	15.501	0	635\\
	8.501	15.501	0	636\\
	9.499	15.501	0	637\\
	9.499	15.501	0	638\\
	nan	nan	0	639\\
	9.501	15.501	0	640\\
	9.501	15.501	0	641\\
	10.499	15.501	0	642\\
	10.499	15.501	0	643\\
	nan	nan	0	644\\
	10.501	15.501	0	645\\
	10.501	15.501	0	646\\
	11.499	15.501	0	647\\
	11.499	15.501	0	648\\
	nan	nan	0	649\\
	11.501	15.501	0	650\\
	11.501	15.501	0	651\\
	12.499	15.501	0	652\\
	12.499	15.501	0	653\\
	nan	nan	0	654\\
	12.501	15.501	0	655\\
	12.501	15.501	0	656\\
	13.499	15.501	0	657\\
	13.499	15.501	0	658\\
	nan	nan	0	659\\
	7.501	16.499	0	660\\
	7.501	16.499	68	661\\
	8.499	16.499	68	662\\
	8.499	16.499	0	663\\
	nan	nan	0	664\\
	8.501	16.499	0	665\\
	8.501	16.499	0	666\\
	9.499	16.499	0	667\\
	9.499	16.499	0	668\\
	nan	nan	0	669\\
	9.501	16.499	0	670\\
	9.501	16.499	0	671\\
	10.499	16.499	0	672\\
	10.499	16.499	0	673\\
	nan	nan	0	674\\
	10.501	16.499	0	675\\
	10.501	16.499	0	676\\
	11.499	16.499	0	677\\
	11.499	16.499	0	678\\
	nan	nan	0	679\\
	11.501	16.499	0	680\\
	11.501	16.499	0	681\\
	12.499	16.499	0	682\\
	12.499	16.499	0	683\\
	nan	nan	0	684\\
	12.501	16.499	0	685\\
	12.501	16.499	0	686\\
	13.499	16.499	0	687\\
	13.499	16.499	0	688\\
	nan	nan	0	689\\
	7.501	16.499	0	690\\
	7.501	16.499	0	691\\
	8.499	16.499	0	692\\
	8.499	16.499	0	693\\
	nan	nan	0	694\\
	8.501	16.499	0	695\\
	8.501	16.499	0	696\\
	9.499	16.499	0	697\\
	9.499	16.499	0	698\\
	nan	nan	0	699\\
	9.501	16.499	0	700\\
	9.501	16.499	0	701\\
	10.499	16.499	0	702\\
	10.499	16.499	0	703\\
	nan	nan	0	704\\
	10.501	16.499	0	705\\
	10.501	16.499	0	706\\
	11.499	16.499	0	707\\
	11.499	16.499	0	708\\
	nan	nan	0	709\\
	11.501	16.499	0	710\\
	11.501	16.499	0	711\\
	12.499	16.499	0	712\\
	12.499	16.499	0	713\\
	nan	nan	0	714\\
	12.501	16.499	0	715\\
	12.501	16.499	0	716\\
	13.499	16.499	0	717\\
	13.499	16.499	0	718\\
	nan	nan	0	719\\
	nan	nan	0	720\\
	nan	nan	0	721\\
	nan	nan	0	722\\
	nan	nan	0	723\\
	nan	nan	0	724\\
	nan	nan	0	725\\
	nan	nan	0	726\\
	nan	nan	0	727\\
	nan	nan	0	728\\
	nan	nan	0	729\\
	nan	nan	0	730\\
	nan	nan	0	731\\
	nan	nan	0	732\\
	nan	nan	0	733\\
	nan	nan	0	734\\
	nan	nan	0	735\\
	nan	nan	0	736\\
	nan	nan	0	737\\
	nan	nan	0	738\\
	nan	nan	0	739\\
	nan	nan	0	740\\
	nan	nan	0	741\\
	nan	nan	0	742\\
	nan	nan	0	743\\
	nan	nan	0	744\\
	nan	nan	0	745\\
	nan	nan	0	746\\
	nan	nan	0	747\\
	nan	nan	0	748\\
	nan	nan	0	749\\
	7.501	16.501	0	750\\
	7.501	16.501	0	751\\
	8.499	16.501	0	752\\
	8.499	16.501	0	753\\
	nan	nan	0	754\\
	8.501	16.501	0	755\\
	8.501	16.501	0	756\\
	9.499	16.501	0	757\\
	9.499	16.501	0	758\\
	nan	nan	0	759\\
	9.501	16.501	0	760\\
	9.501	16.501	0	761\\
	10.499	16.501	0	762\\
	10.499	16.501	0	763\\
	nan	nan	0	764\\
	10.501	16.501	0	765\\
	10.501	16.501	0	766\\
	11.499	16.501	0	767\\
	11.499	16.501	0	768\\
	nan	nan	0	769\\
	11.501	16.501	0	770\\
	11.501	16.501	0	771\\
	12.499	16.501	0	772\\
	12.499	16.501	0	773\\
	nan	nan	0	774\\
	12.501	16.501	0	775\\
	12.501	16.501	0	776\\
	13.499	16.501	0	777\\
	13.499	16.501	0	778\\
	nan	nan	0	779\\
	7.501	16.501	0	780\\
	7.501	16.501	16	781\\
	8.499	16.501	16	782\\
	8.499	16.501	0	783\\
	nan	nan	0	784\\
	8.501	16.501	0	785\\
	8.501	16.501	0	786\\
	9.499	16.501	0	787\\
	9.499	16.501	0	788\\
	nan	nan	0	789\\
	9.501	16.501	0	790\\
	9.501	16.501	0	791\\
	10.499	16.501	0	792\\
	10.499	16.501	0	793\\
	nan	nan	0	794\\
	10.501	16.501	0	795\\
	10.501	16.501	0	796\\
	11.499	16.501	0	797\\
	11.499	16.501	0	798\\
	nan	nan	0	799\\
	11.501	16.501	0	800\\
	11.501	16.501	0	801\\
	12.499	16.501	0	802\\
	12.499	16.501	0	803\\
	nan	nan	0	804\\
	12.501	16.501	0	805\\
	12.501	16.501	0	806\\
	13.499	16.501	0	807\\
	13.499	16.501	0	808\\
	nan	nan	0	809\\
	7.501	17.499	0	810\\
	7.501	17.499	16	811\\
	8.499	17.499	16	812\\
	8.499	17.499	0	813\\
	nan	nan	0	814\\
	8.501	17.499	0	815\\
	8.501	17.499	0	816\\
	9.499	17.499	0	817\\
	9.499	17.499	0	818\\
	nan	nan	0	819\\
	9.501	17.499	0	820\\
	9.501	17.499	0	821\\
	10.499	17.499	0	822\\
	10.499	17.499	0	823\\
	nan	nan	0	824\\
	10.501	17.499	0	825\\
	10.501	17.499	0	826\\
	11.499	17.499	0	827\\
	11.499	17.499	0	828\\
	nan	nan	0	829\\
	11.501	17.499	0	830\\
	11.501	17.499	0	831\\
	12.499	17.499	0	832\\
	12.499	17.499	0	833\\
	nan	nan	0	834\\
	12.501	17.499	0	835\\
	12.501	17.499	0	836\\
	13.499	17.499	0	837\\
	13.499	17.499	0	838\\
	nan	nan	0	839\\
	7.501	17.499	0	840\\
	7.501	17.499	0	841\\
	8.499	17.499	0	842\\
	8.499	17.499	0	843\\
	nan	nan	0	844\\
	8.501	17.499	0	845\\
	8.501	17.499	0	846\\
	9.499	17.499	0	847\\
	9.499	17.499	0	848\\
	nan	nan	0	849\\
	9.501	17.499	0	850\\
	9.501	17.499	0	851\\
	10.499	17.499	0	852\\
	10.499	17.499	0	853\\
	nan	nan	0	854\\
	10.501	17.499	0	855\\
	10.501	17.499	0	856\\
	11.499	17.499	0	857\\
	11.499	17.499	0	858\\
	nan	nan	0	859\\
	11.501	17.499	0	860\\
	11.501	17.499	0	861\\
	12.499	17.499	0	862\\
	12.499	17.499	0	863\\
	nan	nan	0	864\\
	12.501	17.499	0	865\\
	12.501	17.499	0	866\\
	13.499	17.499	0	867\\
	13.499	17.499	0	868\\
	nan	nan	0	869\\
	nan	nan	0	870\\
	nan	nan	0	871\\
	nan	nan	0	872\\
	nan	nan	0	873\\
	nan	nan	0	874\\
	nan	nan	0	875\\
	nan	nan	0	876\\
	nan	nan	0	877\\
	nan	nan	0	878\\
	nan	nan	0	879\\
	nan	nan	0	880\\
	nan	nan	0	881\\
	nan	nan	0	882\\
	nan	nan	0	883\\
	nan	nan	0	884\\
	nan	nan	0	885\\
	nan	nan	0	886\\
	nan	nan	0	887\\
	nan	nan	0	888\\
	nan	nan	0	889\\
	nan	nan	0	890\\
	nan	nan	0	891\\
	nan	nan	0	892\\
	nan	nan	0	893\\
	nan	nan	0	894\\
	nan	nan	0	895\\
	nan	nan	0	896\\
	nan	nan	0	897\\
	nan	nan	0	898\\
	nan	nan	0	899\\
};

\end{axis}

\end{tikzpicture}%

%% file: plots/Revised_Best_Card_Set_delta_d_4_SNR_10.tex
%
%
\begin{tikzpicture}

\begin{axis}[%
scale=0.6,
width=3.284in,
height=3.566in,
at={(0.758in,0.481in)},
scale only axis,
unbounded coords=jump,
colormap={patchmap}{[1pt] rgb(0pt)=(0.75,0.85,0.95); rgb(899pt)=(0.75,0.85,0.95)},
xmin=7.2,
xmax=13.8,
xtick={ 8,  9, 10, 11, 12, 13},
xticklabel style = {font=\footnotesize},
tick align=outside,
xlabel style={font=\color{white!15!black}},
xlabel style = {font=\small},
xlabel={$|\mathcal{P}_{1}^{\star}|$},
ymin=11.2,
ymax=17.8,
ytick={12, 13, 14, 15, 16, 17},
yticklabel style = {font=\footnotesize},
ylabel style={font=\color{white!15!black}},
ylabel style = {font=\small},
ylabel={$|\mathcal{P}_{2}^{\star}|$},
zmin=0,
zmax=2100,
zlabel style={font=\color{white!15!black}},
zlabel style = {font=\small},
zticklabel style = {font=\footnotesize},
zlabel={Frequency},
view={-37.5}{30},
axis background/.style={fill=white},
title style={font=\bfseries},
axis x line*=bottom,
axis y line*=left,
axis z line*=left,
xmajorgrids,
ymajorgrids,
zmajorgrids,
legend style={at={(1.03,1)}, anchor=north west, legend cell align=left, align=left, draw=white!15!black}
]

\addplot3[%
surf,
shader=flat corner, draw=black, mesh/rows=30]
table[row sep=crcr, colormap name=surfmap, point meta=\thisrow{c}] {%
	x	y	z	c\\
	7.501	11.501	0	0\\
	7.501	11.501	0	1\\
	8.499	11.501	0	2\\
	8.499	11.501	0	3\\
	nan	nan	0	4\\
	8.501	11.501	0	5\\
	8.501	11.501	0	6\\
	9.499	11.501	0	7\\
	9.499	11.501	0	8\\
	nan	nan	0	9\\
	9.501	11.501	0	10\\
	9.501	11.501	0	11\\
	10.499	11.501	0	12\\
	10.499	11.501	0	13\\
	nan	nan	0	14\\
	10.501	11.501	0	15\\
	10.501	11.501	0	16\\
	11.499	11.501	0	17\\
	11.499	11.501	0	18\\
	nan	nan	0	19\\
	11.501	11.501	0	20\\
	11.501	11.501	0	21\\
	12.499	11.501	0	22\\
	12.499	11.501	0	23\\
	nan	nan	0	24\\
	12.501	11.501	0	25\\
	12.501	11.501	0	26\\
	13.499	11.501	0	27\\
	13.499	11.501	0	28\\
	nan	nan	0	29\\
	7.501	11.501	0	30\\
	7.501	11.501	0	31\\
	8.499	11.501	0	32\\
	8.499	11.501	0	33\\
	nan	nan	0	34\\
	8.501	11.501	0	35\\
	8.501	11.501	0	36\\
	9.499	11.501	0	37\\
	9.499	11.501	0	38\\
	nan	nan	0	39\\
	9.501	11.501	0	40\\
	9.501	11.501	0	41\\
	10.499	11.501	0	42\\
	10.499	11.501	0	43\\
	nan	nan	0	44\\
	10.501	11.501	0	45\\
	10.501	11.501	0	46\\
	11.499	11.501	0	47\\
	11.499	11.501	0	48\\
	nan	nan	0	49\\
	11.501	11.501	0	50\\
	11.501	11.501	0	51\\
	12.499	11.501	0	52\\
	12.499	11.501	0	53\\
	nan	nan	0	54\\
	12.501	11.501	0	55\\
	12.501	11.501	5	56\\
	13.499	11.501	5	57\\
	13.499	11.501	0	58\\
	nan	nan	0	59\\
	7.501	12.499	0	60\\
	7.501	12.499	0	61\\
	8.499	12.499	0	62\\
	8.499	12.499	0	63\\
	nan	nan	0	64\\
	8.501	12.499	0	65\\
	8.501	12.499	0	66\\
	9.499	12.499	0	67\\
	9.499	12.499	0	68\\
	nan	nan	0	69\\
	9.501	12.499	0	70\\
	9.501	12.499	0	71\\
	10.499	12.499	0	72\\
	10.499	12.499	0	73\\
	nan	nan	0	74\\
	10.501	12.499	0	75\\
	10.501	12.499	0	76\\
	11.499	12.499	0	77\\
	11.499	12.499	0	78\\
	nan	nan	0	79\\
	11.501	12.499	0	80\\
	11.501	12.499	0	81\\
	12.499	12.499	0	82\\
	12.499	12.499	0	83\\
	nan	nan	0	84\\
	12.501	12.499	0	85\\
	12.501	12.499	5	86\\
	13.499	12.499	5	87\\
	13.499	12.499	0	88\\
	nan	nan	0	89\\
	7.501	12.499	0	90\\
	7.501	12.499	0	91\\
	8.499	12.499	0	92\\
	8.499	12.499	0	93\\
	nan	nan	0	94\\
	8.501	12.499	0	95\\
	8.501	12.499	0	96\\
	9.499	12.499	0	97\\
	9.499	12.499	0	98\\
	nan	nan	0	99\\
	9.501	12.499	0	100\\
	9.501	12.499	0	101\\
	10.499	12.499	0	102\\
	10.499	12.499	0	103\\
	nan	nan	0	104\\
	10.501	12.499	0	105\\
	10.501	12.499	0	106\\
	11.499	12.499	0	107\\
	11.499	12.499	0	108\\
	nan	nan	0	109\\
	11.501	12.499	0	110\\
	11.501	12.499	0	111\\
	12.499	12.499	0	112\\
	12.499	12.499	0	113\\
	nan	nan	0	114\\
	12.501	12.499	0	115\\
	12.501	12.499	0	116\\
	13.499	12.499	0	117\\
	13.499	12.499	0	118\\
	nan	nan	0	119\\
	nan	nan	0	120\\
	nan	nan	0	121\\
	nan	nan	0	122\\
	nan	nan	0	123\\
	nan	nan	0	124\\
	nan	nan	0	125\\
	nan	nan	0	126\\
	nan	nan	0	127\\
	nan	nan	0	128\\
	nan	nan	0	129\\
	nan	nan	0	130\\
	nan	nan	0	131\\
	nan	nan	0	132\\
	nan	nan	0	133\\
	nan	nan	0	134\\
	nan	nan	0	135\\
	nan	nan	0	136\\
	nan	nan	0	137\\
	nan	nan	0	138\\
	nan	nan	0	139\\
	nan	nan	0	140\\
	nan	nan	0	141\\
	nan	nan	0	142\\
	nan	nan	0	143\\
	nan	nan	0	144\\
	nan	nan	0	145\\
	nan	nan	0	146\\
	nan	nan	0	147\\
	nan	nan	0	148\\
	nan	nan	0	149\\
	7.501	12.501	0	150\\
	7.501	12.501	0	151\\
	8.499	12.501	0	152\\
	8.499	12.501	0	153\\
	nan	nan	0	154\\
	8.501	12.501	0	155\\
	8.501	12.501	0	156\\
	9.499	12.501	0	157\\
	9.499	12.501	0	158\\
	nan	nan	0	159\\
	9.501	12.501	0	160\\
	9.501	12.501	0	161\\
	10.499	12.501	0	162\\
	10.499	12.501	0	163\\
	nan	nan	0	164\\
	10.501	12.501	0	165\\
	10.501	12.501	0	166\\
	11.499	12.501	0	167\\
	11.499	12.501	0	168\\
	nan	nan	0	169\\
	11.501	12.501	0	170\\
	11.501	12.501	0	171\\
	12.499	12.501	0	172\\
	12.499	12.501	0	173\\
	nan	nan	0	174\\
	12.501	12.501	0	175\\
	12.501	12.501	0	176\\
	13.499	12.501	0	177\\
	13.499	12.501	0	178\\
	nan	nan	0	179\\
	7.501	12.501	0	180\\
	7.501	12.501	0	181\\
	8.499	12.501	0	182\\
	8.499	12.501	0	183\\
	nan	nan	0	184\\
	8.501	12.501	0	185\\
	8.501	12.501	0	186\\
	9.499	12.501	0	187\\
	9.499	12.501	0	188\\
	nan	nan	0	189\\
	9.501	12.501	0	190\\
	9.501	12.501	0	191\\
	10.499	12.501	0	192\\
	10.499	12.501	0	193\\
	nan	nan	0	194\\
	10.501	12.501	0	195\\
	10.501	12.501	1	196\\
	11.499	12.501	1	197\\
	11.499	12.501	0	198\\
	nan	nan	0	199\\
	11.501	12.501	0	200\\
	11.501	12.501	6	201\\
	12.499	12.501	6	202\\
	12.499	12.501	0	203\\
	nan	nan	0	204\\
	12.501	12.501	0	205\\
	12.501	12.501	37	206\\
	13.499	12.501	37	207\\
	13.499	12.501	0	208\\
	nan	nan	0	209\\
	7.501	13.499	0	210\\
	7.501	13.499	0	211\\
	8.499	13.499	0	212\\
	8.499	13.499	0	213\\
	nan	nan	0	214\\
	8.501	13.499	0	215\\
	8.501	13.499	0	216\\
	9.499	13.499	0	217\\
	9.499	13.499	0	218\\
	nan	nan	0	219\\
	9.501	13.499	0	220\\
	9.501	13.499	0	221\\
	10.499	13.499	0	222\\
	10.499	13.499	0	223\\
	nan	nan	0	224\\
	10.501	13.499	0	225\\
	10.501	13.499	1	226\\
	11.499	13.499	1	227\\
	11.499	13.499	0	228\\
	nan	nan	0	229\\
	11.501	13.499	0	230\\
	11.501	13.499	6	231\\
	12.499	13.499	6	232\\
	12.499	13.499	0	233\\
	nan	nan	0	234\\
	12.501	13.499	0	235\\
	12.501	13.499	37	236\\
	13.499	13.499	37	237\\
	13.499	13.499	0	238\\
	nan	nan	0	239\\
	7.501	13.499	0	240\\
	7.501	13.499	0	241\\
	8.499	13.499	0	242\\
	8.499	13.499	0	243\\
	nan	nan	0	244\\
	8.501	13.499	0	245\\
	8.501	13.499	0	246\\
	9.499	13.499	0	247\\
	9.499	13.499	0	248\\
	nan	nan	0	249\\
	9.501	13.499	0	250\\
	9.501	13.499	0	251\\
	10.499	13.499	0	252\\
	10.499	13.499	0	253\\
	nan	nan	0	254\\
	10.501	13.499	0	255\\
	10.501	13.499	0	256\\
	11.499	13.499	0	257\\
	11.499	13.499	0	258\\
	nan	nan	0	259\\
	11.501	13.499	0	260\\
	11.501	13.499	0	261\\
	12.499	13.499	0	262\\
	12.499	13.499	0	263\\
	nan	nan	0	264\\
	12.501	13.499	0	265\\
	12.501	13.499	0	266\\
	13.499	13.499	0	267\\
	13.499	13.499	0	268\\
	nan	nan	0	269\\
	nan	nan	0	270\\
	nan	nan	0	271\\
	nan	nan	0	272\\
	nan	nan	0	273\\
	nan	nan	0	274\\
	nan	nan	0	275\\
	nan	nan	0	276\\
	nan	nan	0	277\\
	nan	nan	0	278\\
	nan	nan	0	279\\
	nan	nan	0	280\\
	nan	nan	0	281\\
	nan	nan	0	282\\
	nan	nan	0	283\\
	nan	nan	0	284\\
	nan	nan	0	285\\
	nan	nan	0	286\\
	nan	nan	0	287\\
	nan	nan	0	288\\
	nan	nan	0	289\\
	nan	nan	0	290\\
	nan	nan	0	291\\
	nan	nan	0	292\\
	nan	nan	0	293\\
	nan	nan	0	294\\
	nan	nan	0	295\\
	nan	nan	0	296\\
	nan	nan	0	297\\
	nan	nan	0	298\\
	nan	nan	0	299\\
	7.501	13.501	0	300\\
	7.501	13.501	0	301\\
	8.499	13.501	0	302\\
	8.499	13.501	0	303\\
	nan	nan	0	304\\
	8.501	13.501	0	305\\
	8.501	13.501	0	306\\
	9.499	13.501	0	307\\
	9.499	13.501	0	308\\
	nan	nan	0	309\\
	9.501	13.501	0	310\\
	9.501	13.501	0	311\\
	10.499	13.501	0	312\\
	10.499	13.501	0	313\\
	nan	nan	0	314\\
	10.501	13.501	0	315\\
	10.501	13.501	0	316\\
	11.499	13.501	0	317\\
	11.499	13.501	0	318\\
	nan	nan	0	319\\
	11.501	13.501	0	320\\
	11.501	13.501	0	321\\
	12.499	13.501	0	322\\
	12.499	13.501	0	323\\
	nan	nan	0	324\\
	12.501	13.501	0	325\\
	12.501	13.501	0	326\\
	13.499	13.501	0	327\\
	13.499	13.501	0	328\\
	nan	nan	0	329\\
	7.501	13.501	0	330\\
	7.501	13.501	0	331\\
	8.499	13.501	0	332\\
	8.499	13.501	0	333\\
	nan	nan	0	334\\
	8.501	13.501	0	335\\
	8.501	13.501	0	336\\
	9.499	13.501	0	337\\
	9.499	13.501	0	338\\
	nan	nan	0	339\\
	9.501	13.501	0	340\\
	9.501	13.501	1	341\\
	10.499	13.501	1	342\\
	10.499	13.501	0	343\\
	nan	nan	0	344\\
	10.501	13.501	0	345\\
	10.501	13.501	11	346\\
	11.499	13.501	11	347\\
	11.499	13.501	0	348\\
	nan	nan	0	349\\
	11.501	13.501	0	350\\
	11.501	13.501	109	351\\
	12.499	13.501	109	352\\
	12.499	13.501	0	353\\
	nan	nan	0	354\\
	12.501	13.501	0	355\\
	12.501	13.501	377	356\\
	13.499	13.501	377	357\\
	13.499	13.501	0	358\\
	nan	nan	0	359\\
	7.501	14.499	0	360\\
	7.501	14.499	0	361\\
	8.499	14.499	0	362\\
	8.499	14.499	0	363\\
	nan	nan	0	364\\
	8.501	14.499	0	365\\
	8.501	14.499	0	366\\
	9.499	14.499	0	367\\
	9.499	14.499	0	368\\
	nan	nan	0	369\\
	9.501	14.499	0	370\\
	9.501	14.499	1	371\\
	10.499	14.499	1	372\\
	10.499	14.499	0	373\\
	nan	nan	0	374\\
	10.501	14.499	0	375\\
	10.501	14.499	11	376\\
	11.499	14.499	11	377\\
	11.499	14.499	0	378\\
	nan	nan	0	379\\
	11.501	14.499	0	380\\
	11.501	14.499	109	381\\
	12.499	14.499	109	382\\
	12.499	14.499	0	383\\
	nan	nan	0	384\\
	12.501	14.499	0	385\\
	12.501	14.499	377	386\\
	13.499	14.499	377	387\\
	13.499	14.499	0	388\\
	nan	nan	0	389\\
	7.501	14.499	0	390\\
	7.501	14.499	0	391\\
	8.499	14.499	0	392\\
	8.499	14.499	0	393\\
	nan	nan	0	394\\
	8.501	14.499	0	395\\
	8.501	14.499	0	396\\
	9.499	14.499	0	397\\
	9.499	14.499	0	398\\
	nan	nan	0	399\\
	9.501	14.499	0	400\\
	9.501	14.499	0	401\\
	10.499	14.499	0	402\\
	10.499	14.499	0	403\\
	nan	nan	0	404\\
	10.501	14.499	0	405\\
	10.501	14.499	0	406\\
	11.499	14.499	0	407\\
	11.499	14.499	0	408\\
	nan	nan	0	409\\
	11.501	14.499	0	410\\
	11.501	14.499	0	411\\
	12.499	14.499	0	412\\
	12.499	14.499	0	413\\
	nan	nan	0	414\\
	12.501	14.499	0	415\\
	12.501	14.499	0	416\\
	13.499	14.499	0	417\\
	13.499	14.499	0	418\\
	nan	nan	0	419\\
	nan	nan	0	420\\
	nan	nan	0	421\\
	nan	nan	0	422\\
	nan	nan	0	423\\
	nan	nan	0	424\\
	nan	nan	0	425\\
	nan	nan	0	426\\
	nan	nan	0	427\\
	nan	nan	0	428\\
	nan	nan	0	429\\
	nan	nan	0	430\\
	nan	nan	0	431\\
	nan	nan	0	432\\
	nan	nan	0	433\\
	nan	nan	0	434\\
	nan	nan	0	435\\
	nan	nan	0	436\\
	nan	nan	0	437\\
	nan	nan	0	438\\
	nan	nan	0	439\\
	nan	nan	0	440\\
	nan	nan	0	441\\
	nan	nan	0	442\\
	nan	nan	0	443\\
	nan	nan	0	444\\
	nan	nan	0	445\\
	nan	nan	0	446\\
	nan	nan	0	447\\
	nan	nan	0	448\\
	nan	nan	0	449\\
	7.501	14.501	0	450\\
	7.501	14.501	0	451\\
	8.499	14.501	0	452\\
	8.499	14.501	0	453\\
	nan	nan	0	454\\
	8.501	14.501	0	455\\
	8.501	14.501	0	456\\
	9.499	14.501	0	457\\
	9.499	14.501	0	458\\
	nan	nan	0	459\\
	9.501	14.501	0	460\\
	9.501	14.501	0	461\\
	10.499	14.501	0	462\\
	10.499	14.501	0	463\\
	nan	nan	0	464\\
	10.501	14.501	0	465\\
	10.501	14.501	0	466\\
	11.499	14.501	0	467\\
	11.499	14.501	0	468\\
	nan	nan	0	469\\
	11.501	14.501	0	470\\
	11.501	14.501	0	471\\
	12.499	14.501	0	472\\
	12.499	14.501	0	473\\
	nan	nan	0	474\\
	12.501	14.501	0	475\\
	12.501	14.501	0	476\\
	13.499	14.501	0	477\\
	13.499	14.501	0	478\\
	nan	nan	0	479\\
	7.501	14.501	0	480\\
	7.501	14.501	0	481\\
	8.499	14.501	0	482\\
	8.499	14.501	0	483\\
	nan	nan	0	484\\
	8.501	14.501	0	485\\
	8.501	14.501	0	486\\
	9.499	14.501	0	487\\
	9.499	14.501	0	488\\
	nan	nan	0	489\\
	9.501	14.501	0	490\\
	9.501	14.501	11	491\\
	10.499	14.501	11	492\\
	10.499	14.501	0	493\\
	nan	nan	0	494\\
	10.501	14.501	0	495\\
	10.501	14.501	134	496\\
	11.499	14.501	134	497\\
	11.499	14.501	0	498\\
	nan	nan	0	499\\
	11.501	14.501	0	500\\
	11.501	14.501	858	501\\
	12.499	14.501	858	502\\
	12.499	14.501	0	503\\
	nan	nan	0	504\\
	12.501	14.501	0	505\\
	12.501	14.501	1302	506\\
	13.499	14.501	1302	507\\
	13.499	14.501	0	508\\
	nan	nan	0	509\\
	7.501	15.499	0	510\\
	7.501	15.499	0	511\\
	8.499	15.499	0	512\\
	8.499	15.499	0	513\\
	nan	nan	0	514\\
	8.501	15.499	0	515\\
	8.501	15.499	0	516\\
	9.499	15.499	0	517\\
	9.499	15.499	0	518\\
	nan	nan	0	519\\
	9.501	15.499	0	520\\
	9.501	15.499	11	521\\
	10.499	15.499	11	522\\
	10.499	15.499	0	523\\
	nan	nan	0	524\\
	10.501	15.499	0	525\\
	10.501	15.499	134	526\\
	11.499	15.499	134	527\\
	11.499	15.499	0	528\\
	nan	nan	0	529\\
	11.501	15.499	0	530\\
	11.501	15.499	858	531\\
	12.499	15.499	858	532\\
	12.499	15.499	0	533\\
	nan	nan	0	534\\
	12.501	15.499	0	535\\
	12.501	15.499	1302	536\\
	13.499	15.499	1302	537\\
	13.499	15.499	0	538\\
	nan	nan	0	539\\
	7.501	15.499	0	540\\
	7.501	15.499	0	541\\
	8.499	15.499	0	542\\
	8.499	15.499	0	543\\
	nan	nan	0	544\\
	8.501	15.499	0	545\\
	8.501	15.499	0	546\\
	9.499	15.499	0	547\\
	9.499	15.499	0	548\\
	nan	nan	0	549\\
	9.501	15.499	0	550\\
	9.501	15.499	0	551\\
	10.499	15.499	0	552\\
	10.499	15.499	0	553\\
	nan	nan	0	554\\
	10.501	15.499	0	555\\
	10.501	15.499	0	556\\
	11.499	15.499	0	557\\
	11.499	15.499	0	558\\
	nan	nan	0	559\\
	11.501	15.499	0	560\\
	11.501	15.499	0	561\\
	12.499	15.499	0	562\\
	12.499	15.499	0	563\\
	nan	nan	0	564\\
	12.501	15.499	0	565\\
	12.501	15.499	0	566\\
	13.499	15.499	0	567\\
	13.499	15.499	0	568\\
	nan	nan	0	569\\
	nan	nan	0	570\\
	nan	nan	0	571\\
	nan	nan	0	572\\
	nan	nan	0	573\\
	nan	nan	0	574\\
	nan	nan	0	575\\
	nan	nan	0	576\\
	nan	nan	0	577\\
	nan	nan	0	578\\
	nan	nan	0	579\\
	nan	nan	0	580\\
	nan	nan	0	581\\
	nan	nan	0	582\\
	nan	nan	0	583\\
	nan	nan	0	584\\
	nan	nan	0	585\\
	nan	nan	0	586\\
	nan	nan	0	587\\
	nan	nan	0	588\\
	nan	nan	0	589\\
	nan	nan	0	590\\
	nan	nan	0	591\\
	nan	nan	0	592\\
	nan	nan	0	593\\
	nan	nan	0	594\\
	nan	nan	0	595\\
	nan	nan	0	596\\
	nan	nan	0	597\\
	nan	nan	0	598\\
	nan	nan	0	599\\
	7.501	15.501	0	600\\
	7.501	15.501	0	601\\
	8.499	15.501	0	602\\
	8.499	15.501	0	603\\
	nan	nan	0	604\\
	8.501	15.501	0	605\\
	8.501	15.501	0	606\\
	9.499	15.501	0	607\\
	9.499	15.501	0	608\\
	nan	nan	0	609\\
	9.501	15.501	0	610\\
	9.501	15.501	0	611\\
	10.499	15.501	0	612\\
	10.499	15.501	0	613\\
	nan	nan	0	614\\
	10.501	15.501	0	615\\
	10.501	15.501	0	616\\
	11.499	15.501	0	617\\
	11.499	15.501	0	618\\
	nan	nan	0	619\\
	11.501	15.501	0	620\\
	11.501	15.501	0	621\\
	12.499	15.501	0	622\\
	12.499	15.501	0	623\\
	nan	nan	0	624\\
	12.501	15.501	0	625\\
	12.501	15.501	0	626\\
	13.499	15.501	0	627\\
	13.499	15.501	0	628\\
	nan	nan	0	629\\
	7.501	15.501	0	630\\
	7.501	15.501	0	631\\
	8.499	15.501	0	632\\
	8.499	15.501	0	633\\
	nan	nan	0	634\\
	8.501	15.501	0	635\\
	8.501	15.501	1	636\\
	9.499	15.501	1	637\\
	9.499	15.501	0	638\\
	nan	nan	0	639\\
	9.501	15.501	0	640\\
	9.501	15.501	55	641\\
	10.499	15.501	55	642\\
	10.499	15.501	0	643\\
	nan	nan	0	644\\
	10.501	15.501	0	645\\
	10.501	15.501	607	646\\
	11.499	15.501	607	647\\
	11.499	15.501	0	648\\
	nan	nan	0	649\\
	11.501	15.501	0	650\\
	11.501	15.501	1910	651\\
	12.499	15.501	1910	652\\
	12.499	15.501	0	653\\
	nan	nan	0	654\\
	12.501	15.501	0	655\\
	12.501	15.501	656	656\\
	13.499	15.501	656	657\\
	13.499	15.501	0	658\\
	nan	nan	0	659\\
	7.501	16.499	0	660\\
	7.501	16.499	0	661\\
	8.499	16.499	0	662\\
	8.499	16.499	0	663\\
	nan	nan	0	664\\
	8.501	16.499	0	665\\
	8.501	16.499	1	666\\
	9.499	16.499	1	667\\
	9.499	16.499	0	668\\
	nan	nan	0	669\\
	9.501	16.499	0	670\\
	9.501	16.499	55	671\\
	10.499	16.499	55	672\\
	10.499	16.499	0	673\\
	nan	nan	0	674\\
	10.501	16.499	0	675\\
	10.501	16.499	607	676\\
	11.499	16.499	607	677\\
	11.499	16.499	0	678\\
	nan	nan	0	679\\
	11.501	16.499	0	680\\
	11.501	16.499	1910	681\\
	12.499	16.499	1910	682\\
	12.499	16.499	0	683\\
	nan	nan	0	684\\
	12.501	16.499	0	685\\
	12.501	16.499	656	686\\
	13.499	16.499	656	687\\
	13.499	16.499	0	688\\
	nan	nan	0	689\\
	7.501	16.499	0	690\\
	7.501	16.499	0	691\\
	8.499	16.499	0	692\\
	8.499	16.499	0	693\\
	nan	nan	0	694\\
	8.501	16.499	0	695\\
	8.501	16.499	0	696\\
	9.499	16.499	0	697\\
	9.499	16.499	0	698\\
	nan	nan	0	699\\
	9.501	16.499	0	700\\
	9.501	16.499	0	701\\
	10.499	16.499	0	702\\
	10.499	16.499	0	703\\
	nan	nan	0	704\\
	10.501	16.499	0	705\\
	10.501	16.499	0	706\\
	11.499	16.499	0	707\\
	11.499	16.499	0	708\\
	nan	nan	0	709\\
	11.501	16.499	0	710\\
	11.501	16.499	0	711\\
	12.499	16.499	0	712\\
	12.499	16.499	0	713\\
	nan	nan	0	714\\
	12.501	16.499	0	715\\
	12.501	16.499	0	716\\
	13.499	16.499	0	717\\
	13.499	16.499	0	718\\
	nan	nan	0	719\\
	nan	nan	0	720\\
	nan	nan	0	721\\
	nan	nan	0	722\\
	nan	nan	0	723\\
	nan	nan	0	724\\
	nan	nan	0	725\\
	nan	nan	0	726\\
	nan	nan	0	727\\
	nan	nan	0	728\\
	nan	nan	0	729\\
	nan	nan	0	730\\
	nan	nan	0	731\\
	nan	nan	0	732\\
	nan	nan	0	733\\
	nan	nan	0	734\\
	nan	nan	0	735\\
	nan	nan	0	736\\
	nan	nan	0	737\\
	nan	nan	0	738\\
	nan	nan	0	739\\
	nan	nan	0	740\\
	nan	nan	0	741\\
	nan	nan	0	742\\
	nan	nan	0	743\\
	nan	nan	0	744\\
	nan	nan	0	745\\
	nan	nan	0	746\\
	nan	nan	0	747\\
	nan	nan	0	748\\
	nan	nan	0	749\\
	7.501	16.501	0	750\\
	7.501	16.501	0	751\\
	8.499	16.501	0	752\\
	8.499	16.501	0	753\\
	nan	nan	0	754\\
	8.501	16.501	0	755\\
	8.501	16.501	0	756\\
	9.499	16.501	0	757\\
	9.499	16.501	0	758\\
	nan	nan	0	759\\
	9.501	16.501	0	760\\
	9.501	16.501	0	761\\
	10.499	16.501	0	762\\
	10.499	16.501	0	763\\
	nan	nan	0	764\\
	10.501	16.501	0	765\\
	10.501	16.501	0	766\\
	11.499	16.501	0	767\\
	11.499	16.501	0	768\\
	nan	nan	0	769\\
	11.501	16.501	0	770\\
	11.501	16.501	0	771\\
	12.499	16.501	0	772\\
	12.499	16.501	0	773\\
	nan	nan	0	774\\
	12.501	16.501	0	775\\
	12.501	16.501	0	776\\
	13.499	16.501	0	777\\
	13.499	16.501	0	778\\
	nan	nan	0	779\\
	7.501	16.501	0	780\\
	7.501	16.501	1	781\\
	8.499	16.501	1	782\\
	8.499	16.501	0	783\\
	nan	nan	0	784\\
	8.501	16.501	0	785\\
	8.501	16.501	10	786\\
	9.499	16.501	10	787\\
	9.499	16.501	0	788\\
	nan	nan	0	789\\
	9.501	16.501	0	790\\
	9.501	16.501	163	791\\
	10.499	16.501	163	792\\
	10.499	16.501	0	793\\
	nan	nan	0	794\\
	10.501	16.501	0	795\\
	10.501	16.501	723	796\\
	11.499	16.501	723	797\\
	11.499	16.501	0	798\\
	nan	nan	0	799\\
	11.501	16.501	0	800\\
	11.501	16.501	522	801\\
	12.499	16.501	522	802\\
	12.499	16.501	0	803\\
	nan	nan	0	804\\
	12.501	16.501	0	805\\
	12.501	16.501	0	806\\
	13.499	16.501	0	807\\
	13.499	16.501	0	808\\
	nan	nan	0	809\\
	7.501	17.499	0	810\\
	7.501	17.499	1	811\\
	8.499	17.499	1	812\\
	8.499	17.499	0	813\\
	nan	nan	0	814\\
	8.501	17.499	0	815\\
	8.501	17.499	10	816\\
	9.499	17.499	10	817\\
	9.499	17.499	0	818\\
	nan	nan	0	819\\
	9.501	17.499	0	820\\
	9.501	17.499	163	821\\
	10.499	17.499	163	822\\
	10.499	17.499	0	823\\
	nan	nan	0	824\\
	10.501	17.499	0	825\\
	10.501	17.499	723	826\\
	11.499	17.499	723	827\\
	11.499	17.499	0	828\\
	nan	nan	0	829\\
	11.501	17.499	0	830\\
	11.501	17.499	522	831\\
	12.499	17.499	522	832\\
	12.499	17.499	0	833\\
	nan	nan	0	834\\
	12.501	17.499	0	835\\
	12.501	17.499	0	836\\
	13.499	17.499	0	837\\
	13.499	17.499	0	838\\
	nan	nan	0	839\\
	7.501	17.499	0	840\\
	7.501	17.499	0	841\\
	8.499	17.499	0	842\\
	8.499	17.499	0	843\\
	nan	nan	0	844\\
	8.501	17.499	0	845\\
	8.501	17.499	0	846\\
	9.499	17.499	0	847\\
	9.499	17.499	0	848\\
	nan	nan	0	849\\
	9.501	17.499	0	850\\
	9.501	17.499	0	851\\
	10.499	17.499	0	852\\
	10.499	17.499	0	853\\
	nan	nan	0	854\\
	10.501	17.499	0	855\\
	10.501	17.499	0	856\\
	11.499	17.499	0	857\\
	11.499	17.499	0	858\\
	nan	nan	0	859\\
	11.501	17.499	0	860\\
	11.501	17.499	0	861\\
	12.499	17.499	0	862\\
	12.499	17.499	0	863\\
	nan	nan	0	864\\
	12.501	17.499	0	865\\
	12.501	17.499	0	866\\
	13.499	17.499	0	867\\
	13.499	17.499	0	868\\
	nan	nan	0	869\\
	nan	nan	0	870\\
	nan	nan	0	871\\
	nan	nan	0	872\\
	nan	nan	0	873\\
	nan	nan	0	874\\
	nan	nan	0	875\\
	nan	nan	0	876\\
	nan	nan	0	877\\
	nan	nan	0	878\\
	nan	nan	0	879\\
	nan	nan	0	880\\
	nan	nan	0	881\\
	nan	nan	0	882\\
	nan	nan	0	883\\
	nan	nan	0	884\\
	nan	nan	0	885\\
	nan	nan	0	886\\
	nan	nan	0	887\\
	nan	nan	0	888\\
	nan	nan	0	889\\
	nan	nan	0	890\\
	nan	nan	0	891\\
	nan	nan	0	892\\
	nan	nan	0	893\\
	nan	nan	0	894\\
	nan	nan	0	895\\
	nan	nan	0	896\\
	nan	nan	0	897\\
	nan	nan	0	898\\
	nan	nan	0	899\\
};

\end{axis}

\end{tikzpicture}%

%% file: plots/Revised_Best_Card_Set_delta_d_4_SNR_30.tex
%
%
\begin{tikzpicture}

\begin{axis}[%
scale=0.6, width=3.284in,
height=3.566in,
at={(0.758in,0.481in)},
scale only axis,
unbounded coords=jump,
colormap={patchmap}{[1pt] rgb(0pt)=(0.75,0.85,0.95); rgb(899pt)=(0.75,0.85,0.95)},
xmin=7.2,
xmax=13.8,
xtick={ 8,  9, 10, 11, 12, 13},
xticklabel style = {font=\footnotesize},
tick align=outside,
xlabel style={font=\color{white!15!black}},
xlabel style = {font=\small},
xlabel={$|\mathcal{P}_{1}^{\star}|$},
ymin=11.2,
ymax=17.8,
ytick={12, 13, 14, 15, 16, 17},
yticklabel style = {font=\footnotesize},
ylabel style={font=\color{white!15!black}},
ylabel style = {font=\small},
ylabel={$|\mathcal{P}_{2}^{\star}|$},
zmin=0,
zmax=4200,
zlabel style={font=\color{white!15!black}},
zlabel style = {font=\small},
zticklabel style = {font=\footnotesize},
zlabel={Frequency},
view={-37.5}{30},
axis background/.style={fill=white},
title style={font=\bfseries},
axis x line*=bottom,
axis y line*=left,
axis z line*=left,
xmajorgrids,
ymajorgrids,
zmajorgrids,
legend style={at={(1.03,1)}, anchor=north west, legend cell align=left, align=left, draw=white!15!black}
]

\addplot3[%
surf,
shader=flat corner, draw=black, mesh/rows=30]
table[row sep=crcr, colormap name=surfmap, point meta=\thisrow{c}] {%
	x	y	z	c\\
	7.501	11.501	0	0\\
	7.501	11.501	0	1\\
	8.499	11.501	0	2\\
	8.499	11.501	0	3\\
	nan	nan	0	4\\
	8.501	11.501	0	5\\
	8.501	11.501	0	6\\
	9.499	11.501	0	7\\
	9.499	11.501	0	8\\
	nan	nan	0	9\\
	9.501	11.501	0	10\\
	9.501	11.501	0	11\\
	10.499	11.501	0	12\\
	10.499	11.501	0	13\\
	nan	nan	0	14\\
	10.501	11.501	0	15\\
	10.501	11.501	0	16\\
	11.499	11.501	0	17\\
	11.499	11.501	0	18\\
	nan	nan	0	19\\
	11.501	11.501	0	20\\
	11.501	11.501	0	21\\
	12.499	11.501	0	22\\
	12.499	11.501	0	23\\
	nan	nan	0	24\\
	12.501	11.501	0	25\\
	12.501	11.501	0	26\\
	13.499	11.501	0	27\\
	13.499	11.501	0	28\\
	nan	nan	0	29\\
	7.501	11.501	0	30\\
	7.501	11.501	0	31\\
	8.499	11.501	0	32\\
	8.499	11.501	0	33\\
	nan	nan	0	34\\
	8.501	11.501	0	35\\
	8.501	11.501	0	36\\
	9.499	11.501	0	37\\
	9.499	11.501	0	38\\
	nan	nan	0	39\\
	9.501	11.501	0	40\\
	9.501	11.501	0	41\\
	10.499	11.501	0	42\\
	10.499	11.501	0	43\\
	nan	nan	0	44\\
	10.501	11.501	0	45\\
	10.501	11.501	0	46\\
	11.499	11.501	0	47\\
	11.499	11.501	0	48\\
	nan	nan	0	49\\
	11.501	11.501	0	50\\
	11.501	11.501	0	51\\
	12.499	11.501	0	52\\
	12.499	11.501	0	53\\
	nan	nan	0	54\\
	12.501	11.501	0	55\\
	12.501	11.501	0	56\\
	13.499	11.501	0	57\\
	13.499	11.501	0	58\\
	nan	nan	0	59\\
	7.501	12.499	0	60\\
	7.501	12.499	0	61\\
	8.499	12.499	0	62\\
	8.499	12.499	0	63\\
	nan	nan	0	64\\
	8.501	12.499	0	65\\
	8.501	12.499	0	66\\
	9.499	12.499	0	67\\
	9.499	12.499	0	68\\
	nan	nan	0	69\\
	9.501	12.499	0	70\\
	9.501	12.499	0	71\\
	10.499	12.499	0	72\\
	10.499	12.499	0	73\\
	nan	nan	0	74\\
	10.501	12.499	0	75\\
	10.501	12.499	0	76\\
	11.499	12.499	0	77\\
	11.499	12.499	0	78\\
	nan	nan	0	79\\
	11.501	12.499	0	80\\
	11.501	12.499	0	81\\
	12.499	12.499	0	82\\
	12.499	12.499	0	83\\
	nan	nan	0	84\\
	12.501	12.499	0	85\\
	12.501	12.499	0	86\\
	13.499	12.499	0	87\\
	13.499	12.499	0	88\\
	nan	nan	0	89\\
	7.501	12.499	0	90\\
	7.501	12.499	0	91\\
	8.499	12.499	0	92\\
	8.499	12.499	0	93\\
	nan	nan	0	94\\
	8.501	12.499	0	95\\
	8.501	12.499	0	96\\
	9.499	12.499	0	97\\
	9.499	12.499	0	98\\
	nan	nan	0	99\\
	9.501	12.499	0	100\\
	9.501	12.499	0	101\\
	10.499	12.499	0	102\\
	10.499	12.499	0	103\\
	nan	nan	0	104\\
	10.501	12.499	0	105\\
	10.501	12.499	0	106\\
	11.499	12.499	0	107\\
	11.499	12.499	0	108\\
	nan	nan	0	109\\
	11.501	12.499	0	110\\
	11.501	12.499	0	111\\
	12.499	12.499	0	112\\
	12.499	12.499	0	113\\
	nan	nan	0	114\\
	12.501	12.499	0	115\\
	12.501	12.499	0	116\\
	13.499	12.499	0	117\\
	13.499	12.499	0	118\\
	nan	nan	0	119\\
	nan	nan	0	120\\
	nan	nan	0	121\\
	nan	nan	0	122\\
	nan	nan	0	123\\
	nan	nan	0	124\\
	nan	nan	0	125\\
	nan	nan	0	126\\
	nan	nan	0	127\\
	nan	nan	0	128\\
	nan	nan	0	129\\
	nan	nan	0	130\\
	nan	nan	0	131\\
	nan	nan	0	132\\
	nan	nan	0	133\\
	nan	nan	0	134\\
	nan	nan	0	135\\
	nan	nan	0	136\\
	nan	nan	0	137\\
	nan	nan	0	138\\
	nan	nan	0	139\\
	nan	nan	0	140\\
	nan	nan	0	141\\
	nan	nan	0	142\\
	nan	nan	0	143\\
	nan	nan	0	144\\
	nan	nan	0	145\\
	nan	nan	0	146\\
	nan	nan	0	147\\
	nan	nan	0	148\\
	nan	nan	0	149\\
	7.501	12.501	0	150\\
	7.501	12.501	0	151\\
	8.499	12.501	0	152\\
	8.499	12.501	0	153\\
	nan	nan	0	154\\
	8.501	12.501	0	155\\
	8.501	12.501	0	156\\
	9.499	12.501	0	157\\
	9.499	12.501	0	158\\
	nan	nan	0	159\\
	9.501	12.501	0	160\\
	9.501	12.501	0	161\\
	10.499	12.501	0	162\\
	10.499	12.501	0	163\\
	nan	nan	0	164\\
	10.501	12.501	0	165\\
	10.501	12.501	0	166\\
	11.499	12.501	0	167\\
	11.499	12.501	0	168\\
	nan	nan	0	169\\
	11.501	12.501	0	170\\
	11.501	12.501	0	171\\
	12.499	12.501	0	172\\
	12.499	12.501	0	173\\
	nan	nan	0	174\\
	12.501	12.501	0	175\\
	12.501	12.501	0	176\\
	13.499	12.501	0	177\\
	13.499	12.501	0	178\\
	nan	nan	0	179\\
	7.501	12.501	0	180\\
	7.501	12.501	0	181\\
	8.499	12.501	0	182\\
	8.499	12.501	0	183\\
	nan	nan	0	184\\
	8.501	12.501	0	185\\
	8.501	12.501	0	186\\
	9.499	12.501	0	187\\
	9.499	12.501	0	188\\
	nan	nan	0	189\\
	9.501	12.501	0	190\\
	9.501	12.501	0	191\\
	10.499	12.501	0	192\\
	10.499	12.501	0	193\\
	nan	nan	0	194\\
	10.501	12.501	0	195\\
	10.501	12.501	0	196\\
	11.499	12.501	0	197\\
	11.499	12.501	0	198\\
	nan	nan	0	199\\
	11.501	12.501	0	200\\
	11.501	12.501	0	201\\
	12.499	12.501	0	202\\
	12.499	12.501	0	203\\
	nan	nan	0	204\\
	12.501	12.501	0	205\\
	12.501	12.501	0	206\\
	13.499	12.501	0	207\\
	13.499	12.501	0	208\\
	nan	nan	0	209\\
	7.501	13.499	0	210\\
	7.501	13.499	0	211\\
	8.499	13.499	0	212\\
	8.499	13.499	0	213\\
	nan	nan	0	214\\
	8.501	13.499	0	215\\
	8.501	13.499	0	216\\
	9.499	13.499	0	217\\
	9.499	13.499	0	218\\
	nan	nan	0	219\\
	9.501	13.499	0	220\\
	9.501	13.499	0	221\\
	10.499	13.499	0	222\\
	10.499	13.499	0	223\\
	nan	nan	0	224\\
	10.501	13.499	0	225\\
	10.501	13.499	0	226\\
	11.499	13.499	0	227\\
	11.499	13.499	0	228\\
	nan	nan	0	229\\
	11.501	13.499	0	230\\
	11.501	13.499	0	231\\
	12.499	13.499	0	232\\
	12.499	13.499	0	233\\
	nan	nan	0	234\\
	12.501	13.499	0	235\\
	12.501	13.499	0	236\\
	13.499	13.499	0	237\\
	13.499	13.499	0	238\\
	nan	nan	0	239\\
	7.501	13.499	0	240\\
	7.501	13.499	0	241\\
	8.499	13.499	0	242\\
	8.499	13.499	0	243\\
	nan	nan	0	244\\
	8.501	13.499	0	245\\
	8.501	13.499	0	246\\
	9.499	13.499	0	247\\
	9.499	13.499	0	248\\
	nan	nan	0	249\\
	9.501	13.499	0	250\\
	9.501	13.499	0	251\\
	10.499	13.499	0	252\\
	10.499	13.499	0	253\\
	nan	nan	0	254\\
	10.501	13.499	0	255\\
	10.501	13.499	0	256\\
	11.499	13.499	0	257\\
	11.499	13.499	0	258\\
	nan	nan	0	259\\
	11.501	13.499	0	260\\
	11.501	13.499	0	261\\
	12.499	13.499	0	262\\
	12.499	13.499	0	263\\
	nan	nan	0	264\\
	12.501	13.499	0	265\\
	12.501	13.499	0	266\\
	13.499	13.499	0	267\\
	13.499	13.499	0	268\\
	nan	nan	0	269\\
	nan	nan	0	270\\
	nan	nan	0	271\\
	nan	nan	0	272\\
	nan	nan	0	273\\
	nan	nan	0	274\\
	nan	nan	0	275\\
	nan	nan	0	276\\
	nan	nan	0	277\\
	nan	nan	0	278\\
	nan	nan	0	279\\
	nan	nan	0	280\\
	nan	nan	0	281\\
	nan	nan	0	282\\
	nan	nan	0	283\\
	nan	nan	0	284\\
	nan	nan	0	285\\
	nan	nan	0	286\\
	nan	nan	0	287\\
	nan	nan	0	288\\
	nan	nan	0	289\\
	nan	nan	0	290\\
	nan	nan	0	291\\
	nan	nan	0	292\\
	nan	nan	0	293\\
	nan	nan	0	294\\
	nan	nan	0	295\\
	nan	nan	0	296\\
	nan	nan	0	297\\
	nan	nan	0	298\\
	nan	nan	0	299\\
	7.501	13.501	0	300\\
	7.501	13.501	0	301\\
	8.499	13.501	0	302\\
	8.499	13.501	0	303\\
	nan	nan	0	304\\
	8.501	13.501	0	305\\
	8.501	13.501	0	306\\
	9.499	13.501	0	307\\
	9.499	13.501	0	308\\
	nan	nan	0	309\\
	9.501	13.501	0	310\\
	9.501	13.501	0	311\\
	10.499	13.501	0	312\\
	10.499	13.501	0	313\\
	nan	nan	0	314\\
	10.501	13.501	0	315\\
	10.501	13.501	0	316\\
	11.499	13.501	0	317\\
	11.499	13.501	0	318\\
	nan	nan	0	319\\
	11.501	13.501	0	320\\
	11.501	13.501	0	321\\
	12.499	13.501	0	322\\
	12.499	13.501	0	323\\
	nan	nan	0	324\\
	12.501	13.501	0	325\\
	12.501	13.501	0	326\\
	13.499	13.501	0	327\\
	13.499	13.501	0	328\\
	nan	nan	0	329\\
	7.501	13.501	0	330\\
	7.501	13.501	0	331\\
	8.499	13.501	0	332\\
	8.499	13.501	0	333\\
	nan	nan	0	334\\
	8.501	13.501	0	335\\
	8.501	13.501	0	336\\
	9.499	13.501	0	337\\
	9.499	13.501	0	338\\
	nan	nan	0	339\\
	9.501	13.501	0	340\\
	9.501	13.501	0	341\\
	10.499	13.501	0	342\\
	10.499	13.501	0	343\\
	nan	nan	0	344\\
	10.501	13.501	0	345\\
	10.501	13.501	0	346\\
	11.499	13.501	0	347\\
	11.499	13.501	0	348\\
	nan	nan	0	349\\
	11.501	13.501	0	350\\
	11.501	13.501	0	351\\
	12.499	13.501	0	352\\
	12.499	13.501	0	353\\
	nan	nan	0	354\\
	12.501	13.501	0	355\\
	12.501	13.501	0	356\\
	13.499	13.501	0	357\\
	13.499	13.501	0	358\\
	nan	nan	0	359\\
	7.501	14.499	0	360\\
	7.501	14.499	0	361\\
	8.499	14.499	0	362\\
	8.499	14.499	0	363\\
	nan	nan	0	364\\
	8.501	14.499	0	365\\
	8.501	14.499	0	366\\
	9.499	14.499	0	367\\
	9.499	14.499	0	368\\
	nan	nan	0	369\\
	9.501	14.499	0	370\\
	9.501	14.499	0	371\\
	10.499	14.499	0	372\\
	10.499	14.499	0	373\\
	nan	nan	0	374\\
	10.501	14.499	0	375\\
	10.501	14.499	0	376\\
	11.499	14.499	0	377\\
	11.499	14.499	0	378\\
	nan	nan	0	379\\
	11.501	14.499	0	380\\
	11.501	14.499	0	381\\
	12.499	14.499	0	382\\
	12.499	14.499	0	383\\
	nan	nan	0	384\\
	12.501	14.499	0	385\\
	12.501	14.499	0	386\\
	13.499	14.499	0	387\\
	13.499	14.499	0	388\\
	nan	nan	0	389\\
	7.501	14.499	0	390\\
	7.501	14.499	0	391\\
	8.499	14.499	0	392\\
	8.499	14.499	0	393\\
	nan	nan	0	394\\
	8.501	14.499	0	395\\
	8.501	14.499	0	396\\
	9.499	14.499	0	397\\
	9.499	14.499	0	398\\
	nan	nan	0	399\\
	9.501	14.499	0	400\\
	9.501	14.499	0	401\\
	10.499	14.499	0	402\\
	10.499	14.499	0	403\\
	nan	nan	0	404\\
	10.501	14.499	0	405\\
	10.501	14.499	0	406\\
	11.499	14.499	0	407\\
	11.499	14.499	0	408\\
	nan	nan	0	409\\
	11.501	14.499	0	410\\
	11.501	14.499	0	411\\
	12.499	14.499	0	412\\
	12.499	14.499	0	413\\
	nan	nan	0	414\\
	12.501	14.499	0	415\\
	12.501	14.499	0	416\\
	13.499	14.499	0	417\\
	13.499	14.499	0	418\\
	nan	nan	0	419\\
	nan	nan	0	420\\
	nan	nan	0	421\\
	nan	nan	0	422\\
	nan	nan	0	423\\
	nan	nan	0	424\\
	nan	nan	0	425\\
	nan	nan	0	426\\
	nan	nan	0	427\\
	nan	nan	0	428\\
	nan	nan	0	429\\
	nan	nan	0	430\\
	nan	nan	0	431\\
	nan	nan	0	432\\
	nan	nan	0	433\\
	nan	nan	0	434\\
	nan	nan	0	435\\
	nan	nan	0	436\\
	nan	nan	0	437\\
	nan	nan	0	438\\
	nan	nan	0	439\\
	nan	nan	0	440\\
	nan	nan	0	441\\
	nan	nan	0	442\\
	nan	nan	0	443\\
	nan	nan	0	444\\
	nan	nan	0	445\\
	nan	nan	0	446\\
	nan	nan	0	447\\
	nan	nan	0	448\\
	nan	nan	0	449\\
	7.501	14.501	0	450\\
	7.501	14.501	0	451\\
	8.499	14.501	0	452\\
	8.499	14.501	0	453\\
	nan	nan	0	454\\
	8.501	14.501	0	455\\
	8.501	14.501	0	456\\
	9.499	14.501	0	457\\
	9.499	14.501	0	458\\
	nan	nan	0	459\\
	9.501	14.501	0	460\\
	9.501	14.501	0	461\\
	10.499	14.501	0	462\\
	10.499	14.501	0	463\\
	nan	nan	0	464\\
	10.501	14.501	0	465\\
	10.501	14.501	0	466\\
	11.499	14.501	0	467\\
	11.499	14.501	0	468\\
	nan	nan	0	469\\
	11.501	14.501	0	470\\
	11.501	14.501	0	471\\
	12.499	14.501	0	472\\
	12.499	14.501	0	473\\
	nan	nan	0	474\\
	12.501	14.501	0	475\\
	12.501	14.501	0	476\\
	13.499	14.501	0	477\\
	13.499	14.501	0	478\\
	nan	nan	0	479\\
	7.501	14.501	0	480\\
	7.501	14.501	0	481\\
	8.499	14.501	0	482\\
	8.499	14.501	0	483\\
	nan	nan	0	484\\
	8.501	14.501	0	485\\
	8.501	14.501	0	486\\
	9.499	14.501	0	487\\
	9.499	14.501	0	488\\
	nan	nan	0	489\\
	9.501	14.501	0	490\\
	9.501	14.501	0	491\\
	10.499	14.501	0	492\\
	10.499	14.501	0	493\\
	nan	nan	0	494\\
	10.501	14.501	0	495\\
	10.501	14.501	0	496\\
	11.499	14.501	0	497\\
	11.499	14.501	0	498\\
	nan	nan	0	499\\
	11.501	14.501	0	500\\
	11.501	14.501	0	501\\
	12.499	14.501	0	502\\
	12.499	14.501	0	503\\
	nan	nan	0	504\\
	12.501	14.501	0	505\\
	12.501	14.501	29	506\\
	13.499	14.501	29	507\\
	13.499	14.501	0	508\\
	nan	nan	0	509\\
	7.501	15.499	0	510\\
	7.501	15.499	0	511\\
	8.499	15.499	0	512\\
	8.499	15.499	0	513\\
	nan	nan	0	514\\
	8.501	15.499	0	515\\
	8.501	15.499	0	516\\
	9.499	15.499	0	517\\
	9.499	15.499	0	518\\
	nan	nan	0	519\\
	9.501	15.499	0	520\\
	9.501	15.499	0	521\\
	10.499	15.499	0	522\\
	10.499	15.499	0	523\\
	nan	nan	0	524\\
	10.501	15.499	0	525\\
	10.501	15.499	0	526\\
	11.499	15.499	0	527\\
	11.499	15.499	0	528\\
	nan	nan	0	529\\
	11.501	15.499	0	530\\
	11.501	15.499	0	531\\
	12.499	15.499	0	532\\
	12.499	15.499	0	533\\
	nan	nan	0	534\\
	12.501	15.499	0	535\\
	12.501	15.499	29	536\\
	13.499	15.499	29	537\\
	13.499	15.499	0	538\\
	nan	nan	0	539\\
	7.501	15.499	0	540\\
	7.501	15.499	0	541\\
	8.499	15.499	0	542\\
	8.499	15.499	0	543\\
	nan	nan	0	544\\
	8.501	15.499	0	545\\
	8.501	15.499	0	546\\
	9.499	15.499	0	547\\
	9.499	15.499	0	548\\
	nan	nan	0	549\\
	9.501	15.499	0	550\\
	9.501	15.499	0	551\\
	10.499	15.499	0	552\\
	10.499	15.499	0	553\\
	nan	nan	0	554\\
	10.501	15.499	0	555\\
	10.501	15.499	0	556\\
	11.499	15.499	0	557\\
	11.499	15.499	0	558\\
	nan	nan	0	559\\
	11.501	15.499	0	560\\
	11.501	15.499	0	561\\
	12.499	15.499	0	562\\
	12.499	15.499	0	563\\
	nan	nan	0	564\\
	12.501	15.499	0	565\\
	12.501	15.499	0	566\\
	13.499	15.499	0	567\\
	13.499	15.499	0	568\\
	nan	nan	0	569\\
	nan	nan	0	570\\
	nan	nan	0	571\\
	nan	nan	0	572\\
	nan	nan	0	573\\
	nan	nan	0	574\\
	nan	nan	0	575\\
	nan	nan	0	576\\
	nan	nan	0	577\\
	nan	nan	0	578\\
	nan	nan	0	579\\
	nan	nan	0	580\\
	nan	nan	0	581\\
	nan	nan	0	582\\
	nan	nan	0	583\\
	nan	nan	0	584\\
	nan	nan	0	585\\
	nan	nan	0	586\\
	nan	nan	0	587\\
	nan	nan	0	588\\
	nan	nan	0	589\\
	nan	nan	0	590\\
	nan	nan	0	591\\
	nan	nan	0	592\\
	nan	nan	0	593\\
	nan	nan	0	594\\
	nan	nan	0	595\\
	nan	nan	0	596\\
	nan	nan	0	597\\
	nan	nan	0	598\\
	nan	nan	0	599\\
	7.501	15.501	0	600\\
	7.501	15.501	0	601\\
	8.499	15.501	0	602\\
	8.499	15.501	0	603\\
	nan	nan	0	604\\
	8.501	15.501	0	605\\
	8.501	15.501	0	606\\
	9.499	15.501	0	607\\
	9.499	15.501	0	608\\
	nan	nan	0	609\\
	9.501	15.501	0	610\\
	9.501	15.501	0	611\\
	10.499	15.501	0	612\\
	10.499	15.501	0	613\\
	nan	nan	0	614\\
	10.501	15.501	0	615\\
	10.501	15.501	0	616\\
	11.499	15.501	0	617\\
	11.499	15.501	0	618\\
	nan	nan	0	619\\
	11.501	15.501	0	620\\
	11.501	15.501	0	621\\
	12.499	15.501	0	622\\
	12.499	15.501	0	623\\
	nan	nan	0	624\\
	12.501	15.501	0	625\\
	12.501	15.501	0	626\\
	13.499	15.501	0	627\\
	13.499	15.501	0	628\\
	nan	nan	0	629\\
	7.501	15.501	0	630\\
	7.501	15.501	0	631\\
	8.499	15.501	0	632\\
	8.499	15.501	0	633\\
	nan	nan	0	634\\
	8.501	15.501	0	635\\
	8.501	15.501	0	636\\
	9.499	15.501	0	637\\
	9.499	15.501	0	638\\
	nan	nan	0	639\\
	9.501	15.501	0	640\\
	9.501	15.501	0	641\\
	10.499	15.501	0	642\\
	10.499	15.501	0	643\\
	nan	nan	0	644\\
	10.501	15.501	0	645\\
	10.501	15.501	0	646\\
	11.499	15.501	0	647\\
	11.499	15.501	0	648\\
	nan	nan	0	649\\
	11.501	15.501	0	650\\
	11.501	15.501	40	651\\
	12.499	15.501	40	652\\
	12.499	15.501	0	653\\
	nan	nan	0	654\\
	12.501	15.501	0	655\\
	12.501	15.501	4160	656\\
	13.499	15.501	4160	657\\
	13.499	15.501	0	658\\
	nan	nan	0	659\\
	7.501	16.499	0	660\\
	7.501	16.499	0	661\\
	8.499	16.499	0	662\\
	8.499	16.499	0	663\\
	nan	nan	0	664\\
	8.501	16.499	0	665\\
	8.501	16.499	0	666\\
	9.499	16.499	0	667\\
	9.499	16.499	0	668\\
	nan	nan	0	669\\
	9.501	16.499	0	670\\
	9.501	16.499	0	671\\
	10.499	16.499	0	672\\
	10.499	16.499	0	673\\
	nan	nan	0	674\\
	10.501	16.499	0	675\\
	10.501	16.499	0	676\\
	11.499	16.499	0	677\\
	11.499	16.499	0	678\\
	nan	nan	0	679\\
	11.501	16.499	0	680\\
	11.501	16.499	40	681\\
	12.499	16.499	40	682\\
	12.499	16.499	0	683\\
	nan	nan	0	684\\
	12.501	16.499	0	685\\
	12.501	16.499	4160	686\\
	13.499	16.499	4160	687\\
	13.499	16.499	0	688\\
	nan	nan	0	689\\
	7.501	16.499	0	690\\
	7.501	16.499	0	691\\
	8.499	16.499	0	692\\
	8.499	16.499	0	693\\
	nan	nan	0	694\\
	8.501	16.499	0	695\\
	8.501	16.499	0	696\\
	9.499	16.499	0	697\\
	9.499	16.499	0	698\\
	nan	nan	0	699\\
	9.501	16.499	0	700\\
	9.501	16.499	0	701\\
	10.499	16.499	0	702\\
	10.499	16.499	0	703\\
	nan	nan	0	704\\
	10.501	16.499	0	705\\
	10.501	16.499	0	706\\
	11.499	16.499	0	707\\
	11.499	16.499	0	708\\
	nan	nan	0	709\\
	11.501	16.499	0	710\\
	11.501	16.499	0	711\\
	12.499	16.499	0	712\\
	12.499	16.499	0	713\\
	nan	nan	0	714\\
	12.501	16.499	0	715\\
	12.501	16.499	0	716\\
	13.499	16.499	0	717\\
	13.499	16.499	0	718\\
	nan	nan	0	719\\
	nan	nan	0	720\\
	nan	nan	0	721\\
	nan	nan	0	722\\
	nan	nan	0	723\\
	nan	nan	0	724\\
	nan	nan	0	725\\
	nan	nan	0	726\\
	nan	nan	0	727\\
	nan	nan	0	728\\
	nan	nan	0	729\\
	nan	nan	0	730\\
	nan	nan	0	731\\
	nan	nan	0	732\\
	nan	nan	0	733\\
	nan	nan	0	734\\
	nan	nan	0	735\\
	nan	nan	0	736\\
	nan	nan	0	737\\
	nan	nan	0	738\\
	nan	nan	0	739\\
	nan	nan	0	740\\
	nan	nan	0	741\\
	nan	nan	0	742\\
	nan	nan	0	743\\
	nan	nan	0	744\\
	nan	nan	0	745\\
	nan	nan	0	746\\
	nan	nan	0	747\\
	nan	nan	0	748\\
	nan	nan	0	749\\
	7.501	16.501	0	750\\
	7.501	16.501	0	751\\
	8.499	16.501	0	752\\
	8.499	16.501	0	753\\
	nan	nan	0	754\\
	8.501	16.501	0	755\\
	8.501	16.501	0	756\\
	9.499	16.501	0	757\\
	9.499	16.501	0	758\\
	nan	nan	0	759\\
	9.501	16.501	0	760\\
	9.501	16.501	0	761\\
	10.499	16.501	0	762\\
	10.499	16.501	0	763\\
	nan	nan	0	764\\
	10.501	16.501	0	765\\
	10.501	16.501	0	766\\
	11.499	16.501	0	767\\
	11.499	16.501	0	768\\
	nan	nan	0	769\\
	11.501	16.501	0	770\\
	11.501	16.501	0	771\\
	12.499	16.501	0	772\\
	12.499	16.501	0	773\\
	nan	nan	0	774\\
	12.501	16.501	0	775\\
	12.501	16.501	0	776\\
	13.499	16.501	0	777\\
	13.499	16.501	0	778\\
	nan	nan	0	779\\
	7.501	16.501	0	780\\
	7.501	16.501	0	781\\
	8.499	16.501	0	782\\
	8.499	16.501	0	783\\
	nan	nan	0	784\\
	8.501	16.501	0	785\\
	8.501	16.501	0	786\\
	9.499	16.501	0	787\\
	9.499	16.501	0	788\\
	nan	nan	0	789\\
	9.501	16.501	0	790\\
	9.501	16.501	0	791\\
	10.499	16.501	0	792\\
	10.499	16.501	0	793\\
	nan	nan	0	794\\
	10.501	16.501	0	795\\
	10.501	16.501	17	796\\
	11.499	16.501	17	797\\
	11.499	16.501	0	798\\
	nan	nan	0	799\\
	11.501	16.501	0	800\\
	11.501	16.501	3254	801\\
	12.499	16.501	3254	802\\
	12.499	16.501	0	803\\
	nan	nan	0	804\\
	12.501	16.501	0	805\\
	12.501	16.501	0	806\\
	13.499	16.501	0	807\\
	13.499	16.501	0	808\\
	nan	nan	0	809\\
	7.501	17.499	0	810\\
	7.501	17.499	0	811\\
	8.499	17.499	0	812\\
	8.499	17.499	0	813\\
	nan	nan	0	814\\
	8.501	17.499	0	815\\
	8.501	17.499	0	816\\
	9.499	17.499	0	817\\
	9.499	17.499	0	818\\
	nan	nan	0	819\\
	9.501	17.499	0	820\\
	9.501	17.499	0	821\\
	10.499	17.499	0	822\\
	10.499	17.499	0	823\\
	nan	nan	0	824\\
	10.501	17.499	0	825\\
	10.501	17.499	17	826\\
	11.499	17.499	17	827\\
	11.499	17.499	0	828\\
	nan	nan	0	829\\
	11.501	17.499	0	830\\
	11.501	17.499	3254	831\\
	12.499	17.499	3254	832\\
	12.499	17.499	0	833\\
	nan	nan	0	834\\
	12.501	17.499	0	835\\
	12.501	17.499	0	836\\
	13.499	17.499	0	837\\
	13.499	17.499	0	838\\
	nan	nan	0	839\\
	7.501	17.499	0	840\\
	7.501	17.499	0	841\\
	8.499	17.499	0	842\\
	8.499	17.499	0	843\\
	nan	nan	0	844\\
	8.501	17.499	0	845\\
	8.501	17.499	0	846\\
	9.499	17.499	0	847\\
	9.499	17.499	0	848\\
	nan	nan	0	849\\
	9.501	17.499	0	850\\
	9.501	17.499	0	851\\
	10.499	17.499	0	852\\
	10.499	17.499	0	853\\
	nan	nan	0	854\\
	10.501	17.499	0	855\\
	10.501	17.499	0	856\\
	11.499	17.499	0	857\\
	11.499	17.499	0	858\\
	nan	nan	0	859\\
	11.501	17.499	0	860\\
	11.501	17.499	0	861\\
	12.499	17.499	0	862\\
	12.499	17.499	0	863\\
	nan	nan	0	864\\
	12.501	17.499	0	865\\
	12.501	17.499	0	866\\
	13.499	17.499	0	867\\
	13.499	17.499	0	868\\
	nan	nan	0	869\\
	nan	nan	0	870\\
	nan	nan	0	871\\
	nan	nan	0	872\\
	nan	nan	0	873\\
	nan	nan	0	874\\
	nan	nan	0	875\\
	nan	nan	0	876\\
	nan	nan	0	877\\
	nan	nan	0	878\\
	nan	nan	0	879\\
	nan	nan	0	880\\
	nan	nan	0	881\\
	nan	nan	0	882\\
	nan	nan	0	883\\
	nan	nan	0	884\\
	nan	nan	0	885\\
	nan	nan	0	886\\
	nan	nan	0	887\\
	nan	nan	0	888\\
	nan	nan	0	889\\
	nan	nan	0	890\\
	nan	nan	0	891\\
	nan	nan	0	892\\
	nan	nan	0	893\\
	nan	nan	0	894\\
	nan	nan	0	895\\
	nan	nan	0	896\\
	nan	nan	0	897\\
	nan	nan	0	898\\
	nan	nan	0	899\\
};

\end{axis}

\end{tikzpicture}%

%% file: plots/Revised_Best_Card_Set_delta_d_4_SNR_50.tex
%
%
\begin{tikzpicture}

\begin{axis}[%
scale=0.6, width=3.284in,
height=3.566in,
at={(0.758in,0.481in)},
scale only axis,
unbounded coords=jump,
colormap={patchmap}{[1pt] rgb(0pt)=(0.75,0.85,0.95); rgb(899pt)=(0.75,0.85,0.95)},
xmin=7.2,
xmax=13.8,
xtick={ 8,  9, 10, 11, 12, 13},
tick align=outside,
xticklabel style = {font=\footnotesize},
xlabel style={font=\color{white!15!black}},
xlabel style = {font=\small},
xlabel={$|\mathcal{P}_{1}^{\star}|$},
ymin=11.2,
ymax=17.8,
ytick={12, 13, 14, 15, 16, 17},
yticklabel style = {font=\footnotesize},
ylabel style={font=\color{white!15!black}},
ylabel style = {font=\small},
ylabel={$|\mathcal{P}_{2}^{\star}|$},
zmin=0,
zmax=4300,
zlabel style={font=\color{white!15!black}},
zlabel style = {font=\small},
zticklabel style = {font=\footnotesize},
zlabel={Frequency},
view={-37.5}{30},
axis background/.style={fill=white},
axis x line*=bottom,
axis y line*=left,
axis z line*=left,
xmajorgrids,
ymajorgrids,
zmajorgrids,
legend style={at={(1.03,1)}, anchor=north west, legend cell align=left, align=left, draw=white!15!black}
]

\addplot3[%
surf,
shader=flat corner, draw=black, mesh/rows=30]
table[row sep=crcr, colormap name=surfmap, point meta=\thisrow{c}] {%
	x	y	z	c\\
	7.501	11.501	0	0\\
	7.501	11.501	0	1\\
	8.499	11.501	0	2\\
	8.499	11.501	0	3\\
	nan	nan	0	4\\
	8.501	11.501	0	5\\
	8.501	11.501	0	6\\
	9.499	11.501	0	7\\
	9.499	11.501	0	8\\
	nan	nan	0	9\\
	9.501	11.501	0	10\\
	9.501	11.501	0	11\\
	10.499	11.501	0	12\\
	10.499	11.501	0	13\\
	nan	nan	0	14\\
	10.501	11.501	0	15\\
	10.501	11.501	0	16\\
	11.499	11.501	0	17\\
	11.499	11.501	0	18\\
	nan	nan	0	19\\
	11.501	11.501	0	20\\
	11.501	11.501	0	21\\
	12.499	11.501	0	22\\
	12.499	11.501	0	23\\
	nan	nan	0	24\\
	12.501	11.501	0	25\\
	12.501	11.501	0	26\\
	13.499	11.501	0	27\\
	13.499	11.501	0	28\\
	nan	nan	0	29\\
	7.501	11.501	0	30\\
	7.501	11.501	0	31\\
	8.499	11.501	0	32\\
	8.499	11.501	0	33\\
	nan	nan	0	34\\
	8.501	11.501	0	35\\
	8.501	11.501	0	36\\
	9.499	11.501	0	37\\
	9.499	11.501	0	38\\
	nan	nan	0	39\\
	9.501	11.501	0	40\\
	9.501	11.501	0	41\\
	10.499	11.501	0	42\\
	10.499	11.501	0	43\\
	nan	nan	0	44\\
	10.501	11.501	0	45\\
	10.501	11.501	0	46\\
	11.499	11.501	0	47\\
	11.499	11.501	0	48\\
	nan	nan	0	49\\
	11.501	11.501	0	50\\
	11.501	11.501	0	51\\
	12.499	11.501	0	52\\
	12.499	11.501	0	53\\
	nan	nan	0	54\\
	12.501	11.501	0	55\\
	12.501	11.501	0	56\\
	13.499	11.501	0	57\\
	13.499	11.501	0	58\\
	nan	nan	0	59\\
	7.501	12.499	0	60\\
	7.501	12.499	0	61\\
	8.499	12.499	0	62\\
	8.499	12.499	0	63\\
	nan	nan	0	64\\
	8.501	12.499	0	65\\
	8.501	12.499	0	66\\
	9.499	12.499	0	67\\
	9.499	12.499	0	68\\
	nan	nan	0	69\\
	9.501	12.499	0	70\\
	9.501	12.499	0	71\\
	10.499	12.499	0	72\\
	10.499	12.499	0	73\\
	nan	nan	0	74\\
	10.501	12.499	0	75\\
	10.501	12.499	0	76\\
	11.499	12.499	0	77\\
	11.499	12.499	0	78\\
	nan	nan	0	79\\
	11.501	12.499	0	80\\
	11.501	12.499	0	81\\
	12.499	12.499	0	82\\
	12.499	12.499	0	83\\
	nan	nan	0	84\\
	12.501	12.499	0	85\\
	12.501	12.499	0	86\\
	13.499	12.499	0	87\\
	13.499	12.499	0	88\\
	nan	nan	0	89\\
	7.501	12.499	0	90\\
	7.501	12.499	0	91\\
	8.499	12.499	0	92\\
	8.499	12.499	0	93\\
	nan	nan	0	94\\
	8.501	12.499	0	95\\
	8.501	12.499	0	96\\
	9.499	12.499	0	97\\
	9.499	12.499	0	98\\
	nan	nan	0	99\\
	9.501	12.499	0	100\\
	9.501	12.499	0	101\\
	10.499	12.499	0	102\\
	10.499	12.499	0	103\\
	nan	nan	0	104\\
	10.501	12.499	0	105\\
	10.501	12.499	0	106\\
	11.499	12.499	0	107\\
	11.499	12.499	0	108\\
	nan	nan	0	109\\
	11.501	12.499	0	110\\
	11.501	12.499	0	111\\
	12.499	12.499	0	112\\
	12.499	12.499	0	113\\
	nan	nan	0	114\\
	12.501	12.499	0	115\\
	12.501	12.499	0	116\\
	13.499	12.499	0	117\\
	13.499	12.499	0	118\\
	nan	nan	0	119\\
	nan	nan	0	120\\
	nan	nan	0	121\\
	nan	nan	0	122\\
	nan	nan	0	123\\
	nan	nan	0	124\\
	nan	nan	0	125\\
	nan	nan	0	126\\
	nan	nan	0	127\\
	nan	nan	0	128\\
	nan	nan	0	129\\
	nan	nan	0	130\\
	nan	nan	0	131\\
	nan	nan	0	132\\
	nan	nan	0	133\\
	nan	nan	0	134\\
	nan	nan	0	135\\
	nan	nan	0	136\\
	nan	nan	0	137\\
	nan	nan	0	138\\
	nan	nan	0	139\\
	nan	nan	0	140\\
	nan	nan	0	141\\
	nan	nan	0	142\\
	nan	nan	0	143\\
	nan	nan	0	144\\
	nan	nan	0	145\\
	nan	nan	0	146\\
	nan	nan	0	147\\
	nan	nan	0	148\\
	nan	nan	0	149\\
	7.501	12.501	0	150\\
	7.501	12.501	0	151\\
	8.499	12.501	0	152\\
	8.499	12.501	0	153\\
	nan	nan	0	154\\
	8.501	12.501	0	155\\
	8.501	12.501	0	156\\
	9.499	12.501	0	157\\
	9.499	12.501	0	158\\
	nan	nan	0	159\\
	9.501	12.501	0	160\\
	9.501	12.501	0	161\\
	10.499	12.501	0	162\\
	10.499	12.501	0	163\\
	nan	nan	0	164\\
	10.501	12.501	0	165\\
	10.501	12.501	0	166\\
	11.499	12.501	0	167\\
	11.499	12.501	0	168\\
	nan	nan	0	169\\
	11.501	12.501	0	170\\
	11.501	12.501	0	171\\
	12.499	12.501	0	172\\
	12.499	12.501	0	173\\
	nan	nan	0	174\\
	12.501	12.501	0	175\\
	12.501	12.501	0	176\\
	13.499	12.501	0	177\\
	13.499	12.501	0	178\\
	nan	nan	0	179\\
	7.501	12.501	0	180\\
	7.501	12.501	0	181\\
	8.499	12.501	0	182\\
	8.499	12.501	0	183\\
	nan	nan	0	184\\
	8.501	12.501	0	185\\
	8.501	12.501	0	186\\
	9.499	12.501	0	187\\
	9.499	12.501	0	188\\
	nan	nan	0	189\\
	9.501	12.501	0	190\\
	9.501	12.501	0	191\\
	10.499	12.501	0	192\\
	10.499	12.501	0	193\\
	nan	nan	0	194\\
	10.501	12.501	0	195\\
	10.501	12.501	0	196\\
	11.499	12.501	0	197\\
	11.499	12.501	0	198\\
	nan	nan	0	199\\
	11.501	12.501	0	200\\
	11.501	12.501	0	201\\
	12.499	12.501	0	202\\
	12.499	12.501	0	203\\
	nan	nan	0	204\\
	12.501	12.501	0	205\\
	12.501	12.501	0	206\\
	13.499	12.501	0	207\\
	13.499	12.501	0	208\\
	nan	nan	0	209\\
	7.501	13.499	0	210\\
	7.501	13.499	0	211\\
	8.499	13.499	0	212\\
	8.499	13.499	0	213\\
	nan	nan	0	214\\
	8.501	13.499	0	215\\
	8.501	13.499	0	216\\
	9.499	13.499	0	217\\
	9.499	13.499	0	218\\
	nan	nan	0	219\\
	9.501	13.499	0	220\\
	9.501	13.499	0	221\\
	10.499	13.499	0	222\\
	10.499	13.499	0	223\\
	nan	nan	0	224\\
	10.501	13.499	0	225\\
	10.501	13.499	0	226\\
	11.499	13.499	0	227\\
	11.499	13.499	0	228\\
	nan	nan	0	229\\
	11.501	13.499	0	230\\
	11.501	13.499	0	231\\
	12.499	13.499	0	232\\
	12.499	13.499	0	233\\
	nan	nan	0	234\\
	12.501	13.499	0	235\\
	12.501	13.499	0	236\\
	13.499	13.499	0	237\\
	13.499	13.499	0	238\\
	nan	nan	0	239\\
	7.501	13.499	0	240\\
	7.501	13.499	0	241\\
	8.499	13.499	0	242\\
	8.499	13.499	0	243\\
	nan	nan	0	244\\
	8.501	13.499	0	245\\
	8.501	13.499	0	246\\
	9.499	13.499	0	247\\
	9.499	13.499	0	248\\
	nan	nan	0	249\\
	9.501	13.499	0	250\\
	9.501	13.499	0	251\\
	10.499	13.499	0	252\\
	10.499	13.499	0	253\\
	nan	nan	0	254\\
	10.501	13.499	0	255\\
	10.501	13.499	0	256\\
	11.499	13.499	0	257\\
	11.499	13.499	0	258\\
	nan	nan	0	259\\
	11.501	13.499	0	260\\
	11.501	13.499	0	261\\
	12.499	13.499	0	262\\
	12.499	13.499	0	263\\
	nan	nan	0	264\\
	12.501	13.499	0	265\\
	12.501	13.499	0	266\\
	13.499	13.499	0	267\\
	13.499	13.499	0	268\\
	nan	nan	0	269\\
	nan	nan	0	270\\
	nan	nan	0	271\\
	nan	nan	0	272\\
	nan	nan	0	273\\
	nan	nan	0	274\\
	nan	nan	0	275\\
	nan	nan	0	276\\
	nan	nan	0	277\\
	nan	nan	0	278\\
	nan	nan	0	279\\
	nan	nan	0	280\\
	nan	nan	0	281\\
	nan	nan	0	282\\
	nan	nan	0	283\\
	nan	nan	0	284\\
	nan	nan	0	285\\
	nan	nan	0	286\\
	nan	nan	0	287\\
	nan	nan	0	288\\
	nan	nan	0	289\\
	nan	nan	0	290\\
	nan	nan	0	291\\
	nan	nan	0	292\\
	nan	nan	0	293\\
	nan	nan	0	294\\
	nan	nan	0	295\\
	nan	nan	0	296\\
	nan	nan	0	297\\
	nan	nan	0	298\\
	nan	nan	0	299\\
	7.501	13.501	0	300\\
	7.501	13.501	0	301\\
	8.499	13.501	0	302\\
	8.499	13.501	0	303\\
	nan	nan	0	304\\
	8.501	13.501	0	305\\
	8.501	13.501	0	306\\
	9.499	13.501	0	307\\
	9.499	13.501	0	308\\
	nan	nan	0	309\\
	9.501	13.501	0	310\\
	9.501	13.501	0	311\\
	10.499	13.501	0	312\\
	10.499	13.501	0	313\\
	nan	nan	0	314\\
	10.501	13.501	0	315\\
	10.501	13.501	0	316\\
	11.499	13.501	0	317\\
	11.499	13.501	0	318\\
	nan	nan	0	319\\
	11.501	13.501	0	320\\
	11.501	13.501	0	321\\
	12.499	13.501	0	322\\
	12.499	13.501	0	323\\
	nan	nan	0	324\\
	12.501	13.501	0	325\\
	12.501	13.501	0	326\\
	13.499	13.501	0	327\\
	13.499	13.501	0	328\\
	nan	nan	0	329\\
	7.501	13.501	0	330\\
	7.501	13.501	0	331\\
	8.499	13.501	0	332\\
	8.499	13.501	0	333\\
	nan	nan	0	334\\
	8.501	13.501	0	335\\
	8.501	13.501	0	336\\
	9.499	13.501	0	337\\
	9.499	13.501	0	338\\
	nan	nan	0	339\\
	9.501	13.501	0	340\\
	9.501	13.501	0	341\\
	10.499	13.501	0	342\\
	10.499	13.501	0	343\\
	nan	nan	0	344\\
	10.501	13.501	0	345\\
	10.501	13.501	0	346\\
	11.499	13.501	0	347\\
	11.499	13.501	0	348\\
	nan	nan	0	349\\
	11.501	13.501	0	350\\
	11.501	13.501	0	351\\
	12.499	13.501	0	352\\
	12.499	13.501	0	353\\
	nan	nan	0	354\\
	12.501	13.501	0	355\\
	12.501	13.501	0	356\\
	13.499	13.501	0	357\\
	13.499	13.501	0	358\\
	nan	nan	0	359\\
	7.501	14.499	0	360\\
	7.501	14.499	0	361\\
	8.499	14.499	0	362\\
	8.499	14.499	0	363\\
	nan	nan	0	364\\
	8.501	14.499	0	365\\
	8.501	14.499	0	366\\
	9.499	14.499	0	367\\
	9.499	14.499	0	368\\
	nan	nan	0	369\\
	9.501	14.499	0	370\\
	9.501	14.499	0	371\\
	10.499	14.499	0	372\\
	10.499	14.499	0	373\\
	nan	nan	0	374\\
	10.501	14.499	0	375\\
	10.501	14.499	0	376\\
	11.499	14.499	0	377\\
	11.499	14.499	0	378\\
	nan	nan	0	379\\
	11.501	14.499	0	380\\
	11.501	14.499	0	381\\
	12.499	14.499	0	382\\
	12.499	14.499	0	383\\
	nan	nan	0	384\\
	12.501	14.499	0	385\\
	12.501	14.499	0	386\\
	13.499	14.499	0	387\\
	13.499	14.499	0	388\\
	nan	nan	0	389\\
	7.501	14.499	0	390\\
	7.501	14.499	0	391\\
	8.499	14.499	0	392\\
	8.499	14.499	0	393\\
	nan	nan	0	394\\
	8.501	14.499	0	395\\
	8.501	14.499	0	396\\
	9.499	14.499	0	397\\
	9.499	14.499	0	398\\
	nan	nan	0	399\\
	9.501	14.499	0	400\\
	9.501	14.499	0	401\\
	10.499	14.499	0	402\\
	10.499	14.499	0	403\\
	nan	nan	0	404\\
	10.501	14.499	0	405\\
	10.501	14.499	0	406\\
	11.499	14.499	0	407\\
	11.499	14.499	0	408\\
	nan	nan	0	409\\
	11.501	14.499	0	410\\
	11.501	14.499	0	411\\
	12.499	14.499	0	412\\
	12.499	14.499	0	413\\
	nan	nan	0	414\\
	12.501	14.499	0	415\\
	12.501	14.499	0	416\\
	13.499	14.499	0	417\\
	13.499	14.499	0	418\\
	nan	nan	0	419\\
	nan	nan	0	420\\
	nan	nan	0	421\\
	nan	nan	0	422\\
	nan	nan	0	423\\
	nan	nan	0	424\\
	nan	nan	0	425\\
	nan	nan	0	426\\
	nan	nan	0	427\\
	nan	nan	0	428\\
	nan	nan	0	429\\
	nan	nan	0	430\\
	nan	nan	0	431\\
	nan	nan	0	432\\
	nan	nan	0	433\\
	nan	nan	0	434\\
	nan	nan	0	435\\
	nan	nan	0	436\\
	nan	nan	0	437\\
	nan	nan	0	438\\
	nan	nan	0	439\\
	nan	nan	0	440\\
	nan	nan	0	441\\
	nan	nan	0	442\\
	nan	nan	0	443\\
	nan	nan	0	444\\
	nan	nan	0	445\\
	nan	nan	0	446\\
	nan	nan	0	447\\
	nan	nan	0	448\\
	nan	nan	0	449\\
	7.501	14.501	0	450\\
	7.501	14.501	0	451\\
	8.499	14.501	0	452\\
	8.499	14.501	0	453\\
	nan	nan	0	454\\
	8.501	14.501	0	455\\
	8.501	14.501	0	456\\
	9.499	14.501	0	457\\
	9.499	14.501	0	458\\
	nan	nan	0	459\\
	9.501	14.501	0	460\\
	9.501	14.501	0	461\\
	10.499	14.501	0	462\\
	10.499	14.501	0	463\\
	nan	nan	0	464\\
	10.501	14.501	0	465\\
	10.501	14.501	0	466\\
	11.499	14.501	0	467\\
	11.499	14.501	0	468\\
	nan	nan	0	469\\
	11.501	14.501	0	470\\
	11.501	14.501	0	471\\
	12.499	14.501	0	472\\
	12.499	14.501	0	473\\
	nan	nan	0	474\\
	12.501	14.501	0	475\\
	12.501	14.501	0	476\\
	13.499	14.501	0	477\\
	13.499	14.501	0	478\\
	nan	nan	0	479\\
	7.501	14.501	0	480\\
	7.501	14.501	0	481\\
	8.499	14.501	0	482\\
	8.499	14.501	0	483\\
	nan	nan	0	484\\
	8.501	14.501	0	485\\
	8.501	14.501	0	486\\
	9.499	14.501	0	487\\
	9.499	14.501	0	488\\
	nan	nan	0	489\\
	9.501	14.501	0	490\\
	9.501	14.501	0	491\\
	10.499	14.501	0	492\\
	10.499	14.501	0	493\\
	nan	nan	0	494\\
	10.501	14.501	0	495\\
	10.501	14.501	0	496\\
	11.499	14.501	0	497\\
	11.499	14.501	0	498\\
	nan	nan	0	499\\
	11.501	14.501	0	500\\
	11.501	14.501	0	501\\
	12.499	14.501	0	502\\
	12.499	14.501	0	503\\
	nan	nan	0	504\\
	12.501	14.501	0	505\\
	12.501	14.501	0	506\\
	13.499	14.501	0	507\\
	13.499	14.501	0	508\\
	nan	nan	0	509\\
	7.501	15.499	0	510\\
	7.501	15.499	0	511\\
	8.499	15.499	0	512\\
	8.499	15.499	0	513\\
	nan	nan	0	514\\
	8.501	15.499	0	515\\
	8.501	15.499	0	516\\
	9.499	15.499	0	517\\
	9.499	15.499	0	518\\
	nan	nan	0	519\\
	9.501	15.499	0	520\\
	9.501	15.499	0	521\\
	10.499	15.499	0	522\\
	10.499	15.499	0	523\\
	nan	nan	0	524\\
	10.501	15.499	0	525\\
	10.501	15.499	0	526\\
	11.499	15.499	0	527\\
	11.499	15.499	0	528\\
	nan	nan	0	529\\
	11.501	15.499	0	530\\
	11.501	15.499	0	531\\
	12.499	15.499	0	532\\
	12.499	15.499	0	533\\
	nan	nan	0	534\\
	12.501	15.499	0	535\\
	12.501	15.499	0	536\\
	13.499	15.499	0	537\\
	13.499	15.499	0	538\\
	nan	nan	0	539\\
	7.501	15.499	0	540\\
	7.501	15.499	0	541\\
	8.499	15.499	0	542\\
	8.499	15.499	0	543\\
	nan	nan	0	544\\
	8.501	15.499	0	545\\
	8.501	15.499	0	546\\
	9.499	15.499	0	547\\
	9.499	15.499	0	548\\
	nan	nan	0	549\\
	9.501	15.499	0	550\\
	9.501	15.499	0	551\\
	10.499	15.499	0	552\\
	10.499	15.499	0	553\\
	nan	nan	0	554\\
	10.501	15.499	0	555\\
	10.501	15.499	0	556\\
	11.499	15.499	0	557\\
	11.499	15.499	0	558\\
	nan	nan	0	559\\
	11.501	15.499	0	560\\
	11.501	15.499	0	561\\
	12.499	15.499	0	562\\
	12.499	15.499	0	563\\
	nan	nan	0	564\\
	12.501	15.499	0	565\\
	12.501	15.499	0	566\\
	13.499	15.499	0	567\\
	13.499	15.499	0	568\\
	nan	nan	0	569\\
	nan	nan	0	570\\
	nan	nan	0	571\\
	nan	nan	0	572\\
	nan	nan	0	573\\
	nan	nan	0	574\\
	nan	nan	0	575\\
	nan	nan	0	576\\
	nan	nan	0	577\\
	nan	nan	0	578\\
	nan	nan	0	579\\
	nan	nan	0	580\\
	nan	nan	0	581\\
	nan	nan	0	582\\
	nan	nan	0	583\\
	nan	nan	0	584\\
	nan	nan	0	585\\
	nan	nan	0	586\\
	nan	nan	0	587\\
	nan	nan	0	588\\
	nan	nan	0	589\\
	nan	nan	0	590\\
	nan	nan	0	591\\
	nan	nan	0	592\\
	nan	nan	0	593\\
	nan	nan	0	594\\
	nan	nan	0	595\\
	nan	nan	0	596\\
	nan	nan	0	597\\
	nan	nan	0	598\\
	nan	nan	0	599\\
	7.501	15.501	0	600\\
	7.501	15.501	0	601\\
	8.499	15.501	0	602\\
	8.499	15.501	0	603\\
	nan	nan	0	604\\
	8.501	15.501	0	605\\
	8.501	15.501	0	606\\
	9.499	15.501	0	607\\
	9.499	15.501	0	608\\
	nan	nan	0	609\\
	9.501	15.501	0	610\\
	9.501	15.501	0	611\\
	10.499	15.501	0	612\\
	10.499	15.501	0	613\\
	nan	nan	0	614\\
	10.501	15.501	0	615\\
	10.501	15.501	0	616\\
	11.499	15.501	0	617\\
	11.499	15.501	0	618\\
	nan	nan	0	619\\
	11.501	15.501	0	620\\
	11.501	15.501	0	621\\
	12.499	15.501	0	622\\
	12.499	15.501	0	623\\
	nan	nan	0	624\\
	12.501	15.501	0	625\\
	12.501	15.501	0	626\\
	13.499	15.501	0	627\\
	13.499	15.501	0	628\\
	nan	nan	0	629\\
	7.501	15.501	0	630\\
	7.501	15.501	0	631\\
	8.499	15.501	0	632\\
	8.499	15.501	0	633\\
	nan	nan	0	634\\
	8.501	15.501	0	635\\
	8.501	15.501	0	636\\
	9.499	15.501	0	637\\
	9.499	15.501	0	638\\
	nan	nan	0	639\\
	9.501	15.501	0	640\\
	9.501	15.501	0	641\\
	10.499	15.501	0	642\\
	10.499	15.501	0	643\\
	nan	nan	0	644\\
	10.501	15.501	0	645\\
	10.501	15.501	0	646\\
	11.499	15.501	0	647\\
	11.499	15.501	0	648\\
	nan	nan	0	649\\
	11.501	15.501	0	650\\
	11.501	15.501	0	651\\
	12.499	15.501	0	652\\
	12.499	15.501	0	653\\
	nan	nan	0	654\\
	12.501	15.501	0	655\\
	12.501	15.501	4215	656\\
	13.499	15.501	4215	657\\
	13.499	15.501	0	658\\
	nan	nan	0	659\\
	7.501	16.499	0	660\\
	7.501	16.499	0	661\\
	8.499	16.499	0	662\\
	8.499	16.499	0	663\\
	nan	nan	0	664\\
	8.501	16.499	0	665\\
	8.501	16.499	0	666\\
	9.499	16.499	0	667\\
	9.499	16.499	0	668\\
	nan	nan	0	669\\
	9.501	16.499	0	670\\
	9.501	16.499	0	671\\
	10.499	16.499	0	672\\
	10.499	16.499	0	673\\
	nan	nan	0	674\\
	10.501	16.499	0	675\\
	10.501	16.499	0	676\\
	11.499	16.499	0	677\\
	11.499	16.499	0	678\\
	nan	nan	0	679\\
	11.501	16.499	0	680\\
	11.501	16.499	0	681\\
	12.499	16.499	0	682\\
	12.499	16.499	0	683\\
	nan	nan	0	684\\
	12.501	16.499	0	685\\
	12.501	16.499	4215	686\\
	13.499	16.499	4215	687\\
	13.499	16.499	0	688\\
	nan	nan	0	689\\
	7.501	16.499	0	690\\
	7.501	16.499	0	691\\
	8.499	16.499	0	692\\
	8.499	16.499	0	693\\
	nan	nan	0	694\\
	8.501	16.499	0	695\\
	8.501	16.499	0	696\\
	9.499	16.499	0	697\\
	9.499	16.499	0	698\\
	nan	nan	0	699\\
	9.501	16.499	0	700\\
	9.501	16.499	0	701\\
	10.499	16.499	0	702\\
	10.499	16.499	0	703\\
	nan	nan	0	704\\
	10.501	16.499	0	705\\
	10.501	16.499	0	706\\
	11.499	16.499	0	707\\
	11.499	16.499	0	708\\
	nan	nan	0	709\\
	11.501	16.499	0	710\\
	11.501	16.499	0	711\\
	12.499	16.499	0	712\\
	12.499	16.499	0	713\\
	nan	nan	0	714\\
	12.501	16.499	0	715\\
	12.501	16.499	0	716\\
	13.499	16.499	0	717\\
	13.499	16.499	0	718\\
	nan	nan	0	719\\
	nan	nan	0	720\\
	nan	nan	0	721\\
	nan	nan	0	722\\
	nan	nan	0	723\\
	nan	nan	0	724\\
	nan	nan	0	725\\
	nan	nan	0	726\\
	nan	nan	0	727\\
	nan	nan	0	728\\
	nan	nan	0	729\\
	nan	nan	0	730\\
	nan	nan	0	731\\
	nan	nan	0	732\\
	nan	nan	0	733\\
	nan	nan	0	734\\
	nan	nan	0	735\\
	nan	nan	0	736\\
	nan	nan	0	737\\
	nan	nan	0	738\\
	nan	nan	0	739\\
	nan	nan	0	740\\
	nan	nan	0	741\\
	nan	nan	0	742\\
	nan	nan	0	743\\
	nan	nan	0	744\\
	nan	nan	0	745\\
	nan	nan	0	746\\
	nan	nan	0	747\\
	nan	nan	0	748\\
	nan	nan	0	749\\
	7.501	16.501	0	750\\
	7.501	16.501	0	751\\
	8.499	16.501	0	752\\
	8.499	16.501	0	753\\
	nan	nan	0	754\\
	8.501	16.501	0	755\\
	8.501	16.501	0	756\\
	9.499	16.501	0	757\\
	9.499	16.501	0	758\\
	nan	nan	0	759\\
	9.501	16.501	0	760\\
	9.501	16.501	0	761\\
	10.499	16.501	0	762\\
	10.499	16.501	0	763\\
	nan	nan	0	764\\
	10.501	16.501	0	765\\
	10.501	16.501	0	766\\
	11.499	16.501	0	767\\
	11.499	16.501	0	768\\
	nan	nan	0	769\\
	11.501	16.501	0	770\\
	11.501	16.501	0	771\\
	12.499	16.501	0	772\\
	12.499	16.501	0	773\\
	nan	nan	0	774\\
	12.501	16.501	0	775\\
	12.501	16.501	0	776\\
	13.499	16.501	0	777\\
	13.499	16.501	0	778\\
	nan	nan	0	779\\
	7.501	16.501	0	780\\
	7.501	16.501	0	781\\
	8.499	16.501	0	782\\
	8.499	16.501	0	783\\
	nan	nan	0	784\\
	8.501	16.501	0	785\\
	8.501	16.501	0	786\\
	9.499	16.501	0	787\\
	9.499	16.501	0	788\\
	nan	nan	0	789\\
	9.501	16.501	0	790\\
	9.501	16.501	0	791\\
	10.499	16.501	0	792\\
	10.499	16.501	0	793\\
	nan	nan	0	794\\
	10.501	16.501	0	795\\
	10.501	16.501	0	796\\
	11.499	16.501	0	797\\
	11.499	16.501	0	798\\
	nan	nan	0	799\\
	11.501	16.501	0	800\\
	11.501	16.501	3285	801\\
	12.499	16.501	3285	802\\
	12.499	16.501	0	803\\
	nan	nan	0	804\\
	12.501	16.501	0	805\\
	12.501	16.501	0	806\\
	13.499	16.501	0	807\\
	13.499	16.501	0	808\\
	nan	nan	0	809\\
	7.501	17.499	0	810\\
	7.501	17.499	0	811\\
	8.499	17.499	0	812\\
	8.499	17.499	0	813\\
	nan	nan	0	814\\
	8.501	17.499	0	815\\
	8.501	17.499	0	816\\
	9.499	17.499	0	817\\
	9.499	17.499	0	818\\
	nan	nan	0	819\\
	9.501	17.499	0	820\\
	9.501	17.499	0	821\\
	10.499	17.499	0	822\\
	10.499	17.499	0	823\\
	nan	nan	0	824\\
	10.501	17.499	0	825\\
	10.501	17.499	0	826\\
	11.499	17.499	0	827\\
	11.499	17.499	0	828\\
	nan	nan	0	829\\
	11.501	17.499	0	830\\
	11.501	17.499	3285	831\\
	12.499	17.499	3285	832\\
	12.499	17.499	0	833\\
	nan	nan	0	834\\
	12.501	17.499	0	835\\
	12.501	17.499	0	836\\
	13.499	17.499	0	837\\
	13.499	17.499	0	838\\
	nan	nan	0	839\\
	7.501	17.499	0	840\\
	7.501	17.499	0	841\\
	8.499	17.499	0	842\\
	8.499	17.499	0	843\\
	nan	nan	0	844\\
	8.501	17.499	0	845\\
	8.501	17.499	0	846\\
	9.499	17.499	0	847\\
	9.499	17.499	0	848\\
	nan	nan	0	849\\
	9.501	17.499	0	850\\
	9.501	17.499	0	851\\
	10.499	17.499	0	852\\
	10.499	17.499	0	853\\
	nan	nan	0	854\\
	10.501	17.499	0	855\\
	10.501	17.499	0	856\\
	11.499	17.499	0	857\\
	11.499	17.499	0	858\\
	nan	nan	0	859\\
	11.501	17.499	0	860\\
	11.501	17.499	0	861\\
	12.499	17.499	0	862\\
	12.499	17.499	0	863\\
	nan	nan	0	864\\
	12.501	17.499	0	865\\
	12.501	17.499	0	866\\
	13.499	17.499	0	867\\
	13.499	17.499	0	868\\
	nan	nan	0	869\\
	nan	nan	0	870\\
	nan	nan	0	871\\
	nan	nan	0	872\\
	nan	nan	0	873\\
	nan	nan	0	874\\
	nan	nan	0	875\\
	nan	nan	0	876\\
	nan	nan	0	877\\
	nan	nan	0	878\\
	nan	nan	0	879\\
	nan	nan	0	880\\
	nan	nan	0	881\\
	nan	nan	0	882\\
	nan	nan	0	883\\
	nan	nan	0	884\\
	nan	nan	0	885\\
	nan	nan	0	886\\
	nan	nan	0	887\\
	nan	nan	0	888\\
	nan	nan	0	889\\
	nan	nan	0	890\\
	nan	nan	0	891\\
	nan	nan	0	892\\
	nan	nan	0	893\\
	nan	nan	0	894\\
	nan	nan	0	895\\
	nan	nan	0	896\\
	nan	nan	0	897\\
	nan	nan	0	898\\
	nan	nan	0	899\\
};

\end{axis}

\end{tikzpicture}%

%% file: plots/Revised_Best_Card_Set_delta_d_-14_SNR_-5.tex
%
%
\begin{tikzpicture}

\begin{axis}[%
scale=0.6, width=3.284in,
height=3.566in,
at={(0.758in,0.481in)},
scale only axis,
unbounded coords=jump,
colormap={patchmap}{[1pt] rgb(0pt)=(0.75,0.85,0.95); rgb(899pt)=(0.75,0.85,0.95)},
xmin=16.2,
xmax=22.8,
xtick={17, 18, 19, 20, 21, 22},
xticklabel style = {font=\footnotesize},
tick align=outside,
xlabel style={font=\color{white!15!black}},
xlabel style = {font=\small},
xlabel={$|\mathcal{P}_{1}^{\star}|$},
ymin=2.2,
ymax=8.8,
ytick={3, 4, 5, 6, 7, 8},
yticklabel style = {font=\footnotesize},
ylabel style={font=\color{white!15!black}},
ylabel style = {font=\small},
ylabel={$|\mathcal{P}_{2}^{\star}|$},
zmin=0,
zmax=2200,
zlabel style={font=\color{white!15!black}},
zlabel style = {font=\small},
zticklabel style = {font=\footnotesize},
zlabel={Frequency},
view={-37.5}{30},
axis background/.style={fill=white},
title style={font=\bfseries},
axis x line*=bottom,
axis y line*=left,
axis z line*=left,
xmajorgrids,
ymajorgrids,
zmajorgrids,
legend style={at={(1.03,1)}, anchor=north west, legend cell align=left, align=left, draw=white!15!black}
]

\addplot3[%
surf,
shader=flat corner, draw=black, mesh/rows=30]
table[row sep=crcr, colormap name=surfmap, point meta=\thisrow{c}] {%
	x	y	z	c\\
	16.501	2.501	0	0\\
	16.501	2.501	0	1\\
	17.499	2.501	0	2\\
	17.499	2.501	0	3\\
	nan	nan	0	4\\
	17.501	2.501	0	5\\
	17.501	2.501	0	6\\
	18.499	2.501	0	7\\
	18.499	2.501	0	8\\
	nan	nan	0	9\\
	18.501	2.501	0	10\\
	18.501	2.501	0	11\\
	19.499	2.501	0	12\\
	19.499	2.501	0	13\\
	nan	nan	0	14\\
	19.501	2.501	0	15\\
	19.501	2.501	0	16\\
	20.499	2.501	0	17\\
	20.499	2.501	0	18\\
	nan	nan	0	19\\
	20.501	2.501	0	20\\
	20.501	2.501	0	21\\
	21.499	2.501	0	22\\
	21.499	2.501	0	23\\
	nan	nan	0	24\\
	21.501	2.501	0	25\\
	21.501	2.501	0	26\\
	22.499	2.501	0	27\\
	22.499	2.501	0	28\\
	nan	nan	0	29\\
	16.501	2.501	0	30\\
	16.501	2.501	232	31\\
	17.499	2.501	232	32\\
	17.499	2.501	0	33\\
	nan	nan	0	34\\
	17.501	2.501	0	35\\
	17.501	2.501	63	36\\
	18.499	2.501	63	37\\
	18.499	2.501	0	38\\
	nan	nan	0	39\\
	18.501	2.501	0	40\\
	18.501	2.501	46	41\\
	19.499	2.501	46	42\\
	19.499	2.501	0	43\\
	nan	nan	0	44\\
	19.501	2.501	0	45\\
	19.501	2.501	10	46\\
	20.499	2.501	10	47\\
	20.499	2.501	0	48\\
	nan	nan	0	49\\
	20.501	2.501	0	50\\
	20.501	2.501	7	51\\
	21.499	2.501	7	52\\
	21.499	2.501	0	53\\
	nan	nan	0	54\\
	21.501	2.501	0	55\\
	21.501	2.501	0	56\\
	22.499	2.501	0	57\\
	22.499	2.501	0	58\\
	nan	nan	0	59\\
	16.501	3.499	0	60\\
	16.501	3.499	232	61\\
	17.499	3.499	232	62\\
	17.499	3.499	0	63\\
	nan	nan	0	64\\
	17.501	3.499	0	65\\
	17.501	3.499	63	66\\
	18.499	3.499	63	67\\
	18.499	3.499	0	68\\
	nan	nan	0	69\\
	18.501	3.499	0	70\\
	18.501	3.499	46	71\\
	19.499	3.499	46	72\\
	19.499	3.499	0	73\\
	nan	nan	0	74\\
	19.501	3.499	0	75\\
	19.501	3.499	10	76\\
	20.499	3.499	10	77\\
	20.499	3.499	0	78\\
	nan	nan	0	79\\
	20.501	3.499	0	80\\
	20.501	3.499	7	81\\
	21.499	3.499	7	82\\
	21.499	3.499	0	83\\
	nan	nan	0	84\\
	21.501	3.499	0	85\\
	21.501	3.499	0	86\\
	22.499	3.499	0	87\\
	22.499	3.499	0	88\\
	nan	nan	0	89\\
	16.501	3.499	0	90\\
	16.501	3.499	0	91\\
	17.499	3.499	0	92\\
	17.499	3.499	0	93\\
	nan	nan	0	94\\
	17.501	3.499	0	95\\
	17.501	3.499	0	96\\
	18.499	3.499	0	97\\
	18.499	3.499	0	98\\
	nan	nan	0	99\\
	18.501	3.499	0	100\\
	18.501	3.499	0	101\\
	19.499	3.499	0	102\\
	19.499	3.499	0	103\\
	nan	nan	0	104\\
	19.501	3.499	0	105\\
	19.501	3.499	0	106\\
	20.499	3.499	0	107\\
	20.499	3.499	0	108\\
	nan	nan	0	109\\
	20.501	3.499	0	110\\
	20.501	3.499	0	111\\
	21.499	3.499	0	112\\
	21.499	3.499	0	113\\
	nan	nan	0	114\\
	21.501	3.499	0	115\\
	21.501	3.499	0	116\\
	22.499	3.499	0	117\\
	22.499	3.499	0	118\\
	nan	nan	0	119\\
	nan	nan	0	120\\
	nan	nan	0	121\\
	nan	nan	0	122\\
	nan	nan	0	123\\
	nan	nan	0	124\\
	nan	nan	0	125\\
	nan	nan	0	126\\
	nan	nan	0	127\\
	nan	nan	0	128\\
	nan	nan	0	129\\
	nan	nan	0	130\\
	nan	nan	0	131\\
	nan	nan	0	132\\
	nan	nan	0	133\\
	nan	nan	0	134\\
	nan	nan	0	135\\
	nan	nan	0	136\\
	nan	nan	0	137\\
	nan	nan	0	138\\
	nan	nan	0	139\\
	nan	nan	0	140\\
	nan	nan	0	141\\
	nan	nan	0	142\\
	nan	nan	0	143\\
	nan	nan	0	144\\
	nan	nan	0	145\\
	nan	nan	0	146\\
	nan	nan	0	147\\
	nan	nan	0	148\\
	nan	nan	0	149\\
	16.501	3.501	0	150\\
	16.501	3.501	0	151\\
	17.499	3.501	0	152\\
	17.499	3.501	0	153\\
	nan	nan	0	154\\
	17.501	3.501	0	155\\
	17.501	3.501	0	156\\
	18.499	3.501	0	157\\
	18.499	3.501	0	158\\
	nan	nan	0	159\\
	18.501	3.501	0	160\\
	18.501	3.501	0	161\\
	19.499	3.501	0	162\\
	19.499	3.501	0	163\\
	nan	nan	0	164\\
	19.501	3.501	0	165\\
	19.501	3.501	0	166\\
	20.499	3.501	0	167\\
	20.499	3.501	0	168\\
	nan	nan	0	169\\
	20.501	3.501	0	170\\
	20.501	3.501	0	171\\
	21.499	3.501	0	172\\
	21.499	3.501	0	173\\
	nan	nan	0	174\\
	21.501	3.501	0	175\\
	21.501	3.501	0	176\\
	22.499	3.501	0	177\\
	22.499	3.501	0	178\\
	nan	nan	0	179\\
	16.501	3.501	0	180\\
	16.501	3.501	675	181\\
	17.499	3.501	675	182\\
	17.499	3.501	0	183\\
	nan	nan	0	184\\
	17.501	3.501	0	185\\
	17.501	3.501	57	186\\
	18.499	3.501	57	187\\
	18.499	3.501	0	188\\
	nan	nan	0	189\\
	18.501	3.501	0	190\\
	18.501	3.501	15	191\\
	19.499	3.501	15	192\\
	19.499	3.501	0	193\\
	nan	nan	0	194\\
	19.501	3.501	0	195\\
	19.501	3.501	1	196\\
	20.499	3.501	1	197\\
	20.499	3.501	0	198\\
	nan	nan	0	199\\
	20.501	3.501	0	200\\
	20.501	3.501	0	201\\
	21.499	3.501	0	202\\
	21.499	3.501	0	203\\
	nan	nan	0	204\\
	21.501	3.501	0	205\\
	21.501	3.501	0	206\\
	22.499	3.501	0	207\\
	22.499	3.501	0	208\\
	nan	nan	0	209\\
	16.501	4.499	0	210\\
	16.501	4.499	675	211\\
	17.499	4.499	675	212\\
	17.499	4.499	0	213\\
	nan	nan	0	214\\
	17.501	4.499	0	215\\
	17.501	4.499	57	216\\
	18.499	4.499	57	217\\
	18.499	4.499	0	218\\
	nan	nan	0	219\\
	18.501	4.499	0	220\\
	18.501	4.499	15	221\\
	19.499	4.499	15	222\\
	19.499	4.499	0	223\\
	nan	nan	0	224\\
	19.501	4.499	0	225\\
	19.501	4.499	1	226\\
	20.499	4.499	1	227\\
	20.499	4.499	0	228\\
	nan	nan	0	229\\
	20.501	4.499	0	230\\
	20.501	4.499	0	231\\
	21.499	4.499	0	232\\
	21.499	4.499	0	233\\
	nan	nan	0	234\\
	21.501	4.499	0	235\\
	21.501	4.499	0	236\\
	22.499	4.499	0	237\\
	22.499	4.499	0	238\\
	nan	nan	0	239\\
	16.501	4.499	0	240\\
	16.501	4.499	0	241\\
	17.499	4.499	0	242\\
	17.499	4.499	0	243\\
	nan	nan	0	244\\
	17.501	4.499	0	245\\
	17.501	4.499	0	246\\
	18.499	4.499	0	247\\
	18.499	4.499	0	248\\
	nan	nan	0	249\\
	18.501	4.499	0	250\\
	18.501	4.499	0	251\\
	19.499	4.499	0	252\\
	19.499	4.499	0	253\\
	nan	nan	0	254\\
	19.501	4.499	0	255\\
	19.501	4.499	0	256\\
	20.499	4.499	0	257\\
	20.499	4.499	0	258\\
	nan	nan	0	259\\
	20.501	4.499	0	260\\
	20.501	4.499	0	261\\
	21.499	4.499	0	262\\
	21.499	4.499	0	263\\
	nan	nan	0	264\\
	21.501	4.499	0	265\\
	21.501	4.499	0	266\\
	22.499	4.499	0	267\\
	22.499	4.499	0	268\\
	nan	nan	0	269\\
	nan	nan	0	270\\
	nan	nan	0	271\\
	nan	nan	0	272\\
	nan	nan	0	273\\
	nan	nan	0	274\\
	nan	nan	0	275\\
	nan	nan	0	276\\
	nan	nan	0	277\\
	nan	nan	0	278\\
	nan	nan	0	279\\
	nan	nan	0	280\\
	nan	nan	0	281\\
	nan	nan	0	282\\
	nan	nan	0	283\\
	nan	nan	0	284\\
	nan	nan	0	285\\
	nan	nan	0	286\\
	nan	nan	0	287\\
	nan	nan	0	288\\
	nan	nan	0	289\\
	nan	nan	0	290\\
	nan	nan	0	291\\
	nan	nan	0	292\\
	nan	nan	0	293\\
	nan	nan	0	294\\
	nan	nan	0	295\\
	nan	nan	0	296\\
	nan	nan	0	297\\
	nan	nan	0	298\\
	nan	nan	0	299\\
	16.501	4.501	0	300\\
	16.501	4.501	0	301\\
	17.499	4.501	0	302\\
	17.499	4.501	0	303\\
	nan	nan	0	304\\
	17.501	4.501	0	305\\
	17.501	4.501	0	306\\
	18.499	4.501	0	307\\
	18.499	4.501	0	308\\
	nan	nan	0	309\\
	18.501	4.501	0	310\\
	18.501	4.501	0	311\\
	19.499	4.501	0	312\\
	19.499	4.501	0	313\\
	nan	nan	0	314\\
	19.501	4.501	0	315\\
	19.501	4.501	0	316\\
	20.499	4.501	0	317\\
	20.499	4.501	0	318\\
	nan	nan	0	319\\
	20.501	4.501	0	320\\
	20.501	4.501	0	321\\
	21.499	4.501	0	322\\
	21.499	4.501	0	323\\
	nan	nan	0	324\\
	21.501	4.501	0	325\\
	21.501	4.501	0	326\\
	22.499	4.501	0	327\\
	22.499	4.501	0	328\\
	nan	nan	0	329\\
	16.501	4.501	0	330\\
	16.501	4.501	1359	331\\
	17.499	4.501	1359	332\\
	17.499	4.501	0	333\\
	nan	nan	0	334\\
	17.501	4.501	0	335\\
	17.501	4.501	24	336\\
	18.499	4.501	24	337\\
	18.499	4.501	0	338\\
	nan	nan	0	339\\
	18.501	4.501	0	340\\
	18.501	4.501	5	341\\
	19.499	4.501	5	342\\
	19.499	4.501	0	343\\
	nan	nan	0	344\\
	19.501	4.501	0	345\\
	19.501	4.501	0	346\\
	20.499	4.501	0	347\\
	20.499	4.501	0	348\\
	nan	nan	0	349\\
	20.501	4.501	0	350\\
	20.501	4.501	0	351\\
	21.499	4.501	0	352\\
	21.499	4.501	0	353\\
	nan	nan	0	354\\
	21.501	4.501	0	355\\
	21.501	4.501	0	356\\
	22.499	4.501	0	357\\
	22.499	4.501	0	358\\
	nan	nan	0	359\\
	16.501	5.499	0	360\\
	16.501	5.499	1359	361\\
	17.499	5.499	1359	362\\
	17.499	5.499	0	363\\
	nan	nan	0	364\\
	17.501	5.499	0	365\\
	17.501	5.499	24	366\\
	18.499	5.499	24	367\\
	18.499	5.499	0	368\\
	nan	nan	0	369\\
	18.501	5.499	0	370\\
	18.501	5.499	5	371\\
	19.499	5.499	5	372\\
	19.499	5.499	0	373\\
	nan	nan	0	374\\
	19.501	5.499	0	375\\
	19.501	5.499	0	376\\
	20.499	5.499	0	377\\
	20.499	5.499	0	378\\
	nan	nan	0	379\\
	20.501	5.499	0	380\\
	20.501	5.499	0	381\\
	21.499	5.499	0	382\\
	21.499	5.499	0	383\\
	nan	nan	0	384\\
	21.501	5.499	0	385\\
	21.501	5.499	0	386\\
	22.499	5.499	0	387\\
	22.499	5.499	0	388\\
	nan	nan	0	389\\
	16.501	5.499	0	390\\
	16.501	5.499	0	391\\
	17.499	5.499	0	392\\
	17.499	5.499	0	393\\
	nan	nan	0	394\\
	17.501	5.499	0	395\\
	17.501	5.499	0	396\\
	18.499	5.499	0	397\\
	18.499	5.499	0	398\\
	nan	nan	0	399\\
	18.501	5.499	0	400\\
	18.501	5.499	0	401\\
	19.499	5.499	0	402\\
	19.499	5.499	0	403\\
	nan	nan	0	404\\
	19.501	5.499	0	405\\
	19.501	5.499	0	406\\
	20.499	5.499	0	407\\
	20.499	5.499	0	408\\
	nan	nan	0	409\\
	20.501	5.499	0	410\\
	20.501	5.499	0	411\\
	21.499	5.499	0	412\\
	21.499	5.499	0	413\\
	nan	nan	0	414\\
	21.501	5.499	0	415\\
	21.501	5.499	0	416\\
	22.499	5.499	0	417\\
	22.499	5.499	0	418\\
	nan	nan	0	419\\
	nan	nan	0	420\\
	nan	nan	0	421\\
	nan	nan	0	422\\
	nan	nan	0	423\\
	nan	nan	0	424\\
	nan	nan	0	425\\
	nan	nan	0	426\\
	nan	nan	0	427\\
	nan	nan	0	428\\
	nan	nan	0	429\\
	nan	nan	0	430\\
	nan	nan	0	431\\
	nan	nan	0	432\\
	nan	nan	0	433\\
	nan	nan	0	434\\
	nan	nan	0	435\\
	nan	nan	0	436\\
	nan	nan	0	437\\
	nan	nan	0	438\\
	nan	nan	0	439\\
	nan	nan	0	440\\
	nan	nan	0	441\\
	nan	nan	0	442\\
	nan	nan	0	443\\
	nan	nan	0	444\\
	nan	nan	0	445\\
	nan	nan	0	446\\
	nan	nan	0	447\\
	nan	nan	0	448\\
	nan	nan	0	449\\
	16.501	5.501	0	450\\
	16.501	5.501	0	451\\
	17.499	5.501	0	452\\
	17.499	5.501	0	453\\
	nan	nan	0	454\\
	17.501	5.501	0	455\\
	17.501	5.501	0	456\\
	18.499	5.501	0	457\\
	18.499	5.501	0	458\\
	nan	nan	0	459\\
	18.501	5.501	0	460\\
	18.501	5.501	0	461\\
	19.499	5.501	0	462\\
	19.499	5.501	0	463\\
	nan	nan	0	464\\
	19.501	5.501	0	465\\
	19.501	5.501	0	466\\
	20.499	5.501	0	467\\
	20.499	5.501	0	468\\
	nan	nan	0	469\\
	20.501	5.501	0	470\\
	20.501	5.501	0	471\\
	21.499	5.501	0	472\\
	21.499	5.501	0	473\\
	nan	nan	0	474\\
	21.501	5.501	0	475\\
	21.501	5.501	0	476\\
	22.499	5.501	0	477\\
	22.499	5.501	0	478\\
	nan	nan	0	479\\
	16.501	5.501	0	480\\
	16.501	5.501	1944	481\\
	17.499	5.501	1944	482\\
	17.499	5.501	0	483\\
	nan	nan	0	484\\
	17.501	5.501	0	485\\
	17.501	5.501	12	486\\
	18.499	5.501	12	487\\
	18.499	5.501	0	488\\
	nan	nan	0	489\\
	18.501	5.501	0	490\\
	18.501	5.501	0	491\\
	19.499	5.501	0	492\\
	19.499	5.501	0	493\\
	nan	nan	0	494\\
	19.501	5.501	0	495\\
	19.501	5.501	0	496\\
	20.499	5.501	0	497\\
	20.499	5.501	0	498\\
	nan	nan	0	499\\
	20.501	5.501	0	500\\
	20.501	5.501	0	501\\
	21.499	5.501	0	502\\
	21.499	5.501	0	503\\
	nan	nan	0	504\\
	21.501	5.501	0	505\\
	21.501	5.501	0	506\\
	22.499	5.501	0	507\\
	22.499	5.501	0	508\\
	nan	nan	0	509\\
	16.501	6.499	0	510\\
	16.501	6.499	1944	511\\
	17.499	6.499	1944	512\\
	17.499	6.499	0	513\\
	nan	nan	0	514\\
	17.501	6.499	0	515\\
	17.501	6.499	12	516\\
	18.499	6.499	12	517\\
	18.499	6.499	0	518\\
	nan	nan	0	519\\
	18.501	6.499	0	520\\
	18.501	6.499	0	521\\
	19.499	6.499	0	522\\
	19.499	6.499	0	523\\
	nan	nan	0	524\\
	19.501	6.499	0	525\\
	19.501	6.499	0	526\\
	20.499	6.499	0	527\\
	20.499	6.499	0	528\\
	nan	nan	0	529\\
	20.501	6.499	0	530\\
	20.501	6.499	0	531\\
	21.499	6.499	0	532\\
	21.499	6.499	0	533\\
	nan	nan	0	534\\
	21.501	6.499	0	535\\
	21.501	6.499	0	536\\
	22.499	6.499	0	537\\
	22.499	6.499	0	538\\
	nan	nan	0	539\\
	16.501	6.499	0	540\\
	16.501	6.499	0	541\\
	17.499	6.499	0	542\\
	17.499	6.499	0	543\\
	nan	nan	0	544\\
	17.501	6.499	0	545\\
	17.501	6.499	0	546\\
	18.499	6.499	0	547\\
	18.499	6.499	0	548\\
	nan	nan	0	549\\
	18.501	6.499	0	550\\
	18.501	6.499	0	551\\
	19.499	6.499	0	552\\
	19.499	6.499	0	553\\
	nan	nan	0	554\\
	19.501	6.499	0	555\\
	19.501	6.499	0	556\\
	20.499	6.499	0	557\\
	20.499	6.499	0	558\\
	nan	nan	0	559\\
	20.501	6.499	0	560\\
	20.501	6.499	0	561\\
	21.499	6.499	0	562\\
	21.499	6.499	0	563\\
	nan	nan	0	564\\
	21.501	6.499	0	565\\
	21.501	6.499	0	566\\
	22.499	6.499	0	567\\
	22.499	6.499	0	568\\
	nan	nan	0	569\\
	nan	nan	0	570\\
	nan	nan	0	571\\
	nan	nan	0	572\\
	nan	nan	0	573\\
	nan	nan	0	574\\
	nan	nan	0	575\\
	nan	nan	0	576\\
	nan	nan	0	577\\
	nan	nan	0	578\\
	nan	nan	0	579\\
	nan	nan	0	580\\
	nan	nan	0	581\\
	nan	nan	0	582\\
	nan	nan	0	583\\
	nan	nan	0	584\\
	nan	nan	0	585\\
	nan	nan	0	586\\
	nan	nan	0	587\\
	nan	nan	0	588\\
	nan	nan	0	589\\
	nan	nan	0	590\\
	nan	nan	0	591\\
	nan	nan	0	592\\
	nan	nan	0	593\\
	nan	nan	0	594\\
	nan	nan	0	595\\
	nan	nan	0	596\\
	nan	nan	0	597\\
	nan	nan	0	598\\
	nan	nan	0	599\\
	16.501	6.501	0	600\\
	16.501	6.501	0	601\\
	17.499	6.501	0	602\\
	17.499	6.501	0	603\\
	nan	nan	0	604\\
	17.501	6.501	0	605\\
	17.501	6.501	0	606\\
	18.499	6.501	0	607\\
	18.499	6.501	0	608\\
	nan	nan	0	609\\
	18.501	6.501	0	610\\
	18.501	6.501	0	611\\
	19.499	6.501	0	612\\
	19.499	6.501	0	613\\
	nan	nan	0	614\\
	19.501	6.501	0	615\\
	19.501	6.501	0	616\\
	20.499	6.501	0	617\\
	20.499	6.501	0	618\\
	nan	nan	0	619\\
	20.501	6.501	0	620\\
	20.501	6.501	0	621\\
	21.499	6.501	0	622\\
	21.499	6.501	0	623\\
	nan	nan	0	624\\
	21.501	6.501	0	625\\
	21.501	6.501	0	626\\
	22.499	6.501	0	627\\
	22.499	6.501	0	628\\
	nan	nan	0	629\\
	16.501	6.501	0	630\\
	16.501	6.501	1962	631\\
	17.499	6.501	1962	632\\
	17.499	6.501	0	633\\
	nan	nan	0	634\\
	17.501	6.501	0	635\\
	17.501	6.501	0	636\\
	18.499	6.501	0	637\\
	18.499	6.501	0	638\\
	nan	nan	0	639\\
	18.501	6.501	0	640\\
	18.501	6.501	0	641\\
	19.499	6.501	0	642\\
	19.499	6.501	0	643\\
	nan	nan	0	644\\
	19.501	6.501	0	645\\
	19.501	6.501	0	646\\
	20.499	6.501	0	647\\
	20.499	6.501	0	648\\
	nan	nan	0	649\\
	20.501	6.501	0	650\\
	20.501	6.501	0	651\\
	21.499	6.501	0	652\\
	21.499	6.501	0	653\\
	nan	nan	0	654\\
	21.501	6.501	0	655\\
	21.501	6.501	0	656\\
	22.499	6.501	0	657\\
	22.499	6.501	0	658\\
	nan	nan	0	659\\
	16.501	7.499	0	660\\
	16.501	7.499	1962	661\\
	17.499	7.499	1962	662\\
	17.499	7.499	0	663\\
	nan	nan	0	664\\
	17.501	7.499	0	665\\
	17.501	7.499	0	666\\
	18.499	7.499	0	667\\
	18.499	7.499	0	668\\
	nan	nan	0	669\\
	18.501	7.499	0	670\\
	18.501	7.499	0	671\\
	19.499	7.499	0	672\\
	19.499	7.499	0	673\\
	nan	nan	0	674\\
	19.501	7.499	0	675\\
	19.501	7.499	0	676\\
	20.499	7.499	0	677\\
	20.499	7.499	0	678\\
	nan	nan	0	679\\
	20.501	7.499	0	680\\
	20.501	7.499	0	681\\
	21.499	7.499	0	682\\
	21.499	7.499	0	683\\
	nan	nan	0	684\\
	21.501	7.499	0	685\\
	21.501	7.499	0	686\\
	22.499	7.499	0	687\\
	22.499	7.499	0	688\\
	nan	nan	0	689\\
	16.501	7.499	0	690\\
	16.501	7.499	0	691\\
	17.499	7.499	0	692\\
	17.499	7.499	0	693\\
	nan	nan	0	694\\
	17.501	7.499	0	695\\
	17.501	7.499	0	696\\
	18.499	7.499	0	697\\
	18.499	7.499	0	698\\
	nan	nan	0	699\\
	18.501	7.499	0	700\\
	18.501	7.499	0	701\\
	19.499	7.499	0	702\\
	19.499	7.499	0	703\\
	nan	nan	0	704\\
	19.501	7.499	0	705\\
	19.501	7.499	0	706\\
	20.499	7.499	0	707\\
	20.499	7.499	0	708\\
	nan	nan	0	709\\
	20.501	7.499	0	710\\
	20.501	7.499	0	711\\
	21.499	7.499	0	712\\
	21.499	7.499	0	713\\
	nan	nan	0	714\\
	21.501	7.499	0	715\\
	21.501	7.499	0	716\\
	22.499	7.499	0	717\\
	22.499	7.499	0	718\\
	nan	nan	0	719\\
	nan	nan	0	720\\
	nan	nan	0	721\\
	nan	nan	0	722\\
	nan	nan	0	723\\
	nan	nan	0	724\\
	nan	nan	0	725\\
	nan	nan	0	726\\
	nan	nan	0	727\\
	nan	nan	0	728\\
	nan	nan	0	729\\
	nan	nan	0	730\\
	nan	nan	0	731\\
	nan	nan	0	732\\
	nan	nan	0	733\\
	nan	nan	0	734\\
	nan	nan	0	735\\
	nan	nan	0	736\\
	nan	nan	0	737\\
	nan	nan	0	738\\
	nan	nan	0	739\\
	nan	nan	0	740\\
	nan	nan	0	741\\
	nan	nan	0	742\\
	nan	nan	0	743\\
	nan	nan	0	744\\
	nan	nan	0	745\\
	nan	nan	0	746\\
	nan	nan	0	747\\
	nan	nan	0	748\\
	nan	nan	0	749\\
	16.501	7.501	0	750\\
	16.501	7.501	0	751\\
	17.499	7.501	0	752\\
	17.499	7.501	0	753\\
	nan	nan	0	754\\
	17.501	7.501	0	755\\
	17.501	7.501	0	756\\
	18.499	7.501	0	757\\
	18.499	7.501	0	758\\
	nan	nan	0	759\\
	18.501	7.501	0	760\\
	18.501	7.501	0	761\\
	19.499	7.501	0	762\\
	19.499	7.501	0	763\\
	nan	nan	0	764\\
	19.501	7.501	0	765\\
	19.501	7.501	0	766\\
	20.499	7.501	0	767\\
	20.499	7.501	0	768\\
	nan	nan	0	769\\
	20.501	7.501	0	770\\
	20.501	7.501	0	771\\
	21.499	7.501	0	772\\
	21.499	7.501	0	773\\
	nan	nan	0	774\\
	21.501	7.501	0	775\\
	21.501	7.501	0	776\\
	22.499	7.501	0	777\\
	22.499	7.501	0	778\\
	nan	nan	0	779\\
	16.501	7.501	0	780\\
	16.501	7.501	1088	781\\
	17.499	7.501	1088	782\\
	17.499	7.501	0	783\\
	nan	nan	0	784\\
	17.501	7.501	0	785\\
	17.501	7.501	0	786\\
	18.499	7.501	0	787\\
	18.499	7.501	0	788\\
	nan	nan	0	789\\
	18.501	7.501	0	790\\
	18.501	7.501	0	791\\
	19.499	7.501	0	792\\
	19.499	7.501	0	793\\
	nan	nan	0	794\\
	19.501	7.501	0	795\\
	19.501	7.501	0	796\\
	20.499	7.501	0	797\\
	20.499	7.501	0	798\\
	nan	nan	0	799\\
	20.501	7.501	0	800\\
	20.501	7.501	0	801\\
	21.499	7.501	0	802\\
	21.499	7.501	0	803\\
	nan	nan	0	804\\
	21.501	7.501	0	805\\
	21.501	7.501	0	806\\
	22.499	7.501	0	807\\
	22.499	7.501	0	808\\
	nan	nan	0	809\\
	16.501	8.499	0	810\\
	16.501	8.499	1088	811\\
	17.499	8.499	1088	812\\
	17.499	8.499	0	813\\
	nan	nan	0	814\\
	17.501	8.499	0	815\\
	17.501	8.499	0	816\\
	18.499	8.499	0	817\\
	18.499	8.499	0	818\\
	nan	nan	0	819\\
	18.501	8.499	0	820\\
	18.501	8.499	0	821\\
	19.499	8.499	0	822\\
	19.499	8.499	0	823\\
	nan	nan	0	824\\
	19.501	8.499	0	825\\
	19.501	8.499	0	826\\
	20.499	8.499	0	827\\
	20.499	8.499	0	828\\
	nan	nan	0	829\\
	20.501	8.499	0	830\\
	20.501	8.499	0	831\\
	21.499	8.499	0	832\\
	21.499	8.499	0	833\\
	nan	nan	0	834\\
	21.501	8.499	0	835\\
	21.501	8.499	0	836\\
	22.499	8.499	0	837\\
	22.499	8.499	0	838\\
	nan	nan	0	839\\
	16.501	8.499	0	840\\
	16.501	8.499	0	841\\
	17.499	8.499	0	842\\
	17.499	8.499	0	843\\
	nan	nan	0	844\\
	17.501	8.499	0	845\\
	17.501	8.499	0	846\\
	18.499	8.499	0	847\\
	18.499	8.499	0	848\\
	nan	nan	0	849\\
	18.501	8.499	0	850\\
	18.501	8.499	0	851\\
	19.499	8.499	0	852\\
	19.499	8.499	0	853\\
	nan	nan	0	854\\
	19.501	8.499	0	855\\
	19.501	8.499	0	856\\
	20.499	8.499	0	857\\
	20.499	8.499	0	858\\
	nan	nan	0	859\\
	20.501	8.499	0	860\\
	20.501	8.499	0	861\\
	21.499	8.499	0	862\\
	21.499	8.499	0	863\\
	nan	nan	0	864\\
	21.501	8.499	0	865\\
	21.501	8.499	0	866\\
	22.499	8.499	0	867\\
	22.499	8.499	0	868\\
	nan	nan	0	869\\
	nan	nan	0	870\\
	nan	nan	0	871\\
	nan	nan	0	872\\
	nan	nan	0	873\\
	nan	nan	0	874\\
	nan	nan	0	875\\
	nan	nan	0	876\\
	nan	nan	0	877\\
	nan	nan	0	878\\
	nan	nan	0	879\\
	nan	nan	0	880\\
	nan	nan	0	881\\
	nan	nan	0	882\\
	nan	nan	0	883\\
	nan	nan	0	884\\
	nan	nan	0	885\\
	nan	nan	0	886\\
	nan	nan	0	887\\
	nan	nan	0	888\\
	nan	nan	0	889\\
	nan	nan	0	890\\
	nan	nan	0	891\\
	nan	nan	0	892\\
	nan	nan	0	893\\
	nan	nan	0	894\\
	nan	nan	0	895\\
	nan	nan	0	896\\
	nan	nan	0	897\\
	nan	nan	0	898\\
	nan	nan	0	899\\
};

\end{axis}

\end{tikzpicture}%

%% file: plots/Revised_Best_Card_Set_delta_d_-14_SNR_10.tex
%
%
\begin{tikzpicture}

\begin{axis}[%
scale=0.6, width=3.284in,
height=3.566in,
at={(0.758in,0.481in)},
scale only axis,
unbounded coords=jump,
colormap={patchmap}{[1pt] rgb(0pt)=(0.75,0.85,0.95); rgb(899pt)=(0.75,0.85,0.95)},
xmin=16.2,
xmax=22.8,
xtick={17, 18, 19, 20, 21, 22},
xticklabel style = {font=\footnotesize},
tick align=outside,
xlabel style={font=\color{white!15!black}},
xlabel style = {font=\small},
xlabel={$|\mathcal{P}_{1}^{\star}|$},
ymin=2.2,
ymax=8.8,
ytick={3, 4, 5, 6, 7, 8},
yticklabel style = {font=\footnotesize},
ylabel style={font=\color{white!15!black}},
ylabel style = {font=\small},
ylabel={$|\mathcal{P}_{2}^{\star}|$},
zmin=0,
zmax=2200,
zlabel style={font=\color{white!15!black}},
zlabel style = {font=\small},
zticklabel style = {font=\footnotesize},
zlabel={Frequency},
view={-37.5}{30},
axis background/.style={fill=white},
title style={font=\bfseries},
axis x line*=bottom,
axis y line*=left,
axis z line*=left,
xmajorgrids,
ymajorgrids,
zmajorgrids,
legend style={at={(1.03,1)}, anchor=north west, legend cell align=left, align=left, draw=white!15!black}
]

\addplot3[%
surf,
shader=flat corner, draw=black, mesh/rows=30]
table[row sep=crcr, colormap name=surfmap, point meta=\thisrow{c}] {%
	x	y	z	c\\
	16.501	2.501	0	0\\
	16.501	2.501	0	1\\
	17.499	2.501	0	2\\
	17.499	2.501	0	3\\
	nan	nan	0	4\\
	17.501	2.501	0	5\\
	17.501	2.501	0	6\\
	18.499	2.501	0	7\\
	18.499	2.501	0	8\\
	nan	nan	0	9\\
	18.501	2.501	0	10\\
	18.501	2.501	0	11\\
	19.499	2.501	0	12\\
	19.499	2.501	0	13\\
	nan	nan	0	14\\
	19.501	2.501	0	15\\
	19.501	2.501	0	16\\
	20.499	2.501	0	17\\
	20.499	2.501	0	18\\
	nan	nan	0	19\\
	20.501	2.501	0	20\\
	20.501	2.501	0	21\\
	21.499	2.501	0	22\\
	21.499	2.501	0	23\\
	nan	nan	0	24\\
	21.501	2.501	0	25\\
	21.501	2.501	0	26\\
	22.499	2.501	0	27\\
	22.499	2.501	0	28\\
	nan	nan	0	29\\
	16.501	2.501	0	30\\
	16.501	2.501	0	31\\
	17.499	2.501	0	32\\
	17.499	2.501	0	33\\
	nan	nan	0	34\\
	17.501	2.501	0	35\\
	17.501	2.501	0	36\\
	18.499	2.501	0	37\\
	18.499	2.501	0	38\\
	nan	nan	0	39\\
	18.501	2.501	0	40\\
	18.501	2.501	0	41\\
	19.499	2.501	0	42\\
	19.499	2.501	0	43\\
	nan	nan	0	44\\
	19.501	2.501	0	45\\
	19.501	2.501	0	46\\
	20.499	2.501	0	47\\
	20.499	2.501	0	48\\
	nan	nan	0	49\\
	20.501	2.501	0	50\\
	20.501	2.501	0	51\\
	21.499	2.501	0	52\\
	21.499	2.501	0	53\\
	nan	nan	0	54\\
	21.501	2.501	0	55\\
	21.501	2.501	0	56\\
	22.499	2.501	0	57\\
	22.499	2.501	0	58\\
	nan	nan	0	59\\
	16.501	3.499	0	60\\
	16.501	3.499	0	61\\
	17.499	3.499	0	62\\
	17.499	3.499	0	63\\
	nan	nan	0	64\\
	17.501	3.499	0	65\\
	17.501	3.499	0	66\\
	18.499	3.499	0	67\\
	18.499	3.499	0	68\\
	nan	nan	0	69\\
	18.501	3.499	0	70\\
	18.501	3.499	0	71\\
	19.499	3.499	0	72\\
	19.499	3.499	0	73\\
	nan	nan	0	74\\
	19.501	3.499	0	75\\
	19.501	3.499	0	76\\
	20.499	3.499	0	77\\
	20.499	3.499	0	78\\
	nan	nan	0	79\\
	20.501	3.499	0	80\\
	20.501	3.499	0	81\\
	21.499	3.499	0	82\\
	21.499	3.499	0	83\\
	nan	nan	0	84\\
	21.501	3.499	0	85\\
	21.501	3.499	0	86\\
	22.499	3.499	0	87\\
	22.499	3.499	0	88\\
	nan	nan	0	89\\
	16.501	3.499	0	90\\
	16.501	3.499	0	91\\
	17.499	3.499	0	92\\
	17.499	3.499	0	93\\
	nan	nan	0	94\\
	17.501	3.499	0	95\\
	17.501	3.499	0	96\\
	18.499	3.499	0	97\\
	18.499	3.499	0	98\\
	nan	nan	0	99\\
	18.501	3.499	0	100\\
	18.501	3.499	0	101\\
	19.499	3.499	0	102\\
	19.499	3.499	0	103\\
	nan	nan	0	104\\
	19.501	3.499	0	105\\
	19.501	3.499	0	106\\
	20.499	3.499	0	107\\
	20.499	3.499	0	108\\
	nan	nan	0	109\\
	20.501	3.499	0	110\\
	20.501	3.499	0	111\\
	21.499	3.499	0	112\\
	21.499	3.499	0	113\\
	nan	nan	0	114\\
	21.501	3.499	0	115\\
	21.501	3.499	0	116\\
	22.499	3.499	0	117\\
	22.499	3.499	0	118\\
	nan	nan	0	119\\
	nan	nan	0	120\\
	nan	nan	0	121\\
	nan	nan	0	122\\
	nan	nan	0	123\\
	nan	nan	0	124\\
	nan	nan	0	125\\
	nan	nan	0	126\\
	nan	nan	0	127\\
	nan	nan	0	128\\
	nan	nan	0	129\\
	nan	nan	0	130\\
	nan	nan	0	131\\
	nan	nan	0	132\\
	nan	nan	0	133\\
	nan	nan	0	134\\
	nan	nan	0	135\\
	nan	nan	0	136\\
	nan	nan	0	137\\
	nan	nan	0	138\\
	nan	nan	0	139\\
	nan	nan	0	140\\
	nan	nan	0	141\\
	nan	nan	0	142\\
	nan	nan	0	143\\
	nan	nan	0	144\\
	nan	nan	0	145\\
	nan	nan	0	146\\
	nan	nan	0	147\\
	nan	nan	0	148\\
	nan	nan	0	149\\
	16.501	3.501	0	150\\
	16.501	3.501	0	151\\
	17.499	3.501	0	152\\
	17.499	3.501	0	153\\
	nan	nan	0	154\\
	17.501	3.501	0	155\\
	17.501	3.501	0	156\\
	18.499	3.501	0	157\\
	18.499	3.501	0	158\\
	nan	nan	0	159\\
	18.501	3.501	0	160\\
	18.501	3.501	0	161\\
	19.499	3.501	0	162\\
	19.499	3.501	0	163\\
	nan	nan	0	164\\
	19.501	3.501	0	165\\
	19.501	3.501	0	166\\
	20.499	3.501	0	167\\
	20.499	3.501	0	168\\
	nan	nan	0	169\\
	20.501	3.501	0	170\\
	20.501	3.501	0	171\\
	21.499	3.501	0	172\\
	21.499	3.501	0	173\\
	nan	nan	0	174\\
	21.501	3.501	0	175\\
	21.501	3.501	0	176\\
	22.499	3.501	0	177\\
	22.499	3.501	0	178\\
	nan	nan	0	179\\
	16.501	3.501	0	180\\
	16.501	3.501	0	181\\
	17.499	3.501	0	182\\
	17.499	3.501	0	183\\
	nan	nan	0	184\\
	17.501	3.501	0	185\\
	17.501	3.501	0	186\\
	18.499	3.501	0	187\\
	18.499	3.501	0	188\\
	nan	nan	0	189\\
	18.501	3.501	0	190\\
	18.501	3.501	0	191\\
	19.499	3.501	0	192\\
	19.499	3.501	0	193\\
	nan	nan	0	194\\
	19.501	3.501	0	195\\
	19.501	3.501	0	196\\
	20.499	3.501	0	197\\
	20.499	3.501	0	198\\
	nan	nan	0	199\\
	20.501	3.501	0	200\\
	20.501	3.501	2	201\\
	21.499	3.501	2	202\\
	21.499	3.501	0	203\\
	nan	nan	0	204\\
	21.501	3.501	0	205\\
	21.501	3.501	2	206\\
	22.499	3.501	2	207\\
	22.499	3.501	0	208\\
	nan	nan	0	209\\
	16.501	4.499	0	210\\
	16.501	4.499	0	211\\
	17.499	4.499	0	212\\
	17.499	4.499	0	213\\
	nan	nan	0	214\\
	17.501	4.499	0	215\\
	17.501	4.499	0	216\\
	18.499	4.499	0	217\\
	18.499	4.499	0	218\\
	nan	nan	0	219\\
	18.501	4.499	0	220\\
	18.501	4.499	0	221\\
	19.499	4.499	0	222\\
	19.499	4.499	0	223\\
	nan	nan	0	224\\
	19.501	4.499	0	225\\
	19.501	4.499	0	226\\
	20.499	4.499	0	227\\
	20.499	4.499	0	228\\
	nan	nan	0	229\\
	20.501	4.499	0	230\\
	20.501	4.499	2	231\\
	21.499	4.499	2	232\\
	21.499	4.499	0	233\\
	nan	nan	0	234\\
	21.501	4.499	0	235\\
	21.501	4.499	2	236\\
	22.499	4.499	2	237\\
	22.499	4.499	0	238\\
	nan	nan	0	239\\
	16.501	4.499	0	240\\
	16.501	4.499	0	241\\
	17.499	4.499	0	242\\
	17.499	4.499	0	243\\
	nan	nan	0	244\\
	17.501	4.499	0	245\\
	17.501	4.499	0	246\\
	18.499	4.499	0	247\\
	18.499	4.499	0	248\\
	nan	nan	0	249\\
	18.501	4.499	0	250\\
	18.501	4.499	0	251\\
	19.499	4.499	0	252\\
	19.499	4.499	0	253\\
	nan	nan	0	254\\
	19.501	4.499	0	255\\
	19.501	4.499	0	256\\
	20.499	4.499	0	257\\
	20.499	4.499	0	258\\
	nan	nan	0	259\\
	20.501	4.499	0	260\\
	20.501	4.499	0	261\\
	21.499	4.499	0	262\\
	21.499	4.499	0	263\\
	nan	nan	0	264\\
	21.501	4.499	0	265\\
	21.501	4.499	0	266\\
	22.499	4.499	0	267\\
	22.499	4.499	0	268\\
	nan	nan	0	269\\
	nan	nan	0	270\\
	nan	nan	0	271\\
	nan	nan	0	272\\
	nan	nan	0	273\\
	nan	nan	0	274\\
	nan	nan	0	275\\
	nan	nan	0	276\\
	nan	nan	0	277\\
	nan	nan	0	278\\
	nan	nan	0	279\\
	nan	nan	0	280\\
	nan	nan	0	281\\
	nan	nan	0	282\\
	nan	nan	0	283\\
	nan	nan	0	284\\
	nan	nan	0	285\\
	nan	nan	0	286\\
	nan	nan	0	287\\
	nan	nan	0	288\\
	nan	nan	0	289\\
	nan	nan	0	290\\
	nan	nan	0	291\\
	nan	nan	0	292\\
	nan	nan	0	293\\
	nan	nan	0	294\\
	nan	nan	0	295\\
	nan	nan	0	296\\
	nan	nan	0	297\\
	nan	nan	0	298\\
	nan	nan	0	299\\
	16.501	4.501	0	300\\
	16.501	4.501	0	301\\
	17.499	4.501	0	302\\
	17.499	4.501	0	303\\
	nan	nan	0	304\\
	17.501	4.501	0	305\\
	17.501	4.501	0	306\\
	18.499	4.501	0	307\\
	18.499	4.501	0	308\\
	nan	nan	0	309\\
	18.501	4.501	0	310\\
	18.501	4.501	0	311\\
	19.499	4.501	0	312\\
	19.499	4.501	0	313\\
	nan	nan	0	314\\
	19.501	4.501	0	315\\
	19.501	4.501	0	316\\
	20.499	4.501	0	317\\
	20.499	4.501	0	318\\
	nan	nan	0	319\\
	20.501	4.501	0	320\\
	20.501	4.501	0	321\\
	21.499	4.501	0	322\\
	21.499	4.501	0	323\\
	nan	nan	0	324\\
	21.501	4.501	0	325\\
	21.501	4.501	0	326\\
	22.499	4.501	0	327\\
	22.499	4.501	0	328\\
	nan	nan	0	329\\
	16.501	4.501	0	330\\
	16.501	4.501	0	331\\
	17.499	4.501	0	332\\
	17.499	4.501	0	333\\
	nan	nan	0	334\\
	17.501	4.501	0	335\\
	17.501	4.501	0	336\\
	18.499	4.501	0	337\\
	18.499	4.501	0	338\\
	nan	nan	0	339\\
	18.501	4.501	0	340\\
	18.501	4.501	0	341\\
	19.499	4.501	0	342\\
	19.499	4.501	0	343\\
	nan	nan	0	344\\
	19.501	4.501	0	345\\
	19.501	4.501	2	346\\
	20.499	4.501	2	347\\
	20.499	4.501	0	348\\
	nan	nan	0	349\\
	20.501	4.501	0	350\\
	20.501	4.501	16	351\\
	21.499	4.501	16	352\\
	21.499	4.501	0	353\\
	nan	nan	0	354\\
	21.501	4.501	0	355\\
	21.501	4.501	28	356\\
	22.499	4.501	28	357\\
	22.499	4.501	0	358\\
	nan	nan	0	359\\
	16.501	5.499	0	360\\
	16.501	5.499	0	361\\
	17.499	5.499	0	362\\
	17.499	5.499	0	363\\
	nan	nan	0	364\\
	17.501	5.499	0	365\\
	17.501	5.499	0	366\\
	18.499	5.499	0	367\\
	18.499	5.499	0	368\\
	nan	nan	0	369\\
	18.501	5.499	0	370\\
	18.501	5.499	0	371\\
	19.499	5.499	0	372\\
	19.499	5.499	0	373\\
	nan	nan	0	374\\
	19.501	5.499	0	375\\
	19.501	5.499	2	376\\
	20.499	5.499	2	377\\
	20.499	5.499	0	378\\
	nan	nan	0	379\\
	20.501	5.499	0	380\\
	20.501	5.499	16	381\\
	21.499	5.499	16	382\\
	21.499	5.499	0	383\\
	nan	nan	0	384\\
	21.501	5.499	0	385\\
	21.501	5.499	28	386\\
	22.499	5.499	28	387\\
	22.499	5.499	0	388\\
	nan	nan	0	389\\
	16.501	5.499	0	390\\
	16.501	5.499	0	391\\
	17.499	5.499	0	392\\
	17.499	5.499	0	393\\
	nan	nan	0	394\\
	17.501	5.499	0	395\\
	17.501	5.499	0	396\\
	18.499	5.499	0	397\\
	18.499	5.499	0	398\\
	nan	nan	0	399\\
	18.501	5.499	0	400\\
	18.501	5.499	0	401\\
	19.499	5.499	0	402\\
	19.499	5.499	0	403\\
	nan	nan	0	404\\
	19.501	5.499	0	405\\
	19.501	5.499	0	406\\
	20.499	5.499	0	407\\
	20.499	5.499	0	408\\
	nan	nan	0	409\\
	20.501	5.499	0	410\\
	20.501	5.499	0	411\\
	21.499	5.499	0	412\\
	21.499	5.499	0	413\\
	nan	nan	0	414\\
	21.501	5.499	0	415\\
	21.501	5.499	0	416\\
	22.499	5.499	0	417\\
	22.499	5.499	0	418\\
	nan	nan	0	419\\
	nan	nan	0	420\\
	nan	nan	0	421\\
	nan	nan	0	422\\
	nan	nan	0	423\\
	nan	nan	0	424\\
	nan	nan	0	425\\
	nan	nan	0	426\\
	nan	nan	0	427\\
	nan	nan	0	428\\
	nan	nan	0	429\\
	nan	nan	0	430\\
	nan	nan	0	431\\
	nan	nan	0	432\\
	nan	nan	0	433\\
	nan	nan	0	434\\
	nan	nan	0	435\\
	nan	nan	0	436\\
	nan	nan	0	437\\
	nan	nan	0	438\\
	nan	nan	0	439\\
	nan	nan	0	440\\
	nan	nan	0	441\\
	nan	nan	0	442\\
	nan	nan	0	443\\
	nan	nan	0	444\\
	nan	nan	0	445\\
	nan	nan	0	446\\
	nan	nan	0	447\\
	nan	nan	0	448\\
	nan	nan	0	449\\
	16.501	5.501	0	450\\
	16.501	5.501	0	451\\
	17.499	5.501	0	452\\
	17.499	5.501	0	453\\
	nan	nan	0	454\\
	17.501	5.501	0	455\\
	17.501	5.501	0	456\\
	18.499	5.501	0	457\\
	18.499	5.501	0	458\\
	nan	nan	0	459\\
	18.501	5.501	0	460\\
	18.501	5.501	0	461\\
	19.499	5.501	0	462\\
	19.499	5.501	0	463\\
	nan	nan	0	464\\
	19.501	5.501	0	465\\
	19.501	5.501	0	466\\
	20.499	5.501	0	467\\
	20.499	5.501	0	468\\
	nan	nan	0	469\\
	20.501	5.501	0	470\\
	20.501	5.501	0	471\\
	21.499	5.501	0	472\\
	21.499	5.501	0	473\\
	nan	nan	0	474\\
	21.501	5.501	0	475\\
	21.501	5.501	0	476\\
	22.499	5.501	0	477\\
	22.499	5.501	0	478\\
	nan	nan	0	479\\
	16.501	5.501	0	480\\
	16.501	5.501	0	481\\
	17.499	5.501	0	482\\
	17.499	5.501	0	483\\
	nan	nan	0	484\\
	17.501	5.501	0	485\\
	17.501	5.501	1	486\\
	18.499	5.501	1	487\\
	18.499	5.501	0	488\\
	nan	nan	0	489\\
	18.501	5.501	0	490\\
	18.501	5.501	8	491\\
	19.499	5.501	8	492\\
	19.499	5.501	0	493\\
	nan	nan	0	494\\
	19.501	5.501	0	495\\
	19.501	5.501	79	496\\
	20.499	5.501	79	497\\
	20.499	5.501	0	498\\
	nan	nan	0	499\\
	20.501	5.501	0	500\\
	20.501	5.501	295	501\\
	21.499	5.501	295	502\\
	21.499	5.501	0	503\\
	nan	nan	0	504\\
	21.501	5.501	0	505\\
	21.501	5.501	304	506\\
	22.499	5.501	304	507\\
	22.499	5.501	0	508\\
	nan	nan	0	509\\
	16.501	6.499	0	510\\
	16.501	6.499	0	511\\
	17.499	6.499	0	512\\
	17.499	6.499	0	513\\
	nan	nan	0	514\\
	17.501	6.499	0	515\\
	17.501	6.499	1	516\\
	18.499	6.499	1	517\\
	18.499	6.499	0	518\\
	nan	nan	0	519\\
	18.501	6.499	0	520\\
	18.501	6.499	8	521\\
	19.499	6.499	8	522\\
	19.499	6.499	0	523\\
	nan	nan	0	524\\
	19.501	6.499	0	525\\
	19.501	6.499	79	526\\
	20.499	6.499	79	527\\
	20.499	6.499	0	528\\
	nan	nan	0	529\\
	20.501	6.499	0	530\\
	20.501	6.499	295	531\\
	21.499	6.499	295	532\\
	21.499	6.499	0	533\\
	nan	nan	0	534\\
	21.501	6.499	0	535\\
	21.501	6.499	304	536\\
	22.499	6.499	304	537\\
	22.499	6.499	0	538\\
	nan	nan	0	539\\
	16.501	6.499	0	540\\
	16.501	6.499	0	541\\
	17.499	6.499	0	542\\
	17.499	6.499	0	543\\
	nan	nan	0	544\\
	17.501	6.499	0	545\\
	17.501	6.499	0	546\\
	18.499	6.499	0	547\\
	18.499	6.499	0	548\\
	nan	nan	0	549\\
	18.501	6.499	0	550\\
	18.501	6.499	0	551\\
	19.499	6.499	0	552\\
	19.499	6.499	0	553\\
	nan	nan	0	554\\
	19.501	6.499	0	555\\
	19.501	6.499	0	556\\
	20.499	6.499	0	557\\
	20.499	6.499	0	558\\
	nan	nan	0	559\\
	20.501	6.499	0	560\\
	20.501	6.499	0	561\\
	21.499	6.499	0	562\\
	21.499	6.499	0	563\\
	nan	nan	0	564\\
	21.501	6.499	0	565\\
	21.501	6.499	0	566\\
	22.499	6.499	0	567\\
	22.499	6.499	0	568\\
	nan	nan	0	569\\
	nan	nan	0	570\\
	nan	nan	0	571\\
	nan	nan	0	572\\
	nan	nan	0	573\\
	nan	nan	0	574\\
	nan	nan	0	575\\
	nan	nan	0	576\\
	nan	nan	0	577\\
	nan	nan	0	578\\
	nan	nan	0	579\\
	nan	nan	0	580\\
	nan	nan	0	581\\
	nan	nan	0	582\\
	nan	nan	0	583\\
	nan	nan	0	584\\
	nan	nan	0	585\\
	nan	nan	0	586\\
	nan	nan	0	587\\
	nan	nan	0	588\\
	nan	nan	0	589\\
	nan	nan	0	590\\
	nan	nan	0	591\\
	nan	nan	0	592\\
	nan	nan	0	593\\
	nan	nan	0	594\\
	nan	nan	0	595\\
	nan	nan	0	596\\
	nan	nan	0	597\\
	nan	nan	0	598\\
	nan	nan	0	599\\
	16.501	6.501	0	600\\
	16.501	6.501	0	601\\
	17.499	6.501	0	602\\
	17.499	6.501	0	603\\
	nan	nan	0	604\\
	17.501	6.501	0	605\\
	17.501	6.501	0	606\\
	18.499	6.501	0	607\\
	18.499	6.501	0	608\\
	nan	nan	0	609\\
	18.501	6.501	0	610\\
	18.501	6.501	0	611\\
	19.499	6.501	0	612\\
	19.499	6.501	0	613\\
	nan	nan	0	614\\
	19.501	6.501	0	615\\
	19.501	6.501	0	616\\
	20.499	6.501	0	617\\
	20.499	6.501	0	618\\
	nan	nan	0	619\\
	20.501	6.501	0	620\\
	20.501	6.501	0	621\\
	21.499	6.501	0	622\\
	21.499	6.501	0	623\\
	nan	nan	0	624\\
	21.501	6.501	0	625\\
	21.501	6.501	0	626\\
	22.499	6.501	0	627\\
	22.499	6.501	0	628\\
	nan	nan	0	629\\
	16.501	6.501	0	630\\
	16.501	6.501	1	631\\
	17.499	6.501	1	632\\
	17.499	6.501	0	633\\
	nan	nan	0	634\\
	17.501	6.501	0	635\\
	17.501	6.501	10	636\\
	18.499	6.501	10	637\\
	18.499	6.501	0	638\\
	nan	nan	0	639\\
	18.501	6.501	0	640\\
	18.501	6.501	160	641\\
	19.499	6.501	160	642\\
	19.499	6.501	0	643\\
	nan	nan	0	644\\
	19.501	6.501	0	645\\
	19.501	6.501	837	646\\
	20.499	6.501	837	647\\
	20.499	6.501	0	648\\
	nan	nan	0	649\\
	20.501	6.501	0	650\\
	20.501	6.501	1461	651\\
	21.499	6.501	1461	652\\
	21.499	6.501	0	653\\
	nan	nan	0	654\\
	21.501	6.501	0	655\\
	21.501	6.501	311	656\\
	22.499	6.501	311	657\\
	22.499	6.501	0	658\\
	nan	nan	0	659\\
	16.501	7.499	0	660\\
	16.501	7.499	1	661\\
	17.499	7.499	1	662\\
	17.499	7.499	0	663\\
	nan	nan	0	664\\
	17.501	7.499	0	665\\
	17.501	7.499	10	666\\
	18.499	7.499	10	667\\
	18.499	7.499	0	668\\
	nan	nan	0	669\\
	18.501	7.499	0	670\\
	18.501	7.499	160	671\\
	19.499	7.499	160	672\\
	19.499	7.499	0	673\\
	nan	nan	0	674\\
	19.501	7.499	0	675\\
	19.501	7.499	837	676\\
	20.499	7.499	837	677\\
	20.499	7.499	0	678\\
	nan	nan	0	679\\
	20.501	7.499	0	680\\
	20.501	7.499	1461	681\\
	21.499	7.499	1461	682\\
	21.499	7.499	0	683\\
	nan	nan	0	684\\
	21.501	7.499	0	685\\
	21.501	7.499	311	686\\
	22.499	7.499	311	687\\
	22.499	7.499	0	688\\
	nan	nan	0	689\\
	16.501	7.499	0	690\\
	16.501	7.499	0	691\\
	17.499	7.499	0	692\\
	17.499	7.499	0	693\\
	nan	nan	0	694\\
	17.501	7.499	0	695\\
	17.501	7.499	0	696\\
	18.499	7.499	0	697\\
	18.499	7.499	0	698\\
	nan	nan	0	699\\
	18.501	7.499	0	700\\
	18.501	7.499	0	701\\
	19.499	7.499	0	702\\
	19.499	7.499	0	703\\
	nan	nan	0	704\\
	19.501	7.499	0	705\\
	19.501	7.499	0	706\\
	20.499	7.499	0	707\\
	20.499	7.499	0	708\\
	nan	nan	0	709\\
	20.501	7.499	0	710\\
	20.501	7.499	0	711\\
	21.499	7.499	0	712\\
	21.499	7.499	0	713\\
	nan	nan	0	714\\
	21.501	7.499	0	715\\
	21.501	7.499	0	716\\
	22.499	7.499	0	717\\
	22.499	7.499	0	718\\
	nan	nan	0	719\\
	nan	nan	0	720\\
	nan	nan	0	721\\
	nan	nan	0	722\\
	nan	nan	0	723\\
	nan	nan	0	724\\
	nan	nan	0	725\\
	nan	nan	0	726\\
	nan	nan	0	727\\
	nan	nan	0	728\\
	nan	nan	0	729\\
	nan	nan	0	730\\
	nan	nan	0	731\\
	nan	nan	0	732\\
	nan	nan	0	733\\
	nan	nan	0	734\\
	nan	nan	0	735\\
	nan	nan	0	736\\
	nan	nan	0	737\\
	nan	nan	0	738\\
	nan	nan	0	739\\
	nan	nan	0	740\\
	nan	nan	0	741\\
	nan	nan	0	742\\
	nan	nan	0	743\\
	nan	nan	0	744\\
	nan	nan	0	745\\
	nan	nan	0	746\\
	nan	nan	0	747\\
	nan	nan	0	748\\
	nan	nan	0	749\\
	16.501	7.501	0	750\\
	16.501	7.501	0	751\\
	17.499	7.501	0	752\\
	17.499	7.501	0	753\\
	nan	nan	0	754\\
	17.501	7.501	0	755\\
	17.501	7.501	0	756\\
	18.499	7.501	0	757\\
	18.499	7.501	0	758\\
	nan	nan	0	759\\
	18.501	7.501	0	760\\
	18.501	7.501	0	761\\
	19.499	7.501	0	762\\
	19.499	7.501	0	763\\
	nan	nan	0	764\\
	19.501	7.501	0	765\\
	19.501	7.501	0	766\\
	20.499	7.501	0	767\\
	20.499	7.501	0	768\\
	nan	nan	0	769\\
	20.501	7.501	0	770\\
	20.501	7.501	0	771\\
	21.499	7.501	0	772\\
	21.499	7.501	0	773\\
	nan	nan	0	774\\
	21.501	7.501	0	775\\
	21.501	7.501	0	776\\
	22.499	7.501	0	777\\
	22.499	7.501	0	778\\
	nan	nan	0	779\\
	16.501	7.501	0	780\\
	16.501	7.501	8	781\\
	17.499	7.501	8	782\\
	17.499	7.501	0	783\\
	nan	nan	0	784\\
	17.501	7.501	0	785\\
	17.501	7.501	105	786\\
	18.499	7.501	105	787\\
	18.499	7.501	0	788\\
	nan	nan	0	789\\
	18.501	7.501	0	790\\
	18.501	7.501	843	791\\
	19.499	7.501	843	792\\
	19.499	7.501	0	793\\
	nan	nan	0	794\\
	19.501	7.501	0	795\\
	19.501	7.501	2179	796\\
	20.499	7.501	2179	797\\
	20.499	7.501	0	798\\
	nan	nan	0	799\\
	20.501	7.501	0	800\\
	20.501	7.501	848	801\\
	21.499	7.501	848	802\\
	21.499	7.501	0	803\\
	nan	nan	0	804\\
	21.501	7.501	0	805\\
	21.501	7.501	0	806\\
	22.499	7.501	0	807\\
	22.499	7.501	0	808\\
	nan	nan	0	809\\
	16.501	8.499	0	810\\
	16.501	8.499	8	811\\
	17.499	8.499	8	812\\
	17.499	8.499	0	813\\
	nan	nan	0	814\\
	17.501	8.499	0	815\\
	17.501	8.499	105	816\\
	18.499	8.499	105	817\\
	18.499	8.499	0	818\\
	nan	nan	0	819\\
	18.501	8.499	0	820\\
	18.501	8.499	843	821\\
	19.499	8.499	843	822\\
	19.499	8.499	0	823\\
	nan	nan	0	824\\
	19.501	8.499	0	825\\
	19.501	8.499	2179	826\\
	20.499	8.499	2179	827\\
	20.499	8.499	0	828\\
	nan	nan	0	829\\
	20.501	8.499	0	830\\
	20.501	8.499	848	831\\
	21.499	8.499	848	832\\
	21.499	8.499	0	833\\
	nan	nan	0	834\\
	21.501	8.499	0	835\\
	21.501	8.499	0	836\\
	22.499	8.499	0	837\\
	22.499	8.499	0	838\\
	nan	nan	0	839\\
	16.501	8.499	0	840\\
	16.501	8.499	0	841\\
	17.499	8.499	0	842\\
	17.499	8.499	0	843\\
	nan	nan	0	844\\
	17.501	8.499	0	845\\
	17.501	8.499	0	846\\
	18.499	8.499	0	847\\
	18.499	8.499	0	848\\
	nan	nan	0	849\\
	18.501	8.499	0	850\\
	18.501	8.499	0	851\\
	19.499	8.499	0	852\\
	19.499	8.499	0	853\\
	nan	nan	0	854\\
	19.501	8.499	0	855\\
	19.501	8.499	0	856\\
	20.499	8.499	0	857\\
	20.499	8.499	0	858\\
	nan	nan	0	859\\
	20.501	8.499	0	860\\
	20.501	8.499	0	861\\
	21.499	8.499	0	862\\
	21.499	8.499	0	863\\
	nan	nan	0	864\\
	21.501	8.499	0	865\\
	21.501	8.499	0	866\\
	22.499	8.499	0	867\\
	22.499	8.499	0	868\\
	nan	nan	0	869\\
	nan	nan	0	870\\
	nan	nan	0	871\\
	nan	nan	0	872\\
	nan	nan	0	873\\
	nan	nan	0	874\\
	nan	nan	0	875\\
	nan	nan	0	876\\
	nan	nan	0	877\\
	nan	nan	0	878\\
	nan	nan	0	879\\
	nan	nan	0	880\\
	nan	nan	0	881\\
	nan	nan	0	882\\
	nan	nan	0	883\\
	nan	nan	0	884\\
	nan	nan	0	885\\
	nan	nan	0	886\\
	nan	nan	0	887\\
	nan	nan	0	888\\
	nan	nan	0	889\\
	nan	nan	0	890\\
	nan	nan	0	891\\
	nan	nan	0	892\\
	nan	nan	0	893\\
	nan	nan	0	894\\
	nan	nan	0	895\\
	nan	nan	0	896\\
	nan	nan	0	897\\
	nan	nan	0	898\\
	nan	nan	0	899\\
};

\end{axis}

\end{tikzpicture}%

%% file: plots/Revised_Best_Card_Set_delta_d_-14_SNR_30.tex
%
%
\begin{tikzpicture}

\begin{axis}[%
scale=0.6, width=3.284in,
height=3.566in,
at={(0.758in,0.481in)},
scale only axis,
unbounded coords=jump,
colormap={patchmap}{[1pt] rgb(0pt)=(0.75,0.85,0.95); rgb(899pt)=(0.75,0.85,0.95)},
xmin=16.2,
xmax=22.8,
xtick={17, 18, 19, 20, 21, 22},
xticklabel style = {font=\footnotesize},
tick align=outside,
xlabel style={font=\color{white!15!black}},
xlabel style = {font=\small},
xlabel={$|\mathcal{P}_{1}^{\star}|$},
ymin=2.2,
ymax=8.8,
ytick={3, 4, 5, 6, 7, 8},
yticklabel style = {font=\footnotesize},
ylabel style={font=\color{white!15!black}},
ylabel style = {font=\small},
ylabel={$|\mathcal{P}_{2}^{\star}|$},
zmin=0,
zmax=6200,
zlabel style={font=\color{white!15!black}},
zlabel style = {font=\small},
zticklabel style = {font=\footnotesize},
zlabel={Frequency},
view={-37.5}{30},
axis background/.style={fill=white},
title style={font=\bfseries},
axis x line*=bottom,
axis y line*=left,
axis z line*=left,
xmajorgrids,
ymajorgrids,
zmajorgrids,
legend style={at={(1.03,1)}, anchor=north west, legend cell align=left, align=left, draw=white!15!black}
]

\addplot3[%
surf,
shader=flat corner, draw=black, mesh/rows=30]
table[row sep=crcr, colormap name=surfmap, point meta=\thisrow{c}] {%
	x	y	z	c\\
	16.501	2.501	0	0\\
	16.501	2.501	0	1\\
	17.499	2.501	0	2\\
	17.499	2.501	0	3\\
	nan	nan	0	4\\
	17.501	2.501	0	5\\
	17.501	2.501	0	6\\
	18.499	2.501	0	7\\
	18.499	2.501	0	8\\
	nan	nan	0	9\\
	18.501	2.501	0	10\\
	18.501	2.501	0	11\\
	19.499	2.501	0	12\\
	19.499	2.501	0	13\\
	nan	nan	0	14\\
	19.501	2.501	0	15\\
	19.501	2.501	0	16\\
	20.499	2.501	0	17\\
	20.499	2.501	0	18\\
	nan	nan	0	19\\
	20.501	2.501	0	20\\
	20.501	2.501	0	21\\
	21.499	2.501	0	22\\
	21.499	2.501	0	23\\
	nan	nan	0	24\\
	21.501	2.501	0	25\\
	21.501	2.501	0	26\\
	22.499	2.501	0	27\\
	22.499	2.501	0	28\\
	nan	nan	0	29\\
	16.501	2.501	0	30\\
	16.501	2.501	0	31\\
	17.499	2.501	0	32\\
	17.499	2.501	0	33\\
	nan	nan	0	34\\
	17.501	2.501	0	35\\
	17.501	2.501	0	36\\
	18.499	2.501	0	37\\
	18.499	2.501	0	38\\
	nan	nan	0	39\\
	18.501	2.501	0	40\\
	18.501	2.501	0	41\\
	19.499	2.501	0	42\\
	19.499	2.501	0	43\\
	nan	nan	0	44\\
	19.501	2.501	0	45\\
	19.501	2.501	0	46\\
	20.499	2.501	0	47\\
	20.499	2.501	0	48\\
	nan	nan	0	49\\
	20.501	2.501	0	50\\
	20.501	2.501	0	51\\
	21.499	2.501	0	52\\
	21.499	2.501	0	53\\
	nan	nan	0	54\\
	21.501	2.501	0	55\\
	21.501	2.501	0	56\\
	22.499	2.501	0	57\\
	22.499	2.501	0	58\\
	nan	nan	0	59\\
	16.501	3.499	0	60\\
	16.501	3.499	0	61\\
	17.499	3.499	0	62\\
	17.499	3.499	0	63\\
	nan	nan	0	64\\
	17.501	3.499	0	65\\
	17.501	3.499	0	66\\
	18.499	3.499	0	67\\
	18.499	3.499	0	68\\
	nan	nan	0	69\\
	18.501	3.499	0	70\\
	18.501	3.499	0	71\\
	19.499	3.499	0	72\\
	19.499	3.499	0	73\\
	nan	nan	0	74\\
	19.501	3.499	0	75\\
	19.501	3.499	0	76\\
	20.499	3.499	0	77\\
	20.499	3.499	0	78\\
	nan	nan	0	79\\
	20.501	3.499	0	80\\
	20.501	3.499	0	81\\
	21.499	3.499	0	82\\
	21.499	3.499	0	83\\
	nan	nan	0	84\\
	21.501	3.499	0	85\\
	21.501	3.499	0	86\\
	22.499	3.499	0	87\\
	22.499	3.499	0	88\\
	nan	nan	0	89\\
	16.501	3.499	0	90\\
	16.501	3.499	0	91\\
	17.499	3.499	0	92\\
	17.499	3.499	0	93\\
	nan	nan	0	94\\
	17.501	3.499	0	95\\
	17.501	3.499	0	96\\
	18.499	3.499	0	97\\
	18.499	3.499	0	98\\
	nan	nan	0	99\\
	18.501	3.499	0	100\\
	18.501	3.499	0	101\\
	19.499	3.499	0	102\\
	19.499	3.499	0	103\\
	nan	nan	0	104\\
	19.501	3.499	0	105\\
	19.501	3.499	0	106\\
	20.499	3.499	0	107\\
	20.499	3.499	0	108\\
	nan	nan	0	109\\
	20.501	3.499	0	110\\
	20.501	3.499	0	111\\
	21.499	3.499	0	112\\
	21.499	3.499	0	113\\
	nan	nan	0	114\\
	21.501	3.499	0	115\\
	21.501	3.499	0	116\\
	22.499	3.499	0	117\\
	22.499	3.499	0	118\\
	nan	nan	0	119\\
	nan	nan	0	120\\
	nan	nan	0	121\\
	nan	nan	0	122\\
	nan	nan	0	123\\
	nan	nan	0	124\\
	nan	nan	0	125\\
	nan	nan	0	126\\
	nan	nan	0	127\\
	nan	nan	0	128\\
	nan	nan	0	129\\
	nan	nan	0	130\\
	nan	nan	0	131\\
	nan	nan	0	132\\
	nan	nan	0	133\\
	nan	nan	0	134\\
	nan	nan	0	135\\
	nan	nan	0	136\\
	nan	nan	0	137\\
	nan	nan	0	138\\
	nan	nan	0	139\\
	nan	nan	0	140\\
	nan	nan	0	141\\
	nan	nan	0	142\\
	nan	nan	0	143\\
	nan	nan	0	144\\
	nan	nan	0	145\\
	nan	nan	0	146\\
	nan	nan	0	147\\
	nan	nan	0	148\\
	nan	nan	0	149\\
	16.501	3.501	0	150\\
	16.501	3.501	0	151\\
	17.499	3.501	0	152\\
	17.499	3.501	0	153\\
	nan	nan	0	154\\
	17.501	3.501	0	155\\
	17.501	3.501	0	156\\
	18.499	3.501	0	157\\
	18.499	3.501	0	158\\
	nan	nan	0	159\\
	18.501	3.501	0	160\\
	18.501	3.501	0	161\\
	19.499	3.501	0	162\\
	19.499	3.501	0	163\\
	nan	nan	0	164\\
	19.501	3.501	0	165\\
	19.501	3.501	0	166\\
	20.499	3.501	0	167\\
	20.499	3.501	0	168\\
	nan	nan	0	169\\
	20.501	3.501	0	170\\
	20.501	3.501	0	171\\
	21.499	3.501	0	172\\
	21.499	3.501	0	173\\
	nan	nan	0	174\\
	21.501	3.501	0	175\\
	21.501	3.501	0	176\\
	22.499	3.501	0	177\\
	22.499	3.501	0	178\\
	nan	nan	0	179\\
	16.501	3.501	0	180\\
	16.501	3.501	0	181\\
	17.499	3.501	0	182\\
	17.499	3.501	0	183\\
	nan	nan	0	184\\
	17.501	3.501	0	185\\
	17.501	3.501	0	186\\
	18.499	3.501	0	187\\
	18.499	3.501	0	188\\
	nan	nan	0	189\\
	18.501	3.501	0	190\\
	18.501	3.501	0	191\\
	19.499	3.501	0	192\\
	19.499	3.501	0	193\\
	nan	nan	0	194\\
	19.501	3.501	0	195\\
	19.501	3.501	0	196\\
	20.499	3.501	0	197\\
	20.499	3.501	0	198\\
	nan	nan	0	199\\
	20.501	3.501	0	200\\
	20.501	3.501	0	201\\
	21.499	3.501	0	202\\
	21.499	3.501	0	203\\
	nan	nan	0	204\\
	21.501	3.501	0	205\\
	21.501	3.501	0	206\\
	22.499	3.501	0	207\\
	22.499	3.501	0	208\\
	nan	nan	0	209\\
	16.501	4.499	0	210\\
	16.501	4.499	0	211\\
	17.499	4.499	0	212\\
	17.499	4.499	0	213\\
	nan	nan	0	214\\
	17.501	4.499	0	215\\
	17.501	4.499	0	216\\
	18.499	4.499	0	217\\
	18.499	4.499	0	218\\
	nan	nan	0	219\\
	18.501	4.499	0	220\\
	18.501	4.499	0	221\\
	19.499	4.499	0	222\\
	19.499	4.499	0	223\\
	nan	nan	0	224\\
	19.501	4.499	0	225\\
	19.501	4.499	0	226\\
	20.499	4.499	0	227\\
	20.499	4.499	0	228\\
	nan	nan	0	229\\
	20.501	4.499	0	230\\
	20.501	4.499	0	231\\
	21.499	4.499	0	232\\
	21.499	4.499	0	233\\
	nan	nan	0	234\\
	21.501	4.499	0	235\\
	21.501	4.499	0	236\\
	22.499	4.499	0	237\\
	22.499	4.499	0	238\\
	nan	nan	0	239\\
	16.501	4.499	0	240\\
	16.501	4.499	0	241\\
	17.499	4.499	0	242\\
	17.499	4.499	0	243\\
	nan	nan	0	244\\
	17.501	4.499	0	245\\
	17.501	4.499	0	246\\
	18.499	4.499	0	247\\
	18.499	4.499	0	248\\
	nan	nan	0	249\\
	18.501	4.499	0	250\\
	18.501	4.499	0	251\\
	19.499	4.499	0	252\\
	19.499	4.499	0	253\\
	nan	nan	0	254\\
	19.501	4.499	0	255\\
	19.501	4.499	0	256\\
	20.499	4.499	0	257\\
	20.499	4.499	0	258\\
	nan	nan	0	259\\
	20.501	4.499	0	260\\
	20.501	4.499	0	261\\
	21.499	4.499	0	262\\
	21.499	4.499	0	263\\
	nan	nan	0	264\\
	21.501	4.499	0	265\\
	21.501	4.499	0	266\\
	22.499	4.499	0	267\\
	22.499	4.499	0	268\\
	nan	nan	0	269\\
	nan	nan	0	270\\
	nan	nan	0	271\\
	nan	nan	0	272\\
	nan	nan	0	273\\
	nan	nan	0	274\\
	nan	nan	0	275\\
	nan	nan	0	276\\
	nan	nan	0	277\\
	nan	nan	0	278\\
	nan	nan	0	279\\
	nan	nan	0	280\\
	nan	nan	0	281\\
	nan	nan	0	282\\
	nan	nan	0	283\\
	nan	nan	0	284\\
	nan	nan	0	285\\
	nan	nan	0	286\\
	nan	nan	0	287\\
	nan	nan	0	288\\
	nan	nan	0	289\\
	nan	nan	0	290\\
	nan	nan	0	291\\
	nan	nan	0	292\\
	nan	nan	0	293\\
	nan	nan	0	294\\
	nan	nan	0	295\\
	nan	nan	0	296\\
	nan	nan	0	297\\
	nan	nan	0	298\\
	nan	nan	0	299\\
	16.501	4.501	0	300\\
	16.501	4.501	0	301\\
	17.499	4.501	0	302\\
	17.499	4.501	0	303\\
	nan	nan	0	304\\
	17.501	4.501	0	305\\
	17.501	4.501	0	306\\
	18.499	4.501	0	307\\
	18.499	4.501	0	308\\
	nan	nan	0	309\\
	18.501	4.501	0	310\\
	18.501	4.501	0	311\\
	19.499	4.501	0	312\\
	19.499	4.501	0	313\\
	nan	nan	0	314\\
	19.501	4.501	0	315\\
	19.501	4.501	0	316\\
	20.499	4.501	0	317\\
	20.499	4.501	0	318\\
	nan	nan	0	319\\
	20.501	4.501	0	320\\
	20.501	4.501	0	321\\
	21.499	4.501	0	322\\
	21.499	4.501	0	323\\
	nan	nan	0	324\\
	21.501	4.501	0	325\\
	21.501	4.501	0	326\\
	22.499	4.501	0	327\\
	22.499	4.501	0	328\\
	nan	nan	0	329\\
	16.501	4.501	0	330\\
	16.501	4.501	0	331\\
	17.499	4.501	0	332\\
	17.499	4.501	0	333\\
	nan	nan	0	334\\
	17.501	4.501	0	335\\
	17.501	4.501	0	336\\
	18.499	4.501	0	337\\
	18.499	4.501	0	338\\
	nan	nan	0	339\\
	18.501	4.501	0	340\\
	18.501	4.501	0	341\\
	19.499	4.501	0	342\\
	19.499	4.501	0	343\\
	nan	nan	0	344\\
	19.501	4.501	0	345\\
	19.501	4.501	0	346\\
	20.499	4.501	0	347\\
	20.499	4.501	0	348\\
	nan	nan	0	349\\
	20.501	4.501	0	350\\
	20.501	4.501	0	351\\
	21.499	4.501	0	352\\
	21.499	4.501	0	353\\
	nan	nan	0	354\\
	21.501	4.501	0	355\\
	21.501	4.501	0	356\\
	22.499	4.501	0	357\\
	22.499	4.501	0	358\\
	nan	nan	0	359\\
	16.501	5.499	0	360\\
	16.501	5.499	0	361\\
	17.499	5.499	0	362\\
	17.499	5.499	0	363\\
	nan	nan	0	364\\
	17.501	5.499	0	365\\
	17.501	5.499	0	366\\
	18.499	5.499	0	367\\
	18.499	5.499	0	368\\
	nan	nan	0	369\\
	18.501	5.499	0	370\\
	18.501	5.499	0	371\\
	19.499	5.499	0	372\\
	19.499	5.499	0	373\\
	nan	nan	0	374\\
	19.501	5.499	0	375\\
	19.501	5.499	0	376\\
	20.499	5.499	0	377\\
	20.499	5.499	0	378\\
	nan	nan	0	379\\
	20.501	5.499	0	380\\
	20.501	5.499	0	381\\
	21.499	5.499	0	382\\
	21.499	5.499	0	383\\
	nan	nan	0	384\\
	21.501	5.499	0	385\\
	21.501	5.499	0	386\\
	22.499	5.499	0	387\\
	22.499	5.499	0	388\\
	nan	nan	0	389\\
	16.501	5.499	0	390\\
	16.501	5.499	0	391\\
	17.499	5.499	0	392\\
	17.499	5.499	0	393\\
	nan	nan	0	394\\
	17.501	5.499	0	395\\
	17.501	5.499	0	396\\
	18.499	5.499	0	397\\
	18.499	5.499	0	398\\
	nan	nan	0	399\\
	18.501	5.499	0	400\\
	18.501	5.499	0	401\\
	19.499	5.499	0	402\\
	19.499	5.499	0	403\\
	nan	nan	0	404\\
	19.501	5.499	0	405\\
	19.501	5.499	0	406\\
	20.499	5.499	0	407\\
	20.499	5.499	0	408\\
	nan	nan	0	409\\
	20.501	5.499	0	410\\
	20.501	5.499	0	411\\
	21.499	5.499	0	412\\
	21.499	5.499	0	413\\
	nan	nan	0	414\\
	21.501	5.499	0	415\\
	21.501	5.499	0	416\\
	22.499	5.499	0	417\\
	22.499	5.499	0	418\\
	nan	nan	0	419\\
	nan	nan	0	420\\
	nan	nan	0	421\\
	nan	nan	0	422\\
	nan	nan	0	423\\
	nan	nan	0	424\\
	nan	nan	0	425\\
	nan	nan	0	426\\
	nan	nan	0	427\\
	nan	nan	0	428\\
	nan	nan	0	429\\
	nan	nan	0	430\\
	nan	nan	0	431\\
	nan	nan	0	432\\
	nan	nan	0	433\\
	nan	nan	0	434\\
	nan	nan	0	435\\
	nan	nan	0	436\\
	nan	nan	0	437\\
	nan	nan	0	438\\
	nan	nan	0	439\\
	nan	nan	0	440\\
	nan	nan	0	441\\
	nan	nan	0	442\\
	nan	nan	0	443\\
	nan	nan	0	444\\
	nan	nan	0	445\\
	nan	nan	0	446\\
	nan	nan	0	447\\
	nan	nan	0	448\\
	nan	nan	0	449\\
	16.501	5.501	0	450\\
	16.501	5.501	0	451\\
	17.499	5.501	0	452\\
	17.499	5.501	0	453\\
	nan	nan	0	454\\
	17.501	5.501	0	455\\
	17.501	5.501	0	456\\
	18.499	5.501	0	457\\
	18.499	5.501	0	458\\
	nan	nan	0	459\\
	18.501	5.501	0	460\\
	18.501	5.501	0	461\\
	19.499	5.501	0	462\\
	19.499	5.501	0	463\\
	nan	nan	0	464\\
	19.501	5.501	0	465\\
	19.501	5.501	0	466\\
	20.499	5.501	0	467\\
	20.499	5.501	0	468\\
	nan	nan	0	469\\
	20.501	5.501	0	470\\
	20.501	5.501	0	471\\
	21.499	5.501	0	472\\
	21.499	5.501	0	473\\
	nan	nan	0	474\\
	21.501	5.501	0	475\\
	21.501	5.501	0	476\\
	22.499	5.501	0	477\\
	22.499	5.501	0	478\\
	nan	nan	0	479\\
	16.501	5.501	0	480\\
	16.501	5.501	0	481\\
	17.499	5.501	0	482\\
	17.499	5.501	0	483\\
	nan	nan	0	484\\
	17.501	5.501	0	485\\
	17.501	5.501	0	486\\
	18.499	5.501	0	487\\
	18.499	5.501	0	488\\
	nan	nan	0	489\\
	18.501	5.501	0	490\\
	18.501	5.501	0	491\\
	19.499	5.501	0	492\\
	19.499	5.501	0	493\\
	nan	nan	0	494\\
	19.501	5.501	0	495\\
	19.501	5.501	0	496\\
	20.499	5.501	0	497\\
	20.499	5.501	0	498\\
	nan	nan	0	499\\
	20.501	5.501	0	500\\
	20.501	5.501	0	501\\
	21.499	5.501	0	502\\
	21.499	5.501	0	503\\
	nan	nan	0	504\\
	21.501	5.501	0	505\\
	21.501	5.501	5	506\\
	22.499	5.501	5	507\\
	22.499	5.501	0	508\\
	nan	nan	0	509\\
	16.501	6.499	0	510\\
	16.501	6.499	0	511\\
	17.499	6.499	0	512\\
	17.499	6.499	0	513\\
	nan	nan	0	514\\
	17.501	6.499	0	515\\
	17.501	6.499	0	516\\
	18.499	6.499	0	517\\
	18.499	6.499	0	518\\
	nan	nan	0	519\\
	18.501	6.499	0	520\\
	18.501	6.499	0	521\\
	19.499	6.499	0	522\\
	19.499	6.499	0	523\\
	nan	nan	0	524\\
	19.501	6.499	0	525\\
	19.501	6.499	0	526\\
	20.499	6.499	0	527\\
	20.499	6.499	0	528\\
	nan	nan	0	529\\
	20.501	6.499	0	530\\
	20.501	6.499	0	531\\
	21.499	6.499	0	532\\
	21.499	6.499	0	533\\
	nan	nan	0	534\\
	21.501	6.499	0	535\\
	21.501	6.499	5	536\\
	22.499	6.499	5	537\\
	22.499	6.499	0	538\\
	nan	nan	0	539\\
	16.501	6.499	0	540\\
	16.501	6.499	0	541\\
	17.499	6.499	0	542\\
	17.499	6.499	0	543\\
	nan	nan	0	544\\
	17.501	6.499	0	545\\
	17.501	6.499	0	546\\
	18.499	6.499	0	547\\
	18.499	6.499	0	548\\
	nan	nan	0	549\\
	18.501	6.499	0	550\\
	18.501	6.499	0	551\\
	19.499	6.499	0	552\\
	19.499	6.499	0	553\\
	nan	nan	0	554\\
	19.501	6.499	0	555\\
	19.501	6.499	0	556\\
	20.499	6.499	0	557\\
	20.499	6.499	0	558\\
	nan	nan	0	559\\
	20.501	6.499	0	560\\
	20.501	6.499	0	561\\
	21.499	6.499	0	562\\
	21.499	6.499	0	563\\
	nan	nan	0	564\\
	21.501	6.499	0	565\\
	21.501	6.499	0	566\\
	22.499	6.499	0	567\\
	22.499	6.499	0	568\\
	nan	nan	0	569\\
	nan	nan	0	570\\
	nan	nan	0	571\\
	nan	nan	0	572\\
	nan	nan	0	573\\
	nan	nan	0	574\\
	nan	nan	0	575\\
	nan	nan	0	576\\
	nan	nan	0	577\\
	nan	nan	0	578\\
	nan	nan	0	579\\
	nan	nan	0	580\\
	nan	nan	0	581\\
	nan	nan	0	582\\
	nan	nan	0	583\\
	nan	nan	0	584\\
	nan	nan	0	585\\
	nan	nan	0	586\\
	nan	nan	0	587\\
	nan	nan	0	588\\
	nan	nan	0	589\\
	nan	nan	0	590\\
	nan	nan	0	591\\
	nan	nan	0	592\\
	nan	nan	0	593\\
	nan	nan	0	594\\
	nan	nan	0	595\\
	nan	nan	0	596\\
	nan	nan	0	597\\
	nan	nan	0	598\\
	nan	nan	0	599\\
	16.501	6.501	0	600\\
	16.501	6.501	0	601\\
	17.499	6.501	0	602\\
	17.499	6.501	0	603\\
	nan	nan	0	604\\
	17.501	6.501	0	605\\
	17.501	6.501	0	606\\
	18.499	6.501	0	607\\
	18.499	6.501	0	608\\
	nan	nan	0	609\\
	18.501	6.501	0	610\\
	18.501	6.501	0	611\\
	19.499	6.501	0	612\\
	19.499	6.501	0	613\\
	nan	nan	0	614\\
	19.501	6.501	0	615\\
	19.501	6.501	0	616\\
	20.499	6.501	0	617\\
	20.499	6.501	0	618\\
	nan	nan	0	619\\
	20.501	6.501	0	620\\
	20.501	6.501	0	621\\
	21.499	6.501	0	622\\
	21.499	6.501	0	623\\
	nan	nan	0	624\\
	21.501	6.501	0	625\\
	21.501	6.501	0	626\\
	22.499	6.501	0	627\\
	22.499	6.501	0	628\\
	nan	nan	0	629\\
	16.501	6.501	0	630\\
	16.501	6.501	0	631\\
	17.499	6.501	0	632\\
	17.499	6.501	0	633\\
	nan	nan	0	634\\
	17.501	6.501	0	635\\
	17.501	6.501	0	636\\
	18.499	6.501	0	637\\
	18.499	6.501	0	638\\
	nan	nan	0	639\\
	18.501	6.501	0	640\\
	18.501	6.501	0	641\\
	19.499	6.501	0	642\\
	19.499	6.501	0	643\\
	nan	nan	0	644\\
	19.501	6.501	0	645\\
	19.501	6.501	0	646\\
	20.499	6.501	0	647\\
	20.499	6.501	0	648\\
	nan	nan	0	649\\
	20.501	6.501	0	650\\
	20.501	6.501	34	651\\
	21.499	6.501	34	652\\
	21.499	6.501	0	653\\
	nan	nan	0	654\\
	21.501	6.501	0	655\\
	21.501	6.501	1991	656\\
	22.499	6.501	1991	657\\
	22.499	6.501	0	658\\
	nan	nan	0	659\\
	16.501	7.499	0	660\\
	16.501	7.499	0	661\\
	17.499	7.499	0	662\\
	17.499	7.499	0	663\\
	nan	nan	0	664\\
	17.501	7.499	0	665\\
	17.501	7.499	0	666\\
	18.499	7.499	0	667\\
	18.499	7.499	0	668\\
	nan	nan	0	669\\
	18.501	7.499	0	670\\
	18.501	7.499	0	671\\
	19.499	7.499	0	672\\
	19.499	7.499	0	673\\
	nan	nan	0	674\\
	19.501	7.499	0	675\\
	19.501	7.499	0	676\\
	20.499	7.499	0	677\\
	20.499	7.499	0	678\\
	nan	nan	0	679\\
	20.501	7.499	0	680\\
	20.501	7.499	34	681\\
	21.499	7.499	34	682\\
	21.499	7.499	0	683\\
	nan	nan	0	684\\
	21.501	7.499	0	685\\
	21.501	7.499	1991	686\\
	22.499	7.499	1991	687\\
	22.499	7.499	0	688\\
	nan	nan	0	689\\
	16.501	7.499	0	690\\
	16.501	7.499	0	691\\
	17.499	7.499	0	692\\
	17.499	7.499	0	693\\
	nan	nan	0	694\\
	17.501	7.499	0	695\\
	17.501	7.499	0	696\\
	18.499	7.499	0	697\\
	18.499	7.499	0	698\\
	nan	nan	0	699\\
	18.501	7.499	0	700\\
	18.501	7.499	0	701\\
	19.499	7.499	0	702\\
	19.499	7.499	0	703\\
	nan	nan	0	704\\
	19.501	7.499	0	705\\
	19.501	7.499	0	706\\
	20.499	7.499	0	707\\
	20.499	7.499	0	708\\
	nan	nan	0	709\\
	20.501	7.499	0	710\\
	20.501	7.499	0	711\\
	21.499	7.499	0	712\\
	21.499	7.499	0	713\\
	nan	nan	0	714\\
	21.501	7.499	0	715\\
	21.501	7.499	0	716\\
	22.499	7.499	0	717\\
	22.499	7.499	0	718\\
	nan	nan	0	719\\
	nan	nan	0	720\\
	nan	nan	0	721\\
	nan	nan	0	722\\
	nan	nan	0	723\\
	nan	nan	0	724\\
	nan	nan	0	725\\
	nan	nan	0	726\\
	nan	nan	0	727\\
	nan	nan	0	728\\
	nan	nan	0	729\\
	nan	nan	0	730\\
	nan	nan	0	731\\
	nan	nan	0	732\\
	nan	nan	0	733\\
	nan	nan	0	734\\
	nan	nan	0	735\\
	nan	nan	0	736\\
	nan	nan	0	737\\
	nan	nan	0	738\\
	nan	nan	0	739\\
	nan	nan	0	740\\
	nan	nan	0	741\\
	nan	nan	0	742\\
	nan	nan	0	743\\
	nan	nan	0	744\\
	nan	nan	0	745\\
	nan	nan	0	746\\
	nan	nan	0	747\\
	nan	nan	0	748\\
	nan	nan	0	749\\
	16.501	7.501	0	750\\
	16.501	7.501	0	751\\
	17.499	7.501	0	752\\
	17.499	7.501	0	753\\
	nan	nan	0	754\\
	17.501	7.501	0	755\\
	17.501	7.501	0	756\\
	18.499	7.501	0	757\\
	18.499	7.501	0	758\\
	nan	nan	0	759\\
	18.501	7.501	0	760\\
	18.501	7.501	0	761\\
	19.499	7.501	0	762\\
	19.499	7.501	0	763\\
	nan	nan	0	764\\
	19.501	7.501	0	765\\
	19.501	7.501	0	766\\
	20.499	7.501	0	767\\
	20.499	7.501	0	768\\
	nan	nan	0	769\\
	20.501	7.501	0	770\\
	20.501	7.501	0	771\\
	21.499	7.501	0	772\\
	21.499	7.501	0	773\\
	nan	nan	0	774\\
	21.501	7.501	0	775\\
	21.501	7.501	0	776\\
	22.499	7.501	0	777\\
	22.499	7.501	0	778\\
	nan	nan	0	779\\
	16.501	7.501	0	780\\
	16.501	7.501	0	781\\
	17.499	7.501	0	782\\
	17.499	7.501	0	783\\
	nan	nan	0	784\\
	17.501	7.501	0	785\\
	17.501	7.501	0	786\\
	18.499	7.501	0	787\\
	18.499	7.501	0	788\\
	nan	nan	0	789\\
	18.501	7.501	0	790\\
	18.501	7.501	0	791\\
	19.499	7.501	0	792\\
	19.499	7.501	0	793\\
	nan	nan	0	794\\
	19.501	7.501	0	795\\
	19.501	7.501	60	796\\
	20.499	7.501	60	797\\
	20.499	7.501	0	798\\
	nan	nan	0	799\\
	20.501	7.501	0	800\\
	20.501	7.501	5410	801\\
	21.499	7.501	5410	802\\
	21.499	7.501	0	803\\
	nan	nan	0	804\\
	21.501	7.501	0	805\\
	21.501	7.501	0	806\\
	22.499	7.501	0	807\\
	22.499	7.501	0	808\\
	nan	nan	0	809\\
	16.501	8.499	0	810\\
	16.501	8.499	0	811\\
	17.499	8.499	0	812\\
	17.499	8.499	0	813\\
	nan	nan	0	814\\
	17.501	8.499	0	815\\
	17.501	8.499	0	816\\
	18.499	8.499	0	817\\
	18.499	8.499	0	818\\
	nan	nan	0	819\\
	18.501	8.499	0	820\\
	18.501	8.499	0	821\\
	19.499	8.499	0	822\\
	19.499	8.499	0	823\\
	nan	nan	0	824\\
	19.501	8.499	0	825\\
	19.501	8.499	60	826\\
	20.499	8.499	60	827\\
	20.499	8.499	0	828\\
	nan	nan	0	829\\
	20.501	8.499	0	830\\
	20.501	8.499	5410	831\\
	21.499	8.499	5410	832\\
	21.499	8.499	0	833\\
	nan	nan	0	834\\
	21.501	8.499	0	835\\
	21.501	8.499	0	836\\
	22.499	8.499	0	837\\
	22.499	8.499	0	838\\
	nan	nan	0	839\\
	16.501	8.499	0	840\\
	16.501	8.499	0	841\\
	17.499	8.499	0	842\\
	17.499	8.499	0	843\\
	nan	nan	0	844\\
	17.501	8.499	0	845\\
	17.501	8.499	0	846\\
	18.499	8.499	0	847\\
	18.499	8.499	0	848\\
	nan	nan	0	849\\
	18.501	8.499	0	850\\
	18.501	8.499	0	851\\
	19.499	8.499	0	852\\
	19.499	8.499	0	853\\
	nan	nan	0	854\\
	19.501	8.499	0	855\\
	19.501	8.499	0	856\\
	20.499	8.499	0	857\\
	20.499	8.499	0	858\\
	nan	nan	0	859\\
	20.501	8.499	0	860\\
	20.501	8.499	0	861\\
	21.499	8.499	0	862\\
	21.499	8.499	0	863\\
	nan	nan	0	864\\
	21.501	8.499	0	865\\
	21.501	8.499	0	866\\
	22.499	8.499	0	867\\
	22.499	8.499	0	868\\
	nan	nan	0	869\\
	nan	nan	0	870\\
	nan	nan	0	871\\
	nan	nan	0	872\\
	nan	nan	0	873\\
	nan	nan	0	874\\
	nan	nan	0	875\\
	nan	nan	0	876\\
	nan	nan	0	877\\
	nan	nan	0	878\\
	nan	nan	0	879\\
	nan	nan	0	880\\
	nan	nan	0	881\\
	nan	nan	0	882\\
	nan	nan	0	883\\
	nan	nan	0	884\\
	nan	nan	0	885\\
	nan	nan	0	886\\
	nan	nan	0	887\\
	nan	nan	0	888\\
	nan	nan	0	889\\
	nan	nan	0	890\\
	nan	nan	0	891\\
	nan	nan	0	892\\
	nan	nan	0	893\\
	nan	nan	0	894\\
	nan	nan	0	895\\
	nan	nan	0	896\\
	nan	nan	0	897\\
	nan	nan	0	898\\
	nan	nan	0	899\\
};

\end{axis}

\end{tikzpicture}%

%% file: plots/Revised_Best_Card_Set_delta_d_-14_SNR_50.tex
%
%
\begin{tikzpicture}

\begin{axis}[%
scale=0.6, width=3.284in,
height=3.566in,
at={(0.758in,0.481in)},
scale only axis,
unbounded coords=jump,
colormap={patchmap}{[1pt] rgb(0pt)=(0.75,0.85,0.95); rgb(899pt)=(0.75,0.85,0.95)},
xmin=16.2,
xmax=22.8,
xtick={17, 18, 19, 20, 21, 22},
xticklabel style = {font=\footnotesize},
tick align=outside,
xlabel style={font=\color{white!15!black}},
xlabel style = {font=\small},
xlabel={$|\mathcal{P}_{1}^{\star}|$},
ymin=2.2,
ymax=8.8,
ytick={3, 4, 5, 6, 7, 8},
yticklabel style = {font=\footnotesize},
ylabel style={font=\color{white!15!black}},
ylabel style = {font=\small},
ylabel={$|\mathcal{P}_{2}^{\star}|$},
zmin=0,
zmax=6200,
zlabel style={font=\color{white!15!black}},
zlabel style = {font=\small},
zticklabel style = {font=\footnotesize},
zlabel={Frequency},
view={-37.5}{30},
axis background/.style={fill=white},
title style={font=\bfseries},
axis x line*=bottom,
axis y line*=left,
axis z line*=left,
xmajorgrids,
ymajorgrids,
zmajorgrids,
legend style={at={(1.03,1)}, anchor=north west, legend cell align=left, align=left, draw=white!15!black}
]

\addplot3[%
surf,
shader=flat corner, draw=black, mesh/rows=30]
table[row sep=crcr, colormap name=surfmap, point meta=\thisrow{c}] {%
	x	y	z	c\\
	16.501	2.501	0	0\\
	16.501	2.501	0	1\\
	17.499	2.501	0	2\\
	17.499	2.501	0	3\\
	nan	nan	0	4\\
	17.501	2.501	0	5\\
	17.501	2.501	0	6\\
	18.499	2.501	0	7\\
	18.499	2.501	0	8\\
	nan	nan	0	9\\
	18.501	2.501	0	10\\
	18.501	2.501	0	11\\
	19.499	2.501	0	12\\
	19.499	2.501	0	13\\
	nan	nan	0	14\\
	19.501	2.501	0	15\\
	19.501	2.501	0	16\\
	20.499	2.501	0	17\\
	20.499	2.501	0	18\\
	nan	nan	0	19\\
	20.501	2.501	0	20\\
	20.501	2.501	0	21\\
	21.499	2.501	0	22\\
	21.499	2.501	0	23\\
	nan	nan	0	24\\
	21.501	2.501	0	25\\
	21.501	2.501	0	26\\
	22.499	2.501	0	27\\
	22.499	2.501	0	28\\
	nan	nan	0	29\\
	16.501	2.501	0	30\\
	16.501	2.501	0	31\\
	17.499	2.501	0	32\\
	17.499	2.501	0	33\\
	nan	nan	0	34\\
	17.501	2.501	0	35\\
	17.501	2.501	0	36\\
	18.499	2.501	0	37\\
	18.499	2.501	0	38\\
	nan	nan	0	39\\
	18.501	2.501	0	40\\
	18.501	2.501	0	41\\
	19.499	2.501	0	42\\
	19.499	2.501	0	43\\
	nan	nan	0	44\\
	19.501	2.501	0	45\\
	19.501	2.501	0	46\\
	20.499	2.501	0	47\\
	20.499	2.501	0	48\\
	nan	nan	0	49\\
	20.501	2.501	0	50\\
	20.501	2.501	0	51\\
	21.499	2.501	0	52\\
	21.499	2.501	0	53\\
	nan	nan	0	54\\
	21.501	2.501	0	55\\
	21.501	2.501	0	56\\
	22.499	2.501	0	57\\
	22.499	2.501	0	58\\
	nan	nan	0	59\\
	16.501	3.499	0	60\\
	16.501	3.499	0	61\\
	17.499	3.499	0	62\\
	17.499	3.499	0	63\\
	nan	nan	0	64\\
	17.501	3.499	0	65\\
	17.501	3.499	0	66\\
	18.499	3.499	0	67\\
	18.499	3.499	0	68\\
	nan	nan	0	69\\
	18.501	3.499	0	70\\
	18.501	3.499	0	71\\
	19.499	3.499	0	72\\
	19.499	3.499	0	73\\
	nan	nan	0	74\\
	19.501	3.499	0	75\\
	19.501	3.499	0	76\\
	20.499	3.499	0	77\\
	20.499	3.499	0	78\\
	nan	nan	0	79\\
	20.501	3.499	0	80\\
	20.501	3.499	0	81\\
	21.499	3.499	0	82\\
	21.499	3.499	0	83\\
	nan	nan	0	84\\
	21.501	3.499	0	85\\
	21.501	3.499	0	86\\
	22.499	3.499	0	87\\
	22.499	3.499	0	88\\
	nan	nan	0	89\\
	16.501	3.499	0	90\\
	16.501	3.499	0	91\\
	17.499	3.499	0	92\\
	17.499	3.499	0	93\\
	nan	nan	0	94\\
	17.501	3.499	0	95\\
	17.501	3.499	0	96\\
	18.499	3.499	0	97\\
	18.499	3.499	0	98\\
	nan	nan	0	99\\
	18.501	3.499	0	100\\
	18.501	3.499	0	101\\
	19.499	3.499	0	102\\
	19.499	3.499	0	103\\
	nan	nan	0	104\\
	19.501	3.499	0	105\\
	19.501	3.499	0	106\\
	20.499	3.499	0	107\\
	20.499	3.499	0	108\\
	nan	nan	0	109\\
	20.501	3.499	0	110\\
	20.501	3.499	0	111\\
	21.499	3.499	0	112\\
	21.499	3.499	0	113\\
	nan	nan	0	114\\
	21.501	3.499	0	115\\
	21.501	3.499	0	116\\
	22.499	3.499	0	117\\
	22.499	3.499	0	118\\
	nan	nan	0	119\\
	nan	nan	0	120\\
	nan	nan	0	121\\
	nan	nan	0	122\\
	nan	nan	0	123\\
	nan	nan	0	124\\
	nan	nan	0	125\\
	nan	nan	0	126\\
	nan	nan	0	127\\
	nan	nan	0	128\\
	nan	nan	0	129\\
	nan	nan	0	130\\
	nan	nan	0	131\\
	nan	nan	0	132\\
	nan	nan	0	133\\
	nan	nan	0	134\\
	nan	nan	0	135\\
	nan	nan	0	136\\
	nan	nan	0	137\\
	nan	nan	0	138\\
	nan	nan	0	139\\
	nan	nan	0	140\\
	nan	nan	0	141\\
	nan	nan	0	142\\
	nan	nan	0	143\\
	nan	nan	0	144\\
	nan	nan	0	145\\
	nan	nan	0	146\\
	nan	nan	0	147\\
	nan	nan	0	148\\
	nan	nan	0	149\\
	16.501	3.501	0	150\\
	16.501	3.501	0	151\\
	17.499	3.501	0	152\\
	17.499	3.501	0	153\\
	nan	nan	0	154\\
	17.501	3.501	0	155\\
	17.501	3.501	0	156\\
	18.499	3.501	0	157\\
	18.499	3.501	0	158\\
	nan	nan	0	159\\
	18.501	3.501	0	160\\
	18.501	3.501	0	161\\
	19.499	3.501	0	162\\
	19.499	3.501	0	163\\
	nan	nan	0	164\\
	19.501	3.501	0	165\\
	19.501	3.501	0	166\\
	20.499	3.501	0	167\\
	20.499	3.501	0	168\\
	nan	nan	0	169\\
	20.501	3.501	0	170\\
	20.501	3.501	0	171\\
	21.499	3.501	0	172\\
	21.499	3.501	0	173\\
	nan	nan	0	174\\
	21.501	3.501	0	175\\
	21.501	3.501	0	176\\
	22.499	3.501	0	177\\
	22.499	3.501	0	178\\
	nan	nan	0	179\\
	16.501	3.501	0	180\\
	16.501	3.501	0	181\\
	17.499	3.501	0	182\\
	17.499	3.501	0	183\\
	nan	nan	0	184\\
	17.501	3.501	0	185\\
	17.501	3.501	0	186\\
	18.499	3.501	0	187\\
	18.499	3.501	0	188\\
	nan	nan	0	189\\
	18.501	3.501	0	190\\
	18.501	3.501	0	191\\
	19.499	3.501	0	192\\
	19.499	3.501	0	193\\
	nan	nan	0	194\\
	19.501	3.501	0	195\\
	19.501	3.501	0	196\\
	20.499	3.501	0	197\\
	20.499	3.501	0	198\\
	nan	nan	0	199\\
	20.501	3.501	0	200\\
	20.501	3.501	0	201\\
	21.499	3.501	0	202\\
	21.499	3.501	0	203\\
	nan	nan	0	204\\
	21.501	3.501	0	205\\
	21.501	3.501	0	206\\
	22.499	3.501	0	207\\
	22.499	3.501	0	208\\
	nan	nan	0	209\\
	16.501	4.499	0	210\\
	16.501	4.499	0	211\\
	17.499	4.499	0	212\\
	17.499	4.499	0	213\\
	nan	nan	0	214\\
	17.501	4.499	0	215\\
	17.501	4.499	0	216\\
	18.499	4.499	0	217\\
	18.499	4.499	0	218\\
	nan	nan	0	219\\
	18.501	4.499	0	220\\
	18.501	4.499	0	221\\
	19.499	4.499	0	222\\
	19.499	4.499	0	223\\
	nan	nan	0	224\\
	19.501	4.499	0	225\\
	19.501	4.499	0	226\\
	20.499	4.499	0	227\\
	20.499	4.499	0	228\\
	nan	nan	0	229\\
	20.501	4.499	0	230\\
	20.501	4.499	0	231\\
	21.499	4.499	0	232\\
	21.499	4.499	0	233\\
	nan	nan	0	234\\
	21.501	4.499	0	235\\
	21.501	4.499	0	236\\
	22.499	4.499	0	237\\
	22.499	4.499	0	238\\
	nan	nan	0	239\\
	16.501	4.499	0	240\\
	16.501	4.499	0	241\\
	17.499	4.499	0	242\\
	17.499	4.499	0	243\\
	nan	nan	0	244\\
	17.501	4.499	0	245\\
	17.501	4.499	0	246\\
	18.499	4.499	0	247\\
	18.499	4.499	0	248\\
	nan	nan	0	249\\
	18.501	4.499	0	250\\
	18.501	4.499	0	251\\
	19.499	4.499	0	252\\
	19.499	4.499	0	253\\
	nan	nan	0	254\\
	19.501	4.499	0	255\\
	19.501	4.499	0	256\\
	20.499	4.499	0	257\\
	20.499	4.499	0	258\\
	nan	nan	0	259\\
	20.501	4.499	0	260\\
	20.501	4.499	0	261\\
	21.499	4.499	0	262\\
	21.499	4.499	0	263\\
	nan	nan	0	264\\
	21.501	4.499	0	265\\
	21.501	4.499	0	266\\
	22.499	4.499	0	267\\
	22.499	4.499	0	268\\
	nan	nan	0	269\\
	nan	nan	0	270\\
	nan	nan	0	271\\
	nan	nan	0	272\\
	nan	nan	0	273\\
	nan	nan	0	274\\
	nan	nan	0	275\\
	nan	nan	0	276\\
	nan	nan	0	277\\
	nan	nan	0	278\\
	nan	nan	0	279\\
	nan	nan	0	280\\
	nan	nan	0	281\\
	nan	nan	0	282\\
	nan	nan	0	283\\
	nan	nan	0	284\\
	nan	nan	0	285\\
	nan	nan	0	286\\
	nan	nan	0	287\\
	nan	nan	0	288\\
	nan	nan	0	289\\
	nan	nan	0	290\\
	nan	nan	0	291\\
	nan	nan	0	292\\
	nan	nan	0	293\\
	nan	nan	0	294\\
	nan	nan	0	295\\
	nan	nan	0	296\\
	nan	nan	0	297\\
	nan	nan	0	298\\
	nan	nan	0	299\\
	16.501	4.501	0	300\\
	16.501	4.501	0	301\\
	17.499	4.501	0	302\\
	17.499	4.501	0	303\\
	nan	nan	0	304\\
	17.501	4.501	0	305\\
	17.501	4.501	0	306\\
	18.499	4.501	0	307\\
	18.499	4.501	0	308\\
	nan	nan	0	309\\
	18.501	4.501	0	310\\
	18.501	4.501	0	311\\
	19.499	4.501	0	312\\
	19.499	4.501	0	313\\
	nan	nan	0	314\\
	19.501	4.501	0	315\\
	19.501	4.501	0	316\\
	20.499	4.501	0	317\\
	20.499	4.501	0	318\\
	nan	nan	0	319\\
	20.501	4.501	0	320\\
	20.501	4.501	0	321\\
	21.499	4.501	0	322\\
	21.499	4.501	0	323\\
	nan	nan	0	324\\
	21.501	4.501	0	325\\
	21.501	4.501	0	326\\
	22.499	4.501	0	327\\
	22.499	4.501	0	328\\
	nan	nan	0	329\\
	16.501	4.501	0	330\\
	16.501	4.501	0	331\\
	17.499	4.501	0	332\\
	17.499	4.501	0	333\\
	nan	nan	0	334\\
	17.501	4.501	0	335\\
	17.501	4.501	0	336\\
	18.499	4.501	0	337\\
	18.499	4.501	0	338\\
	nan	nan	0	339\\
	18.501	4.501	0	340\\
	18.501	4.501	0	341\\
	19.499	4.501	0	342\\
	19.499	4.501	0	343\\
	nan	nan	0	344\\
	19.501	4.501	0	345\\
	19.501	4.501	0	346\\
	20.499	4.501	0	347\\
	20.499	4.501	0	348\\
	nan	nan	0	349\\
	20.501	4.501	0	350\\
	20.501	4.501	0	351\\
	21.499	4.501	0	352\\
	21.499	4.501	0	353\\
	nan	nan	0	354\\
	21.501	4.501	0	355\\
	21.501	4.501	0	356\\
	22.499	4.501	0	357\\
	22.499	4.501	0	358\\
	nan	nan	0	359\\
	16.501	5.499	0	360\\
	16.501	5.499	0	361\\
	17.499	5.499	0	362\\
	17.499	5.499	0	363\\
	nan	nan	0	364\\
	17.501	5.499	0	365\\
	17.501	5.499	0	366\\
	18.499	5.499	0	367\\
	18.499	5.499	0	368\\
	nan	nan	0	369\\
	18.501	5.499	0	370\\
	18.501	5.499	0	371\\
	19.499	5.499	0	372\\
	19.499	5.499	0	373\\
	nan	nan	0	374\\
	19.501	5.499	0	375\\
	19.501	5.499	0	376\\
	20.499	5.499	0	377\\
	20.499	5.499	0	378\\
	nan	nan	0	379\\
	20.501	5.499	0	380\\
	20.501	5.499	0	381\\
	21.499	5.499	0	382\\
	21.499	5.499	0	383\\
	nan	nan	0	384\\
	21.501	5.499	0	385\\
	21.501	5.499	0	386\\
	22.499	5.499	0	387\\
	22.499	5.499	0	388\\
	nan	nan	0	389\\
	16.501	5.499	0	390\\
	16.501	5.499	0	391\\
	17.499	5.499	0	392\\
	17.499	5.499	0	393\\
	nan	nan	0	394\\
	17.501	5.499	0	395\\
	17.501	5.499	0	396\\
	18.499	5.499	0	397\\
	18.499	5.499	0	398\\
	nan	nan	0	399\\
	18.501	5.499	0	400\\
	18.501	5.499	0	401\\
	19.499	5.499	0	402\\
	19.499	5.499	0	403\\
	nan	nan	0	404\\
	19.501	5.499	0	405\\
	19.501	5.499	0	406\\
	20.499	5.499	0	407\\
	20.499	5.499	0	408\\
	nan	nan	0	409\\
	20.501	5.499	0	410\\
	20.501	5.499	0	411\\
	21.499	5.499	0	412\\
	21.499	5.499	0	413\\
	nan	nan	0	414\\
	21.501	5.499	0	415\\
	21.501	5.499	0	416\\
	22.499	5.499	0	417\\
	22.499	5.499	0	418\\
	nan	nan	0	419\\
	nan	nan	0	420\\
	nan	nan	0	421\\
	nan	nan	0	422\\
	nan	nan	0	423\\
	nan	nan	0	424\\
	nan	nan	0	425\\
	nan	nan	0	426\\
	nan	nan	0	427\\
	nan	nan	0	428\\
	nan	nan	0	429\\
	nan	nan	0	430\\
	nan	nan	0	431\\
	nan	nan	0	432\\
	nan	nan	0	433\\
	nan	nan	0	434\\
	nan	nan	0	435\\
	nan	nan	0	436\\
	nan	nan	0	437\\
	nan	nan	0	438\\
	nan	nan	0	439\\
	nan	nan	0	440\\
	nan	nan	0	441\\
	nan	nan	0	442\\
	nan	nan	0	443\\
	nan	nan	0	444\\
	nan	nan	0	445\\
	nan	nan	0	446\\
	nan	nan	0	447\\
	nan	nan	0	448\\
	nan	nan	0	449\\
	16.501	5.501	0	450\\
	16.501	5.501	0	451\\
	17.499	5.501	0	452\\
	17.499	5.501	0	453\\
	nan	nan	0	454\\
	17.501	5.501	0	455\\
	17.501	5.501	0	456\\
	18.499	5.501	0	457\\
	18.499	5.501	0	458\\
	nan	nan	0	459\\
	18.501	5.501	0	460\\
	18.501	5.501	0	461\\
	19.499	5.501	0	462\\
	19.499	5.501	0	463\\
	nan	nan	0	464\\
	19.501	5.501	0	465\\
	19.501	5.501	0	466\\
	20.499	5.501	0	467\\
	20.499	5.501	0	468\\
	nan	nan	0	469\\
	20.501	5.501	0	470\\
	20.501	5.501	0	471\\
	21.499	5.501	0	472\\
	21.499	5.501	0	473\\
	nan	nan	0	474\\
	21.501	5.501	0	475\\
	21.501	5.501	0	476\\
	22.499	5.501	0	477\\
	22.499	5.501	0	478\\
	nan	nan	0	479\\
	16.501	5.501	0	480\\
	16.501	5.501	0	481\\
	17.499	5.501	0	482\\
	17.499	5.501	0	483\\
	nan	nan	0	484\\
	17.501	5.501	0	485\\
	17.501	5.501	0	486\\
	18.499	5.501	0	487\\
	18.499	5.501	0	488\\
	nan	nan	0	489\\
	18.501	5.501	0	490\\
	18.501	5.501	0	491\\
	19.499	5.501	0	492\\
	19.499	5.501	0	493\\
	nan	nan	0	494\\
	19.501	5.501	0	495\\
	19.501	5.501	0	496\\
	20.499	5.501	0	497\\
	20.499	5.501	0	498\\
	nan	nan	0	499\\
	20.501	5.501	0	500\\
	20.501	5.501	0	501\\
	21.499	5.501	0	502\\
	21.499	5.501	0	503\\
	nan	nan	0	504\\
	21.501	5.501	0	505\\
	21.501	5.501	0	506\\
	22.499	5.501	0	507\\
	22.499	5.501	0	508\\
	nan	nan	0	509\\
	16.501	6.499	0	510\\
	16.501	6.499	0	511\\
	17.499	6.499	0	512\\
	17.499	6.499	0	513\\
	nan	nan	0	514\\
	17.501	6.499	0	515\\
	17.501	6.499	0	516\\
	18.499	6.499	0	517\\
	18.499	6.499	0	518\\
	nan	nan	0	519\\
	18.501	6.499	0	520\\
	18.501	6.499	0	521\\
	19.499	6.499	0	522\\
	19.499	6.499	0	523\\
	nan	nan	0	524\\
	19.501	6.499	0	525\\
	19.501	6.499	0	526\\
	20.499	6.499	0	527\\
	20.499	6.499	0	528\\
	nan	nan	0	529\\
	20.501	6.499	0	530\\
	20.501	6.499	0	531\\
	21.499	6.499	0	532\\
	21.499	6.499	0	533\\
	nan	nan	0	534\\
	21.501	6.499	0	535\\
	21.501	6.499	0	536\\
	22.499	6.499	0	537\\
	22.499	6.499	0	538\\
	nan	nan	0	539\\
	16.501	6.499	0	540\\
	16.501	6.499	0	541\\
	17.499	6.499	0	542\\
	17.499	6.499	0	543\\
	nan	nan	0	544\\
	17.501	6.499	0	545\\
	17.501	6.499	0	546\\
	18.499	6.499	0	547\\
	18.499	6.499	0	548\\
	nan	nan	0	549\\
	18.501	6.499	0	550\\
	18.501	6.499	0	551\\
	19.499	6.499	0	552\\
	19.499	6.499	0	553\\
	nan	nan	0	554\\
	19.501	6.499	0	555\\
	19.501	6.499	0	556\\
	20.499	6.499	0	557\\
	20.499	6.499	0	558\\
	nan	nan	0	559\\
	20.501	6.499	0	560\\
	20.501	6.499	0	561\\
	21.499	6.499	0	562\\
	21.499	6.499	0	563\\
	nan	nan	0	564\\
	21.501	6.499	0	565\\
	21.501	6.499	0	566\\
	22.499	6.499	0	567\\
	22.499	6.499	0	568\\
	nan	nan	0	569\\
	nan	nan	0	570\\
	nan	nan	0	571\\
	nan	nan	0	572\\
	nan	nan	0	573\\
	nan	nan	0	574\\
	nan	nan	0	575\\
	nan	nan	0	576\\
	nan	nan	0	577\\
	nan	nan	0	578\\
	nan	nan	0	579\\
	nan	nan	0	580\\
	nan	nan	0	581\\
	nan	nan	0	582\\
	nan	nan	0	583\\
	nan	nan	0	584\\
	nan	nan	0	585\\
	nan	nan	0	586\\
	nan	nan	0	587\\
	nan	nan	0	588\\
	nan	nan	0	589\\
	nan	nan	0	590\\
	nan	nan	0	591\\
	nan	nan	0	592\\
	nan	nan	0	593\\
	nan	nan	0	594\\
	nan	nan	0	595\\
	nan	nan	0	596\\
	nan	nan	0	597\\
	nan	nan	0	598\\
	nan	nan	0	599\\
	16.501	6.501	0	600\\
	16.501	6.501	0	601\\
	17.499	6.501	0	602\\
	17.499	6.501	0	603\\
	nan	nan	0	604\\
	17.501	6.501	0	605\\
	17.501	6.501	0	606\\
	18.499	6.501	0	607\\
	18.499	6.501	0	608\\
	nan	nan	0	609\\
	18.501	6.501	0	610\\
	18.501	6.501	0	611\\
	19.499	6.501	0	612\\
	19.499	6.501	0	613\\
	nan	nan	0	614\\
	19.501	6.501	0	615\\
	19.501	6.501	0	616\\
	20.499	6.501	0	617\\
	20.499	6.501	0	618\\
	nan	nan	0	619\\
	20.501	6.501	0	620\\
	20.501	6.501	0	621\\
	21.499	6.501	0	622\\
	21.499	6.501	0	623\\
	nan	nan	0	624\\
	21.501	6.501	0	625\\
	21.501	6.501	0	626\\
	22.499	6.501	0	627\\
	22.499	6.501	0	628\\
	nan	nan	0	629\\
	16.501	6.501	0	630\\
	16.501	6.501	0	631\\
	17.499	6.501	0	632\\
	17.499	6.501	0	633\\
	nan	nan	0	634\\
	17.501	6.501	0	635\\
	17.501	6.501	0	636\\
	18.499	6.501	0	637\\
	18.499	6.501	0	638\\
	nan	nan	0	639\\
	18.501	6.501	0	640\\
	18.501	6.501	0	641\\
	19.499	6.501	0	642\\
	19.499	6.501	0	643\\
	nan	nan	0	644\\
	19.501	6.501	0	645\\
	19.501	6.501	0	646\\
	20.499	6.501	0	647\\
	20.499	6.501	0	648\\
	nan	nan	0	649\\
	20.501	6.501	0	650\\
	20.501	6.501	0	651\\
	21.499	6.501	0	652\\
	21.499	6.501	0	653\\
	nan	nan	0	654\\
	21.501	6.501	0	655\\
	21.501	6.501	2053	656\\
	22.499	6.501	2053	657\\
	22.499	6.501	0	658\\
	nan	nan	0	659\\
	16.501	7.499	0	660\\
	16.501	7.499	0	661\\
	17.499	7.499	0	662\\
	17.499	7.499	0	663\\
	nan	nan	0	664\\
	17.501	7.499	0	665\\
	17.501	7.499	0	666\\
	18.499	7.499	0	667\\
	18.499	7.499	0	668\\
	nan	nan	0	669\\
	18.501	7.499	0	670\\
	18.501	7.499	0	671\\
	19.499	7.499	0	672\\
	19.499	7.499	0	673\\
	nan	nan	0	674\\
	19.501	7.499	0	675\\
	19.501	7.499	0	676\\
	20.499	7.499	0	677\\
	20.499	7.499	0	678\\
	nan	nan	0	679\\
	20.501	7.499	0	680\\
	20.501	7.499	0	681\\
	21.499	7.499	0	682\\
	21.499	7.499	0	683\\
	nan	nan	0	684\\
	21.501	7.499	0	685\\
	21.501	7.499	2053	686\\
	22.499	7.499	2053	687\\
	22.499	7.499	0	688\\
	nan	nan	0	689\\
	16.501	7.499	0	690\\
	16.501	7.499	0	691\\
	17.499	7.499	0	692\\
	17.499	7.499	0	693\\
	nan	nan	0	694\\
	17.501	7.499	0	695\\
	17.501	7.499	0	696\\
	18.499	7.499	0	697\\
	18.499	7.499	0	698\\
	nan	nan	0	699\\
	18.501	7.499	0	700\\
	18.501	7.499	0	701\\
	19.499	7.499	0	702\\
	19.499	7.499	0	703\\
	nan	nan	0	704\\
	19.501	7.499	0	705\\
	19.501	7.499	0	706\\
	20.499	7.499	0	707\\
	20.499	7.499	0	708\\
	nan	nan	0	709\\
	20.501	7.499	0	710\\
	20.501	7.499	0	711\\
	21.499	7.499	0	712\\
	21.499	7.499	0	713\\
	nan	nan	0	714\\
	21.501	7.499	0	715\\
	21.501	7.499	0	716\\
	22.499	7.499	0	717\\
	22.499	7.499	0	718\\
	nan	nan	0	719\\
	nan	nan	0	720\\
	nan	nan	0	721\\
	nan	nan	0	722\\
	nan	nan	0	723\\
	nan	nan	0	724\\
	nan	nan	0	725\\
	nan	nan	0	726\\
	nan	nan	0	727\\
	nan	nan	0	728\\
	nan	nan	0	729\\
	nan	nan	0	730\\
	nan	nan	0	731\\
	nan	nan	0	732\\
	nan	nan	0	733\\
	nan	nan	0	734\\
	nan	nan	0	735\\
	nan	nan	0	736\\
	nan	nan	0	737\\
	nan	nan	0	738\\
	nan	nan	0	739\\
	nan	nan	0	740\\
	nan	nan	0	741\\
	nan	nan	0	742\\
	nan	nan	0	743\\
	nan	nan	0	744\\
	nan	nan	0	745\\
	nan	nan	0	746\\
	nan	nan	0	747\\
	nan	nan	0	748\\
	nan	nan	0	749\\
	16.501	7.501	0	750\\
	16.501	7.501	0	751\\
	17.499	7.501	0	752\\
	17.499	7.501	0	753\\
	nan	nan	0	754\\
	17.501	7.501	0	755\\
	17.501	7.501	0	756\\
	18.499	7.501	0	757\\
	18.499	7.501	0	758\\
	nan	nan	0	759\\
	18.501	7.501	0	760\\
	18.501	7.501	0	761\\
	19.499	7.501	0	762\\
	19.499	7.501	0	763\\
	nan	nan	0	764\\
	19.501	7.501	0	765\\
	19.501	7.501	0	766\\
	20.499	7.501	0	767\\
	20.499	7.501	0	768\\
	nan	nan	0	769\\
	20.501	7.501	0	770\\
	20.501	7.501	0	771\\
	21.499	7.501	0	772\\
	21.499	7.501	0	773\\
	nan	nan	0	774\\
	21.501	7.501	0	775\\
	21.501	7.501	0	776\\
	22.499	7.501	0	777\\
	22.499	7.501	0	778\\
	nan	nan	0	779\\
	16.501	7.501	0	780\\
	16.501	7.501	0	781\\
	17.499	7.501	0	782\\
	17.499	7.501	0	783\\
	nan	nan	0	784\\
	17.501	7.501	0	785\\
	17.501	7.501	0	786\\
	18.499	7.501	0	787\\
	18.499	7.501	0	788\\
	nan	nan	0	789\\
	18.501	7.501	0	790\\
	18.501	7.501	0	791\\
	19.499	7.501	0	792\\
	19.499	7.501	0	793\\
	nan	nan	0	794\\
	19.501	7.501	0	795\\
	19.501	7.501	0	796\\
	20.499	7.501	0	797\\
	20.499	7.501	0	798\\
	nan	nan	0	799\\
	20.501	7.501	0	800\\
	20.501	7.501	5447	801\\
	21.499	7.501	5447	802\\
	21.499	7.501	0	803\\
	nan	nan	0	804\\
	21.501	7.501	0	805\\
	21.501	7.501	0	806\\
	22.499	7.501	0	807\\
	22.499	7.501	0	808\\
	nan	nan	0	809\\
	16.501	8.499	0	810\\
	16.501	8.499	0	811\\
	17.499	8.499	0	812\\
	17.499	8.499	0	813\\
	nan	nan	0	814\\
	17.501	8.499	0	815\\
	17.501	8.499	0	816\\
	18.499	8.499	0	817\\
	18.499	8.499	0	818\\
	nan	nan	0	819\\
	18.501	8.499	0	820\\
	18.501	8.499	0	821\\
	19.499	8.499	0	822\\
	19.499	8.499	0	823\\
	nan	nan	0	824\\
	19.501	8.499	0	825\\
	19.501	8.499	0	826\\
	20.499	8.499	0	827\\
	20.499	8.499	0	828\\
	nan	nan	0	829\\
	20.501	8.499	0	830\\
	20.501	8.499	5447	831\\
	21.499	8.499	5447	832\\
	21.499	8.499	0	833\\
	nan	nan	0	834\\
	21.501	8.499	0	835\\
	21.501	8.499	0	836\\
	22.499	8.499	0	837\\
	22.499	8.499	0	838\\
	nan	nan	0	839\\
	16.501	8.499	0	840\\
	16.501	8.499	0	841\\
	17.499	8.499	0	842\\
	17.499	8.499	0	843\\
	nan	nan	0	844\\
	17.501	8.499	0	845\\
	17.501	8.499	0	846\\
	18.499	8.499	0	847\\
	18.499	8.499	0	848\\
	nan	nan	0	849\\
	18.501	8.499	0	850\\
	18.501	8.499	0	851\\
	19.499	8.499	0	852\\
	19.499	8.499	0	853\\
	nan	nan	0	854\\
	19.501	8.499	0	855\\
	19.501	8.499	0	856\\
	20.499	8.499	0	857\\
	20.499	8.499	0	858\\
	nan	nan	0	859\\
	20.501	8.499	0	860\\
	20.501	8.499	0	861\\
	21.499	8.499	0	862\\
	21.499	8.499	0	863\\
	nan	nan	0	864\\
	21.501	8.499	0	865\\
	21.501	8.499	0	866\\
	22.499	8.499	0	867\\
	22.499	8.499	0	868\\
	nan	nan	0	869\\
	nan	nan	0	870\\
	nan	nan	0	871\\
	nan	nan	0	872\\
	nan	nan	0	873\\
	nan	nan	0	874\\
	nan	nan	0	875\\
	nan	nan	0	876\\
	nan	nan	0	877\\
	nan	nan	0	878\\
	nan	nan	0	879\\
	nan	nan	0	880\\
	nan	nan	0	881\\
	nan	nan	0	882\\
	nan	nan	0	883\\
	nan	nan	0	884\\
	nan	nan	0	885\\
	nan	nan	0	886\\
	nan	nan	0	887\\
	nan	nan	0	888\\
	nan	nan	0	889\\
	nan	nan	0	890\\
	nan	nan	0	891\\
	nan	nan	0	892\\
	nan	nan	0	893\\
	nan	nan	0	894\\
	nan	nan	0	895\\
	nan	nan	0	896\\
	nan	nan	0	897\\
	nan	nan	0	898\\
	nan	nan	0	899\\
};

\end{axis}

\end{tikzpicture}%

%% file: plots/MSE_over_SNR_Plot.tex

\definecolor{mycolor1}{rgb}{0.00000,0.44700,0.74100}%
\definecolor{mycolor2}{rgb}{0.85000,0.32500,0.09800}%
\definecolor{mycolor3}{rgb}{0.92900,0.69400,0.12500}%
\definecolor{mycolor4}{rgb}{0.49400,0.18400,0.55600}%
\definecolor{mycolor5}{rgb}{0.46600,0.67400,0.18800}%
\definecolor{mycolor6}{rgb}{0.30100,0.74500,0.93300}%
\begin{tikzpicture}

\begin{axis}[%
scale=0.4, width=4.521in,
height=3.566in,
at={(0.758in,0.481in)},
scale only axis,
xmin=-6,
xmax=51,
xlabel style={font=\color{white!15!black}},
xlabel style = {font=\small},
xlabel={$\mathsf{SNR}$ [dB]},
ymin=0.1,
ymax=0.605,
ylabel style={font=\color{white!15!black}},
ylabel style = {font=\small},
ylabel={$\frac{\mathbb{E}_{\bm h}[\mathsf{MSE}^{\star}]}{\max(|\mathcal{D}_{1}|,|\mathcal{D}_{2}|)}$},
axis background/.style={fill=white},
legend style={legend cell align=left, align=left, draw=white!15!black}, legend pos=outer north east,
]
\addplot [color=mycolor1, line width=1.5pt, mark=square, mark options={solid, mycolor1}]
table[row sep=crcr]{%
	-5	0.147884515274042\\
	0	0.14181659763462\\
	5	0.138280123743333\\
	10	0.13607419219587\\
	15	0.134789295019997\\
	20	0.134022173913227\\
	25	0.133656985519676\\
	30	0.133464074086683\\
	35	0.133385622260073\\
	40	0.133350698040762\\
	45	0.1333419518638\\
	50	0.133335107077591\\
};
\addlegendentry{\footnotesize$\Delta D=22$}

\addplot [color=mycolor2, line width=1.5pt, mark=square, mark options={solid, mycolor2}]
table[row sep=crcr]{%
	-5	0.22634986335996\\
	0	0.214193812298008\\
	5	0.207715141321848\\
	10	0.204102927386462\\
	15	0.202142249903405\\
	20	0.201033569063855\\
	25	0.20043965878195\\
	30	0.200197985542683\\
	35	0.200070829040199\\
	40	0.200034247321115\\
	45	0.200009010764026\\
	50	0.200003425839541\\
};
\addlegendentry{\footnotesize$\Delta D=18$}

\addplot [color=mycolor3, line width=1.5pt, mark=square, mark options={solid, mycolor3}]
table[row sep=crcr]{%
	-5	0.335684200023787\\
	0	0.31994903801064\\
	5	0.310948912627472\\
	10	0.306112755282314\\
	15	0.303200869612389\\
	20	0.301565289689849\\
	25	0.300713366720983\\
	30	0.300281643285347\\
	35	0.300121683122029\\
	40	0.300048285119685\\
	45	0.300015520097381\\
	50	0.300005018443501\\
};
\addlegendentry{\footnotesize$\Delta D=12$}

\addplot [color=mycolor4, line width=1.5pt, mark=square, mark options={solid, mycolor4}]
table[row sep=crcr]{%
	-5	0.558753411414074\\
	0	0.533301913292541\\
	5	0.518410600495214\\
	10	0.510156205729324\\
	15	0.505249510628226\\
	20	0.502671096170617\\
	25	0.501179001531728\\
	30	0.500484522418179\\
	35	0.500206231700909\\
	40	0.500079660105068\\
	45	0.500033209889604\\
	50	0.50001389302895\\
};
\addlegendentry{\footnotesize$\Delta D=0$}

\addplot [color=mycolor5, line width=1.5pt, mark=square, mark options={solid, mycolor5}]
table[row sep=crcr]{%
	-5	0.446567341568818\\
	0	0.42676074207723\\
	5	0.414781081302843\\
	10	0.40794438400176\\
	15	0.404223530892248\\
	20	0.402069176956456\\
	25	0.400892744862772\\
	30	0.400332029229104\\
	35	0.400162149176969\\
	40	0.400049729276801\\
	45	0.400021017333733\\
	50	0.400007026216704\\
};
\addlegendentry{\footnotesize$\Delta D=-6$}

\addplot [color=mycolor6, line width=1.5pt, mark=square, mark options={solid, mycolor6}]
table[row sep=crcr]{%
	-5	0.371662924289133\\
	0	0.355356988486409\\
	5	0.345847863104528\\
	10	0.340129961208034\\
	15	0.336869839396955\\
	20	0.334985533284551\\
	25	0.334092870049338\\
	30	0.333666023969565\\
	35	0.333480105690031\\
	40	0.333386238788148\\
	45	0.333355057668396\\
	50	0.333339989624415\\
};
\addlegendentry{\footnotesize$\Delta D=-10$}

\end{axis}
\end{tikzpicture}%

%% file: plots/MSE_over_SNR_Plot_with_Benchmark_Comp.tex

\definecolor{mycolor1}{rgb}{0.00000,0.44700,0.74100}%
\definecolor{mycolor2}{rgb}{0.85000,0.32500,0.09800}%
\definecolor{mycolor3}{rgb}{0.92900,0.69400,0.12500}%
\definecolor{mycolor4}{rgb}{0.49400,0.18400,0.55600}%
\begin{tikzpicture}
\hspace{0.75cm}
\begin{axis}[%
scale=0.4,width=4.521in,
height=3.566in,
at={(0.758in,0.481in)},
scale only axis,
xmin=-6,
xmax=51,
xlabel style={font=\color{white!15!black}},
xlabel style = {font=\small},
xlabel={$\mathsf{SNR}$ [dB]},
ymin=0.25,
ymax=0.5,
ylabel style={font=\color{white!15!black}},
ylabel style = {font=\small},
ylabel={$\frac{\mathbb{E}_{\bm h}[\max(\mathsf{MSE}_{1},\mathsf{MSE}_{2})]}{\max(|\mathcal{D}_{1}|,|\mathcal{D}_{2}|)}$},
axis background/.style={fill=white},
legend style={legend cell align=left, align=left, draw=white!15!black},  legend pos=outer north east,
]
\addplot [color=mycolor1, line width=1.5pt, mark=o, mark options={solid, mycolor1}]
table[row sep=crcr]{%
	-5	0.476737769336917\\
	0	0.468172848828791\\
	5	0.465844419425617\\
	10	0.46460991719383\\
	15	0.464035624475693\\
	20	0.464524842530489\\
	25	0.464032006863969\\
	30	0.463296121313369\\
	35	0.46425713349179\\
	40	0.463878406768426\\
	45	0.463661068175248\\
	50	0.464345977292549\\
};
\addlegendentry{\footnotesize$\text{Full-Pwr, }\Delta D=10$}

\addplot [color=mycolor2, line width=1.5pt, mark=square, mark options={solid, mycolor2}]
table[row sep=crcr]{%
	-5	0.372713134128393\\
	0	0.355338312431071\\
	5	0.345604193424703\\
	10	0.340082438942197\\
	15	0.336835128098411\\
	20	0.335099099612538\\
	25	0.334194453076007\\
	30	0.333638370731413\\
	35	0.333457387372862\\
	40	0.333385005222976\\
	45	0.333346782446102\\
	50	0.333343093600627\\
};
\addlegendentry{\footnotesize$\text{Robust, }\Delta D=10$}

\addplot [color=mycolor3, line width=1.5pt, mark=o, mark options={solid, mycolor3}]
table[row sep=crcr]{%
	-5	0.483888247453659\\
	0	0.476764289852006\\
	5	0.474068153674864\\
	10	0.475050890964088\\
	15	0.473366307944338\\
	20	0.472203004861213\\
	25	0.473336267983588\\
	30	0.472923960077658\\
	35	0.473197329183692\\
	40	0.472314511288577\\
	45	0.472644936469709\\
	50	0.473486256922309\\
};
\addlegendentry{\footnotesize$\text{Full-Pwr, }\Delta D=-14$}

\addplot [color=mycolor4, line width=1.5pt, mark=square, mark options={solid, mycolor4}]
table[row sep=crcr]{%
	-5	0.299561961734914\\
	0	0.284646116617976\\
	5	0.276613674573752\\
	10	0.272157870326156\\
	15	0.269571434983394\\
	20	0.268019316888667\\
	25	0.267302425349906\\
	30	0.266920363636994\\
	35	0.266762941922031\\
	40	0.266700366672213\\
	45	0.266678326857914\\
	50	0.26667372063769\\
};
\addlegendentry{\footnotesize$\text{Robust, }\Delta D=-14$}

\end{axis}
\end{tikzpicture}%